\documentstyle[epsfig,12pt]{report}

\newcommand{\be}{\begin{equation}}
\newcommand{\ee}{\end{equation}}
\newcommand{\bea}{\begin{eqnarray}}
\newcommand{\eea}{\end{eqnarray}}

\newcommand{\bp}{\mbox{\boldmath $p$}}

\newcommand{\br}{\mbox{\boldmath $r$}}
\newcommand{\bl}{\mbox{\boldmath $l$}}
\newcommand{\Qb}{\overline{Q}}
\newcommand{\vep}{\varepsilon}

\def\Pom{{\bf I\!P}}

\def\lsim{\mathrel{\rlap{\lower4pt\hbox{\hskip1pt$\sim$}}
    \raise1pt\hbox{$<$}}}         

\def\gsim{\mathrel{\rlap{\lower4pt\hbox{\hskip1pt$\sim$}}
    \raise1pt\hbox{$>$}}}         
\newcommand{\dst}{\displaystyle}
\newcommand{\fr}[2]{\frac{{\dst #1}}{{\dst #2}}}

\begin{document}
\thispagestyle{empty}
~\\
\begin{center}
{\Large \bf Diffractive production of vector mesons
in Deep Inelastic Scattering within $k_t$-factorization approach}\\

\vspace{5cm} {\large \bf Dissertation }\\

zur\\

Erlangung des Doktorgrades (Dr. rer. nat.) \\

der\\

Mathematisch-Naturwissenschaftlichen Fakult\"at

der\\

Rheinischen Friedrich-Wilhelms-Universit\"at Bonn

\vspace{6cm}
{vorgelegt von}

\vspace{1cm}
\large {\bf Igor Ivanov} \\  \normalsize 

aus\\

 Russland

\vspace{3cm}
Bonn 2002
\end{center}

\newpage
\thispagestyle{empty}

\noindent Angefertigt am Institut f\"ur Kernphysik\\
des Forschungszentrums J\"ulich GmbH\\

\noindent mit Genehmigung der
Mathematisch-Naturwissenschaftlichen Fakult\"at der\\
Rheinischen Friedrich-Wilhelms-Universit\"at Bonn

\vspace{17cm}
\noindent 1. Referent:  Prof. Dr. J.~Speth \\

\noindent 2. Referent:  Prof. Dr. H.R.~Petry

\vspace{1cm}
\noindent Tag der Promotion:

\newpage
\begin{center}
{\large\bf Abstract}\\
\end{center}
In this work we give a theoretical description of the elastic vector meson
production in diffractive DIS developed within 
the $k_t$-factorization formalism. 
Since the $k_t$-factorization scheme does not require
large values of $Q^2+m_V^2$, we conduct an analysis that is
applicable to all values of $Q^2$ from photo- up to
highly virtual production of vector mesons.
The basic quantity in this approach --- the unintegrated gluon
structure function --- was for the first time extracted
from the experimental data on $F_{2p}$, 
thoroughly investigated, and consistently used
in the vector meson production calculation.
Moreover, by limiting ourselves to the lowest Fock state 
of the vector meson, we were able to construct in a closed form
the theory of spin-angular coupling in the vector meson.
This allowed us for the first time to address
the production of a vector meson in a given spin-angular state.
We performed an extensive analytical and numerical 
investigation of the properties of
$1S$, $2S$, and $D$-wave vector meson production reactions.
Treating the physical ground state vector mesons as purely $1S$ states,
we observed a good overall agreement with all available experimental
data on vector meson production. For the excited states, 
our analysis predicts a picture which is remarkably different from $1S$-state, 
so that such reactions can be regarded as potential sources
of new information on the structure of excited states in vector mesons.

\newpage

\begin{center}
~\\
\vspace{6cm}
{\large to \\ \vspace{1cm}
Claudia I \\\vspace{1cm} and \\\vspace{1cm} Claudia II}
\end{center}

\newpage

\tableofcontents

\newpage

\chapter{Introduction}

 In the past 30 years the particle physics theory has
proved numerous times to provide a good, consistent, unified
description of the great variety of nuclear, low and high energy
particle physics experiments.  Being based on the ideas of QFT
applicability, gauge approach to fundamental interactions,
symmetry and naturalness considerations, the Standard Model
managed to explain virtually all phenomena in electromagnetic,
weak and strong interactions, to predict new particles and
effects. Though the questions of fundamental origin lie beyond the
scope of the Standard Model, its precision in description, for
instance, QED phenomena reaches the magnitude of $10^{-10}$.

However, the current situation is not that optimistic in the
domain of strong interactions. The gauge--based formulation ---
the Quantum Chromodynamics (QCD) --- seems to offer reasonably
good description only of the energetic enough processes (more
accurately: only when every vertex involves at least one highly
virtual particle) thanks to the famous asymptotic freedom. The
major difficulty lies in the behavior of the QCD coupling constant:
$\alpha_s(Q^2)$ exhibits infrared growth and becomes
comparable to unity at $Q^2 \sim$ 1 GeV$^2$. The net result is
that the perturbation theory --- the only prolific universal
treatment of various high-energy processes --- fails to give even
qualitative description of low-energy, essentially
non-perturbative phenomena. Additional difficulties arise from the
non-abelian nature of QCD, chiral symmetry breaking,
non-trivial QCD vacuum, instantons etc.

On the other hand, many separate concepts have been developed,
which do not cling to the perturbative QCD (pQCD) and provide
reasonably good description of phenomena in their applicability
regions. The fundamental problem of the theory of strong
interactions is that these heterogeneous approaches do not
match\footnote{ Just a few examples of poor accordance among
various approaches: the quark generated ladder diagrams do not
appear to correspond uniquely to any of experimentally observed
Regge trajectories. Another example is the vague status of
$\alpha_s =$ const BFKL results in true QCD.}. They do not
comprise a unified picture of strong interactions. Given such a
lack of universal, rigorously derived results, one must admit that
the subject of our investigation belongs to the realm of
phenomenology rather than rigorous theory.

\section{Diffractive processes and Pomeron}

In the light of these problems, the careful examination of regions
where two or more approaches overlap (or visa versa, where neither
of the concepts exhausts the interaction) are of great interest.
Diffractive Deep Inelastic Scattering (DIS) is exactly one of
these fields.

\begin{figure}[!htb]
   \centering
   \epsfig{file=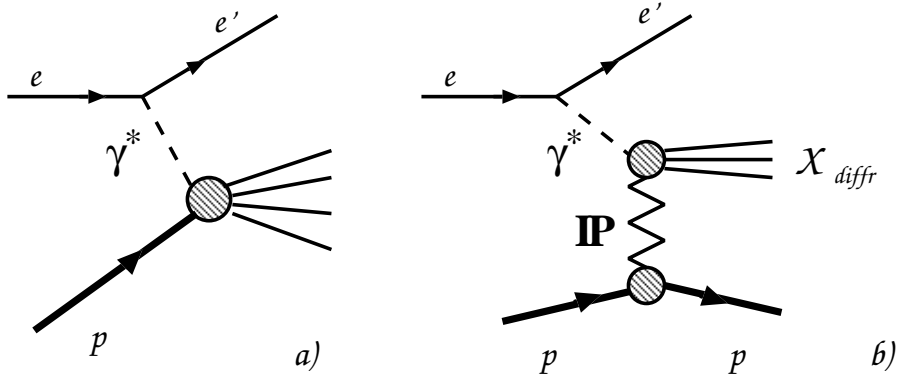,width=120mm}
   \caption{Examples of deep inelastic scattering process:
{\it (a)} hard DIS and {\it (b)} diffractive DIS. In the latter case
$M^2_{diffr} \ll s$ and the process proceeds via pomeron $t$-channel exchange.}
   \label{DIS}
\end{figure}

A typical hard DIS process (Fig.\ref{DIS}a) occurs when a virtual
photon\footnote{We will always imply that the virtual photon is
emitted by an electron, which means the photon is always
space-like: if $q$ is photon momentum, then $Q^2 \equiv -q^2 >0$.}
strikes a proton to produce a hard system $X$ with large invariant
mass\footnote{In hard DIS phenomenology this quantity is usually
labeled as $W^2$. However, for simplicity we will use notation
$s$.} $s$ and large enough multiplicity, final state hadrons being
distributed over whole rapidity range approximately smoothly.

However, as it was noted long ago, sometimes the proton survives,
being only slightly deflected, and a virtual photon turns into a so-called diffractive
system $X_{diffr}$ with invariant mass $M^2_{diffr} \ll s$.
In this process the proton and the diffractive system are naturally
separated by a large rapidity gap and a condition which appears necessary
for the rapidity gap formation is $Q^2 \ll s$, or in terms of Bjorken $x$
\begin{equation}\label{x}
x = {Q^2 \over s} \ll 1\,.
\end{equation}

This is one of the most common cases
of diffractive DIS (DDIS) processes. In fact, the class of diffractive
processes is not confined within DIS; it is much broader.
There are many other reactions which possess the generic features
--- the rapidity gap and smallness of $M^2_{diffr}$ ---
and therefore can be classified as diffractive processes
(for a recent review see \cite{NNN-vocab}).

How can a typical diffractive process occur? Certainly, it must be
kind of a peripheral interaction: if the photon struck directly
one of the valence quarks, the proton would 'explode', providing no
way for the large rapidity gap formation. What remains is the
possibility of the $t$--channel exchange by not-too-energetic
'particle' (Fig.\ref{DIS}b), which would be a natural mechanism of
the experimentally observed  weak proton deflection and small
$M^2_{diffr}$. Further experimental features suggest that this
'particle' should be chargeless and colorless, its interaction
with other particles should be of strong (not EM or weak) nature,
its 'propagation' should be independent of the specific process
($\gamma p$, $\gamma\gamma$, $pp$, $p\bar p$, etc), and it should
be of spin 1 (due to approximately $s$--constant $pp$ cross
section). In the early 60s this 'particle' was dubbed {\bf
Pomeron} (symbol $\Pom$).

Further properties come from combining the Regge picture and BFKL
results with experimental observations (for a detailed review of
Regge theory see \cite{Regge}). They include, first of all, the
asymptotic equality of total $pp$ and $p\bar p$ cross sections
(the Pomeranchuk theorem). Formulated long ago, it was
experimentally verified only recently. Then, the Regge theory
predicts the power-like $s$-dependence of the total $pp$ cross
section $\sigma \propto s^{2(\alpha_\Pom -1)}$, which has also
been experimentally observed, with intercept $\delta_\Pom \equiv
\alpha_\Pom -1 \approx 0.08$. On the other hand, the BFKL equation
\cite{BFKL,BFKLNLO}
succeeded in reproducing such power-like dependence
in QCD, but in a simplified case $\alpha_s = $const. In this
approach the hard pomeron is treated as two reggeized gluons
 --- an ansatz used currently in diffraction phenomenology with great
 success.
However, the predictive power of the BFKL approach for the numerical
value of the pomeron intercept is still limited and
not all issues with sensitivity of the result to  the infrared region
have been understood. For further reading on pomerons,
a topic very intriguing by itself, we refer to \cite{pomerons}.

\section{Vector meson production in diffractive DIS}

There are several possible final states $X$ in a typical
diffractive DIS (DDIS) process $\gamma^* p \to X p$:
system $X$ can be a real photon, a $q \bar q$ continuum pair
forming two jets or $q \bar q$ bound state,
for example,  a vector meson.
Let us now focus specifically on exclusive vector meson production
in diffractive DIS.
This reaction has been studied extensively at fixed target
DIS experiments at CERN and FNAL and more recently
by the H1 and ZEUS collaborations at HERA.

Despite the great deal of theoretical work on vector meson production
in diffractive DIS \cite{IK,KNZ98,early,early1}, there is a number
of issues that have not yet been seriously analyzed 
and still need a closer investigation.

One of them concerns the vector meson production 
in a seeming soft region, namely, at small values of $Q^2$ and $m_V^2$
(or, to put it short, at small values of $Q^2+m_V^2$).
Indeed, the majority of early calculations treated
the vector meson productions within the DGLAP-inspired approach,
the production amplitudes being expressed in terms
of the intergrated gluon density $G(x,Q^2)$. 
Certainly, this line of calculation is not applicable
at small enough values of $Q^2+m_V^2$, say at $Q^2+m_V^2 \lsim 1$ GeV$^2$. 
However it is necessary to understand that the DGLAP-based approach
 not only {\em can} be avoided but also {\em should} be avoided
when one studies diffractive scattering. Indeed, at high energies and
small to moderate values of $Q^2$ the dynamics of the amplitudes 
is governed by large logarithms of $\log(1/x)$ rather than $\log(Q^2)$,
and the correct and the most natural method to treat processes
in this kinematical region is $k_t$-factorization approach.
This approach does not place any restriction on the value of $Q^2$
as long as one works at large enough energies.

Although the strategy of the evaluation of the vector meson 
production amplitudes within the $k_t$-factorization approach
is essentially clear, performing reliable numerical prediction
is not a straghtforward task. The impediment consists in presence
of purely soft, non-perturbative quantities in the 
calculation, namely, the {\em gluon content of the proton} and
the {\em wave function} of the vector meson. 

 The presence of the former quantity is a specific feature of the
particular final state we investigate, however, the
unintegrated gluon density
\be
{\cal F}(x,Q^{2})={\partial G(x,Q^{2})\over \partial \log Q^{2}}
\,,
\ee
is the basic quantity in all 
$k_t$-factorization calculations. Unfortunately, 
no reliable Ansatz or parametrization has been developed,
and this gap needs to be filled.

Another issue that has never been brought under scrutiny
is the spin-angular coupling inside the vector meson.
In an off-forward scattering
$\gamma^*_{\lambda_\gamma} \to V_{\lambda_V}$
the $s$-channel helicity flip amplitudes can be non-vanishing.
Because of the well known quark
helicity conservation in high energy QCD scattering, such
a helicity flip is possible only due to the internal motion and
spin--angular momentum coupling of quarks in a vector meson.
This issue was accurately analyzed only in very recent papers
\cite{IK,KNZ98}, where it was shown that helicity non-conserving
amplitudes are not negligible, as they had been thought before.
Thus, as such, the helicity flip amplitudes would offer
a great deal of unique information of internal constituent motion
and spin--angular momentum structure of vector mesons,
unaccessible in other experiments. In addition, the vector meson
decays are self-analyzing and the full set of helicity amplitudes
can be measured experimentally. For unpolarized incident leptons,
the angular distribution of decay products is parameterized in terms of
15 spin-density matrix elements, which can be calculated via
five --- two helicity conserving plus three helicity violating ---
basic helicity amplitudes \cite{spinmatrix}.

Certainly, the helicity structure of the vector meson
production amplitudes must be analyzed only along with
a careful treatment of the spinorial structure of the $q\bar q \to V$
transition. It is thus rather surprising that the above issue of
sensitivity of the production amplitudes 
to the spin-angular momentum coupling has not been
addressed before. Namely, in a typical
vector meson production calculation, a vector meson
has been implicitly taken as $1S$ state and at the same time
an unjustified ansatz was used for $q\bar q \to V$ transition
spinorial structure, namely, of $\bar u' \gamma_\mu u \cdot V_\mu$
type. Being a mere analogy of $q\bar q \gamma$ vertex, this
ansatz in fact corresponds neither to pure $S$ nor to pure $D$
wave state but rather to their certain mixture.
Only in \cite{NNN1s2s} the cases of $1S$ and $2S$ vector mesons
were compared and the necessity of similar calculation for $D$ wave
states was stressed. Such calculations however have been missing in
literature until now.

In addition to purely theoretical needs, there are more issues
that call upon a thorough analysis of the $D$-wave effects. For
instance, different spin properties of the $S$- and $D$-wave
production may resolve the long standing problem of the $D$-wave
vs. $2S$-wave assignment for the $\rho'(1480)$ and $\rho'(1700)$
mesons (as well as the $\omega'$ and $\phi'$ mesons). Furthermore,
the deuteron which is a spin--1 ground state in the $pn$ system is
known to have a substantial $D$ wave admixture, which mostly
derives from the tensor forces induced by pion exchange between
nucleons. Recently, there has been much discussion \cite{Riska} of
the nonperturbative long-range pion exchange between light quarks
and antiquarks in a vector meson, which is a natural source of the
$S$-$D$ mixing in the ground state $\rho$ and $\omega$ mesons.\\

In the present work we addressed both issues. 
We performed an accurate determination of the unintegrated 
gluon density from the experimental data on the structure function
$F_{2p}$ and gave its convenient and ready-to-use parametrizations.
Besides, we constructed the consistent description of the
vector mesons with spin-angular coupling taken into account,
which enabled us to calculate diffractive production amplitudes
for pure $S$-wave and $D$-wave states as well as for an arbitrary
$S/D$ wave mixture.
This resulted in a complete theory of the vector meson production
in diffractive DIS within the $k_t$-factorization approach.

\section{The strategy of the thesis}

The guideline of the thesis is the following.
The main text is comprised of three Parts. Part I is an 
introduction to the $k_t$-factorization approach.
Here we calculate some basic scattering processes, such as
the virtual Compton scattering, and introduce the concept
of the differential densities of partons. The discussion
on the similarities and distinctions between the DGLAP-motivated
description and $k_t$-factorization description of diffractive
processes can also be found here.

In Part II we turn to the vector meson production amplitudes.
These are preceeded by the theory of vector meson structure
within the truncated Fock space, that is, when
the vector meson is assumed to be a bound state of $q\bar q$ pair only.
Upon obtaining the closed analytical expessions for vector meson
production amplitudes, we perform the twist expansion and
illustrated some of the most salient properties
of the $S$-wave and $D$-wave vector meson amplitudes.

Part III contains the numerical analysis of the expressions obtained
and the concrete prediction to various experimentally observed quantities.
At first we perform an extraction of the differential gluon density
of the proton and thoroughly investigate its properties.
Having brought the differential glue under control, 
we turn to the vector meson production amplitudes and
give a large number of predictions for $1S$, $2S$, and $D$-wave
states. Whenever the experimental results are avaliable, we confront
our predictions with the data. This Part is concluded  with
a detailed consideration of the effect of the Coulomb tail of the 
vector meson wave function and of the $S/D$-wave mixing.

Finally, we summarize our main findings in Conclusions.
Some lengthy calculations can be found in Appendices.

All the results presented in this thesis were derived by the author.
The text is based on publications \cite{IN99} and \cite{IN2001},
and on the works in progress \cite{IN2002}.
Preliminary results have been presented as talks
at \cite{DIS2000,lowx2001}.

\part{Basics of $k_t$-factorization scheme}
\chapter{The virtual Compton scattering}

We start our acquaintance with the $k_t$-factorization scheme
with calculation of the imaginary part of the 
forward virtual Compton scattering amplitude.
By means of optical theorem it is related with the total
photoabsorption cross section and with the structre functions
of the proton.

The purpose to get started with this quantity is twofold.
First, during this calculation we will follow all
steps and discuss all major feature of the $k_t$-factorization
scheme of calculations. Being rather simple, the Compton scattering 
amplitude will keep us from being distracted by inessential
technical complications, which would arise 
in other diffractive reactions.

The second purpose is to derive the well-known expression
for the structure function $F_{2p}$ in terms of unintegrated
gluon density of the proton: the basic quantity in any 
$k_t$-factorization calculation. These expressions will be used
later, when we discuss the determination of the unintegrated glue
from the experimental data.

The third aim is to use the simplicity of this amplitude
to gain as much insight into the dynamics 
of photon-proton peripheral interactions.
This information will be used later in deriving the 
vector meson production amplitudes thanks to a remarkable
similarity between the virtual
Compton scattering and the vector meson electroproduction
processes. Indeed, in the proton rest frame, both can be viewed 
as follows: a photon dissociates into a $q\bar q$ pair, 
which interacts with gluon content of the proton 
and then is projected onto the final
state. The hard dynamics in both cases are the same, only
difference lurking in the final state projection.

\section{Modeling virtual photoabsorption in QCD}

The quantity which is measured in deep inelastic leptoproduction is the
total cross section of photoabsorption $\gamma^{*}_{\mu}p \to X$ summed over
all hadronic final states $X$, where $\mu,\nu=\pm 1,0$ are helicities of
$(T)$ transverse and $(L)$ longitudinal virtual photons. One usually
starts with the imaginary part of the amplitude $A_{\mu\nu}$
of forward Compton scattering $\gamma_{\mu}^{*}p \to \gamma_{\nu}^{*}
p'$, which by optical theorem gives the total cross cross section
of photoabsorption of virtual photons
\bea
\sigma_{T}^{\gamma^{*}p}(x_{bj},Q^2) =
{1\over \sqrt{(W^{2}+Q^{2}-m_{p}^{2})^{2}+4Q^{2}m_{p}^{2}}}
{\rm Im} A_{\pm\pm}\, ,
\label{eq:3.1.1}
\eea
\be
\sigma_{L}^{\gamma^{*}p}(x_{bj},Q^2) =
{1\over \sqrt{(W^{2}+Q^{2}-m_{p}^{2})^{2}+4Q^{2}m_{p}^{2}}}
{\rm Im} A_{00}\, ,
\label{eq:3.1.2}
\ee
where $W$ is the total energy in the $\gamma^{*}p$ {\it c.m.s.}, $m_{p}$ is
the proton mass, $Q^{2}$ is the virtuality of the photon
and  $x_{bj}=Q^{2}/(Q^{2}+W^{2} - m_{p}^{2})$ is the Bjorken variable.
Hereafter we will suppress the subscript $bj$ and use $x\equiv x_{bj}$.

\begin{figure}[!htb]
   \centering
   \epsfig{file=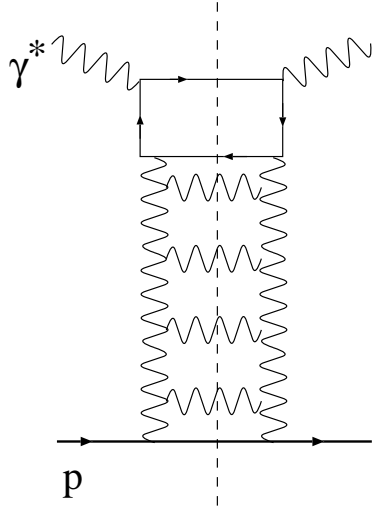,width=5cm}
   \caption{\em The pQCD modeling of DIS in terms of multiproduction
of parton final states.}
   \label{Multiproduction}
\end{figure}
In perturbative QCD (pQCD) one models virtual photoabsorption in terms of the
multiple production of gluons, quarks and antiquarks
(fig.~\ref{Multiproduction}). The experimental
integration over the full phase space of hadronic states $X$ is substituted
in the pQCD calculation by integration over the whole phase space of QCD
partons
\be
\int |M_{{\cal X}}|^{2} d\tau_{{\cal X}} \Rightarrow \sum_{n} |M_{n}|^{2}
\prod\int_{0}^{1} {dx_{i} \over x_{i}} d^{2}{\vec{\kappa}}_{i}\, ,
\label{eq:3.1.3}
\ee
where the integration over the transverse momenta of partons goes over the
whole allowed region
\be
0 \leq \vec{\kappa}_{i}^{2} \leq {1\over 4}W^{2} = {Q^{2} (1-x) \over 4x}\, .
\label{eq:3.1.4}
\ee
The core of the so-called DGLAP approximation \cite{DGLAP} is an observation that
at finite $x$ the dominant contribution to the multiparton production cross
sections comes from a tiny part of the phase space
\bea
&1 \geq x_{1}  \geq x_{2} ... \geq  x_{n-1}  \geq  x_{n}  \geq  x &\, ,
 \nonumber\\
&0 \leq \vec{\kappa}_{1}^{2} \ll \vec{\kappa}_{2}^{2}...\ll\vec{\kappa}_{n-1}^{2}
\ll \vec{k}^{2} \ll  Q^2 &\,,\label{eq:3.1.5}
\eea
in which the upper limit of integration over transverse momenta of partons is
much smaller than the kinematical limit (\ref{eq:3.1.4}). At very small $x$
this limitation of the transverse phase
space becomes much too restrictive and the
DGLAP approximation is doomed to failure.

\begin{figure}[!htb]
   \centering
   \epsfig{file=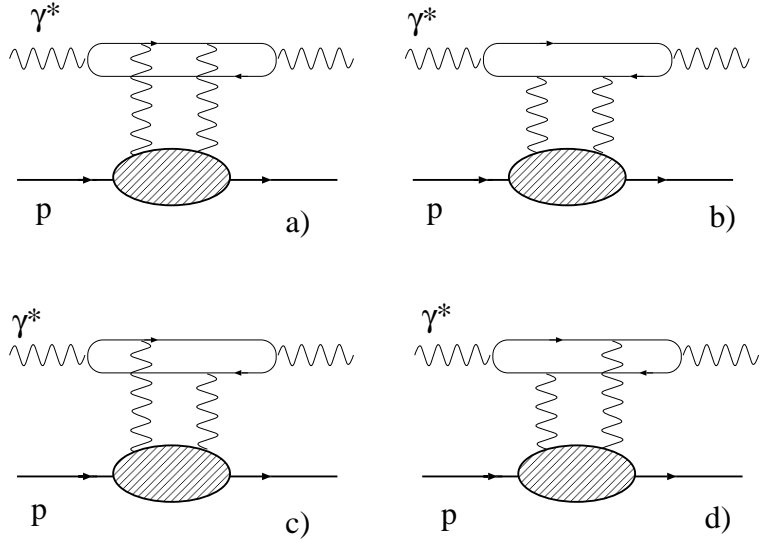,width=10cm}
   \caption{\em The $k_t$-factorization representation for
DIS at small $x$.}
   \label{KappaFactorization}
\end{figure}

Hereafter we focus on how lifting the restrictions on the transverse phase
space changes our understanding of the gluon structure function of the
nucleon at very small $x$, that is, very large ${1\over x}$\,. In this kinematical
region the gluon density $g(x,Q^{2})$ is much higher than the density of
charged partons $q(x,Q^{2}), \bar{q}(x,Q^{2})$. As Fadin, Kuraev and Lipatov
\cite{FKL} have shown, to the leading $\log {1\over x}$
(LL${1\over x}$) approximation the dominant contribution to
photoabsorption comes  in this regime from multigluon
final states of fig.~\ref{Multiproduction};
alternatively, to the LL${1\over x}$ splitting of gluons
into gluons dominates the splitting of gluons into $q\bar{q}$ pairs.
As a matter of fact, for the purposes of the present analysis
we do not need the full BFKL dynamics, in the
$k_t$-factorization only the $q\bar{q}$ loop is treated explicitly
to the  LL${1\over x}$ approximation. In this
regime the Compton scattering can be viewed as an interaction of the nucleon with
the lightcone $q\bar{q}$  Fock states of the photon via the exchange by gluons,
fig.~\ref{KappaFactorization},
and the Compton scattering amplitude takes the form
\be
A_{\nu\mu}=\Psi^{*}_{\nu,\lambda\bar{\lambda}}\otimes A_{q\bar{q}}\otimes
\Psi_{\mu,\lambda\bar{\lambda}}
\label{eq:3.1.6}
\ee
Here $\Psi_{\mu,\lambda\bar{\lambda}}$ is the $Q^{2}$ and $q,\bar{q}$ helicity
$\lambda,\bar{\lambda}$ dependent lightcone wave function of the photon and
the  QCD pomeron exchange $q\bar{q}$-proton scattering kernel $A_{q\bar{q}}$
does not depend on, and conserves exactly, the $q,\bar{q}$ helicities,
summation over which is understood in (\ref{eq:3.1.6}).

The resummation of diagrams of fig.~\ref{Multiproduction}
defines the unintegrated gluon structure function of the target,
which is represented in diagrams of fig.~\ref{KappaFactorization}
as the dashed blob.

\section{Details of calculation}

Suppose that there were no interaction between the gluons 
exchanged in $t$-channel, so that the full calculation of the Compton 
scattering amplitude amounts only to picking up the Born
diagrams.
Consider one of such diagrams, i.e. Diagr.a in 
Fig.~\ref{KappaFactorization}, but without the dashed blob. 
A virtual photon turns into $q\bar q$
pair which interacts with a proton via two-gluon exchange.
The general expression for this amplitude reads:
\begin{eqnarray}
&&iA = \int{d^4k\over (2\pi)^4}\int{d^4\kappa\over (2\pi)^4} \
\bar{u}' (-ig\gamma^{\nu'} t^{B'}) i{\hat p -\hat\kappa +m \over
\left[(p-\kappa)^2 - m^2 + i\epsilon\right]}(-ig\gamma^{\mu'}
t^{A'})u \nonumber\\&& \cdot(-i){g_{\mu\mu'}\delta_{AA'}\over
\kappa^2 - \mu^2 +i\epsilon}
\cdot(-i){g_{\nu\nu'}\delta_{BB'}\over \kappa^2 - \mu^2
+i\epsilon} \cdot e_f^2 \nonumber\\ &&\cdot{ Sp\left\{ (-ie)\hat
e\  i(\hat k -\hat q + m)\ (-ie) \hat e^*\ i(\hat k+m)\
(-ig\gamma^\nu t^B)\ i(\hat k + \hat \kappa + m)\ (-ig\gamma^\mu
t^A)\ i(\hat k+m) \right\} \over \left[k^2 - m^2 +
i\epsilon\right]^2 \left[(k+\kappa)^2 -m^2 +i\epsilon\right]
\left[(k-q)^2-m^2 +i\epsilon\right] }\nonumber\\ \label{photon1}
\end{eqnarray}
Let's first calculate the numerator.

\subsubsection{Color factor}

If we consider photon scattering off a single quark, we have
\begin{equation}
{1 \over N_c}Sp\{t^{B'}t^{A'}\}
\cdot\delta_{AA'}\delta_{BB'}Sp\{t^Bt^A\}= {1 \over N_c}{1 \over
2}\delta_{AB}{1 \over 2}\delta_{AB}= {1 \over 2}{1 \over
2N_c}(N_c^2-1) = {1 \over 2}C_F = {2 \over 3}\label{photon2}
\end{equation}
However, we should take into account that quarks are sitting
inside a colorless proton, whose color structure is
\begin{equation}
\psi_{color} = {1 \over \sqrt{6}} \epsilon^{abc} q^a q^b
q^c\label{photon3}
\end{equation}
In this case there are two ways a pair of gluons can couple 3
quark lines. In the first way both gluons couple to
the same quark. Since the quark momentum does not change after
these two interactions, the nucleon stays in the same state:
$\langle N|N\rangle = 1$. In the second case, gluon legs are
attached to different quark lines, so that extra momentum $\kappa$
circulates between quarks, which gives rise to the factor
 $\langle N|\exp(i\kappa r_1 - i\kappa r_2)|N\rangle$, i.e.
to the two-body formfactor. Therefore, for the lower line instead
of
\begin{equation}
{1 \over N_c}Sp\{t^{B}t^{A}\} = {1 \over N_c} {1 \over 2}
\delta_{AB}\label{photon4}
\end{equation}
one has
\begin{eqnarray}
&&{1 \over 6} \epsilon^{abc} \left(3\delta_{aa'}\delta_{bb'}
t^A_{cc''} t^B_{c''c'} + 6 \delta_{aa'}t^A_{bb'} t^B_{cc'}\langle
N|\exp(i\kappa r_1 - i\kappa r_2)|N\rangle \right)
\epsilon^{a'b'c'}\nonumber\\ &=& Sp\{t^{A}t^{B}\} -
Sp\{t^{A}t^{B}\}\langle N|\exp(i\kappa r_1 - i\kappa r_2)|N\rangle
\nonumber\\ & = &{1 \over 2} \delta_{AB}(1 - \langle
N|\exp(i\kappa r_1 - i\kappa r_2)|N\rangle )\,.\label{photon5}
\end{eqnarray}
Note also that a similar calculation for $N_c$ number of colors
would yeild the same result. Thus, the
overall color factor is
\begin{equation}
2V(\kappa) \equiv 2(1 - \langle N|\exp(i\kappa r_1 - i\kappa
r_2)|N\rangle).\label{photon5a}
\end{equation}

As known, the highest power $s$ contribution comes from so--called
nonsense components of gluon propagator (density matrix)
decomposition:
\begin{equation}
g_{\mu\mu'} = {2 p'_{\mu}q'_{\mu'} \over s} + {2 p'_{\mu}q'_{\mu'}
\over s} + g_{\mu\mu'}^{\bot} \ \approx \ {2 p'_{\mu}q'_{\mu'}
\over s}\,.\label{photon6}
\end{equation}
Therefore, the lower trace is calculated trivially
\begin{equation}
{1 \over 2} Sp\{\hat p' \hat q' \hat p' \hat q'\} =
s^2\,.\label{photon7}
\end{equation}
So, combining all factors, one has for numeractor of
Eq.(\ref{photon1})
\begin{eqnarray}
&&(4\pi\alpha_s)^2\ 4\pi\alpha_{em}\ e_f^2\cdot 2 {4 \over s^2}
s^2 \cdot Sp\left\{ \hat e\  (\hat k -\hat q + m)\ \hat e^*\ (\hat
k+m)\ \hat q'\ (\hat k + \hat \kappa + m)\ \hat q'\ (\hat
k+m)\right\}\nonumber\\ &=&(4\pi\alpha_s)^2\ 4\pi\alpha_{em}\
e_f^2\cdot 8 \cdot 2s^2\cdot I^{(a)}(\gamma^* \to
\gamma^*)\,.\label{photon8}
\end{eqnarray}
Note that we factored out $2s^2$ because it will appear later in
all trace calculations. So, the resulting expression for amplitude
(\ref{photon1}) looks like
\begin{eqnarray}
A &=& (4\pi\alpha_s)^2\ 4\pi\alpha_{em}\ 16 e_f^2\ s^2 \cdot
\int{d^4k\over (2\pi)^4}\int{d^4\kappa\over (2\pi)^4}\cdot {1
\over \left[(p-\kappa)^2 - m^2 + i\epsilon\right] \left[\kappa^2 -
\mu^2 +i\epsilon\right]^2} \label{photon9}\\ && \cdot\,{ \tilde
I(\gamma^* \to \gamma^*) \over \left[k^2 - m^2 +
i\epsilon\right]^2 \left[(k+\kappa)^2 -m^2 +i\epsilon\right]
\left[(k-q)^2-m^2 +i\epsilon\right] }\nonumber
\end{eqnarray}

One can now immediately write similar expressions for three other
diagrams (Fig.~\ref{KappaFactorization} b,c,d). 
Indeed, they will differ from
Eq.(\ref{photon9}) only by the last line. Aside from different
expressions for traces, the quark line propagator structures will
read:
\begin{eqnarray}
(b)&&\quad [k^2 - m^2 + i\epsilon][(k-q)^2-m^2 +i\epsilon]
[(k+\kappa)^2 -m^2 +i\epsilon] [(k+\kappa-q)^2 -m^2
+i\epsilon]\nonumber\\ (c)&&\quad [k^2 - m^2 +
i\epsilon][(k-q)^2-m^2 +i\epsilon] [(k-\kappa)^2 -m^2 +i\epsilon]
[(k-\kappa-q)^2 -m^2 +i\epsilon]\nonumber\\ (d)&&\quad [k^2 - m^2
+ i\epsilon][(k-q)^2-m^2 +i\epsilon] [(k-\kappa)^2 -m^2
+i\epsilon] [(k-q)^2 -m^2 +i\epsilon]\label{photon9a}
\end{eqnarray}

\section{Denominator and trace evaluation}
Now we turn to calculation of denominators. As usual, we implement
Sudakov's decomposition and
\begin{eqnarray}
k&=& yp' + zq' + \vec{k}\,;\nonumber\\ \kappa &=& \alpha p'
+ \beta q' + \vec{\kappa}\;\nonumber\\ q &=& q' -xp'\,,
\label{photon10}
\end{eqnarray}
and make use of relation
\begin{equation}
d^4k = {1\over 2} s dy\,dz\,d^2\vec{k}\,. \label{photon11}
\end{equation}
The complete analysis of denominator hierarchy and
their integrals is performed in Appendix A. We show there that,
for example, for diagram A, the imaginary part of the desired
integral is equal to
\be
Im \left\{\int {dy\ dz\ d\alpha\ d\beta \over \mbox{propagators}}\right\} =
{4\pi^2 \over s^3} \int_0^1 dz\ {1-z \over z}
{1 \over [\vec k^2 + m^2 + z(1-z)Q^2]^2}
{1 \over [\vec \kappa^2 + \mu^2]^2}\,.
\ee
The answers for the other three diagrams differ only by replacements
$\vec k \to \vec k+ \vec\kappa$ in quark propagators whenever appropriate.
The whole expression for the imaginary part of the amplitude is
then
\bea
Im A &=& s {32 \over (2\pi)^2}e_f^2 \cdot \alpha_{em}\cdot 
\int dz d^2\vec{k} \int {d^2 \vec{\kappa} V(\kappa) \alpha_s^2 \over
(\vec{\kappa}^2 + \mu^2)^2} \nonumber\\ &&\times \Biggl[ {1-z \over
z} {I^{(a)} \over [\vec{k}^2 + m^2 + z(1-z)Q^2]^2}
+ {z\over 1-z}{ I^{(d)} \over 
[(\vec{k}+\vec\kappa)^2 + m^2 + z(1-z)Q^2]^2}\nonumber\\
&& + {I^{(b)}+I^{(c)} \over 
[(\vec{k}+\vec{\kappa})^2 +m^2 + z(1-z)Q^2]
[\vec{k}^2 +m^2 + z(1-z)Q^2]}\Biggr]\,.
\,.\label{photon21}
\eea
The short-hand notation $\alpha_s^2$ should be in fact understood
as
\be
\alpha_s(\mbox{lower})\cdot \alpha_s(\mbox{upper})
\equiv \alpha_s(\vec\kappa^2)\cdot 
\alpha_s(q^2)\,.
\ee
with
\be
q^2 = \mbox{max}[\vec k^2+m^2+z(1-z)Q^2,\vec\kappa^2]\,.
\label{alphasarg}
\ee

Now we calculate the integrands $I$ which enter
Eq.(\ref{photon21}) $Sp\{...\} = 2s^2 \cdot I$. We will do this
via light cone helicity amplitude technique. In the subsequent
discussion we will use the following convention:
$$
q'^\mu = q_+n_+^\mu
$$
i.e. the light cone direction $+$ is taken along photon
propagation.

A crutial point that justifies the usage of the helicity 
amplitude technique for all quarks lines inside the loop is that
{\em in the trace calculation all fermions can be treated 
as on-mass shell} thanks to the presence of $\hat n_-$ vertices. 
This property comes from the following arguments.

Note that every intermediate quark line in any diagram
couples at least to one of the $t$-channel gluons.
Algebraically, it means that every $\hat k + m$ stands
near the factor $\hat n_-$.
Let us apply the Sudakov decomposition to the $\gamma$ matrix:
\begin{eqnarray*}
&&\gamma^\mu = \gamma_+n_+^\mu + \gamma_-n_-^\mu + \vec \gamma^\mu\,;\\
&&\gamma_+ = \hat n_- = {1 \over \sqrt{2}}(\gamma_0 + \gamma_3)\,,\quad
\gamma_- = \hat n_+ = {1 \over \sqrt{2}}(\gamma_0 - \gamma_3)\,.
\end{eqnarray*}
Now decompose the propagator numerator of
the constituent, to which this $\hat n_-$ leg couples:
\begin{equation}
\hat k + m = k_+\gamma_- + k_-\gamma_+ - \vec k\vec \gamma +m
\label{propag}
\end{equation}
and rewrite it using notation of (\ref{LC6}), (\ref{LC6b}) as
\begin{equation}
\hat k + m = k_+\gamma_- + k^*_-\gamma_+ - \vec k\vec \gamma +m
+ (k_- - k_-^*)\gamma_+ =
\hat k^* + m + {k^2 - m^2 \over 2k_+}\gamma_+\,.
\label{propag1}
\end{equation}
In other words, we expressed the virtual quark propagator as the sum of
{\it on shell} quark propagator and an additional "instantaneous
interaction" term.
However, since $\hat n_-$ is inserted between two $(\hat k +m)$ factors,
this item does not work due to identity $\gamma_+\gamma_+ = 0$.
The net result is that
{\it wherever $\hat n_-$ appears, both constituents can be treated
on mass shell in the trace calculation}, which completes the proof.

Having established that the fermion lines in the
trace calculation can indeed be taken as if the quarks
were real, we can now decompose the numerator of each ofthe quark lines as
\begin{equation}
\hat k +m  \to \hat k^* + m =  
\sum_{\lambda} \bar{u}_\lambda u_\lambda\,,
\label{tracetrick4}
\end{equation}
where spinors $u_\lambda$ are for an on-mass shell fermion.

In the case of antiquark line the above derived property
is valid as well. The only
thing to remember here is that antiquark propagates upstream the
fermion arrow, so that
\begin{equation}
(-\hat k) +m = - \sum_{\lambda} \bar{v}_\lambda v_\lambda\,,
\label{tracetrick5}
\end{equation}
i.e. each antiquark propagator gives rise to factor $-1$.\\

The derivation is given in Appendix B in full detail and yields
\bea
Im A^T &=& s {32 \over (2\pi)^2}\sum_i e_i^2 \cdot
\alpha_{em}\cdot \int dz\;d^2\vec{k} \int {d^2 \vec{\kappa} V(\kappa)
\alpha_S^2\over (\vec{\kappa}^2 + \mu^2)^2} \nonumber\\ 
&&\times \Biggl\{ m^2\left[{1\over (\vec{k}+\vec{\kappa})^2 
+m^2 + z(1-z)Q^2} - {1 \over \vec{k}^2 + m^2 + z(1-z)Q^2} \right]^2
\label{imat1}\\&&\times
+ [z^2+(1-z)^2] \left[{\vec{k}+\vec{\kappa}\over (\vec{k}+\vec{\kappa})^2 
+m^2 + z(1-z)Q^2} - {\vec k \over \vec{k}^2 + m^2 + z(1-z)Q^2} \right]^2
\Biggr\}\nonumber\\[5mm]
Im A^L &=& s {32 \over (2\pi)^2}\sum_i e_i^2 \cdot
\alpha_{em}\cdot \int dz\;d^2\vec{k} \int {d^2 \vec{\kappa} V(\kappa)
\alpha_S^2\over (\vec{\kappa}^2 + \mu^2)^2} \nonumber\\ 
&&\times 4z^2(1-z)^2Q^2\left[{1\over (\vec{k}+\vec{\kappa})^2 
+m^2 + z(1-z)Q^2} - {1 \over \vec{k}^2 + m^2 + z(1-z)Q^2} \right]^2\,.
\label{imal1}
\eea

\section{Gluon density}\label{sectgluon}

It is obvious that Eqs.(\ref{imat1}) and (\ref{imal1}) are not
directly related to the real experimental situation, for
up to now we assumed that the two exchanged gluons do not interact.
Such interaction will definitely change the properties of
the entire $t$-channel exchange, and in fact, 
as predicted by the BFKL equation, the resultant Pomeron
has rather little in common with the initial two perturbative gluons.

Since the BFKL evolution necessarily involves soft gluons
(see below more on soft-to-hard diffusion), it does not allow
for accurate perturbative calculations.
It must be understood however that although we do not know
what happens "inside the Pomeron" on the way from proton to the 
quark-antiquark pair, we nevertheless know 
--- and the knowlegde is based on the leading order BFKL analysis ---
that eventually the $q\bar q$ pair will interact 
with nothing else but {\em two gluons}.
We underline that this conclusion does not require the 
gluons to be hard, but rather it relies on the fact that 
higher Fock states of the $t$-channel can be, 
to the leading $\log{1\over x}$ approximation, absorbed in the 
two gluon state \cite{NNPZsigmadipole}.

Thus, the only thing we need to know is the momentum distribution
of the {\em uppermost gluons}, or, to put it exactly, 
the probability distribution to find a gluon with given lightcone
momentum fraction $x_g$ and the transverse momentum $\vec\kappa$.
\be
dn_g = {\cal F}(\vec\kappa, x_g) {d\vec\kappa^2\over \vec\kappa^2}
{dx_g\over x_g}\,.\label{dgsfdefine}
\ee
This distribution is called the {\em unintegrated (or differential)
gluon structure function}, DGSF, or simply the 
{\em unintegrated gluon density}.

Since the differential gluon density is uncalculable with pQCD,
a reasonable way to proceed in our computation of the compton
scattering amplitude further consists in finding out 
the correspondence $V(\kappa) \leftrightarrow {\cal F}(\vec\kappa,x_g)$.
Namely, we will calculate the unintegrated gluon density {\em at the Born
level} ${\cal F}_{Born}$ in terms of $V(\kappa)$, and then
postulate that the BFKL dynamics amounts to replacement 
${\cal F}_{Born} \to {\cal F}$. This procedure will give us a unique
prescription how to correctly incorporate the unintegrated gluon density
into the $k_t$-factorization calculations.

In order to provide a gentle introduction into the concept
of the unintegarated parton densities, we start with
famous Fermi-Weizs\"acker-Williams approximation in QED.
We will find the expression for the unintegrated photon densities
in the case of a single charged particle and charge neutral
positronium and then translate the results to the case of color
forces.

\subsection{Differential density of gauge bosons: the QED primer}

\begin{figure}[!htb]
   \centering
   \epsfig{file=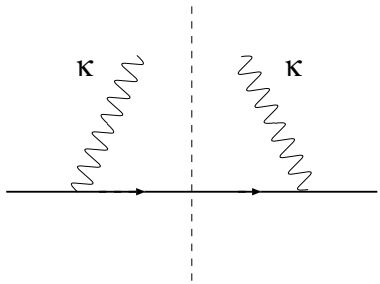,width=5cm}
   \caption{\em The Fermi-Weizs\"acker-Williams diagram for
calculation of the flux of equivalent photons}
   \label{WWFermi}
\end{figure}

For the pedagogical introduction we recall the
celebrated Fermi-Weizs\"acker-Williams approximation in QED, which is the
well known precursor of the parton model (for
the review see \cite{Budnev}). Here
high energy reactions in the  Coulomb field of a charged particle are treated
as collisions with equivalent transversely polarized photons --- partons of
the charged particle, Fig.\ref{WWFermi}.
The familiar flux of comoving equivalent transverse
soft photons carrying a lightcone fraction $x_{\gamma} \ll 1$ of the momentum of a
relativistic particle, let it be the electron, reads
\be
dn^\gamma_e = {\alpha_{em} \over \pi }
{\vec{\kappa}^2 d\vec{\kappa}^2 \over (\vec{\kappa}^2 +\kappa_{z}^{2})^{2}}
{dx_{\gamma} \over x_{\gamma}}
\approx {\alpha_{em} \over \pi }
{d\vec{\kappa}^2  \over \vec{\kappa}^2 }
{dx_{\gamma} \over x_{\gamma}}\,,
\label{eq:2.1}
\ee
Here $\vec{\kappa}$ is photon transverse momentum and $\kappa_{z}=m_{e}x_{\gamma}$
is the photon longitudinal momentum in the electron Breit frame.
The origin of $\vec{\kappa}^2$ in the numerator is in the current conservation,
i.e. gauge invariance. Then the unintegrated photon structure function
of the electron is by definition
\be
{\cal F}_{\gamma}(x_{\gamma},\vec{\kappa}^2)
 = { \partial G_\gamma \over \partial \log \vec{\kappa}^2} =
x_{\gamma}{dn^\gamma_e \over dx_{\gamma} d\log\vec{\kappa}^{2}}
= {\alpha_{em} \over \pi}\left( {\vec{\kappa}^2 \over
\vec{\kappa}^2 +\kappa_{z}^{2}}\right)^{2}
\,.
\label{eq:2.2}
\ee

\subsection{Differential density of photons in a positronium}

\begin{figure}[!htb]
   \centering
   \epsfig{file=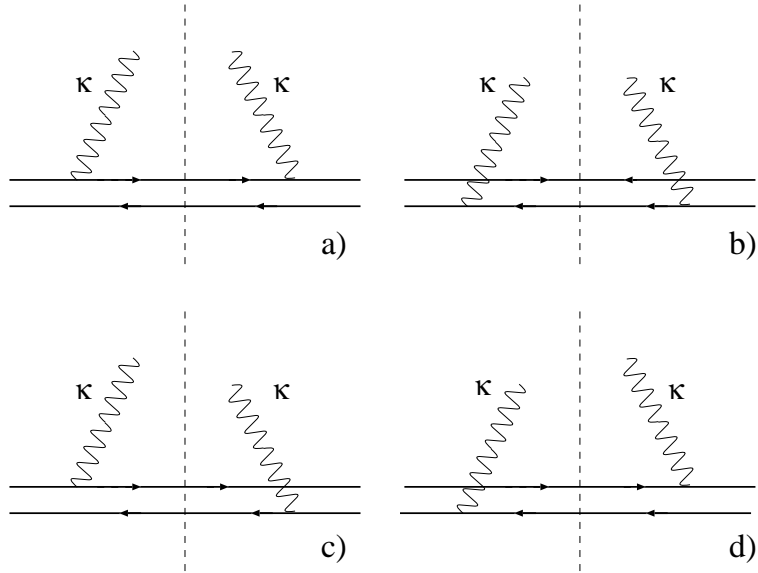,width=10cm}
   \caption{\em The Fermi-Weizs\"acker-Williams diagrams for
calculation of the flux of equivalent photons in positronium.}
   \label{WWpositronium}
\end{figure}

If the relativistic particle is a positronium, Fig.~\ref{WWpositronium},
destructive interference
of electromagnetic fields of the electron and positron must be taken
into account. Specifically, for soft photons with the wavelength
$\lambda = {1\over \kappa} \gg a_{{\rm P}}$, where  $a_{{\rm P}}$ is the
positronium Bohr radius, the electromagnetic fields of an electron
and positron cancel each other and the flux of photons vanishes,
whereas for $\lambda \ll a_{{\rm P}}$ the flux of photons will be twice
that for a single electron.
The above properties are quantified by the formula
\be
{\cal F}_{\gamma}^{{\rm P}}(x_{\gamma},\vec{\kappa}^2) =
N_c{\alpha_{em} \over \pi} \left( {\vec{\kappa}^2 \over
\vec{\kappa}^2 +\kappa_{z}^{2}}\right)^{2} V(\kappa)\,,
\label{eq:2.3}
\ee
where factor $N_c=2$ is a number of charged particles in the positronium
and corresponds to the Feynman diagrams of
Fig.~\ref{WWpositronium}a,~\ref{WWpositronium}b.
The vertex function $V(\kappa)$ is expressed in terms of the two-body
formfactor of the positronium,
\be
V(\kappa) = 1 - F_2(\vec{\kappa},-\vec{\kappa}) =
1 - \langle{\rm P}|\exp(i\vec{\kappa} (\br_{-} - \br_{+}))|{\rm P}\rangle\,,
\label{eq:2.4}
\ee
where $\br_{-} - \br_{+}$ is the spatial separation of $e^+$ and $e^{-}$
in the positronium.
The two-body formfactor $F_{2}(\vec{\kappa},-\vec{\kappa})$
describes the destructive interference of electromagnetic fields of the
electron and positron and corresponds to the Feynman diagrams
of Fig.~\ref{WWpositronium}c,~\ref{WWpositronium}d.
It vanishes for large enough $\kappa \gsim a_{{\rm P}}^{-1}$,
leaving us with $V(\kappa)=1$, whereas for soft gluons one has
\be
V(\kappa) \propto \vec{\kappa}^{2}a_{{\rm P}}^{2}
\label{eq:2.5}
\ee
One can say that the law (\ref{eq:2.5}) is driven by electromagnetic gauge
invariance, which guarantees that long wave photons decouple
from the charge neutral system.

Finally, recall that the derivation of the differential flux
of transverse polarized photons would equally hold if the
photons were massive vector bosons interacting with the conserved current,
the only change being in the propagator. For instance, for the charge
neutral source one finds
\be
{\cal F}_{V}^{{\rm P}}(x_{V},\vec{\kappa}^2) =
N_c{\alpha_{em} \over \pi} \left( {\vec{\kappa}^2 \over
\vec{\kappa}^2 +m_{V}^{2}}\right)^{2} V(\kappa)\,.
\label{eq:2.6}
\ee
Recall that the massive vector fields are Yukawa-Debye screened
with the screening radius
\be
R_{c}={1\over m_{V}}\, .
\label{eq:2.7}
\ee
To the lowest order in QED perturbation theory the two exchanged photons in
figs.\ref{WWFermi},~\ref{WWpositronium} do not interact
and we shall often refer to (\ref{eq:2.6}) as
the Born approximation for the differential vector boson structure function.
One can regard (\ref{eq:2.6}) as a minimal model for soft $\vec{\kappa}$
behavior of differential structure function for Yukawa-Debye
screened vector bosons.

\subsection{Differential gluon density in a proton}

The expression for the Born level unintegrated gluon density
in color neutral proton can be obtained immediately by
generalization of (\ref{eq:2.3}) and (\ref{eq:2.6}). The only
thing one as to do is an accurate color algebra, which leads to
\begin{equation}
{\cal F}_g^{(Born)} = C_F N_c {\alpha_{s}(\vec\kappa^2) \over \pi} 
V(\kappa)\,.\label{density4}
\end{equation}
Therefore, a prescription how to include unintegrated gluon
density is as follows:
\begin{equation}
C_F N_c {\alpha_{s}(\vec\kappa^2) \over \pi} V(\kappa) 
\equiv {\cal F}^{(Born)} \to{\cal F}\label{replacerule}\,.
\end{equation}
Note that in this prescription one of the two strong coupling constants
in (\ref{photon21}) (the lower one) is absorbed into the definition
of ${\cal F}$.

\section{Final expressions}

With all pieces brought together,
the answer for the total photoabsorption cross section reads
\bea
\sigma_T(x,Q^2) &=& {\alpha_{em} \over \pi}\sum_i e_i^2 \int_0^1 dz
\int d^2\vec k\int {d^2\vec\kappa \over
\vec\kappa^4}\alpha_s(q^2)\nonumber\\ &&\times {\cal F}(x_g,\vec\kappa^2)
\left[m_i^2\Phi_0^2 + [z^2+(1-z)^2]\vec\Phi_1^2\right]\,;\label{sigmat}\\[3mm]
\sigma_L(x,Q^2) &=& {\alpha_{em} \over \pi}\sum_i e_i^2 \int_0^1 dz
\int d^2\vec k\int {d^2\vec\kappa \over
\vec\kappa^4}\alpha_s(q^2)\cdot{\cal F}(x_g,\vec\kappa^2)
4Q^2z^2(1-z)^2\Phi_0^2\,,\label{sigmal}
\eea
where
\be
\Phi_0 = {1   \over  \vec{k}^{2}+\varepsilon^{2}}-
{1 \over  (\vec{k}-\vec{\kappa})^{2}+\varepsilon^{2}}\,;\quad
\vec\Phi_1 = {\vec{k}   \over  \vec{k}^{2}+\varepsilon^{2}}-
{\vec{k} - \vec{\kappa} \over  (\vec{k}-\vec{\kappa})^{2}+\varepsilon^{2}}
\,.
\ee
Here
\be
\varepsilon^2=z(1-z)Q^{2}+m_{f}^{2}\,, 
\ee
and the density of gluons enters at
\be
x_{g}={Q^{2} +M_{t}^{2} \over W^{2}+Q^{2}}=
x\left(1 +{M_{t}^{2} \over Q^{2}}\right) \,.
\label{eq:3.1.11}
\ee
Here $M_{t}$ is the transverse mass of the produced $q\bar{q}$
pair in the photon-gluon fusion $\gamma^{*}g\rightarrow q\bar{q}$:
\be
M_{t}^{2} = {m_{f}^{2}+\vec{k}^{2} \over 1-z}+
{m_{f}^{2}+(\vec{k}-\vec{\kappa})^{2} \over z}   \, .
\label{eq:3.1.12}
\ee

No restrictions on the transverse momentum in the $q\bar{q}$ loop,
$\vec{k}$, and gluon momentum, $\vec{\kappa}$, are imposed in the above
representations. The above used BFKL scheme defines
DGSF uniquely in terms of physical observables.

We note that equations obtained are for forward
diagonal Compton scattering, but similar representation 
in terms of the unintegrated gluons structure
function holds also for the off-forward Compton scattering at finite momentum
transfer ${\bf \Delta}$, for off-diagonal Compton scattering
when the virtualities of the initial and final state photons are different,
$Q_{f}^{2}\neq Q_{i}^{2}$, including the timelike photons and vector mesons,
$Q_{f}^{2}=-m_{V}^{2}$, in the final state.

The photoabsorption cross sections define the dimensionless structure functions
\be
F_{T,L}(x,Q^{2})={Q^{2}\over 4\pi^{2}\alpha_{em}}\sigma_{T,L}
\label{eq:3.1.13}
\ee
and $F_{2}=F_{T}+F_{L}$, which admit the familiar 
pQCD parton model interpretation
\be
F_{T}(x,Q^{2})= \sum_{f=u,d,s,c,b,..} e_{f}^{2}
[q_{f}(x,Q^{2})+\bar{q}_{f}(x,Q^{2})]\, ,
\label{eq:3.1.14}
\ee
where $q_{f}(x,Q^{2}), \bar{q}_{f}(x,Q^{2})$ are the integrated
densities of quarks and antiquarks carrying the fraction $x$
of the lightcone momentum of the target and
transverse momenta $\leq Q$.

\section{The virtual Compton amplitude in the impact parameter space}

A deep further insight into the Compton amplitude --- and into the
diffractive processes in general --- can be gained by switching to
the impact parameter (the transverse coordinate) representation.
We will see that the answer will allow for a simple probabilistic
form
\be
A = \int dz \int d^2\vec r \sigma(\vec r) |\Psi_\gamma(z,\vec r
)|^2\,.\label{comptonimpact}
\ee
The quantity $\sigma(\vec r)$ has the meaning of the dipole cross
section, while the $\Psi_\gamma(z,\vec r)$ represents the photon
wave function.

In order to switch to the impact parameter space, we perform the
2-dimensional Fourier transform. We use the following equalities:
\bea
{1 \over \vec k^2 + \vep^2} = {1 \over 2\pi}\int d^2\vec r e^{i\vec k\vec r}
f_0(r)\,; &\Rightarrow& f_0(r) \equiv \int d^2\vec k {e^{-i\vec k\vec
r}\over \vec k^2 \vep^2} = K_0(\vep r)\,;\nonumber\\
{\vec k \over \vec k^2 + \vep^2} = {1 \over 2\pi}\int d^2\vec r e^{i\vec k\vec r}
\vec f_1(\vec r)\,; &\Rightarrow& f_1(\vec r) \equiv
\left(-i{\partial \over \partial \vec r}\right)\int d^2\vec k {e^{-i\vec k\vec
r}\over \vec k^2 \vep^2}\nonumber\\
&=& \left(-i{\partial \over \partial \vec r}\right) K_0(\vep r)
= -i \vep {\vec r\over r}K_1(\vep r)\,.
\eea
This leads to representation
\bea
\Phi_0 &=& {1\over 2\pi}\int d^2\vec r e^{i\vec k\vec r}K_0(\vep
r)\left(1 - e^{i\vec\kappa \vec r}\right)\,;\nonumber\\
\vec\Phi_1 &=& {1\over 2\pi}\int d^2\vec r e^{i\vec k\vec r}
(-i\vep){\vec r\over r}K_1(\vep r)
\left(1 - e^{i\vec\kappa \vec r}\right)\,.
\eea

Let us now take an approximation that there is no other $\vec k$
dependence in the photoabsorption cross section. Then one will
immediately have
\bea
\int d^2\ \vec k \Phi_0 \Phi_0^* &=& \int d^2\vec k {1 \over 4\pi^2}
\int d^2 \vec r_1 d^2\vec r_2\ e^{i\vec k\vec r_1 - i \vec k\vec r_2}
K_0(\vep r_1)K_0(\vep r_2)\left(1 - e^{i\vec\kappa \vec r_1}\right)
\left(1 - e^{-i\vec\kappa \vec r_2}\right) \nonumber\\
&=& \int d^2 \vec r\ K^2_0(\vep r)\ 2[1-\cos(\vec\kappa\vec r)]\,.
\eea
Substituting these expressions into photoproduction cross
sections, one gets
\bea
\sigma_T(x,Q^2) &=& {\alpha_{em} \over \pi}\sum_i e_i^2 \int_0^1
dz\int d^2\vec r
\left[m_i^2 K_0^2(\vep r) + [z^2 + (1-z)^2]\vep^2 K_1^2(\vep r)\right]
\nonumber\\&\times& \int {d^2\vec\kappa \over \vec\kappa^4}
\alpha_s{\cal F}(x_g,\vec\kappa^2)\ 2[1-\cos(\vec\kappa\vec
r)]\,;\nonumber\\[2mm]
\sigma_L(x,Q^2) &=& {\alpha_{em} \over \pi}\sum_i e_i^2 \int_0^1 dz
\int d^2\vec r \cdot 4Q^2z^2(1-z)^2 \cdot K_0^2(\vep r)
\nonumber\\&\times& \int {d^2\vec\kappa \over \vec\kappa^4}
\alpha_s{\cal F}(x_g,\vec\kappa^2)\ 2[1-\cos(\vec\kappa\vec r)]\,.\label{siglt}
\eea

\subsection{Dipole cross section}

The above results can put into the form (\ref{comptonimpact})
by breaking Eqs.~(\ref{siglt}) into some positively defined
cross section and the square of the photo wave function.
This is done in an unambiguous way by defining the {\em dipole
cross section}.

Let us first consider the total quark-proton cross section
\be
\sigma_{qp} = {2\pi\over 3}\int {d^2\vec \kappa\over \vec\kappa^4}
\alpha_s(\vec\kappa^2){\cal F}(x,\vec\kappa^2)\,.\label{sigmaqp}
\ee
Note that this expression does not depend on the quark
transverse momentum $\vec k$. This means that precisely this
cross section corresponds not only to the plane wave, but also
to any transverse wave packet. In particular, a localized state
in the impact parameter space (that is, a quark with a fixed
separation $\vec \rho_{qp}$ from the proton) would be described by the
same formula.

Given this cross section, we can now ask for interaction 
of color dipole with the proton. 
In this each extra gluon that is attached to the antiquark
rather then to the quark gives rise of extra phase factor
$\exp(\pm i\vec \kappa\vec r)$ as well as extra minus sign.
These factors for the four diagrams will then sum up to produce
$$
1 - e^{i\vec \kappa\vec r} - e^{-i\vec \kappa\vec r} + 1 =
2[1-\cos(\vec\kappa\vec r)]\,.
$$
The result for the dipole cross section is then 
\bea
\sigma_{dip}(\vec r) &=& {2 \pi\over 3}\int {d^2\vec \kappa\over \vec\kappa^4}
\alpha_s\left(\mbox{max}\left[\vec\kappa^2,{A \over r^2}\right]\right){\cal F}(x,\vec\kappa^2)
2[1-\cos(\vec\kappa\vec r)] \nonumber\\
&=&
{4 \pi^2 \over 3} \int {d\vec\kappa^2\over \vec\kappa^4}
\alpha_s\left(\mbox{max}\left[\vec\kappa^2,{A \over r^2}\right]\right)
{\cal F}(x,\vec\kappa^2)[1-J_0(\kappa r)]\,.
\eea
Note that in contrast to (\ref{sigmaqp}) 
the argument of $\alpha_s$ contains now the effect
of a possible screening from the complementary charge.
Indeed, even when $\vec\kappa^2$ is small, the strong coupling
constant does not boost up, for such a boost requires the presence
of soft gluon vertex correction loops, which are strongly suppressed
by the color anti-charge of the antiquark.

Having defined the dipole color cross section, we can now  
return to the photoabsorption cross section and cast it into form
\bea
&&\sigma_T =\int_0^1 dz \int d^2\vec r \ \sigma_{dip}(\vec r)
\cdot |\Psi_\gamma^T|^2(z,\vec r)\,, \quad
\sigma_L =\int_0^1 dz \int d^2\vec r \ \sigma_{dip}(\vec r)
\cdot |\Psi_\gamma^L|^2(z,\vec r)\,; \nonumber\\
&&\quad|\Psi_\gamma^T(z,\vec r)|^2 = {3\alpha_{em} \over \pi^2}
\sum_i e_i^2 \left[m_i^2K_0^2(\vep r) + [z^2+(1-z)^2]\vep^2K_1^2(\vep r)
\right]\,;\nonumber\\
&&\quad|\Psi_\gamma^L(z,\vec r)|^2 = {3\alpha_{em} \over \pi^2}
\sum_i e_i^2 \ 4Q^2 z^2(1-z)^2 K_0^2(\vep r)\,.
\eea
Note that this representation literally represents 
the probabilistic form (\ref{eq:3.1.6}) 
of the forward scattering amplitude.

\chapter{DGLAP vs. $k_t$ factorization}

The calculation of the forward Compton scattering amplitude
conducted in the previous chapter within the framework of the 
$k_t$-factoriation approach can be used now to investigate
the major similarities and the gross differences in comparison
with the widely used DGLAP approach to the computation of
high energy reactions.

It turns out that in the double logarithmic regime --- that is,
when both $\log{1\over x}$ and $\log{Q^2}$ are large,
we might expect that both approaches are applicable
and their predictions should asymptotically converge.
We are now going to demostrate that it is indeed the case,
and during this analysis we will also show
what sort of phase space restrictions DGLAP approach contains
and what it leads to.

\section{How DGLAP and $k_t$ factorization approaches
meet at high $Q^2$}

Recall the familiar DGLAP equation \cite{DGLAP} 
for scaling violations at small $x$,
\be
{d F_{2}(x,Q^{2}) \over d\log Q^{2}}= \sum_{f} e_{f}^{2}
{\alpha_{S}(Q^{2}) \over 2\pi} \int_{x}^{1} dy [y^2+(1-y)^{2}]
G\left({x\over y},Q^{2}\right)
\approx {\alpha_{S}(Q^{2}) \over 3\pi} G(2x,Q^{2})\sum_{f} e_{f}^{2}\, ,
\label{eq:3.2.1}
\ee
where for the sake of simplicity we consider only light flavours.
Upon integration we find
\be
F_{2}(x,Q^{2})\approx \sum_{f} e_{f}^{2}\int_{\mu^2}^{Q^{2}}
{d\Qb^{2} \over \Qb^{2}} {\alpha_{S}(\Qb^{2}) \over 3\pi}
G(2x,\Qb^{2})\, ,
\label{eq:3.2.2}
\ee
with $\mu^2$ being the proper cut-off.
In order to see the correspondence between the $k_t$-factorization and
DGLAP factorization
it is instructive to follow the derivation of (\ref{eq:3.2.2}) from the
$\vec{\kappa}_{\perp}$-representation.

First, separate the
$\vec{\kappa}^{2}$-integration into the DGLAP part
of the gluon phase space
$\vec{\kappa}^{2} \lsim \Qb^{2}=\epsilon^2 +\vec{k}^2$ and beyond-DGLAP region
$\vec{\kappa}^{2} \gsim \Qb^{2}$. One readily finds
\be
\left({\vec{k} \over  \vec{k}^{2}+\varepsilon^{2}} -
{\vec{k}-\vec{\kappa} \over  (\vec{k}-\vec{\kappa})^{2}+\varepsilon^{2}}\right)^{2}
\to
\left\{\begin{array}{lcr}
\left({2z^2(1-z)^2Q^4 \over \Qb^8}-{2z(1-z)Q^2 \over \Qb^{6}} +
{1\over \Qb^{4}}\right)\vec{\kappa}^2
& \mbox{if}& \vec{\kappa}^2 \ll \Qb^{2} \\[2mm]
\left({1\over \Qb^{2}}-{z(1-z)Q^2 \over \Qb^{4}}\right),&
\mbox{if}& \vec{\kappa}^{2} \gsim \Qb^{2}
\end{array}\right.
\label{eq:3.2.3}
\ee

Consider first the contribution from the DGLAP part of the phase space
$\vec{\kappa}^{2} \lsim \Qb^{2}$.
Notice that because of the factor $\vec{\kappa}^{2}$ in (\ref{eq:3.2.3}),
the straightforward $\vec{\kappa}^2$ integration of the DGLAP component
yields $G(x_{g},\Qb^{2})$ and $\Qb^2$ is precisely the pQCD hard scale
for the gluonic transverse momentum scale:
\bea
\int^{\Qb^2}_{0}
{ d\vec{\kappa}^2
\over \vec{\kappa}^{4}}\alpha_{S}(q^{2})
{\cal F}(x_{g},\vec{\kappa}^{2})
\left({\vec{k} \over  \vec{k}^{2}+\varepsilon^{2}} -
{\vec{k}-\vec{\kappa} \over  (\vec{k}-\vec{\kappa})^{2}+\varepsilon^{2}}\right)^{2}
\nonumber\\
=\left({2z^2(1-z)^2Q^4 \over \Qb^8}-{2z(1-z)Q^2 \over \Qb^{6}} +
{1\over \Qb^{4}}\right) \alpha_s(\Qb^2)G(x_{g},\Qb^{2})
\label{eq:3.2.4}
\eea
The contribution from the beyond-DGLAP region of the phase space
can be evaluated in terms of ${\cal F}(x_{g},\Qb^{2})$ and the
rescaling factor $C_2$:
\bea
\int_{\Qb^2}^{\infty}
{ d\vec{\kappa}^2
\over \vec{\kappa}^{4}}\alpha_{S}(q^{2})
{\cal F}(x_{g},\vec{\kappa}^{2})\left({1\over \Qb^{2}}-{z(1-z)Q^2
\over \Qb^{4}}\right)
=\left({1\over \Qb^{4}}-{z(1-z)Q^2 \over \Qb^{6}}\right)
\alpha_s(\Qb^2){\cal F}(x_{g},\Qb^{2})I(x_{g},\Qb^2)\nonumber\\
=
\left({2z^2(1-z)^2Q^4 \over \Qb^8}-{2z(1-z)Q^2 \over \Qb^{6}} +
{1\over \Qb^{4}}\right)\alpha_s(\Qb^2){\cal F}(x_{g},\Qb^{2})
\log C_{2}(x_{g},\Qb^2,z)\,.\label{eq:3.2.5}
\eea
The latter form of (\ref{eq:3.2.5}) allows for convenient
combination (\ref{eq:3.2.4}) and (\ref{eq:3.2.5})
rescaling the hard scale in the GSF
\be
G(x_{g},\Qb^{2})+{\cal F}(x_{g},\Qb^{2})\log C_2(x_{g},\Qb^2,z)
\approx G(x_{g},C_{2}(x_{g},\Qb^2,z)\Qb^2)\, .
\label{eq:3.2.6}
\ee
Here the exact value of $I(x_{g},\Qb^2)$ depends on
the rate of the $\vec{\kappa}^2$-rise of ${\cal F}(x_{g},\vec{\kappa}^{2})$.
At small $x_{g}$ and small to moderate $\Qb^{2}$ one finds
$I(x_{g},\Qb^2)$  substantially larger than 1 and
$C_{2}(x_{g},\Qb^2,z)\gg 1$, see more discussion below in section 9.

Now change from $d\vec{k}^{2}$ integration to $d\Qb^2$ and again split the
$z$,$Q^2$ integration into the DGLAP part of the phase space
$\Qb^{2} \ll {1 \over 4}Q^2$, where
either $z < {\Qb^2 \over Q^{2}}$ or $1-z < {\Qb^2 \over Q^{2}}$,  and
the beyond-DGLAP region  $\Qb^{2} \gsim {1\over 4}Q^2$, where $0< z < 1$.
As a result one finds
\bea
\int dz [z^2+(1-z)^2]\left({2z^2(1-z)^2Q^4 \over \Qb^8}-{2z(1-z)Q^2 \over \Qb^{6}} +
{1\over \Qb^{4}}\right)\nonumber\\
 =
\left\{\begin{array}{lcr}
{4 \over 3\Qb^2 Q^2},
& \mbox{if}&  \Qb^{2}\ll {1\over 4}Q^{2} \\[2mm]
\left(2A_{2}{Q^4\over \Qb^{8}}-2A_{1}{Q^{2} \over \Qb^6} +A_{0}{1\over \Qb^{4}}\right),&
\mbox{if}& \Qb^{2} \gsim {1\over 4}Q^{2}
\end{array}\right.
\label{eq:3.2.7}
\eea
where
\be
A_{m}=\int_{0}^{1} dz [z^2+(1-z)^2]z^m(1-z)^m
\label{eq:3.2.8}
\ee
Let $\overline{C}_2$ be $C_{2}(x_{g},\Qb^2,z)$ at a mean point. Notice also
that $M_{t}^{2} \sim Q^{2}$, so that $x_{g}\sim 2x$. Then the
contribution from the DGLAP phase space of $\Qb^2$ can be cast in precisely
the form (\ref{eq:3.2.2})
\be
\left. F_{2}(x,Q^{2})\right|_{DGLAP}\approx \sum_{f}
e_{f}^{2}\int_{0}^{{\overline{C_2}\over 4}Q^{2}} {d\Qb^{2} \over \Qb^{2}}
{\alpha_{S}(\Qb^{2}) \over 3\pi}
G(2x,\Qb^{2})\, .
\label{eq:3.2.9}
\ee

The beyond-DGLAP region of the phase space gives the extra contribution of
the form
\bea
\left. \Delta F_{2}(x,Q^{2})\right|_{non-DGLAP}\sim
\sum_{f} e_{f}^{2}
{\alpha_{S}(Q^2) \over 3\pi}
\int_{Q^{2}}^{\infty}
{d\Qb^{2} \over \Qb^{2}}{Q^2 \over \Qb^2}G(2x,\Qb^{2}) \nonumber \\
\sim
\sum_{f} e_{f}^{2}
{\alpha_{S}(Q^2) \over 3\pi}
G(2x,Q^{2})
\, .
\label{eq:3.2.10}
\eea
Eqs.(\ref{eq:3.2.9}) and (\ref{eq:3.2.10}) immediately reveal
the phenomenological consequences of lifting the DGLAP restrictions
in the transverse momenta integration.
Indeed, the DGLAP approach respects the following strict inequalities
\be
\vec{\kappa}^2 \ll \vec{k}^2 \quad \mbox{and} \quad
\vec{k}^2 \ll Q^2\,.
\ee
As we just saw, removing the first limitation effectively shifted
the upper limit in the $\Qb^2$ integral to
${{\overline C_2} \over 4} Q^2 \not = Q^2$, while lifting the second constraint
led to an additional, purely non-DGLAP contribution.
Although both of these corrections lack one leading log-$Q^2$ factor
they are numerically substantial.

The above analysis suggests that the DGLAP and $k_t$-factorization
schemes converge logarithmically at large $Q^{2}$.
However, in order to reproduce the
result (\ref{eq:3.2.9}) and (\ref{eq:3.2.10}) for the full phase space
by the conventional DGLAP contribution (\ref{eq:3.2.2}) from the restricted
phase space (\ref{eq:3.1.5}) one has to ask for DGLAP gluon density
$G_{pt}(x,Q^{2})$ larger than the integrated GSF
in the $k_t$-factorization scheme and the
difference may be quite substantial in the domain of strong scaling
violations.

\section{The different evolution paths: soft-to-hard diffusion and
vice versa}

\begin{figure}[!htb]
   \centering
  \epsfig{file=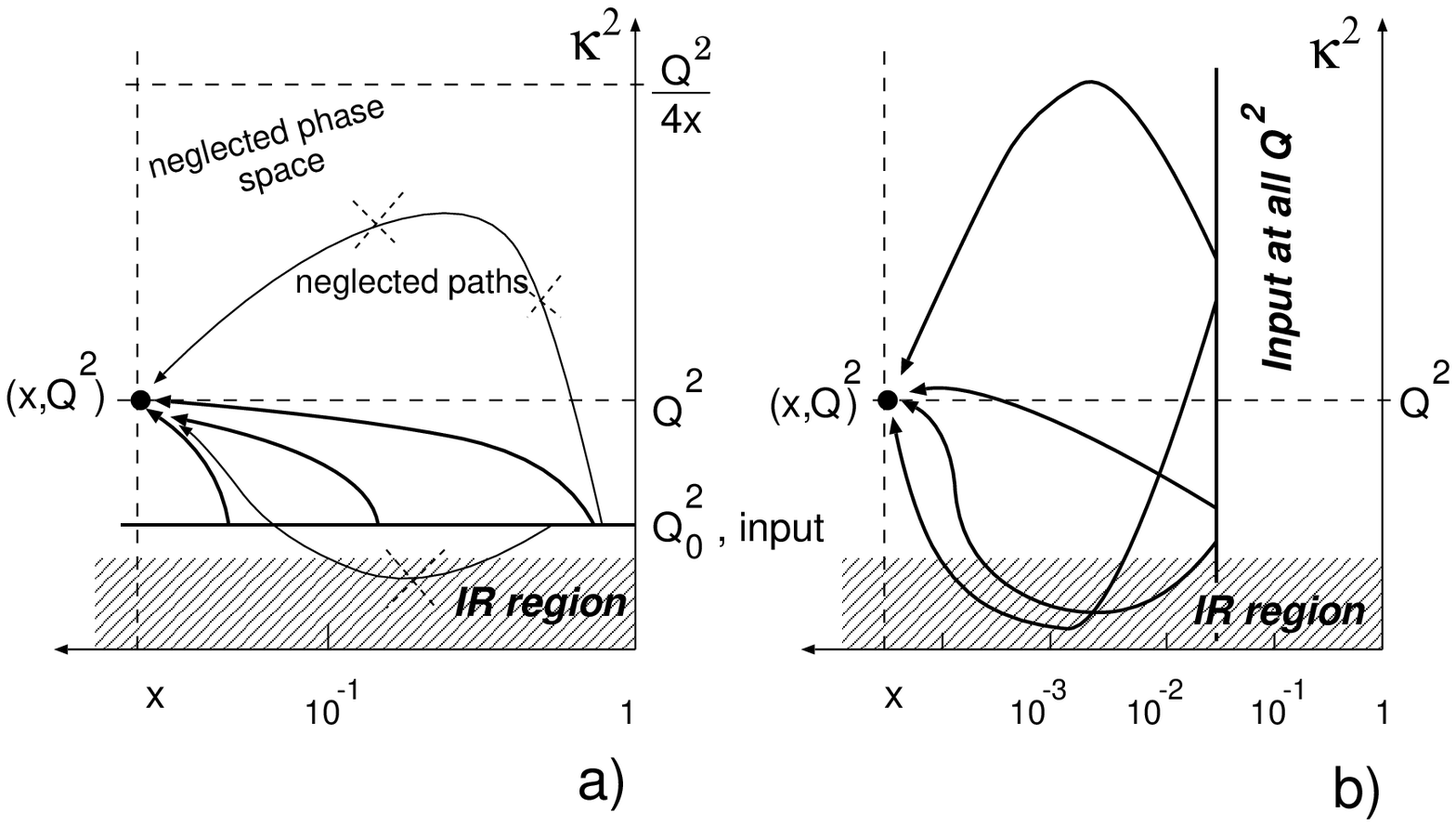,width=16cm}
   \caption{\em The Huygens principle for $Q^{2},x$
evolution of DIS structure
functions with {\rm (a)} DGLAP restricted transverse phase space and {\rm (b)}
for the BFKL
$x$ evolution without restrictions on the transverse phase space
and hard-to-soft {\rm \&} soft-to-hard diffusion.}
   \label{Huygens}
\end{figure}

The above discussion of the contributions to the total cross section
from the DGLAP and non-DGLAP parts of the phase space can conveniently
be cast in the form of the Huygens principle. To the standard DGLAP
leading log$Q^{2}$ (LL$Q^{2}$) approximation one only considers
the contribution from the restricted part of the
available transverse phase space (\ref{eq:3.1.5}).
The familiar Huygens principle for the homogeneous DGLAP LL$Q^{2}$ evaluation
of parton densities in the $x$-$Q^{2}$ plane is illustrated in
Fig.~\ref{Huygens}a: one starts with the boundary condition
$p(x,Q_{0}^{2})$ as a function of $x$ at fixed $Q_{0}^{2}$,
the evolution paths $(z,\tilde{Q}^{2})$ for the calculation
of $p(x,Q^{2})$ shown in Fig.~\ref{Huygens}a are confined 
to a rectangle $x \leq z \leq 1,~ Q_{0}^{2} \leq \tilde{Q}^{2}\leq Q^{2}$, 
the evolution is unidirectional in the sense that there is no feedback 
on the $x$-dependence of $p(x,Q_{1}^{2})$
from the $x$-dependence of $p(x,Q_{2}^{2})$ at $Q_{2}^{2} \geq Q_{1}^{2}$.
In Fig.~\ref{Huygens}a we show some examples of evolution paths which are
kinematically allowed but neglected in the DGLAP approximation.
Starting with about flat or slowly rising $G(x,Q_{0}^{2})$, one finds that
the larger $Q^{2}$, the steeper the small-$x$ rise of $G(x,Q^{2})$.

At $x\ll 1$ the DGLAP contribution from the restricted transverse
phase space (\ref{eq:3.1.5}) no longer dominates
the multiparton production cross sections, the restriction
(\ref{eq:3.1.5}) must be lifted and the contribution to the
cross section from small $\vec{\kappa}_{i}^2$ and large $\vec{\kappa}_{i}^2\gsim Q^2$
can no longer be neglected. The Huygens principle for the
homogeneous BFKL evolution is illustrated in Fig.~\ref{Huygens}b: one starts with
the boundary condition ${\cal F}(x_{0},Q^{2})$ as a function of $Q^{2}$
at fixed $x_{0}\ll 1$, the evolution paths $(z,\tilde{Q}^{2})$
for the calculation
of $p(x,Q^{2})$ are confined to a stripe $x \leq z \leq x_{0}$, in contrast
the the unidirectional DGLAP evolution one can say that under BFKL evolution
the small-$x$ behaviour of $p(x,Q^{2})$ at large $Q^{2}$ is fed partly by
the $x$-dependence of soft  $p(x,Q^{2})$ at larger $x$ and vice versa.
The most dramatic consequence of this soft-to-hard and
hard-to-soft diffusion, which can
not be eliminated, is that  at very small $x$ the $x$-dependence
of the gluon structure
in the soft and hard regions will eventually be the same.
The rate of such a hard-to-soft diffusion is evidently
sensitive to the infrared regularization of pQCD,
 the model estimates show that in the HERA range
of $x$ it is very slow \cite{NZZ94,NZBFKL}.

\part{Derivation of vector meson production amplitudes}

\chapter{Description of a vector meson}

In this chapter we first introduce the vector meson
light cone wave function (LCWF) and show how it emerges
in diagrammatic calculations. Then, describing $S$ and $D$
wave type vector particles, we give at once expressions for
$S$ and $D$ wave vector meson spinorial structures, which we
then prove by computing the normalization condition for LCWF.
Finally, we also calculate $V \to e^+e^-$ decay constants
to be used afterwards.

\section{Bound states in QFT}

While describing particle motion in non-relativistic Quantum Mechanics,
one usually deals with a configuration space particle wave function,
which is a good description because the number of particles is conserved.
So, when one has a system of particles and shows that the wave function
corresponding to their relative motion descreases at large relative
distances at least exponentially, one can speak of a bound state.

In Quantum Field Theory (QFT) this approach needs an update,
since the field function becomes an operator in Fock space.
Besides, since a bound state always implies the presence of interaction,
the projection of a physical bound state onto
the Fock space of {\it free, non--interacting, plane-wave} state vectors
has a rather complicated structure:
\begin{equation}
|V_{phys}\rangle = c_0|q\bar{q}\rangle + c_1 |q\bar{q} g\rangle + c_2 |q\bar{q} gg\rangle
+ c_3 |q\bar{q} q\bar{q}\rangle + ...  \label{fock}
\end{equation}
We emphasize that in this decomposition
quarks and gluons are assumed free, i.e. {\bf on mass shell}.
Coefficients $c_i$ can be called
'wave functions' of the given projection of a physical vector meson,
with $|c_i|^2$ being the probability of finding a vector meson
in a given state.

The exact treatment of any reaction involving the vector meson
must account for all terms in the above expansion.
Demostration of a method that would account for all these terms 
is however still an unresolved task, and currently one is
bound to the term-by-term analysis of the vector meson reactions.

Given a large number of papers devoted to the high-energy reactions
involving vector mesons, and in particular, the process of
difractive vector meson production in DIS, one might expect
that the lowest Fock state in the above decomposition has
been already thoroughly studied. It turns out however
that it is not so, for in all early calculations the importance of
the spin-angular coupling inside the vector meson and dramatic
effects it entails was heavily overlooked.

In this Chapter we close this gap. We construct
a full and exact theory of the vector meson stucture,
provided the vector meson Fock space is saturated
only to the lowest $q\bar q$ state. Being only an approximation,
this approach still is of vital importance to the whole field,
for {\em it results in a complete, self-consistent and self-contained
spin-angular description of the vector meson}.
To our best knowlegde, our work represent the only satisfactory
theory of spin structure of the vector meson.

\section{LCWF and vertex factor}\label{sectionLCWF}

Let us now outline how a wave function of a bound state appears in
the diagrammatic language.

In the non-relativistic quantum mechanics, the two-particle
bound state problem can be immediately reformulated
as a problem for one particle of reduced mass $\mu$, moving
in the external potential. This reformulation allows one to
split the wave function into two factors: the wave function
of the motion of the composite particle as whole and the wave function
corresponding to the internal motion of constituents.
The former part factors out trivially, while the latter
wave function obeys the following Schodinger equation
\begin{equation}
\left[{\hat p^2 \over 2\mu} + V(r)\right]\psi(r) = E \psi(r)\,.\label{LCWF1}
\end{equation}
Since the wave function $\psi(r)$ and the interaction operator $V(r)$
exhibit good infinity behavior, one can rewrite this equation
in the momentum representation
\begin{eqnarray}
&&{p^2 \over 2\mu}\psi(p) + {1 \over (2\pi)^3} \int d^3k V(k)\psi(p-k)
= E\psi(p)\,;\nonumber\\
&&\left({p^2 \over m} - E\right)\psi(p) =
- {1 \over (2\pi)^3} \int d^3k V(k)\psi(p-k)\,.\label{LCWF2}
\end{eqnarray}
In this notation, this equation can be viewed as a homogeneous
non-relativistic Bethe-Salpeter equation for the wave function
$\psi(p)$ that describes the relative motion of constituents inside
a composite particle.

\begin{figure}[!htb]
   \centering
   \epsfig{file=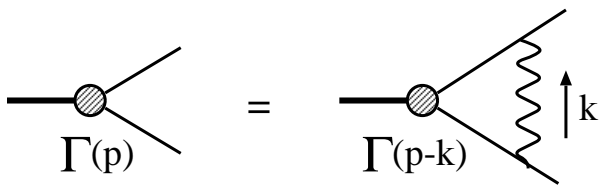,width=80mm}
   \caption{The diagrammatic interpretation of the integral equation
for vertex function $\Gamma(p)$ ($p$ is the relative constituents momentum). }
   \label{LCWFfig}
\end{figure}

Let us now introduce
\begin{equation}
\Gamma(p) \equiv \left({p^2 \over m} - E\right)\psi(p) \label{LCWF3}
\end{equation}
Then Eq.(\ref{LCWF2}) can be rewritten as
\begin{equation}
\Gamma(p) = - {1 \over (2\pi)^3} \int d^3k V(k)
\fr{1}{\fr{(p-k)^2}{m} - E} \Gamma(p-k)\label{LCWF4}
\end{equation}
This equation has an absolutely straightforward diagrammatic interpretation
(Fig.\ref{LCWFfig}). One sees that $\Gamma(p)$ stands for
bound-state $\to$ constituents transition  vertex,
with $p$ being the relative momentum of the constituents,
factor $ 1/[{(p-k)^2 \over m} - E]$ describes propagation of $q\bar q$
pair and $V(k)$ stands for the interaction between constituents.
Of course, the kinetic energy $p^2/2\mu \not = E$, the total energy,
which is in fact negative, so no pole arises in the propagator.

In the relativistic case, i.e. in QFT, it is not clear
{\it a priori} whether the whole picture that involves
somehow defined wave function and representation of
the vector meson as free non-interacting constituents
would work at all. So, in our approach we will be aiming at
introducing an appropriately defined wave function
and demonstrating that hard processes involving vector mesons
can be expressed in terms of expectation
values of $q \bar q$ amplitudes between wave functions,
i.e. we intend to treat a hard process in a
{\it probabilistic, quantum mechanics-like manner}.

In the following we will show that this program succeeds.
Namely, we will introduce the {\it radial} wave function
of the $q \bar q$ state of a vector meson as
\begin{equation}
\psi(q)\equiv  {\Gamma(q) \over M^2 - m_V^2}\,
\label{LCWF6}
\end{equation}
(the angular dependence of the wave function will be treated
separately, see Sect.\ref{sectspin})
Here $\Gamma(q)$ is the vertex factor,
$M^2$ is the eigenvalue of the relativistic kinetic operator
of the on mass shell $q \bar q$ state and $m_V^2$ is eigenvalue
of the total relativisitic Hamiltonian, which is of course
equal to the mass of the vector meson squared.
Then, during an {\it accurate and honest} analysis of a hard process Feynman
diagrams, we will always make sure that wave function (\ref{LCWF6}) automatically appears
in calculations and {\it the rest} looks the same as if both fermions were
on mass shell. If we see that fermion virtualities modify the results,
or if different Fock states mix during hard interactions of the vector meson,
it would signal the invalidity of the free particle parton language
and consequenly the breakdown of the whole approach. This restriction
must always be taken into account when obtaining and interpreting the
parton model-based results.

\section{Light cone formalism}\label{sectLC}

The term "light cone approach" to high--energy process
calculations can have different meanings. Some prefer
to re-formulate the whole QFT within light cone dynamics,
introduce light cone quantization and derive light cone Feynman
rules (on Light Cone Field Theory see \cite{LCQFT0,LCQFT}).
 However, one should keep in mind that even
within the usual QFT the light cone formalism can be freely used
as a means to greatly simplify intermediate calculations.
This is exactly the way we will use it.

It was noted long ago \cite{sudak}
that the calculation of a high energy collision
is simplied if one decomposes all momenta in terms of light cone
$n_+^\mu, n_-^\mu$ and transversal components, which we will
always mark with the vector sign over a letter
(so called Sudakov's decomposition):
\begin{eqnarray}
&&n_+^\mu = {1 \over \sqrt{2}}(1, \vec 0, 1)\,;
\quad n_-^\mu = {1 \over \sqrt{2}}(1, \vec 0, -1)\,;
\quad (n_+n_-)=1\,,\ (n_+n_+) = (n_-n_-) = 0\nonumber\\
&&p^\mu = p_+n_+^\mu + p_-n_-^\mu + \vec p^\mu\,;
\quad p^2 = 2p_+p_- - \vec p^2\,. \label{LC1}
\end{eqnarray}
Indeed, imagine two high energy particles colliding with momenta
$p^\mu$ and $q^\mu$ respectively and equal masses $m$.
Then one can choose such a frame of reference that
in the Sudakov's decomposition
\begin{eqnarray}
&&p^\mu = p_+n_+^\mu + p_-n_-^\mu + \vec p^\mu\,,\nonumber\\
&&q^\mu = q_+n_+^\mu + q_-n_-^\mu + \vec q^\mu\nonumber\label{LC2}
\end{eqnarray}
quantities $p_-$ and $q_+$ are large $p_-,\, q_+ \gg m$ while
$p_+ = (\vec p^2 + m^2)/(2p_-)$ and $q_- = (\vec q^2 + m^2)/(2q_+)$
are small. The total energy squared of these two particles
will be defined as
\begin{equation}
s \equiv 2q_+p_-\,. \label{LC3}
\end{equation}
Note that our definition of $s$ is somewhat different from
the more familiar one $(p+q)^2$ by terms
$\propto m^2, \vec p^2$. However, it is not of any imporatnce
for us, since in the course of all calculations
we will keep track only of the highest power $s$ contributions,
i.e. everything will be calculated in the leading power $s$ approximation.

\begin{figure}[!htb]
   \centering
   \epsfig{file=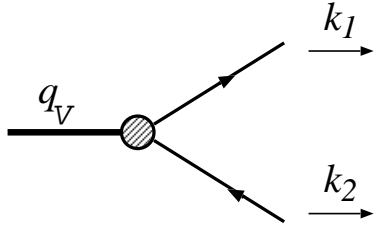,width=50mm}
   \caption{Kinematics of $V \to q\bar q$ vertex on the light cone.
Vector meson momentum $q_V$ is taken incoming, constituents momenta are
outgoing.}
   \label{LCWFkinem}
\end{figure}

Let us now go further and examine the kinematics of
a typical $q \bar q V$ vertex (Fig.\ref{LCWFkinem}).
The Sudakov's decomposition of all momenta reads:
\begin{eqnarray}
&&q_V^\mu = q_{V+}n_+^\mu + q_{V-}n_-^\mu\,;\nonumber\\
&&k^\mu_1 = k_{1+}n_+^\mu + k_{1-}n_-^\mu + \vec k^\mu =
zq_{V+}n_+^\mu + y q_{V-}n_-^\mu + \vec k^\mu\,;\nonumber\\
&&k^\mu_2 =  k_{2+}n_+^\mu + k_{2-}n_-^\mu - \vec k^\mu =
(1-z)q_{V+}n_+^\mu + (1-y)q_{V-}n_-^\mu - \vec k^\mu\,\label{LC6}
\end{eqnarray}
so that
\begin{equation}
q_V^2 = 2q_{V+}q_{V-} = m_V^2\,;\quad
k_1^\mu + k_2^\mu = q_V^\mu\,;\quad k_i^2 \not = m^2\,,
  \label{LC6a}
\end{equation}
i.e. quarks can be off mass shell.
Now let us introduce momenta $k_i^*$, which would correspond to
on mass shell fermions. The only component in $k_i$ subject to modification
is $k_{i-}$ component, or absolutely equivalently, the energy.
Large $k_{i+}$ components are insensitive to (reasonable) quark virtuality
variations. So, to obtain the on mass shell momenta, one has to replace
\begin{equation}
k_{i-} = {k_i^2 + \vec k^2 \over 2k_{i+}} \to
k^*_{i-} = {m^2 + \vec k^2 \over 2k_{i+}}\,.
  \label{LC6b}
\end{equation}
Then the 4-vector
\begin{equation}
  \label{LC6c}
  q^\mu = k^{*\mu}_1 + k^{*\mu}_2
\end{equation}
squared is equal to
\begin{equation}
  \label{LC6d}
  M^2 = q^2 = 2q_+q_- = 2q_{V+}\left(k^*_{1-} + k^*_{2-}\right) =
{\vec k^2 + m^2 \over z(1-z)}\,.
\end{equation}
And again we emphasize that the {\it Feynman invariant mass}
(i.e. the total 4-momentum squared) of the virtual
quark-antiquark pair is $m_V^2$. The quantity $M^2$ is the invariant
mass of the free, non-interacting $q\bar q$ state (see \ref{LCWF6}).
However, it is precisely $M$, not $m_V$ that will govern
the hard interaction of $q\bar q$ pair with gluons.

Finally, it is useful to introduce the relative momentum of
free $q\bar q$ system:
\begin{equation}
2p_\mu = (k^*_1 - k^*_2)_\mu\;.\quad
\end{equation}
Then, trivial algebra leads to
\begin{equation}
 M^2 = 4m^2 + 4{\bf p}^2\;;\quad p^2 = - {\bf p}^2\,;
\quad (pq) =0\,,\label{LC10}
\end{equation}
where ${\bf p}$ is the 3--dimensional relative momentum in the
$q\bar q$ pair rest frame of reference.
Its components are
\begin{equation}
{\bf p} = (\vec p, p_z)\,;\quad \vec p = \vec k\,; \quad
p_z = {1 \over 2} (2z-1)M\,.\label{LC11}
\end{equation}

\section{Spin structure of a vector particle}\label{sectspin}

Let us start with a well known example of a deuteron, which
is a non-relativistic analogy of a vector meson: they are both
vector particles built up of two fermions.
To have the correct $P$-parity, proton and neutron must sit
in the spin--triplet state, thus leaving us with two possible
values of their angular momenta: $L =0$ and 2.

In the conventional non-relativistic language one describes
the spin-angular coupling by the Clebsh-Gordan technique.
The non-relativistic Feynman diagram calculations
can be best performed in an alternative approach.
Here a deuteron, being a vector particle, is described by
a 3 dimensional polarization vector ${\bf V}$.
So, while calculating high energy processes
involving $d \to pn$ transitions,
one can use the following spin structure of
deuteron-nucleon-nucleon vertex:
\begin{equation}
\phi_n^+\,{\bf \Gamma}\, \phi_p \cdot {\bf V} \label{deuteron1}
\end{equation}
Since both nucleons can be treated on mass shell,
only two terms enter $\Gamma_i$, which can be written as:
\begin{equation}
\phi^+_n\,\left[u(p)\sigma^i +
w(p)(3p^ip^j - \delta^{ij} p^2) \sigma^j\right]
\, \phi_p \cdot V^i \label{deuteron2}
\end{equation}
Here $\sigma^i$ are Pauli matrices and ${\bf p}$ is
the relative proton--neutron momentum.
One immediately recognizes here spin structures
corresponding to $pn$ pair sitting in
$S$ and $D$ waves respectively.
In particular, squaring the above expression
gives
\begin{eqnarray}
&&({\bf V}{\bf V}^*) \quad \mbox{for $|S|^2$} \nonumber\\
&&3({\bf p}{\bf V})({\bf p}{\bf V}^*) -({\bf V}{\bf V}^*){\bf p}^2
\quad \mbox{for $SD$ interference}\nonumber\\
&&3{\bf p^2}({\bf p}{\bf V})({\bf p}{\bf V}^*) +({\bf V}{\bf V}^*){\bf p}^4
\quad \mbox{for $|D|^2$} \label{deuteron3}
\end{eqnarray}

Now, let us go relativistic and turn to vector mesons.
The polarization state of a vector particle is described by
a four-vector $V_\mu$. Therefore, a general form of $q\bar q V$ vertex
has the form
$$
\bar u' \Gamma_\mu u \cdot V_\mu \cdot \Gamma(p)\,,
$$
where $\Gamma(p)$ is the familiar vertex factor.
Up to now, it has been customary in literature
to choose the simplest form
of the spinorial structure $\Gamma_\mu$:
\begin{equation}
\bar u' \gamma_\mu u \cdot V_\mu \cdot \Gamma(p)\,.\label{naive}
\end{equation}
However, one must admit that (\ref{naive}) is simply an analogy of
$q \bar q \gamma$ vertex and does not reflect the true internal structure
of a vector meson.
It is known \cite{SD} that the correct spinorial structure
corresponding to pure the $S$ wave $q\bar q$ state reads
\begin{equation}
  S_\mu = \gamma_\mu - { 2 p_\mu \over M + 2m} =
\left( g_{\mu\nu} - {2p_\mu p_\nu \over m (M+2m)}\right) \gamma_\nu
\equiv {\cal S}_{\mu\nu}\gamma_\nu\,.\label{Swave}
\end{equation}
It is implied here that spinorial structures are inserted between
{\it on mass shell spinors} in accordance with
our principal guideline (see discussion in Sect.\ref{sectionLCWF}).

Once $S$ wave spinorial structure
is established, the expression for $D$ wave can be obtained by
contracting $S$ wave with the symmetric traceless tensor of rank two
$3p_ip_j - \delta_{ij}{\bf p}^2$, rewritten in the Lorenz notation.
To do so, one should replace
$$
p_i \to p_\mu\,;\quad
\delta_{ij} \to - g_{\mu\nu} + {q_\mu q_\nu \over M^2}
$$
(in the $q\bar q$ pair rest frame of reference $q_\mu = (M,\,0,\,0,\,0)$).
However, since $q_\mu$ inserted between on mass shell spinors gives zero
due to the Ward identity, one obtains the required tensor
in the form $3p_\mu p_\nu + g_{\mu\nu}{\bf p}^2$.
Its contraction with ${\cal S}_\mu$ yields
\begin{equation}
  D_\mu = (3p_\mu p_\nu + g_{\mu\nu}{\bf p}^2) \cdot {\cal S}_{\nu\rho}\gamma_\rho
= {\bf p}^2\gamma_\mu + (M+m)p_\mu =
\left( {\bf p}^2 g_{\mu\nu} + {M+m \over m}p_\mu p_\nu\right) \gamma_\nu
\equiv {\cal D}_{\mu\nu}\gamma_\nu\,.\label{Dwave}
\end{equation}
We will prove below that structures (\ref{Swave}), (\ref{Dwave})
after being squared indeed perfectly reproduce
(\ref{deuteron3}), i.e. they indeed correspond to pure $S$ and $D$ waves.\\

The quantities ${\cal S}_{\mu\nu}$ and ${\cal D}_{\mu\nu}$
used in (\ref{Swave}), (\ref{Dwave})  have the meaning of
$S/D$ wave projectors, which will be used in all subsequent calculations.
 Namely, all calculations will be at first
performed for the naive $q\bar q V$ vertex (\ref{naive}) and
then we will apply the projector technique to obtain
expressions for $S$ and $D$ wave states.

\section{Vector meson LCWF normalization}\label{sectnorm}

Before tackling the diffractive vector meson production process,
we first should have a prescription of normalization of the
vector meson wave function.

\subsection{Naive $q\bar q V$ vertex}

\begin{figure}[!htb]
   \centering
   \epsfig{file=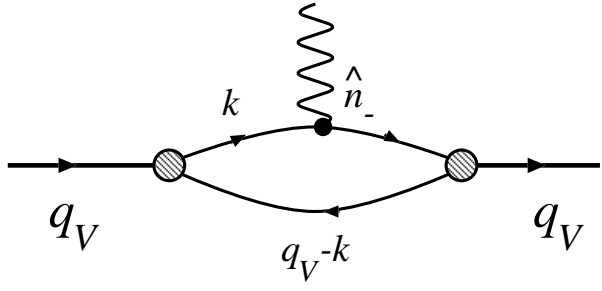,width=80mm}
   \caption{Diagram used for normalizing the vector meson LCWF.
The amplitude of this diagrams is set equal to $2q_+i$.}
   \label{normalization}
\end{figure}

A natural way to normalize the wave function of a composite system is
to put the amplitude given by this diagram in Fig.\ref{normalization}
equal to $2 q_+ i$. Here extra leg carries zero momentum but couples
to the fermion line as $\gamma_\mu n_-^\mu$.
Note that for a charged composite particle
(a deuteron) this is precisely setting the electric formfactor
equal to unity in the soft photon limit. \\

As described above, we first treat $q \bar q V$ vertex as
 $\bar{u}' \gamma_\mu u \cdot \Gamma(p)$. In this case the general expression
for this amplitude is
\begin{eqnarray}
A = { (-1) \over (2\pi)^4} N_c \cdot
\int d^4 k { Sp\{ i \hat V_1 \Gamma \cdot i(\hat k - \hat q_V + m) \cdot
i \hat V^*_2 \Gamma^* \cdot i(\hat k + m) \cdot i \hat n_- \cdot
i(\hat k + m)\} \over [k^2 -m^2 +i \epsilon] \cdot [k^2 -m^2 +i \epsilon]
\cdot [(k-q_V)^2 -m^2 +i \epsilon]} \nonumber\\
= {N_c \over (2\pi)^4}\cdot \int d^4 k { |\Gamma|^2
Sp\{...\} \over [k^2 -m^2 +i \epsilon]^2 [(k-q_V)^2 -m^2 +i \epsilon]}\,,
\end{eqnarray}
where $N_c=3$ is a trivial color factor originating from the quark loop.
We deliberately recognized $V_1$ and $V_2$ as distinct entities
just to make sure later that such a loop is indeed diagonal
in polarization states.

The first step is to rewrite this expression in terms of Sudakov's variables.
As usual, one implements decomposition
$$
k^\mu = zq_{V+}n_+^\mu + y q_{V-}n_-^\mu + \vec k^\mu\,;\quad
q_V^2 = 2q_{V+}q_{V-} = m_V^2
$$
and transforms
$$
d^4k = {1 \over 2}m_V^2 d^2\vec k dy dz\,.
$$
Now we note that vertex functions $\Gamma$ do not depend on $y$
(and neither does the trace, as will be shown later),
so we can immediately perform the integrations over $y$
by means of Cauchy theorem. Indeed, since the integral
\begin{equation}
\int_{-\infty}^{\infty} dy
{1 \over [yzm_V^2 - (\vec k^2 +m^2) + i \epsilon]^2}
{1 \over [(1-y)(1-z)m_V^2 - (\vec k^2 +m^2) + i \epsilon]}\,,
\label{int}
\end{equation}
is convergent and has good infinity behavior,
one can close the integration contour in the most convenient way.
To do so, one should analyze the position of all poles on the complex
$y$ plane:
$$
y_{1,2} = {\vec k^2 + m^2 \over zm_V^2} - {i\epsilon \over zm_V^2};\quad
y_3 = 1 - {\vec k^2 + m^2 \over (1-z)m_V^2} + {i\epsilon \over (1-z)m_V^2}\,.
$$
One sees that if $z<0$ or $z>1$, {\it all} poles lie on the same
side of the real axis in the complex $y$ plane, which leads to
zero contribution. The contribution that survives comes from
region $0<z<1$, which has in fact a simple physical meaning:
all constituents must move in the same direction.
In this region, we close the integration contour through the
upper half-plane and take a residue at $y = y_3$.
Physically, it corresponds to putting the antiquark on mass shell.
After this procedure, one gets for (\ref{int}):
\begin{eqnarray}
&&- {2\pi i \over (1-z)m_V^2}
{1 \over [yzm_V^2 - (\vec k^2 +m^2) + i \epsilon]^2}\bigg|_{y=y_3}
= -{2\pi i \over (1-z)m_V^2}
{(1-z)^2 \over [\vec k^2 + m^2 - z(1-z)m_V^2]^2}\nonumber\\
&&= - {2\pi i \over m_V^2}{1 \over z^2(1-z)}
{1 \over [M^2 - m_V^2]^2}\,.\nonumber
\end{eqnarray}
One immediately recognizes here the same two-particle propagator
as in (\ref{LCWF6}). Therefore, the equation for the amplitude reads
\begin{equation}
A = i {N_c \over (2\pi)^3} \cdot \int d^2 \vec k {dz \over z^2 (1-z)} \cdot |\psi|^2
\cdot \left(-{1 \over 2} Sp\{...\}\right)\,.
\label{a1}
\end{equation}

We now turn to the trace calculations. As we found during the 
calculation of the forward compton scattering amplitude,
all the fermions can be treated in the trace calculations
as if they were on mass shell. We will use this property in all subsequent
calculations.

The easiest way to calculate the traces in our case
is to do it covariantly,
without involving further the Sudakov technique.
Since quarks in numerator can be treated on mass shell, we first note that
$$
(\hat k + m) \hat n_- (\hat k + m) = 2 (k^* n_-) (\hat k^* + m)
$$
so that
$$
-{1 \over 2} Sp\{...\} = - z q_+ Sp\{\hat V_1 (\hat k^* - \hat q + m)
\hat V_2^* (\hat k^* +m)\} =
-2 z q_+ \left[ M^2 (V_1 V_2^*) + 4 (V_1 p)(V^*_2 p) \right]\,,
$$
where $p$ is the relative quark-antiquark momentum [see (\ref{LC10})].
Note that in the antiquark propagator we replaced $\hat k - \hat q_V
\to \hat k^* - \hat q$, since the antiquark is now put on mass shell.
Besides, we explicitly used here gauge condition $(qV)=0$, which
means that polarization vectors must be written
for {\it on mass shell $q\bar q$ pair}, not the vector meson,
--- another important consequence of our approach.
Substituting this into (\ref{a1}), one gets
\begin{equation}
1 = {N_c \over (2\pi)^3} \int d^2 \vec k {dz \over z(1-z)} |\psi|^2
\left[-M^2 (V_1 V_2^*) - 4 (V_1 p)(V^*_2 p)\right]\,.
\label{a2}
\end{equation}
A prominent feature of this equation is the orthogonality of
$V_L$ and $V_T$ polarization states --- the necessary condition for any
normalization prescription.

The next step is to realize that the integral can be cast in the form of
$d^3 {\bf p}$ integration by means of
$$
{dz \over z(1-z)} d^2\vec k= {4 \over M} dp_z d^2\vec p
= {4 \over M} d^3 {\bf p}\,.
$$
Tus, the final expression for normalization condition is
\begin{equation}
1 = {N_c \over (2\pi)^3} \int d^3 {\bf p}
 {4 \over M}|\psi|^2 [-M^2 (V_1 V_2^*) - 4 (V_1 p)(V^*_2 p)]\,.
\label{a3}
\end{equation}
We see that the expression being integrated is explicitly spherically
non-symmetric, which is a manifestation of a certain $D$ wave admixture.
Thus we now apply projector technique to obtain results
for $S$ and $D$ wave states.

\subsection{Normalization for $S$ wave vector meson}

The correct expressions for pure $S$/$D$ type vertices
can be readily obtained with the aid of projector technique.
Namely, to obtain an expression for $S$ wave, replace
\begin{equation}
V_\mu \to V_\nu {\cal S}_{\nu\mu}\,.
\end{equation}
Such a replacement for $V_1$ leads to
$$
-M^2(V_1 V_2^*) - 4 (V_1 p)(V^*_2 p) \Rightarrow
 - M^2(V_1 V_2^*) - {4 M \over M+2m} (V_1 p)(V^*_2 p)
$$
Then, one applies the same replacement to $V^*_2$ to obtain
$$
 - M^2(V_1 V_2^*) - {4 M \over M+2m} (V_1 p)(V^*_2 p)
\Rightarrow - M^2(V_1 V_2^*) \,.
$$
Therefore, the answer for $S$ wave states reads
\begin{equation}
\framebox(200,40){$\dst 1 = \fr{N_c}{(2\pi)^3} \int d^3 {\bf p}\
4M |\psi^S({\bf p}^2)|^2\,$}
\label{a6}
\end{equation}
which is manifestly spherically symmetric.

\subsection{Normalization for $D$ wave vector meson}

Results for $D$ wave states are derived in the same way. The
replacements $V_\mu \to V_\nu {\cal D}_{\nu\mu}$ lead to
\begin{equation}
-M^2{\bf p}^4(V_1 V_2^*) + 3M^2{\bf p}^2 (V_1 p)(V^*_2 p)\,.
\label{a8}
\end{equation}
After angular averaging
$$
\langle p_i p_j\rangle \to {1\over 3} {\bf p}^2 \delta_{ij},
$$
one gets the normalization formula for $D$ wave state:
\begin{equation}
\framebox(200,40){$ \dst 1 = \fr{N_c}{(2\pi)^3} \int d^3 {\bf p}\ 8M {\bf p}^4
|\psi^D({\bf p}^2)|^2\,$}\label{a9}
\end{equation}

Several remarks are in order. First, $S$ wave $\to$ $D$ wave transitions
are forbidden. Indeed, such an amplitude
will be proportional to
\begin{equation}
-M^2[{\bf p}^2(V_1 V_2^*) + 3(V_1 p)(V^*_2 p)]\,.\label{a10}
\end{equation}
which vanishes after angular integration.
Then, we emphasize that the structure of results
(\ref{a6}), (\ref{a8}), (\ref{a10}) is absolutely identical to Eq.(\ref{deuteron3}).
This fact can be viewed as the {\it proof} that spinorial structures
(\ref{Swave}), (\ref{Dwave}) indeed correspond to pure $S$ and $D$ wave states.

\section{Decay constant}

\begin{figure}[!htb]
   \centering
   \epsfig{file=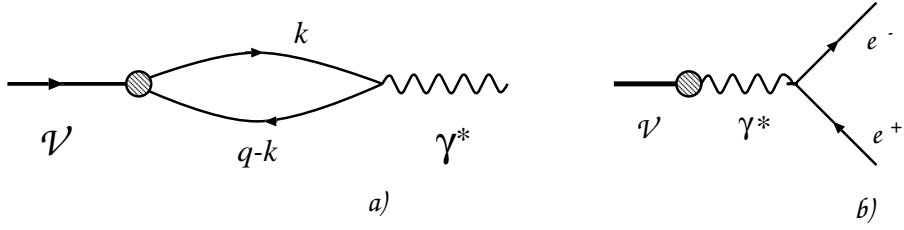,width=120mm}
   \caption{Normalizing LCWF to $\Gamma(V\to e^+e^-)$ decay width:
 (a) the diagram for $V \to \gamma^*$ transition, (b) the diagram for
$V\to e^+e^-$ decay.}
   \label{decay}
\end{figure}

An additional normalization condition consists in relating
the vector meson wave function to the experimentally measurable
physical quantity --- $V \to e^+e^-$ decay width (Fig.\ref{decay}).
The loop at Fig.\ref{decay}a describes transition $V \to \gamma^*$ and enters
the amplitude of the decay $V \to e^+e^-$ (Fig.\ref{decay}b).
Let us define the decay constant via relation
\begin{equation}
{\cal A} = i\langle 0|J_\mu^{em}|V\rangle = - i f_V c_V \sqrt{4\pi\alpha} V_\mu \,.
\label{b1}
\end{equation}
So defined $f_V$ has dimension dim$[f_V] = m^2$.
The quantity $c_V$ reflects the flavor content of a vector meson
(in the previous calculations it simply gave unity) and is equal to
\begin{equation}
c_V = {1 \over \sqrt{2}}, {1 \over 3\sqrt{2}}, - {1 \over 3}, {2 \over 3}\label{b2}
\end{equation}
for $\rho, \omega, \phi, J/\psi$ mesons correspondingly.

Knowing that such a loop does not mix polarization states, we can multiply both sides of
eq.(\ref{b1}) by $V^*$ and get the expression
\begin{equation}
i f_V = {(-1) \over (2\pi)^4} N_c \cdot \int d^4 k
{ Sp\{i \hat V^* \cdot i(\hat k + m) \cdot i \hat V \Gamma
\cdot i(\hat k - \hat q_V + m)\}
\over (k^2-m^2+i\epsilon) \cdot ((k-q_V)^2 -m^2 +i\epsilon)}\,. \label{b3}
\end{equation}

Calculations similar to the above normalization condition
derivation yield (for the naive type of vertex)
\begin{equation}
f_V = {N_c \over (2\pi)^3} \cdot \int {dz \over z(1-z)} d^2 \vec k
\ \psi_V [-M^2(V V^*) - 4 (V p)(V^* p)]\,.
\label{b4}
\end{equation}

Applying now the projector technique, one gets in the case of $S$ states
(after proper angular averaging)
\begin{equation}
\framebox(200,40){$ \dst f^{(S)} = \fr{N_c}{(2\pi)^3} \cdot
\int d^3 {\bf p}\ \psi_S \fr{8}{3} (M+m)\,$}
\label{b5}
\end{equation}
and in the case of $D$ wave states
\begin{equation}
\framebox(200,40){$ \dst f^{(D)} = \fr{N_c}{(2\pi)^3} \cdot
\int d^3 {\bf p}\ \psi_D \fr{32}{3} \fr{{\bf p}^4}{M+2m}\,.$}
\label{b6}
\end{equation}

Finally, one can write down the expression for the decay width in terms of
$f_V$:
\begin{equation}
\Gamma(V\to e^+e^-) = {1 \over 32 \pi^2 m_V^2} \cdot {m_V \over 2} 4\pi |A|^2 =
{4 \pi \alpha^2 \over 3 m_V^3} \cdot f_V^2 c_V^2\,.\label{b8}
\end{equation}
This formula can be used to extract the numerical value of $f_V$ from experimental data.

\section{Ansatz for LCWF}

Later on, we will be presenting numerical analyses
of vector meson production cross sections, for which
we will need some wave function Ansatz. Here we describe
two forms of the wave function that will be exploited
there. By no means should they be expected to accurately
represent the true radial wave functions in a vector meson.
Our Ans\"atze are pure guesses, based on non-relativistic
quantum mechanical experience, of how the wave function 
{\em might} look like. Undoubtedly, such an approach
involved a certain degree of ambiguity to the numerical results,
and in our subsequent analysis we will study this
ambiguity in detail.

\subsection{Suppressed Coulomb wave functions}

A first guess for the vector meson wave function,
especially in the case of heavy non-relativistic meson, would be
a Coulomb-like form, similar to wave function of a positronium:
\be
\psi({\bf p}) \propto {1 \over (1+{\bf p}^2a^2)^2}\,,
\ee
where $a$ is a typical size of the meson.

However, such a {\em hard wave function} will not fit our
course of calculations, since the expression for
the decay constant will be ultra-violet divergent.
Furthermore, as we will see later, this hard wave function
will lead to the vector meson production amplitudes saturated
not at the scale ${\bf p}^2 \lsim 1/a^2$, but will extend to
$1/a^2 \ll {\bf p}^2 \lsim Q^2$.

Thus, it appears that the hard wave function Ansatz
leads to complications, which do not seem to be resolvable within
the lowest Fock state only. Therefore, starting from now,
we will limit ourselves to the {\em soft wave function} Ansatz only,
"soft wave function" meaning that all integrals of 
physically relevant amplitudes involving the wave function
will be saturated by ${\bf p}^2 \lsim 1/a^2$.

If we still prefer to have a Coulomb-like wave function,
we can consider its slightly regularized form, which
we will call the "suppressed Coulomb" wave function.
Besides, in order to be able to conduct simple estimates,
we will take as simple form as possible.
So, in this ansatz the normalized wave functions read
\begin{eqnarray}
\psi_{1S}({\bf p}^2) &=& {c_1 \over \sqrt{M}}
{1 \over  (1 + a_1^2{\bf p}^2)^2}; \nonumber\\[2mm]
\psi_{2S}({\bf p}^2) &=& {c_2 \over \sqrt{M}}
{(\xi_{node} - a_2^2{\bf p}^2) \over (1 + a_2^2{\bf p}^2)^3}; \nonumber\\[2mm]
\psi_{D}({\bf p}^2) &=& {c_D \over \sqrt{M}}
{1 \over (1+ a_D^2{\bf p}^2)^4} \,.\label{coulomb1}
\end{eqnarray}
with normalization constants to be determined from Eqs.(\ref{a6})
and (\ref{a9}). Here parameters $a_i$ are connected to the
size of a bound system: in the coordinate representation $\psi_i
\propto \exp(- r/a_i)$. For strict Coulomb functions one would have 
$a_D  = 3a_2/2 = 3a_1 = R_{Bohr}$, where $R_{Bohr}$
is the Bohr radius. However, this relation should be treated with
care in the case of $q\bar{q}$ quarkonia, where
the quark-antiquark potential is quite complicated and therefore
$a_i$ should rather be considered as free parameters.
Value of  $\xi_{node}$ pinpoints the position 
of the node in the $2S$ radial wave function. For the pure Coulombic system
$\xi_{node} = 1$, but in our case the exact value
of $\xi_{node}$ should be obtained from the requirement
of the orthogognality between $1S$ and $2S$ states with $a_1$ and 
$a_2$ fixed from other requirements.

\subsection{Oscillator type LCWF}

By oscillator-type wave function we mean
$$
\psi({\bf p}) \propto exp\left(- {{\bf p}^2a^2 \over 2}\right)\,,
$$
with $a$ again being a typical size of the wave function.
This wave function Ansatz corresponds to the case of 
a strong confinement. Although the approximately quadratic potential
that leads to such an abrupt descrease at ${\bf p}^2 > 1/a^2$
is not exactly what is suspected about the color-singlet static
quark-antiquark potential (in the quenched approximation),
these wave function  still possess the main confinement-like
 properties.

In this ansatz one has
\begin{eqnarray}
  \psi_{1S} & = & c_1\exp\left(-{{\bf p}^2 a_1^2 \over 2}\right)\,;
\nonumber\\
  \psi_{2S} & = & c_2 \left(\xi_{node} - {\bf p}^2 a_2^2 \right)
\exp\left(-{{\bf p}^2 a_2^2 \over 2}\right); \nonumber\\
  \psi_{D} & = & c_D  \exp\left(-{{\bf p}^2 a_D^2 \over 2}\right)
\,.\label{oscillator1}
\end{eqnarray}
Note again that for purely oscillator potential one also has relation
$a_D = a_2 = a_1$, which might not hold in our case, since
the oscillator type potential is also a crude approximation
of the true quark-antiquark interaction. The position of node
$\xi_{node}$ would be equal to 3/2 for pure oscillator model, but again in our
case its value can turn out different.

\chapter{Vector meson production amplitudes}

\section{Preliminary notes}

\begin{figure}[!htb]
   \centering
   \epsfig{file=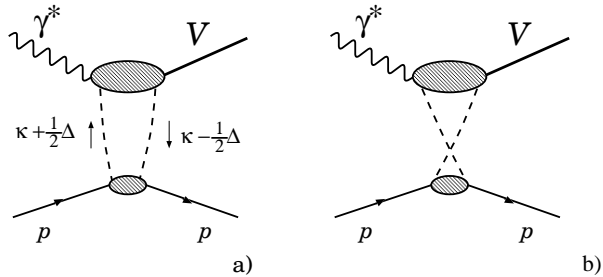,width=80mm}
   \caption{The QCD--inspired diagrams for $\gamma^* p \to Vp$ process
with two gluon $t$--channel. Only Diagr.(a) does contribute to the
imaginary part of amplitudes.}
   \label{main1}
\end{figure}

Having set up the notation and defined and described a vector meson
by itself, we are now ready to evaluate the full set of amplitudes
of its off-forward virtual diffractive photoproduction.

\begin{figure}[!htb]
   \centering
   \epsfig{file=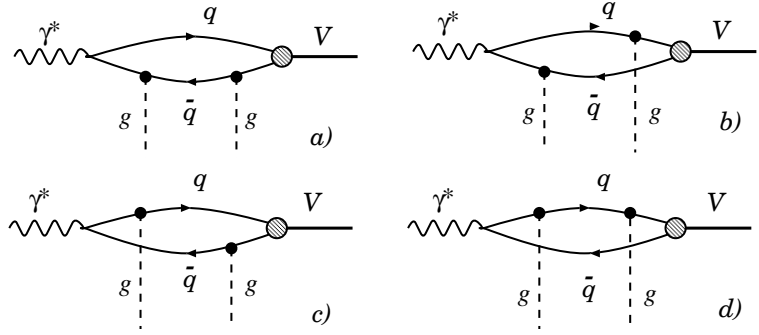,width=100mm}
   \caption{The content of the upper blob in Fig.\ref{main1}a in the
pQCD approach. The true vector meson internal structure is approximated
by $q\bar q$ Fock state.}
   \label{main2}
\end{figure}

In the pQCD motivated approach to this process the pomeron
exchange is viewed as a two-gluon exchange as it is shown in
Fig.\ref{main1}a. Using the scalarization procedure, we will split
the diagram into 2 pieces and will treat each of them separately.
The upper blob describes the pomeron-assisted transition of the
virtual photon into a vector meson. In the perturbative QCD
approach, which is legitimate here due to the presence of the
relevant hard scale $\overline Q^2 = m_q^2 + z(1-z)Q^2$, the
$q\bar q$ fluctuation of the virtual photon interacts with two
hard gluons and then fuses to produce a vector meson. This
interaction is described by four diagrams given in
Fig.\ref{main2}, with all possible two-gluons attachments to $q
\bar q$ pair taken into account. All of them are equally important
and needed for maintaining gauge--invariance and color
transparency. The latter property means that in the case of very
soft gluons the upper blob must yield zero, for the $q \bar q$
pair is colorless.

The lower part of the general diagram Fig.\ref{main1}a
is of course not computable in pQCD. The physically meaningful procedure is
to relate it to the experimentally measurable gluon density.
To do so, we will first calculate this lower blob in the Born approximation
and then give a prescription how to introduce the unintergrated gluon density.
In the course of this procedure, we will neglect in the intermediate
calculations proton off-forwardness  and take it into account only
at the very end, as a certain factor to the unintegrated gluon density.

\begin{figure}[!htb]
   \centering
   \epsfig{file=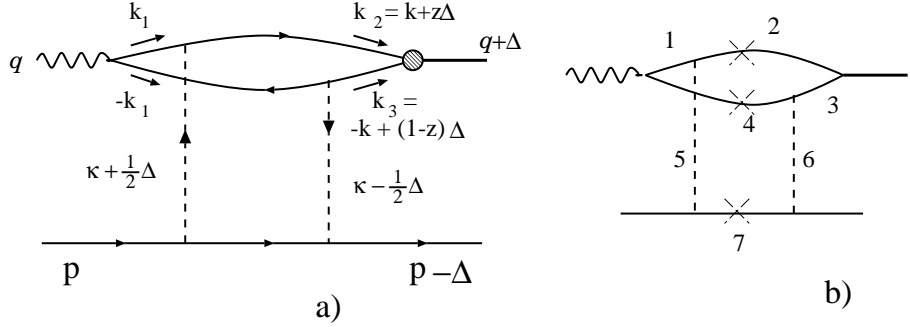,width=120mm}
   \caption{(a): A particularly useful convention of the loop momenta
(only transverse components of quark momenta are shown).
(b): the propagator notation used while calculating denominators.
The crosses denote on mass shell particles.}
   \label{main3}
\end{figure}

\section{Notation and helicity amplitudes}

In our calculation we will use the following Sudakov's
decomposition (see also Fig.\ref{main3}a)
\begin{eqnarray}
k_\mu= y{p_\mu}' + z{q_\mu}' + \vec k_\mu\,;\nonumber\\
\kappa_\mu = \alpha {p_\mu}' + \beta {q_\mu}' + \vec \kappa_\mu\,;\nonumber\\
\Delta_\mu = \delta {p_\mu}' + \sigma {q_\mu}' + \vec \Delta_\mu\label{c1}
\end{eqnarray}
Here $k$ and $\kappa$ are momenta that circulate in the quark
and gluon loops respectively; $\Delta$ is the momentum transfer.
Vectors ${p_\mu}'$ and ${q_\mu}'$ denote the light-cone momenta: they are
 related to the proton and virtual photon momenta as
\begin{equation}
p_\mu = {p_\mu}' + {m_p^2 \over s}{q_\mu}';
\quad {q_\mu} = {q_\mu}'-x{p_\mu}';\quad q'^2 = p'^2 = 0;
\quad x = {Q^2 \over s}\ll 1; \quad s = 2(p'q')\,.
\label{c2}
\end{equation}
As it was mentioned in the Introduction, the condition $x\ll 1$ is
necessary to speak about diffractive processes. The longitudinal
momentum transfer can be readily found from kinematics
(see Fig.\ref{main1}). To the higher power $s$ terms it reads
\begin{eqnarray}
m_p^2 =p^2=(p-\Delta)^2=m_p^2-2(p\Delta) + \Delta^2;
&\Rightarrow &\sigma = - {\vec \Delta^2 \over s}\;;\nonumber\\
m_V^2 = (q +\Delta)^2 = -Q^2 + 2(q\Delta) +\Delta^2
&\Rightarrow&  \delta = x + {m_V^2 + \vec \Delta^2 \over s}\;.
\label{c3}
\end{eqnarray}
The final vector meson momentum reads:
\begin{equation}
q_{V\mu} = {q_\mu}' + {m_V^2 + \vec \Delta^2 \over s}{p_\mu}' + \vec \Delta_\mu\,.
  \label{c2a}
\end{equation}
Finally, throughout the text transverse momenta will be marked by the vector sign
as $\vec k$ and 3D vectors will be written in bold.

There are several possible helicity amplitudes in
the transition $\gamma^*_{\lambda_\gamma} \to V_{\lambda_V}$.
First of all, both photon and vector meson can be transversely polarized.
The polarization vectors are
\begin{equation}
e_{T\mu} = \vec e_\mu\,; \quad
V_{T\mu} = \vec V_\mu + {2 (\vec \Delta \vec V) \over s}(p' -q')_\mu\,.
\label{polart}
\end{equation}
Note that we took into account the fact that
the vector meson momentum has finite transverse component
$\vec \Delta$.
Then, the virtual photon can have the scalar polarization
(which is often called longitudinal;
we will use both terms)
with polarization vector
\begin{equation}
e_{0\mu} = {1 \over Q}(q' + xp')_\mu\,.\label{e0}
\end{equation}
Finally, the longitudinal polarization state of a vector meson
is described by
\begin{equation}
V_{L\mu} = {1 \over M}\left( {q_\mu}' + {\vec \Delta^2 -M^2 \over s}{p_\mu}'
+ \vec \Delta_\mu \right)\,.\label{VL}
\end{equation}
Note that, as we already mentioned, in the self-consistent approach
we must take the {\it running polarization vector} for
the longitudinal polarization state. It depends on $M$, not $m_V$,
which reflects the fact that in our approach we first calculate the
production of {\it an on-shell $q\bar q$ pair with}
(whose dynamics is governed by $M$)
and then projects it onto the physical vector meson. We
 stress that this projection will automatically arise
in the course of usual Feynman diagram evaluation.

Thus, there are 5 different amplitudes:
\begin{eqnarray}
&&L \to L\nonumber\\
&&T \to T\  (\lambda_\gamma = \lambda_V)\nonumber\\
&&T\to L\nonumber\\
&&L\to T\nonumber\\
&&T \to T\  (\lambda_\gamma = -\lambda_V)\label{five}
\end{eqnarray}
The first two are helicity conserving amplitudes.
They are dominant and almost insensitive
to the momentum transfer $\vec \Delta$.
The next two are single helicity flipping amplitudes.
They are unavoidably proportional to $|\vec \Delta|$
in the combination $(\vec e\vec \Delta)$ or $(\vec V^*\vec \Delta)$
and would be vanishing for the strictly forward scattering.
Finally, the last amplitude corresponds to the double helicity flip
and will be proportional to $(\vec e\vec \Delta)(\vec V^*\vec \Delta)$.\\

\section{General amplitude}
We will take diagr.(c) at Fig.\ref{main2}
(it is shown in Fig.\ref{main3}) as a generic diagram
and perform a thorough analysis for it. It turns out that
the other diagrams are calculated in the same fashion.

The general expression for the amplitude given by diagr.(c) reads:
\begin{eqnarray}
&&iA = \int{d^4k\over (2\pi)^4}\int{d^4\kappa\over (2\pi)^4}
\ \bar{u}_p' (-ig\gamma^{\nu'} t^{B'})
i{\hat p -\hat\kappa_1 +m_p \over \left[(p-\kappa)^2 - m_p^2 +
i\epsilon\right]}(-ig\gamma^{\mu'} t^{A'})u_p
\nonumber\\&&
\cdot(-i){g_{\mu\mu'}\delta_{AA'}\over \kappa_1^2 - \mu^2 +i\epsilon}
\cdot(-i){g_{\nu\nu'}\delta_{BB'}\over \kappa_2^2 - \mu^2 +i\epsilon}
\cdot c_V \cdot \Gamma^*
\nonumber\\
&&\cdot{
Sp\left\{ i e\hat e\  i(\hat k_4 + m)\ (-ig\gamma^\nu t^B)\ i(\hat k_3+m)
i \hat V^*\ i(\hat k_2+m)\ (-ig\gamma^\mu t^A)\ i(\hat k_1+m)
\right\}
\over \left[k_1^2 - m^2 + i\epsilon\right]
\left[k_2^2 - m^2 + i\epsilon\right]
\left[k_3^2 - m^2 + i\epsilon\right]
\left[k_4^2 - m^2 + i\epsilon\right]
}\nonumber\\
\label{gv1}
\end{eqnarray}
Here $c_V$ is the same as in (\ref{b2}) and $\Gamma$ is the
familiar $q \bar q \to V$ vertex function. Note that we introduced
'gluon mass' $\mu$ in gluon propagators to account for confinement
at a phenomenological level.

Let's first calculate the numerator.

\section{Color factor}
\begin{figure}[!htb]
   \centering
   \epsfig{file=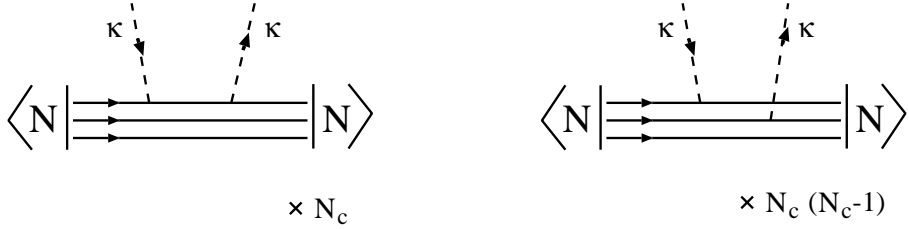,width=120mm}
   \caption{The ways two glons can couple a colorless nucleon.}
   \label{color}
\end{figure}

If we consider {\it strictly forward} gluon scattering off a single
quark, we have
\begin{equation}
{1 \over N_c}Sp\{t^{B'}t^{A'}\} \cdot\delta_{AA'}\delta_{BB'}Sp\{t^Bt^A\}=
{1 \over N_c}{1 \over 2}\delta_{AB}{1 \over 2}\delta_{AB}=
{1 \over 2}{N_c^2-1 \over 2N_c} = {1 \over 2}C_F = {2 \over 3}\label{color1}
\end{equation}
However, we should take into account that quarks are sitting inside
a colorless proton, whose color structure is
\begin{equation}
\psi_{color} = {1 \over \sqrt{6}} \epsilon^{abc} q^a q^b q^c\label{color2}
\end{equation}
In this case there are two ways a pair of gluons can couple 3 quark lines
(see Fig.\ref{color}). In the first way both gluons couple to the same quark.
Since the quark momentum does not change after these two interactions,
the nucleon stays in the same state: $\langle N|N\rangle = 1$.
In the second case gluon legs are
attached to different quark lines, so that extra momentum $\kappa$
circulates between quarks, which gives rise to the factor
 $\langle N|\exp(i\kappa r_1 - i\kappa r_2)|N\rangle$, i.e.
to the two-body formfactor.
Therefore, for the lower line instead of
\begin{equation}
{1 \over N_c}Sp\{t^{B}t^{A}\} = {1 \over N_c} {1 \over 2} \delta_{AB}\label{color3}
\end{equation}
one has
\begin{eqnarray}
&&{1 \over 6} \epsilon^{abc}
\left(3\delta_{aa'}\delta_{bb'} t^A_{cc''} t^B_{c''c'}
+ 6 \delta_{aa'}t^A_{bb'} t^B_{cc'}\langle N|\exp(i\kappa r_1 - i\kappa r_2)|N\rangle
\right)
\epsilon^{a'b'c'}\nonumber\\
&=& Sp\{t^{A}t^{B}\} - Sp\{t^{A}t^{B}\}\langle N|\exp(i\kappa r_1 - i\kappa r_2)|N\rangle
\nonumber\\
& = &{1 \over 2} \delta_{AB}(1 - \langle N|\exp(i\kappa r_1 - i\kappa r_2)|N\rangle
)\,.\label{color4}
\end{eqnarray}
Note also that a similar calculation for $N_c$ number of colors would yield
the same result. Thus, the overall color factor is
\begin{equation}
{1 \over 2} C_F N_c V(\kappa) = 2V(\kappa) \equiv
{1 \over 2} C_F N_c (1 - \langle N|\exp(i\kappa r_1 - i\kappa r_2)|N\rangle).\label{color5}
\end{equation}

\section{Scalarization of upper and lower parts}

As known, the highest power $s$ contribution comes from so--called nonsense
components of gluon propagator (density matrix) decomposition:
\begin{equation}
g_{\mu\mu'} = {2 p'_{\mu}q'_{\mu'} \over s} + {2 p'_{\mu}q'_{\mu'} \over s} +
g_{\mu\mu'}^{\bot} \ \approx \ {2 p'_{\mu}q'_{\mu'} \over s}\,.\label{gv2}
\end{equation}
The lower (proton) line gives then
\begin{equation}
\bar u'(p) \cdot \hat q' (\hat p -  \hat \kappa_1 + m_p)
\hat q' \cdot u(p-\Delta)\label{gv2a}
\end{equation}
To the highest power $s$ order, it can be rewritten as
\begin{equation}
\bar u_p' \hat q' \hat p' \hat q' u_p \bigg|_{forward} =
{1 \over 2} Sp\{\hat p' \hat q' \hat p' \hat q'\} = s^2\,.\label{gv3}
\end{equation}
As we mentioned, the effect of off--forwardness (skewedness)
will be taken into account later.
So, combining all factors, one has for numerator of Eq.(\ref{gv1})
\begin{eqnarray}
&&(4\pi\alpha_s)^2\ \sqrt{4\pi\alpha_{em}}\ c_V\cdot
{1 \over 2} C_F N_c V(\kappa){4 \over s^2} s^2
\cdot Sp\left\{
\hat e\  (\hat k_4 + m)\ \hat q'\ (\hat k_3 + m)\ \hat V^*\
(\hat k_2+m)\ \hat q'\ (\hat k_1 + m)
\right\}\nonumber\\
&=&(4\pi\alpha_s)^2\ \sqrt{4\pi\alpha_{em}}\ c_V\cdot
2 C_F N_c V(\kappa) \cdot 2s^2\cdot
I^{(c)}(\gamma^* \to V)\,.\label{gv4}
\end{eqnarray}
Note that we factored out $2s^2$ from the trace
because it will appear later in all trace calculations.
So, the resulting expression for amplitude (\ref{gv1}) looks like
\begin{eqnarray}
A &=& \ \sqrt{4\pi\alpha_{em}}\ 4 C_F N_c s^2 \ c_V\cdot
\int{d^4k\over (2\pi)^4}\int{d^4\kappa\over (2\pi)^4}\cdot
{(4\pi\alpha_s)^2 V(\kappa) \over \left[(p-\kappa_1)^2 - m_p^2 + i\epsilon\right]
\left[\kappa_1^2 - \mu^2 +i\epsilon\right]
\left[\kappa_2^2 - \mu^2 +i\epsilon\right]}
\nonumber\\
&& \cdot\,{ \Gamma^* I^{(c)}(\gamma^* \to V) \over
\left[k_1^2 - m^2 + i\epsilon\right]
\left[k_2^2 - m^2 + i\epsilon\right]
\left[k_3^2 - m^2 + i\epsilon\right]
\left[k_4^2 - m^2 + i\epsilon\right]
}\label{gv5}
\end{eqnarray}
One can now immediately write similar expressions for the other three
diagrams (Fig.\ref{main2} a,b,d). Indeed, they will differ from
Eq.(\ref{gv5}) only by the last line: they will have different
expressions for traces and propagator structures (i.e. the exact
values for $k_i$.).

\section{Denominator evaluation}
Now we turn to the calculation of denominators.
As usual, we implement Sudakov's decomposition (\ref{c1})
and make use of relation
$$
d^4k = {1\over 2} s dy\,dz\,d^2\vec k\,.
$$
The physical picture of the way we will do the
resulting integrals is the following.
We are interested only in the imaginary part of these diagrams.
In fact, it can be shown that at the level of accuracy used here
the diagram in Fig.\ref{main1}a gives rise only to the imaginary part
of the amplitude. The real part is given by Fig.\ref{main1}b
and can be readily found from analiticity (so that here is no need
for additional calculations), however it turns out small due to
smallness of pomeron intercept, so we will neglect it in our
subsequent calculations.

The imaginary part is computed by setting three
particles in the $s$-channel cut on mass shell
(this is illustrated in Fig.\ref{main3}b). One way to do so is to apply
Cutkosky rule to modify our expression. Another, more straightforward
and 'honest' way is to calculate three of the integrals
(namely, over $\alpha, \beta, y$) via residues.
That's what we are going to do.

The details of this calculations are given in Appendix \ref{apa}.
Here we cite the result:
\begin{eqnarray}
&&Im \int dy\ dz\ d\alpha\ d\beta\ {\Gamma \over [\mbox{all propagators}]}
\nonumber\\
&&= \left(-{\pi i \over s}\right)\left(-{2\pi i \over s}\right)^2\cdot
\int {dz \over z(1-z)} \psi_V(z,\vec k^2) \cdot{1 \over
[\vec k_1^2 +m^2 +z(1-z)Q^2]} { 1 \over (\vec \kappa^2 + \mu^2)^2}\label{gv14a}
\end{eqnarray}
Here $\vec k_1$ is the transverse momentum flowing through photon vertex
along the fermion line.
Particularly, for diagr.(c) it is equal to
$\vec k_1 = \vec k - (1-z)\vec \Delta - \vec \kappa_2$
(with the specific quark loop momentum choice given at Fig.(\ref{main3}a)).

Thus, the amplitude for diagr.(c) has the form
\begin{eqnarray}
A &=& \sqrt{4\pi\alpha_{em}}\ 4C_F N_c s^2 \ c_V
\cdot {1 \over 2}s \ {1 \over 2}s \cdot
\left(-{\pi i \over s}\right)\left(-{2\pi i \over s}\right)^2
\cdot {1 \over (2\pi)^8} \nonumber\\
&&\cdot \int {dz \over z(1-z)} d^2\vec k \psi_V(z,\vec k^2)
\int {d^2 \vec \kappa V(\kappa) \over (\vec \kappa^2 + \mu^2)^2}
\ (4\pi\alpha_s)^2\
\cdot {I^{(c)}\over [\vec k^2_{1} + m^2 + z(1-z)Q^2]}
\,.\label{gv15}
\end{eqnarray}
After bringing all coefficients together, one gets
\begin{eqnarray}
A^{(a)} &=& is {C_F N_c c_V\sqrt{4\pi\alpha_{em}}  \over 64\pi^5}\cdot
\int {dz \over z(1-z)} d^2\vec k \psi_V(z,\vec k^2)
\int {d^2 \vec \kappa V(\kappa)\over (\vec \kappa^2 + \mu^2)^2}
(4\pi\alpha_s)^2
\nonumber\\
&& \times {I^{(c)}\over \vec k^2_{1} + m^2 + z(1-z)Q^2}\,.\label{gv16}
\end{eqnarray}

The other diagrams are calculated in the same way. The most
important difference is that for each diagram we will have a
propagator $1/[\vec k_1^2 + m^2 +z(1-z)Q^2]$ with its own
definition of $\vec{k}_1$, the transverse momentum in photon vertex:
\begin{eqnarray}
&\mbox{diagr.a} & \vec{k}_{1a} = \vec{k} - (1-z)\vec\Delta = \vec{r} - {1\over 2}
\vec\Delta\nonumber\\ &\mbox{diagr.b} & \vec{k}_{1b} = \vec{k} - (1-z)\vec\Delta +
\vec\kappa + {1\over 2} \vec\Delta = \vec{r} + \vec\kappa \nonumber\\
&\mbox{diagr.c} & \vec{k}_{1c} = \vec{k} - (1-z)\vec\Delta - \vec\kappa + {1\over 2}
\vec\Delta = \vec{r} - \vec\kappa \nonumber\\ &\mbox{diagr.d} & \vec{k}_{1d} = \vec{k} +
z\vec\Delta = \vec{r} + {1\over 2} \vec\Delta\label{g1}
\end{eqnarray}
Here $\vec{r} \equiv \vec{k} - (1-2z)\vec\Delta/2$.
Thus, the whole expression for the imaginary part of the amplitude
is
\begin{eqnarray}
A &=& is {C_F N_c c_V\sqrt{4\pi\alpha_{em}}  \over 64\pi^5}\cdot
\int {dz \over z(1-z)} d^2\vec k \psi_V(z,\vec k^2)
\int {d^2 \vec \kappa V(\kappa)\over (\vec \kappa^2 + \mu^2)^2}
\,(4\pi\alpha_s)^2\nonumber\\
&&\times \Biggl[
{1-z \over z} {I^{(a)} \over \vec{k}_{1a}^2 + m^2 + z(1-z)Q^2}
 + {I^{(b)} \over \vec{k}_{1b}^2 +m^2 + z(1-z)Q^2}\nonumber\\
&& + {I^{(c)} \over \vec{k}_{1c}^2 +m^2 + z(1-z)Q^2}
+{z \over 1-z}{ I^{(d)} \over \vec{k}_{1d}^2 + m^2 + z(1-z)Q^2}
\Biggr]
\,.\label{gv17}
\end{eqnarray}

\section{Off-forward gluon density}

In Sect.~\ref{sectgluon} we discussed the 
{\em forward unintegrated gluon density} ${\cal F}(x_g,\vec\kappa)$
and developed a prescription (see Eq.(\ref{replacerule}))
how to introduce it into the $k_t$-factorization calculations.
Being devised for forward scattering processes only,
this gluon density bears a clear probabilistic sense,
which is in fact reflected in the word 'density'.

In the present case of vector meson production, the initial
and final states are kinematically distinct, therefore
the forward unintegrated gluon density in its pure form 
is not the relevant quantity; instead, off-forward (or skewed)
gluon structure function \cite{offforwardgluon}
\be
{\cal F}(x_1,x_2,\vec\kappa,\vec\Delta)
\ee
should be used. It depends on the lightcone momenta $x_1$
and $x_2$ carried by the first and the second gluon,
on the transverse momentum $\vec\kappa$ inside the gluon loop,
and on the total transverse momentum transfer $\vec\Delta$.
At $x_1=x_2$ and $\vec\Delta=0$, the forward gluon structure 
function is recovered, which means that even the strictly forward
vector meson production should be described by off-forward gluon
structure function.

The experimental determination of the off-forward
gluon density might be in principle possible from
accurate measurements of the off-forward virtual Compton 
scattering process, but the lack of such measurements
makes this analysis not feasible in the nearest future.
Thus, the original idea to involve in the vector meson calculations
as little ambiguity as possible by determining the gluon content 
of the proton from other reactions does not work.
One might of course expect that replacement 
\be
{\cal F}(x_1,x_2,\vec\kappa,\vec\Delta) \to 
{\cal F}\left({x_1+x_2 \over 2},\vec\kappa\right)
\ee
should be quite legitimate in kinematical regimes when
$x_1 \approx x_2$, $\vec\Delta \ll \vec\kappa$.
However, such a regime takes place only in the photoproduction
of light mesons and is severely violated as we go to higher
$Q^2$ or higher $m_V$, therefore such a replacement
would be a poor option for the most of the case we study.

There is however way around, which allows us
to reduce the unknown off-forward gluon distributions to
the forward ones. As shown in \cite{shuvaev},
if the energy behavior of the gluon densities
is describable by a simple Regge-type behavior
\be
{\cal F} \propto \left({1 \over x}\right)^\lambda\label{simplelaw}
\ee
then in the case $x_1 \gg x_2$ the off-forward unintegrated 
gluon structure functions can be related to the forward 
unintegrated gluon density according to
\be
{\cal F}(x_1,x_2 \to 0,\vec\kappa,\vec\Delta \to 0) = R_g\cdot
{\cal F}(x_1,\vec\kappa)\,;\quad R_g = {2^{2\lambda+3} \over \sqrt{\pi}}
{\Gamma(\lambda+{5\over 2}) \over \Gamma(\lambda+4)}\,.
\label{shuvaev}
\ee
Bearing this exact result in mind, one can hope that
a similar relation will hold for gluon densities with a
somewhat more complicated energy behavior, if the
effective exponent $\lambda$ is calculated in the vicinity 
of the kinematical point $x_1,\,\vec\kappa$.

Note that for the purposes of approximate numerical calculation
correspondence (\ref{shuvaev}) can be further simplified.
Introduce an agrument shift $c(\lambda)$ such that
\be
{\cal F}(x_1,0,\vec\kappa,0) \propto {2^{2\lambda+3} \over \sqrt{\pi}}
{\Gamma(\lambda+{5\over 2}) \over \Gamma(\lambda+4)}\cdot
\left({1 \over x_1}\right)^\lambda = 
\left({1 \over c(\lambda)x_1}\right)^\lambda
\ee
holds for all $x_1$. Simple arithmetics shows that
$c(\lambda)$ changes from $\approx 0.435$ at $\lambda=0$ to 
$0.4$ at $\lambda = 1$. Given this very flat dependence,
we can approximate $c(\lambda)$ by a constant value $0.41$
so that
\be
{\cal F}(x_g,0,\vec\kappa,0) \approx {\cal F}(0.41 x_g,\vec\kappa)\,.
\label{shuvaevsimple}
\ee
This form will be used in our numerical calculations.

The effect of non-zero $\vec\Delta$ comes both from 
the Pomeron-exchange and from the proton impact factor.
Since Pomeron singularity moves in complex $j$ plane as 
$|t|=\vec\Delta^2$ changes, the value of the effective Pomeron 
intercept will be affected by $\vec\Delta^2$, a customary
representation of this effect (in the linear Rogge trajectory
approximation) being
\be
\alpha_\Pom(\vec\Delta^2) = \alpha_\Pom(0) - b_\Pom \vec\Delta^2\,.
\ee
Although in our case the effective intercept
of the gluon density is not an input number, but is generated dynamically,
we still account for the above effect by multiplying
the gluon density by factor
\be
\exp\left[-b_\Pom\vec\Delta^2\log(x_0/x)\right]\,,
\ee
with $x_0=0.03$ and trajectory slope $b_\Pom$ being different
for hard ($x_\Pom^{hard} = 0.07$) and soft 
($x_\Pom^{soft} = 0.15$) parts of the gluon density.
Since the resulting contribution to the slope increases with
energy growth, this effect is called the diffractive cone shrinkage.

The second --- and the most significant effect of non-zero
$\vec\Delta$ --- comes from the proton impact factor.
Effectively, it amounts to introduction of 
a proton formfactor $F(\Delta^2)$, which would be
equal to 1 at $\vec\Delta^2 = 0$ and
would start significantly decreasing when $\vec\Delta^2$ grows
larger than a certain scale $1/\Lambda_p^2$.
In our numerical calculations we used the dipole formfactor
\be
F(\vec\Delta^2) = \left({1 \over 1 + \vec\Delta^2/\Lambda_p^2}\right)^2
\ee
with $\Lambda_p^2 = 1$ GeV$^{-2}$.

\section{Final results for the naive vertex}

Now, with the off-forward gluon structure function
properly defined, the only thing left 
to be computed is integrands $I^i(\gamma\to
V)$. For convenience, their calculation is also given in Appendix
\ref{apb}. It turns out that the results can be written in the
same way for all four diagrams via $\vec k_1$ given by (\ref{g1}),
i.e. all quantities:
$$
-{1-z \over z}I^{(a)}\,,\quad I^{(b)}\,,\quad I^{(c)}\,,\quad -{z
\over 1-z}I^{(d)}
$$
can be written in a similar way:
\begin{eqnarray}
T \to T&& \left[(\vec{e}\vec{V}^*)(m^2 + \vec{k}\vec k_1) + (\vec{V}^*\vec{k})(\vec{e}\vec k_1)(1-2z)^2 -
(\vec{e}\vec{k})(\vec{V}^*\vec k_1)\right]\nonumber\\
L \to L&&-4z^2(1-z)^2QM\nonumber\\
T \to L&&2z(1-z)M(\vec{e}\vec k_1)(1-2z)\nonumber\\
L \to T&&-2z(1-z)Q(1-2z)(\vec{V}^*\vec{k})\nonumber
\end{eqnarray}

Therefore, we can cast amplitude (\ref{gv17}) in a compact form
with aid of functions $\vec \Phi_{1}$ and $\Phi_2$:
\begin{equation}
\Phi_2 = -{1 \over (\vec{r}+\vec\kappa)^2 + \overline Q^2} -{1 \over
(\vec{r}-\vec\kappa)^2 + \overline Q^2} + {1 \over (\vec{r} + \vec\Delta/2)^2 +
\overline Q^2} + {1 \over (\vec{r} - \vec\Delta/2)^2 + \overline
Q^2}\label{g2}
\end{equation}
and
\begin{equation}
\vec{\Phi}_1 = -{\vec{r} + \vec\kappa \over (\vec{r}+\vec\kappa)^2 + \overline Q^2}
-{\vec{r} - \vec\kappa \over (\vec{r}-\vec\kappa)^2 + \overline Q^2}
+ {\vec{r} + \vec\Delta/2 \over (\vec{r} + \vec\Delta/2)^2 + \overline Q^2}
+ {\vec{r} - \vec\Delta/2 \over (\vec{r} - \vec\Delta/2)^2 + \overline Q^2}\label{g5}
\end{equation}
With these functions, for the naive $q \bar q V$ vertex the whole
expression in square brackets in (\ref{gv17}) wit sign minus (which we denote
here as $I_{\lambda_\gamma \lambda_V}$) has the form:
\begin{eqnarray}
I_{LL} &=& - 4 QM z^2 (1-z)^2 \Phi_2\,;\nonumber\\
I_{TT} &=& (\vec{e}\vec{V}^*)[m^2\Phi_2 +
(\vec{k}\vec{\Phi}_1)] + (1-2z)^2(\vec{k}\vec{V}^*)(\vec{e}\vec{\Phi}_1)
- (\vec{e}\vec{k})(\vec{V}^*\vec{\Phi}_1)\,;\nonumber\\
I_{TL} &=& 2Mz(1-z)(1-2z)(\vec{e}\vec{\Phi}_1)\,;\nonumber\\
I_{LT} &=& -2Qz(1-z)(1-2z)(V\vec{k})\Phi_2\,.\label{f3}
\end{eqnarray}

Replacing $V(\kappa)$ by the unintegrated gluon density according to
(\ref{replacerule}) and using approximation for the off-forward 
gluon density (\ref{shuvaevsimple}), we obtain the final expressions
for  a general amplitude of reaction
$\gamma^*_{\lambda_\gamma} \to V_{\lambda_V}$ in the naive 
vertex:
\begin{center}
\framebox(350,80){
\parbox{16cm}{
\begin{eqnarray}
A(x,Q^{2},\vec \Delta)=
- is{c_{V}\sqrt{4\pi\alpha_{em}}
\over 4\pi^{2}}
\int_{0}^{1} {dz\over z(1-z)} \int d^2 \vec k \psi(z,\vec k)
\nonumber\\
\int {d^{2} \vec \kappa
\over
\vec\kappa^{4}}\alpha_{S}{\cal{F}}(x,\vec \kappa,\vec \Delta)
\cdot I(\gamma^{*}\to V)\, ,
\label{f1}
\end{eqnarray}}}
\end{center}

\section{Final results for $S$ and $D$ wave amplitudes}

Now we can use projector technique to obtain results for $S$/$D$
wave states.
\begin{eqnarray}
&&V_\mu \to V_\nu {\cal S}_{\nu\mu}\,; \quad
{\cal S}_{\mu\nu} = g_{\mu\nu} - {2p_\mu p_\nu \over m (M+2m)}
\ \Rightarrow\ I^S = I + {2 ({\bf V}{\bf p}) \over m(M+2m)}
p_\mu \otimes \gamma_\mu\,;\label{f4}\\
&&V_\mu \to V_\nu {\cal D}_{\nu\mu}\,; \quad
{\cal D}_{\mu\nu} = {\bf p}^2 g_{\mu\nu} + {(M+m)p_\mu p_\nu \over m}
\ \Rightarrow\ I^D = I{\bf p}^2 -  {(M+m) ({\bf V}{\bf p}) \over m}
p_\mu \otimes \gamma_\mu\,;\nonumber
\end{eqnarray}
Note that $({\bf V}{\bf p})$ is 3D scalar product.
While contracting, we encounter terms proportional to $p_\mu \otimes \gamma_\mu$
which should be understood as
\begin{equation}
p_\mu \otimes \gamma_\mu = I_{V_T}\{\vec V \to \vec p\}
+ I_{V_L}\{1 \equiv V_z \to p_z \equiv {1 \over 2}(2z-1)M\} .\label{f5}
\end{equation}
The result of this substitution reads:\\
for $e_T$:
\begin{eqnarray}
&&I_{T\to T}\{\vec V \to \vec p\}
+ I_{T\to L}\{1 \equiv V_z \to p_z \equiv {1 \over 2}(2z-1)M\} \nonumber\\
&&=m^2 \left[ (\vec{e}\vec{k})\Phi_2 - (\vec{e}\vec{\Phi}_1)(1-2z)^2\right]\label{f6}
\end{eqnarray}
for $e_0$:
\begin{eqnarray}
&&I_{L\to T}\{\vec V \to \vec p\}
+ I_{L\to L}\{1 \equiv V_z \to p_z \equiv {1 \over 2}(2z-1)M\} \nonumber\\
&&=- 2Qz(1-z)(2z-1)m^2 \Phi_2\label{f7}
\end{eqnarray}

So, the resulting integrands for $S$ wave type mesons are
\begin{center}
\framebox(480,200){
\parbox{16cm}{
\begin{eqnarray}
I^S_{L\to L} &=& - 4 QM z^2 (1-z)^2
\left[ 1 + { (1-2z)^2\over 4z(1-z)} {2m \over M+2m}\right] \Phi_2\,;\nonumber\\[1mm]
I^S_{T\to T} &=& (\vec{e}\vec{V}^*)[m^2\Phi_2 + (\vec{k}\vec{\Phi}_1)] + (1-2z)^2(\vec{k}\vec{V}^*)(\vec{e}\vec{\Phi}_1){M \over M+2m}
\nonumber\\&&- (\vec{e}\vec{k})(\vec{V}^*\vec{\Phi}_1) + {2m \over M+2m}(\vec{k}\vec{e})(\vec{k}\vec{V}^*)\Phi_2\,;
\nonumber\\[1mm]
I^S_{T\to L} &=& 2Mz(1-z)(1-2z)(\vec{e}\vec{\Phi}_1)
\left[ 1 + { (1-2z)^2\over 4z(1-z)} {2m \over M+2m}\right]
- {Mm\over M+2m}(1-2z)(\vec{e}\vec{k})\Phi_2\,;\nonumber\\[1mm]
I^S_{L\to T} &=& -2Qz(1-z)(1-2z)(\vec{V}^*\vec{k}){M \over M+2m}\Phi_2\,.\label{f8}
\end{eqnarray}}}
\end{center}
and for $D$ wave type mesons are
\begin{center}
\framebox(400,150){
\parbox{16cm}{
\begin{eqnarray}
I^D_{L\to L} &=& - QM z (1-z)
\left( \vec{k}^2 - {4m \over M}p_z^2 \right)  \Phi_2\,;\nonumber\\[2mm]
I^D_{T\to T} &=& (\vec{e}\vec{V}^*){\bf p^2}[m^2\Phi_2 + (\vec{k}\vec{\Phi}_1)]
+ (1-2z)^2({\bf p}^2 + m^2 + Mm)(\vec{k}\vec{V}^*)(\vec{e}\vec{\Phi}_1) \nonumber\\
&&- {\bf p}^2(\vec{e}\vec{k})(\vec{V}^*\vec{\Phi}_1) - m(M+m)(\vec{k}\vec{e})(\vec{k}\vec{V}^*)\Phi_2\,;
\nonumber\\[2mm]
I^D_{T\to L} &=& {1 \over 2}M(1-2z)
\left[(\vec{e}\vec{\Phi}_1)\left( \vec{k}^2 - {4m \over M}p_z^2 \right)
+ m(M+m)(\vec{e}\vec{k})\Phi_2 \right] \,;\nonumber\\[2mm]
I^D_{L\to T} &=& -2Qz(1-z)(1-2z)(\vec{V}^*\vec{k})({\bf p}^2 + m^2 + Mm)\Phi_2\,.\label{f9}
\end{eqnarray}}}
\end{center}

Equations (\ref{f8}), (\ref{f9}) together with expression
(\ref{f1}) constitute the ultimate sets of all helicity amplitudes.
They give explicit answers for the vector meson production amplitudes
within leading-log-approximation.

\newpage

\chapter{Analysis for heavy quarkonia}

The general answers (\ref{f8}), (\ref{f9}) are of course incomprehensible
at a quick glance. Therefore, a further analysis is needed to
grasp the most vivid features of the results and to disentangle
$s$-channel helicity conserving and double helicity flip amplitudes.

Since in the heavy vector mesons quarks can be treated
non-relativistically, further simplifications in analytical
formulas (\ref{f8}), (\ref{f9}) are possible due to the presence
of an additional small parameter ${\bf p}^2/m^2$.

In what follows we will first perform the twist expansion and then
relate simplified amplitudes to the decay constants (\ref{b5}),
(\ref{b6}). We will then analyze twist hierarchy of the amplitudes
and compare results for $S$ vs. $D$ wave states. Though we perform
this analysis for heavy mesons, we wish to stress that all
qualitative features ($S$ vs. $D$ difference, $Q^2$ dependence
etc.) will hold for light quarkonia as well.

\section{Twist expansion}
Here we are going to expand the amplitudes (or to be more exact,
the quantities $\vec \Phi_1$ and $\Phi_2$ (\ref{g2}), (\ref{g5}))
in inverse powers of the hard scale $\overline Q^2$ and
then perform azimuthal angular averaging over $\phi_\kappa$.

Expanding $\Phi_2$ in twists in the main
logarithmic region
\begin{equation}
\mu^2, \vec\Delta^2, \vec{k}^2 \ll \vec\kappa^2 \ll \overline Q^2\,,
\label{g3}
\end{equation}
one observes that
twist--1 terms cancel, so one has to retain twist--2 and twist--3 terms
proportional $\vec\kappa^2$:
\begin{equation}
\Phi_2 =
 {2 \vec\kappa^2 \over \overline Q^4} - {8 \vec\kappa^2 \vec{r}^2 \over \overline Q^6}\label{g4}
\end{equation}

The analogous decomposition for $\vec \Phi_{1}$ reads
\begin{equation}
\vec \Phi_{1} =
{4 \vec r \vec\kappa^2 \over \overline Q^4} - {12\vec r
\vec\kappa^2 r^2 \over \overline Q^6}
 - {\vec \Delta (\vec{r}\vec\Delta)\over \overline Q^4} \label{g6}
\end{equation}
Note that the last term does not contain $\vec\kappa^2$.
However, one must track it because it will be important
in double helicity flip amplitudes.

\section{Twist expansion for $S$ wave type mesons}

With the aid of this decomposition one obtains:\\
for amplitude $L\to L$
\begin{equation}
I^S_{L\to L} = -4QMz^2(1-z)^2{2 \vec\kappa^2 \over \overline Q^4}
\left[ 1 + {(1-2z)^2\over 4z(1-z)} {2m \over M+2m}\right]\,,\label{h3}
\end{equation}
for amplitude $T\to T$
\begin{eqnarray}
&&I^S_{T\to T} = (\vec{e}\vec{V}^*)\left[m^2{2 \vec\kappa^2 \over \overline Q^4}
+{4 \vec\kappa^2 \over \overline Q^4} \vec{k}^2\right]
\ +\ {2m \over M+2m}\cdot {1 \over 2}\vec{k}^2 (\vec{e}\vec{V}^*){2 \vec\kappa^2 \over \overline Q^4}
\label{h4}\\
&&+ \left[(1-2z)^2{M \over M+2m} - 1\right]
\left[ \vec{k}^2 (\vec{e}\vec{V}^*){2 \vec\kappa^2 \over \overline Q^4} -
{\vec{k}^2 \over 2\overline Q^4}(\vec{e}\vec\Delta)(\vec{V}^*\vec\Delta)
\left(1 + {6 \vec\kappa^2 (1-2z)^2 \over \overline Q^2}\right) \right] \nonumber
\end{eqnarray}
This amplitude is naturally split into $s$--channel helicity conserving
and double helicity flip parts
\begin{eqnarray}
I^S_{T\to T}(\lambda_\gamma = \lambda_V) &= &
(\vec{e}\vec{V}^*){2 \vec\kappa^2 \over \overline Q^4}
\left[ m^2 + 2 \vec{k}^2(z^2 + (1-z)^2) + {m\over M+2m} \vec{k}^2 (1-2(1-2z)^2)\right]
\nonumber\\
I^S_{T\to T}(\lambda_\gamma = - \lambda_V) &= &
4z(1-z)(\vec{e}\vec\Delta)(\vec{V}^*\vec\Delta){\vec{k}^2 \over 2\overline Q^4}
\left(1 + {6 \vec\kappa^2 (1-2z)^2 \over \overline Q^2}\right)
\left[ 1 + {(1-2z)^2\over 4z(1-z)} {2m \over M+2m}\right]
\nonumber\\ \label{h5}
\end{eqnarray}
Finally, single spin flip amplitudes are
\begin{eqnarray}
I^S_{T\to L} &=& - 2Mz(1-z)(1-2z)^2 {2 \vec\kappa^2 \over \overline Q^4}(\vec{e}\vec\Delta)
\left[ 1 + {(1-2z)^2\over 4z(1-z)} {2m \over M+2m}\right]\,,\nonumber\\
I^S_{L\to T} &=&-2 Qz(1-z)(1-2z)^2{2 \vec\kappa^2 \over \overline Q^4}(\vec{V}^*\vec\Delta)
{2 \vec{k}^2 \over \overline Q^2}{M\over M+2m}\,.\label{h6}
\end{eqnarray}

\section{Twist expansion for $D$--type vector mesons}
Here we will need to track higher--twist terms.
It will turn out later that leading contributions vanish,
so twist--3 terms will be crucial for our results.\\
For amplitude $L\to L$ one has
\begin{equation}
I^D_{L\to L}=-QMz(1-z)\left(\vec{k}^2 - {4m\over M}p_z^2\right)
\cdot {2 \vec\kappa^2 \over \overline Q^4} \left(1 - {4 \vec{k}^2\over \overline Q^2}\right)\,.
\label{h7}
\end{equation}
For $T\to T$ amplitude, one obtains
\begin{eqnarray}
&&I^D_{T\to T}\ =\ (\vec{e}\vec{V}^*){\bf p}^2 \left[m^2{2 \vec\kappa^2 \over \overline Q^4}
\left(1 - {4 \vec{k}^2\over \overline Q^2}\right)
+ {2 \vec\kappa^2 \over \overline Q^4} 2\vec{k}^2  \right]\nonumber\\
&&+\left[-4z(1-z){\bf p}^2 + (1-2z)^2m(M+m)\right]
\left\{ {2 \vec\kappa^2 \over \overline Q^4} \vec{k}^2 (\vec{e}\vec{V}^*) -
{\vec{k}^2 \over 2\overline Q^4}(\vec{e}\vec\Delta)(\vec{V}^*\vec\Delta)
\left[1 + {6 \vec\kappa^2(1-2z)^2 \over \overline Q^2}\right] \right\}\nonumber\\
&&-m(M+m) {1 \over 2}\vec{k}^2 (\vec{e}\vec{V}^*){2 \vec\kappa^2 \over \overline Q^4}
 \left( 1 - {4 \vec{k}^2\over \overline Q^2} \right)\label{h8}
\end{eqnarray}
Note that we kept track of all terms $\propto |p|^4$. Again, one
can separate out $s$-channel helicity conserving and double
helicity flip parts:
\begin{eqnarray}
I^D_{T\to T}(\lambda_\gamma = \lambda_V) &= &
(\vec{e}\vec{V}^*){\vec\kappa^2 \over \overline Q^4}
\Biggr[ 2{\bf p}^2\left( m^2 + 2\vec{k}^2 - 4\vec{k}^2 {m^2 \over \overline Q^2}\right)
-m(M+m)\vec{k}^2 \left(1 - {4 \vec{k}^2 \over \overline Q^2}\right)
\nonumber\\
&&-2\vec{k}^2 \left(\vec{k}^2 - {4m \over M}p_z^2\right)\Biggr]
\nonumber\\
I^D_{T\to T}(\lambda_\gamma = - \lambda_V) &= &
\left(\vec{k}^2 - {4m \over M}p_z^2\right)
(\vec{e}\vec\Delta)(\vec{V}^*\vec\Delta){\vec{k}^2 \over 2\overline Q^4}
\left(1 + {6 \vec\kappa^2 (1-2z)^2 \over \overline Q^2}\right)
\nonumber\\ \label{h9}
\end{eqnarray}

Finally, single helicity flipping amplitudes are
\begin{eqnarray}
I^D_{T\to L} &=& {1 \over 2}M(1-2z)
\left[{-2(1-2z)(\vec{e}\vec\Delta)\vec\kappa^2 \over \overline Q^4}\left(\vec{k}^2 - {4m\over M}p_z^2\right)
+ m(M+m){4\vec\kappa^2 \vec{k}^2 \over \overline Q^6}(1-2z)(\vec{e}\vec\Delta)
\right]\nonumber\\
&=&- {\vec\kappa^2 \over \overline Q^4}(1-2z)^2M(\vec{e}\vec\Delta)
\left[\vec{k}^2 - {4m\over M}p_z^2 - m(M+m){2 \vec{k}^2 \over \overline Q^2}\right]\,,\nonumber\\
I^D_{L\to T} &=& - 8Qz(1-z)(1-2z)^2({\bf p}^2 + m^2 + mM)
{\vec\kappa^2 \over \overline Q^6}\vec{k}^2 (\vec{V}^*\vec\Delta)\,.\label{h10}
\end{eqnarray}

\section{Final results for $S$ wave mesons}

In order to grasp the major features of various $S$ and $D$ wave
amplitudes, further simplifications can be achieved if one
neglects spherically non-symmetric arguments of $\alpha_s$ and
gluon density. First we rewrite general expression (\ref{f1}) in
the more convenient form
\begin{eqnarray}
A(x,Q^{2},\vec \Delta)=
-is{c_{V}\sqrt{4\pi\alpha_{em}}
\over 4\pi^{2}}
\int d^3 {\bf p} {4 \over M} \psi({\bf p}^2)\int {d^{2} \vec \kappa \over \kappa^{4}}
\alpha_{S}{\cal{F}}\cdot I(\gamma^{*}\to V)\, .
\label{i1}
\end{eqnarray}
In this expression everything except for integrands $I(\gamma^{*}\to V)$ is
spherically symmetric, thus making it possible to perform angular
averaging over $\Omega_{\bf p}$ in these integrands.

\subsection{$S$ wave: $\Omega_{\bf p}$ averaging}

Here all the calculations are rather straightforward.
In the non-relativistic case one can everywhere put $z \to 1/2\,; M =2m=m_V$.
The resultant integrands are:
\begin{center}
\framebox(370,200){
\parbox{16cm}{
\begin{eqnarray}
I^S(L\to L) &=& - {8QM \over (Q^2 + M^2)^2} \vec\kappa^2\nonumber\\[2mm]
I^S(T \to T; \lambda_\gamma = \lambda_V) &=& {8 M^2 \over (Q^2 + M^2)^2} \vec\kappa^2
\nonumber\\[2mm]
I^S(T \to T; \lambda_\gamma = -\lambda_V) &=&
{16 \over 3} {(\vec{e}\vec\Delta)(\vec{V}^*\vec\Delta) \over (Q^2 + M^2)^2}
\left[ 1 + {96 \over 5} {\vec\kappa^2 {\bf p}^2 \over M^2 (Q^2
+M^2)}\right]\nonumber\\[2mm]
I^S(T\to L) &=& - {64 \over 3} {M (\vec{e}\vec\Delta) \over (Q^2 +M^2)^2}
{{\bf p}^2 \over M^2} \vec\kappa^2\nonumber\\[2mm]
I^S(L\to T) &=& - {512 \over 15} {Q(\vec{V}^*\vec\Delta) \over (Q^2+M^2)^2}
{{\bf p}^4 \over M^2 (Q^2 +M^2)}\vec\kappa^2
\label{i3}
\end{eqnarray}}}
\end{center}
Note several things: since the accurate $1S$ wave differs from the naive $\gamma_\mu$
spinorial structure only by relativistic corrections,
one would obtain {\it the same results} in the case of naive $q\bar q V$ vertex.
The only difference would be only extra factor 2 for the $L\to T$ amplitude
(which is higher-twist amplitude, anyway).

Thus the only thing left is $|{\bf p}|$ integration.
Note that these amplitudes are naturally expressed in terms of decay constants.
Indeed, in the extremely non-relativistic case expression (\ref{b5})
turns into
\begin{equation}
f^{(S)} = {3m_V \over 2\pi^3} \int d^3{\bf p}\ \psi_S\quad \Rightarrow\quad
\int d^3{\bf p}\ \psi_S = {2\pi^3 \over 3m_V} f^{(S)}
\label{i4}
\end{equation}

\subsection{$S$ wave: answers for $L \to L$ up to differential
cross section}

Here we would like to digress and for the sake of logical
completeness show how one obtains the final result for the
differential cross section with the example of $L\to L$ amplitude. If
needed, the same can be done for the other amplitudes, so we will
do it just once.

One has:
\begin{eqnarray}
&&\int d^3{\bf p}\ \psi_S  {4 \over m_V} {-8Qm_V \over (Q^2 +m_V^2)^2}
\cdot \pi \int^{\overline Q^2} {d\vec\kappa^2 \over \vec\kappa^2}
{\partial G(x,\vec\kappa^{2})\over \partial \log \vec\kappa^{2}}
\alpha_s(\vec\kappa^2)\exp(-{1\over 2}B_{3\Pom}\vec \Delta^2) \nonumber\\
&&=- {32 \pi Q \over (Q^2+m_V^2)^2}
G(x,\overline Q_0^2) \alpha_s(\overline Q_0^2)\exp(-{1\over 2}B_{3\Pom}\vec \Delta^2)
\int d^3{\bf p}\ \psi_S\nonumber\\
&&=- {32 \pi Q \over (Q^2+m_V^2)^2} \exp(-{1\over 2}B_{3\Pom}\vec \Delta^2)
G(x,\overline Q_0^2) \alpha_s(\overline Q_0^2)
{2\pi^3 \over 3m_V} f_V\nonumber
\end{eqnarray}
With this result, (\ref{i1}) becomes
\begin{eqnarray}
A &=& i s {f_V c_V \sqrt{4\pi\alpha_{em}} \over 4\pi^2}
\cdot {64 \pi^4 \over 3} {Q \over m_V} {G\cdot \alpha_s \over  (Q^2 +m_V^2)^2}
\nonumber\\
 &=& is{16 \pi^2 \over 3} {Q \over m_V} \cdot c_V f_V \cdot \sqrt{4\pi\alpha_{em}}
 {G\cdot \alpha_s \exp(-{1\over 2}B_{3\Pom}\vec \Delta^2)\over  (Q^2 +m_V^2)^2}
\nonumber
\end{eqnarray}
The expression for the differential cross section reads ($t \equiv \vec \Delta^2$):
\begin{eqnarray}
{d\sigma \over dt}&=& {1 \over 16\pi s^2}|A|^2\nonumber\\
&=&{16 \pi^3 \over 9} {Q^2 \over m_V^2} \cdot (c_V f_V)^2 4\pi\alpha_{em}
\cdot {G^2 \alpha_s^2  \exp(-B_{3\Pom}t) \over (Q^2 +m_V^2)^4}\,.
\nonumber
\end{eqnarray}
Finally, one can express this cross section through $\Gamma(V \to e^+e^-)$
(see (\ref{b8})):
\begin{equation}
\framebox(330,50){$ \dst
{d\sigma \over dt}={16 \pi^3 \over 3 \alpha_{em}} Q^2 m_V \cdot\Gamma(V \to e^+e^-)
\cdot {G^2 \alpha_s^2  \exp(-B_{3\Pom}t) \over (Q^2 +m_V^2)^4}\,.$}\label{i8}
\end{equation}

\subsection{$S$ wave: the other amplitudes}
\subsubsection{$T \to T, \lambda_\gamma = \lambda_V$}.
In the non-relativistic case, this amplitude is readily obtained from the above formulas
after $Q \to m_V$ replacement in the numerator of the amplitude (see (\ref{i3})).
This means in particular that in this limit
\begin{equation}
R^{(S)} \equiv \left({A_{LL}\over A_{TT}}\right)^2
= {Q^2 \over m_V^2}\,.\label{i10}
\end{equation}

\subsubsection{$T \to T, \lambda_\gamma = -\lambda_V$}.
This amplitude is very interesting because of the competition of two
very different terms --- soft and hard scale contribution. Indeed,
integration over gluon loop gives
\begin{equation}
A \propto {G(x,\mu^2) \over \mu^2} +
{96 \over 5} {G(x,\overline Q^2) {\bf p}^2 \over M^2 (Q^2 +M^2)}\label{i11}
\end{equation}
We see that the soft contribution turns out to be
of leading twist, while the pQCD contribution is of higher twist.
This observation was first made in \cite{IK} for the naive type
of $q \bar q V$ vertex; here we see that it also holds for
accurate $S$ and $D$ wave vector mesons.

\subsubsection{$T \to L$ and $L \to T$}.
In the case of heavy quarkonia
these single spin flip amplitudes have suppressing
non-relativistic factors. Besides, the amplitude $L \to T$ is of
twist 3, which is another source of suppression. Their ratios to
$A(T\to T) \equiv A(T\to T; \lambda_\gamma = \lambda_V^*)$ read

\begin{equation}
{A(T\to L) \over A(T\to T)} = - {8 \over 3} {(\vec{e}\vec\Delta) \over m_V} \cdot w_2\,;\quad
{A(L\to T) \over A(T\to T)}
= - {64 \over 15} {Q(\vec{V}^*\vec\Delta) \over Q^2 +  m_V^2} \cdot w_4\,.\label{i12}
\end{equation}
The model dependent quantities $w_2$ and $w_4$ are defined via
\begin{equation}
w_2 = {1 \over m_V^2}
{\int d^3{\bf p} \ {\bf p}^2\ \psi_S \over \int d^3{\bf p} \ \psi_S}\,;\quad
w_4 = {1 \over m_V^4}
{\int d^3{\bf p} \ {\bf p}^4\ \psi_S \over \int d^3{\bf p} \ \psi_S}\,.\label{i13}
\end{equation}
Within the oscillator ansatz (\ref{oscillator1}) their values are
\begin{equation}
w_2 = {3 \over 2}{1 \over (m_V R)^2}\,;\quad
w_4 = {15 \over 4}{1 \over (m_V R)^4}\,.\label{i14}
\end{equation}

\section{Final results for $D$ wave}

This case is much more tricky. It turns out that the leading terms
in integrands $I^{(D)}$, proportional to $m^2 |p|^2$ cancel out
after angular averaging, so that many new terms, including higher
twist terms come into play. This cancellation is in fact quite
understandable. Indeed, in the very beginning we showed that
vertex $\bar u' \gamma_\mu u$ contains both $S$ and $D$ waves,
with $D$ wave probability being suppressed for heavy quarks due to
non-relativistic motion. This means in particular that the {\it
photon} couples to $q \bar q$ pairs sitting either in $S$ or $D$
wave state. However, at the other end of the quark loop we have a
vector meson in pure $D$ wave state. Therefore, the largest items
in $\langle \gamma_{S+D}|...|V_D\rangle$ cancel out due to
$S$--$D$ orthogonality.

\subsection{$D$ wave: $\Omega_{\bf p}$ averaging for $L\to L$ amplitude}

If we limited ourselves only to the leading ${\bf p}^2/m^2$ terms,
we would get
$$
\int d\Omega_{\bf p} \left(\vec{k}^2 - {4m\over M}p_z^2\right)
= 4\pi\cdot \left({2\over 3}{\bf p}^2 - 2\cdot{1\over 3}{\bf
p}^2\right)= 0\,,
$$
which is the manifestation of $S$--$D$ orthogonality.
Thus, we see that ${\bf p}^2/m^2$ terms vanish after angular
averaging. Therefore, one has to be extremely careful now
and must take into account all possible sources of
${\bf p}^4/m^4$ terms.
To do so, one has to perform the following averaging
\begin{equation}
\left\langle 4z(1-z)\cdot{1 \over \overline Q^4}\cdot
\left(\vec{k}^2 - {4m\over M}p_z^2\right)\cdot \left(1 - {4 \vec{k}^2\over \overline Q^2}
\right)\right\rangle
\label{i15}
\end{equation}

Before performing the averaging, let us generate
a list of useful formulas :
\begin{eqnarray}
&&\langle \vec{k}^2 \rangle = {2 \over 3}{\bf p}^2 \quad
\langle p_z^2 \rangle  =  {1 \over 3}{\bf p}^2 \nonumber\\[1mm]
&&\langle \vec{k}^2 \vec{k}^2 \rangle  =  {8 \over 15}{\bf p}^4 \quad
\langle \vec{k}^2 p_z^2 \rangle  =  {2 \over 15}{\bf p}^4 \quad
\langle p_z^2 p_z^2 \rangle  =  {3 \over 15}{\bf p}^4 \label{i2}
\end{eqnarray}
Finally, remember that $p_z^2 = {1 \over 4}(1-2z)^2M^2$.

One has to perform the following avegaring
\begin{equation}
\left\langle 4z(1-z)\cdot{1 \over \overline Q^4}\cdot
\left(\vec{k}^2 - {4m\over M}p_z^2\right)\cdot 
\left(1 - {4 \vec{k}^2\over \overline Q^2}\right)\right\rangle
\label{i15ap}
\end{equation}
Note that all factors should be carefully examined; 
all four do contribute to the final answer.
Decomposing $\overline Q^2$ as
\begin{equation}
\overline Q^2 = m^2 + z(1-z)Q^2 = m^2 + {1 \over 4}Q^2 - 
{1 \over 4} (1-2z)^2Q^2 \equiv \overline Q^2_0 - 
{p_z^2 \over M^2}Q^2\,,\label{i16ap}
\end{equation}
with $\overline Q_0^2 = m^2 + Q^2/4$, one gets
\begin{equation}
{1 \over \overline Q^4} = {1 \over \overline Q_0^4}
\left( 1 + 2{p_z^2 \over M^2}{Q^2 \over \overline Q_0^2}\right)\,.\label{i17ap}
\end{equation}
So, omitting $\overline Q_0^{-4}$, one has
\begin{equation}
\left\langle \left(1 -{4p_z^2 \over M^2}\right)
\cdot\left(1 + 2{p_z^2 \over M^2}{Q^2 \over \overline Q_0^2}\right)
\cdot\left(\vec{k}^2 - {4m\over M}p_z^2\right)\cdot \left(1 - {4 \vec{k}^2\over \overline Q_0^2}
\right)\right\rangle
\label{i18ap}
\end{equation}
With the aid of (\ref{i1}), one obtains
\begin{eqnarray}
&&\left\langle \vec{k}^2 - {4m\over M}p_z^2 \right\rangle
- {4 \over M^2}\left\langle \vec{k}^2 p_z^2 - 2 p_z^2 p_z^2 \right\rangle
+ 2{Q^2 \over M^2 \overline Q^2_0}\left\langle \vec{k}^2 p_z^2 - 2 p_z^2 p_z^2 \right\rangle
- {4 \over \overline Q_0^2} \left\langle \vec{k}^2 \vec{k}^2 - 2 p_z^2 \vec{k}^2\right\rangle
\nonumber\\[1mm]
&=&\left({2 \over 3}{\bf p}^2  - {4m \over 3M}{\bf p}^2 \right)
- {4 \over M^2} {\bf p}^4 \left( {2 \over 15} - {6 \over 15}\right)
+  2{Q^2 \over M^2 \overline Q^2_0}{\bf p}^4\left( {2 \over 15} - {6 \over 15}\right)
- {4 \over \overline Q_0^2} {\bf p}^4\left( {8 \over 15} - {4 \over 15}\right)
\nonumber\\[1mm]
&=&{2 \over 3}{\bf p}^2{4 {\bf p}^2 \over M+2m} + {16{\bf p}^4 \over 15 M^2}
- { 8Q^2{\bf p}^4 \over 15 M^2 \overline Q^2_0}
-{16{\bf p}^4 \over 15 \overline Q^2_0}\label{i19ap}\\[1mm]
&=&{4 {\bf p}^4 \over 3M^2}
\left( 1 + {4 \over 5} - {8 \over 5}{Q^2 \over Q^2 + M^2}
- {16 \over 5}{M^2 \over Q^2 + M^2} \right)\nonumber\\[1mm]
&=&{4 {\bf p}^4 \over 15M^2}\left(1 - 8{M^2 \over Q^2 + M^2}\right)\nonumber
\end{eqnarray}

One can now again express the integral over quark loop through
the decay constant (see (\ref{b6})):
\begin{equation}
\int d^3{\bf p}\ {\bf p}^4\ \psi_D \Rightarrow f^{(D)}\cdot {\pi^3 m_V \over 2}\,.
\label{i20}
\end{equation}
to give
\begin{equation}
 - {64 \pi^4 \over 15}\; {Q \over m_V} \;
{ G\cdot \alpha_s\cdot \exp(-{1\over 2}B_{3\Pom}\vec \Delta^2)
\over (Q^2+m_V^2)^2}\cdot f^{(D)}\cdot
\left( 1 - 8 {m_V^2 \over Q^2 + m_V^2}\right)\label{i21}
\end{equation}
Comparison with $L\to L$ amplitude reveals that
\begin{equation}
\framebox(220,50){$\dst{A^D_{LL}\over A^S_{LL}} = {1 \over 5}\left( 1 - 8 {m_V^2 \over Q^2 + m_V^2}\right)
\cdot {f^{(D)} \over f^{(S)}}\,.$}\label{i22}
\end{equation}

\subsection{$D$ wave: the other amplitudes}

For the helicity conserving amplitude one has to repeat the same
averaging procedure. The calculation proceeds as follows:
\begin{eqnarray}
&&\Biggl\langle \left(1 + 2{p_z^2 \over M^2}{Q^2 \over \overline Q_0^2}\right)
\cdot
\Biggr[ 2{\bf p}^2\left( m^2 + 2\vec{k}^2 - 4\vec{k}^2 {m^2 \over \overline Q^2}\right)
-m(M+m)\vec{k}^2 \left(1 - {4 \vec{k}^2 \over \overline Q^2}\right)
-2\vec{k}^2 \left(\vec{k}^2 - {4m \over M}p_z^2\right)\Biggr]\Biggr\rangle
\nonumber\\
&& = 2 m^2 {\bf p}^2 - {2 \over 3}m(M+m){\bf p}^2
+ 2{p_z^2 \over M^2}{Q^2 \over \overline Q_0^2}
\left({1\over 2}M^2{1\over 3}- 3 M^2{2 \over 15} \right)
+ {8 \over 3}{\bf p}^4 -   {8 \over 15} {\bf p}^4
+ {m^2 \over \overline Q^2_0}{\bf p}^4{16 \over 15}\nonumber\\
&&= - {2\over 3}{\bf p}^4 - {8\over 15}{\bf p}^4 + {8\over 3}{\bf p}^4
+ {16 \over 15} {M^2 \over M^2 + Q^2}{\bf p}^4
+ {8 \over 15} {Q^2 \over M^2 + Q^2}{\bf p}^4
\nonumber\\
&&= 2{\bf p}^4 \left(1 + {4 \over 15} {M^2 \over M^2 + Q^2}\right)\,.
\label{i23ap}
\end{eqnarray}

The result can be written as
\begin{equation}
\dst { A^D_{LL} \over A^D_{TT}} = {1\over 15}
{ 1 - 8\fr{m_V^2}{Q^2 + m_V^2} \over 1 + \fr{4}{15}
\fr{m_V^2}{Q^2 + m_V^2}}\,. \label{i24}
\end{equation}

In the case of double helicity flip amplitude we again have
contributions from soft and hard scales with the same hierarchy
of twists. Namely,
\begin{equation}
A \propto {G(x,\mu^2) \over \mu^2} -
{96 \over 7} {G(x,\overline Q^2) {\bf p}^2 \over M^2 (Q^2 +M^2)}\label{i24a}
\end{equation}
so we again see the soft domination in the double helicity flip amplitude.

In the case of single spin flip amplitudes, no dangerous calcellations among
leading terms arise. Before giving a list of amplitudes, we wish to emphasize that
in the case of $D$ wave mesons there is no non-relativistic suppressing factors
like $w_2$ and $w_4$ defined in (\ref{i13}). This means that for
moderate momentum transfers helicity non-conserving amplitudes are absolutely
important for that case of $D$ wave mesons.

\section{$S$ wave vs. $D$ wave comparison}

We would like to present our final results in the form which
stresses the remarkable differences between $S$ wave and $D$ wave amplitudes.
Below we give a table of the ratios
\begin{equation}
\rho_{ij} \equiv {A^D(i\to j) \over A^S(i\to j)}\, {f^{(S)} \over f^{(D)}}\label{rho}
\end{equation}
for helicity conserving and single spin flipping amplitudes.
Double spin flip amplitudes are not given due to the presence
of incalculable non-perturbative contributions.
\begin{eqnarray}
\rho_{LL} &=&{1 \over 5}
\left(1 - 8{m_V^2 \over Q^2 + m_V^2}\right)\nonumber\\
\rho_{TT} &=& 3
\left(1 + {4 \over 15}{m_V^2 \over Q^2 + m_V^2}\right)\nonumber\\
\rho_{TL} &=&
- {3 \over 5} {1 \over w_2} \left(1 + 3{m_V^2 \over Q^2 + m_V^2}\right)\nonumber\\
\rho_{LT} &=&
{9 \over 8}{1 \over w_4}\label{i26}
\end{eqnarray}

Thus, we note several things. First, the abnormally large higher
twist contributions to $D$ wave amplitudes are seen here as terms
$\propto m_V^2/(Q^2 + m_V^2)$. They even force the opposite sign
of $L \to L$ amplitude in the moderate $Q^2$ domain. Second, we
see highly non--trivial and even non--monotonous $Q^2$ dependence
of $(A_{LL}/A_{TT})^2$ ratio, which will lead to the presence of a dip
in experimentally measured $\sigma_L/\sigma_T$ for $D$ wave meson
production. Finally, we must stress that in the case of $D$ wave
mesons there is no non-relativistic suppression for single spin
flip amplitudes as it was in $S$ wave mesons. This leads us to a
conclusion that $s$-channel helicity is strongly violated in the
case of $D$ wave meson production.

\part{Numerical analysis}

\chapter{Determination of the unintegrated gluon structure function
of the proton: DGD2000 analysis}

The familiar objects from Gribov-Lipatov-Dokshitzer-Altarelli-Parisi
(DGLAP) evolution description of deep inelastic scattering (DIS) are
quark, antiquark and gluon distribution functions $q_{i}(x,Q^{2}),
\bar{q}_{i}(x,Q^{2}),g(x,Q^{2})$ (hereafter $x,Q^{2}$ are the standard
DIS variables). At small $x$ they describe the integral flux of partons with
the lightcone momentum $x$ in units of the target momentum and
transverse momentum squared $\leq Q^{2}$ and form
the basis of the highly sophisticated description of hard scattering
processes in terms of collinear partons \cite{DGLAP}. On the other hand, at very
small $x$ the object of the Balitskii-Fadin-Kuraev-Lipatov evolution
equation is the differential gluon structure
function (DGSF) of the target \cite{FKL,BL}
\be
{\cal F}(x,Q^{2})={\partial G(x,Q^{2})\over \partial \log Q^{2}}
\label{eq:1.1}
\ee
with $G(x,Q^2)\equiv xg(x,Q^2)$
(evidently the related unintegrated distributions can be defined also
for charged partons). For instance, it is precisely DGSF of the target
proton that emerges in the familiar color dipole picture of inclusive
DIS at small $x$ \cite{NZZ,NZ94,MuellerPatel} and diffractive DIS into dijets
\cite{NZsplit}. Another familiar example is the QCD
calculation of helicity amplitudes of diffractive DIS into
continuum \cite{NPZLT,Twist4} and production of vector
mesons \cite{Vmeson,JETPVM}. DGSF's are custom-tailored
for QCD treatment of hard processes,
when one needs to keep track of the transverse momentum of gluons
neglected in the standard collinear approximation
\cite{PomKperp,Zotov,Forshaw,Stasto}.

In the past two decades DGLAP phenomenology of DIS has become a big industry
and several groups --- notably GRV \cite{GRV}, CTEQ \cite{CTEQ} \& MRS
\cite{MRS} and some other \cite{BaroneF2} --- keep continuously incorporating
new experimental data and providing the high energy community
with updates of the parton distribution functions supplemented with the
interpolation routines facilitating practical applications. On the other hand,
there are several pertinent issues --- the onset of the purely perturbative
QCD treatment of DIS and the impact of soft mechanisms of photoabsorption
on the proton structure function in the region of large $Q^{2}$ being
top ones on the list --- which can not be answered within the DGLAP approach
itself because DGLAP evolution is obviously hampered at moderate to
small $Q^{2}$. The related issue is to what extent the soft mechanisms of
photoabsorption can bias the $Q^{2}$ dependence of the proton structure
function and, consequently, determination of the gluon density from
scaling violations. We recall here the recent dispute
\cite{dF2dLogQ2} on the applicability
of the DGLAP analysis at $Q^{2} \lsim $ 2--4 GeV$^{2}$ triggered by the
so-called Caldwell's plot \cite{Caldwell}. Arguably the
$\vec{\kappa}$-factorization formalism of
DGSF, in which the interesting observables are expanded in interactions
of gluons of transverse momentum $\vec{\kappa}$  changing
from soft to hard is better suited to look into the issue of soft-hard
interface. Last but not least, neglecting the transverse momentum
$\vec{\kappa}$ of gluons is a questionable approximation in evaluation of
production cross sections of jets or hadrons with large transverse
momentum. It is distressing, then, that convenient, ready-to-use,
parameterizations of DGSF are not yet available in the literature.

Here we perform a simple phenomenological determination
of the DGSF at small $x$  based  on the 1978 Baltskii-Lipatov (BL) scheme
\cite{BL}, in which the DGSF is directly related to the physical observable
--- the proton structure function $F_{2p}(x,Q^{2})$. In early 90's the BL
scheme has been extended to other observables and
dubbed $\vec{\kappa}$-factorization \cite{Kfact};
it is also closely related to the color dipole factorization in the
color dipole  BFKL approach \cite{NZZ,NZ94,MuellerPatel}.
Our interest is in producing the ready-to-use  Ansatz for
${\cal F}(x,\vec{\kappa}^2)$, so that we take advantage of large body of the
early work on color dipole BFKL
factorization \cite{NZ94,NZHERA,BFKLRegge} and follow a very pragmatic
strategy first applied in \cite{NPZLT,Twist4}: (i) for hard gluons
with large $\vec{\kappa}$ we make as much use as possible of the existing
DGLAP parameterizations of $G(x,\vec{\kappa}^{2})$, (ii) for the
extrapolation of hard gluon densities to small $\vec{\kappa}^{2}$ we use an
Ansatz \cite{NZsplit} which correctly describes the color gauge
invariance constraints on radiation of soft gluons by
colour singlet targets, (iii) as suggested by color dipole
phenomenology, we supplement the density of gluons with small
$\vec{\kappa}^{2}$ by nonperturbative soft component, (iv) as
suggested by the soft-hard diffusion inherent to the BFKL evolution,
we allow for propagation of the predominantly hard-interaction driven
small-$x$ rise of DGSF into the soft region invoking plausible
soft-to-hard interpolations. The last two components of DGSF are
parameterized following the modern wisdom on the infrared freezing of
the QCD coupling and short propagation radius of perturbative gluons. Having
specified the infrared regularization, we can apply the resulting
${\cal F}(x,\vec{\kappa}^2)$ to evaluation of the photoabsorption cross
 section in the whole range of small to large $Q^{2}$.

\section{The Ansatz for differential gluon structure function}

The major insight into parameterization of DGSF comes from early experience
with color dipole phenomenology of small-$x$ DIS. In color dipole approach,
which is closely related to $\vec{\kappa}$-factorization, the principal
quantity is the total cross section of interaction of the $q\bar{q}$
color dipole $\br$ with the proton target \cite{NZ94,NZglue,BGNPZUnit}
\be
\sigma(x,r)=
\frac{\pi^{2} r^{2}}{3}
\int
\frac{d \vec{\kappa}^{2}}{\vec{\kappa}^{2}}
\frac{4[1-J_{0}(\kappa r)]}{(\kappa r)^{2}}
\alpha_{S}\left({\rm max}\{\vec{\kappa}^{2}, {A\over r^2}\}\right)
{\cal F}(x,\vec{\kappa}^{2}) \,,
\label{eq:4.00}
\ee
which for very small color dipoles can be approximated by
\be
\sigma(x,r)=
\frac{\pi^{2} r^{2}}{3}\alpha_{S}\left({A\over r^2}\right)
G\left(x,{A\over r^2}\right) \,,
\label{eq:4.0}
\ee
where $A\approx 10$ comes from properties of the Bessel function
$J_{0}(z)$.
The phenomenological properties of the dipole cross section
are well understood, for extraction of $\sigma(x,r)$ from the
experimental data see \cite{NNPZsigmadipole,NNPZVM}.
The known dipole size dependence of $\sigma(x,r)$ serves as a
constrain on the possible
$\vec{\kappa}^2$-dependence of ${\cal F}(x,\vec{\kappa}^{2})$.

As we argued in section 3.2, DGLAP fits are likely to overestimate
${\cal F}_{hard}(x,\vec{\kappa}^2)$ at moderate $\vec{\kappa}^{2}$.
Still, approximation (\ref{eq:4.0}) does a good job when
the hardness $A/r^2$ is very large, and at
large $Q^{2}$ we can arguably approximate the DGSF by the direct
differentiation of available fits (GRV, CTEQ, MRS, ...) to the integrated
gluon structure function $G_{pt}(x,Q^{2})$:
\be
{\cal F}_{pt}(x,\vec{\kappa}^{2}) \approx {\partial G_{pt}(x,\vec{\kappa}^{2}) \over
\partial \log \vec{\kappa}^{2}}
\label{eq:4.1}
\ee
Hereafter the subscript $pt$  serves as a reminder that these gluon distributions
were obtained from the pQCD evolution analyses of the proton structure
function and cross sections of related hard processes.

The available DGLAP fits are only applicable at $\vec{\kappa}^{2}\geq Q_{c}^{2}$,
see table 1 for the values of $Q_c^{2}$, in the extrapolation to
soft region $\vec{\kappa}^{2}\leq Q_{c}^{2}$ we are bound to educated guess.
 To this end recall that perturbative
gluons are confined and do not propagate to large distances;
recent fits \cite{LatticeQCD} to the lattice QCD data suggest
Yukawa-Debye screening of perturbative color fields with propagation/screening
radius $R_{c}\approx 0.27 fm$.
Incidentally, precisely this value of $R_{c}$ for Yukawa screened
colour fields has been used since 1994 in the very successful color
dipole phenomenology of small-$x$  DIS \cite{NZHERA,BFKLRegge}.
Furthermore, important finding of \cite{BFKLRegge}
is a good quantitative description
of the rising component of the proton structure function
starting with the Yukawa-screened perturbative two-gluon exchange as a boundary
condition for the color dipole BFKL evolution.

The above suggests that $\vec{\kappa}^{2}$ dependence of perturbative hard
${\cal F}_{hard}(x,\vec{\kappa}^{2})$ in the soft region $\vec{\kappa}^{2}\leq Q_{c}^{2}$
is similar to the Yukawa-screened flux of photons in the
positron, cf. eq.~(\ref{eq:2.3}),
with $\alpha_{em}$ replaced by the running
strong coupling of quarks $C_{F}\alpha_{S}(\vec{\kappa}^{2})$ and
with factor $N_c$ instead of 2 leptons in the positronium,
for the early discussion see \cite{NZsplit},
\be
{\cal F}^{(B)}_{pt}(\vec{\kappa}^2) =
C_F N_c {\alpha_{s}(\vec{\kappa}^2) \over \pi} \left( {\vec{\kappa}^2 \over
\vec{\kappa}^2 +\mu_{pt}^{2}}\right)^{2} V_{N}(\vec{\kappa})\,,
\label{eq:4.2}
\ee
Here $\mu_{pt}={1\over R_{c}}=0.75$ GeV  is the inverse Yukawa
screening radius and must not be interpreted as a gluon mass;
more sophisticated forms of screening can well be considered.
Following \cite{NZHERA,BFKLRegge,NZ91,NZZ94}
we impose also the infrared freezing of strong coupling:
$\alpha_{S}(\vec{\kappa}^2)\leq 0.82$; recently the concept of freezing coupling
has become very popular, for the review see \cite{Freezing}.

The vertex function $V_{N}(\vec{\kappa})$
describes the decoupling of soft gluons, $\vec{\kappa} \ll {1\over R_{p}}$,
from color neutral proton and has the same structure as in
eq.~(\ref{eq:2.4}). In the nonrelativistic oscillator model
for the nucleon one can
relate the two-quark form factor of the nucleon to the
single-quark form factor,
\be
F_{2}(\vec{\kappa},-\vec{\kappa})=F_{1}\left({2N_{c} \over N_{c}-1}
\vec{\kappa}^{2}\right)\, .
\label{eq:4.3}
\ee
To the extent that $R_{c}^{2} \ll R_{p}^{2}$ the detailed functional
form of $F_{2}(\vec{\kappa},-\vec{\kappa})$ is not crucial, the simple
relation (\ref{eq:4.3}) will be used also for a more realistic
dipole approximation
\be
F_{1}(\vec{\kappa}^{2})={1\over (1 + {\vec{\kappa}^2 \over \Lambda^{2}})^{2}}\,.
\label{eq:4.4}
\ee
The gluon probed radius of the proton and the charge radius of the proton
can be somewhat different and $\Lambda \sim 1$ GeV must be regarded as
a free parameter.
Anticipating the forthcoming discussion of the
diffraction slope in vector meson production we put $\Lambda=1$ GeV.

As discussed above, the hard-to-soft diffusion makes the DGSF rising at
small-$x$ even in the soft region. We model this hard-to-soft diffusion
by matching the $\vec{\kappa}^{2}$ dependence (\ref{eq:4.2}) to the DGLAP
fit $ {\cal F}_{pt}(x,Q_{c}^{2})$ at the soft-hard interface $Q_{c}^{2}$
and assigning to ${\cal F}_{hard}(x,\vec{\kappa}^2)$ in the region of
$\vec{\kappa}^{2}\leq Q_{c}^{2}$ the $\vec{\kappa}^2$ dependence of
the Born term (\ref{eq:4.2}) and the $x$-dependence
as shown by the DGLAP fit $ {\cal F}_{pt}(x,Q_{c}^{2})$, i.e.,
\be
{\cal F}_{hard}(x,\vec{\kappa}^2)=
{\cal F}^{(B)}_{pt}(\vec{\kappa}^2){{\cal F}_{pt}(x,Q_{c}^{2})
\over {\cal F}_{pt}^{(B)}(Q_{c}^{2})}
\theta(Q_{c}^{2}-\vec{\kappa}^{2}) +{\cal F}_{pt}(x,\vec{\kappa}^2)
\theta(\vec{\kappa}^{2}-Q_{c}^{2})\,.
\label{eq:4.5}
\ee

Because the accepted propagation radius $R_{c} \sim 0.3$ fm for perturbative
gluons is short compared to a typical range of strong
interaction, the dipole cross section (\ref{eq:4.00}) evaluated with
the DGSF (\ref{eq:4.5}) would miss an interaction strength in the soft region,
for large color dipoles. In Ref.\cite{NZHERA,BFKLRegge} this missing strength for
large dipoles has been modeled
by the non-perturbative, soft mechanism with energy-independent dipole cross
section, whose specific form \cite{NZHERA,JETPVM} has been driven by early
analysis \cite{NZ91} of the nonperturbative two-gluon exchange and tested
against the diffractive vector meson production data \cite{JETPVM}.
More recently several closely related models for $\sigma_{\rm soft}(r)$ have
appeared in the literature, see for instance models for dipole-dipole
scattering via polarization of non-perturbative QCD vacuum \cite{Nachtmann}
and the model of soft-hard two-component pomeron \cite{LANDSH}. In the
spirit of eq.~(\ref{eq:4.1}) one can parameterize interaction of large
color dipoles in terms of the genuinely soft, nonperturbative
component of DGSF. The principal point about this non-perturbative component
of DGSF is that it must not be subjected to pQCD evolution. Thus the
arguments about the hard-to-soft diffusion driven rise of perturbative
DGSF even at small $\vec{\kappa}^{2}$ do not apply to the non-perturbative DGSF
and we take it energy-independent,
\be
{\cal F}^{(B)}_{soft}(x,\vec{\kappa}^2) = a_{soft}
C_F N_c {\alpha_{s}(\vec{\kappa}^2) \over \pi} \left( {\vec{\kappa}^2 \over
\vec{\kappa}^2 +\mu_{soft}^{2}}\right)^2 V_{N}(\vec{\kappa})\,,
\label{eq:4.6}
\ee
where $\mu_{soft}^2 \ll \mu_{pt}^2$. Furthermore, it is natural that the
soft component of DGSF decreases in the perturbative domain of
$\vec{\kappa}^2 \gsim \mu_{pt}^2$ faster than the perturbative Born term
(\ref{eq:4.2}), which is achieved by the extrapolation of the form suggested
in \cite{NPZLT,Twist4}
\be
{\cal F}(x,\vec{\kappa}^2)= {\cal F}^{(B)}_{soft}(x,\vec{\kappa}^2)
{\kappa_{s}^2 \over
\vec{\kappa}^2 +\kappa_{s}^2} + {\cal F}_{hard}(x,\vec{\kappa}^2)
{\vec{\kappa}^2 \over
\vec{\kappa}^2 +\kappa_{h}^2} \, ,
\label{eq:4.7}
\ee
with $\kappa_{s} \sim \mu_{pt}$.

The above described Ansatz for DGSF must be regarded as a poor man's approximation.
The separation of small-$\vec{\kappa}^{2}$ DGSF into the genuine nonperturbative
component and small-$\vec{\kappa}^{2}$ tail of the hard perturbative DGSF
is not unique. Specifically, we attributed to the latter the same small-$x$
rise as in the DGLAP fits at $Q_{c}^{2}$, though one can not exclude that the
hard DGSF has a small $x$-independent component. The issues of soft-hard
separation and whether the non-perturbative component of DGSF enters
different observables in a universal manner must be addressed in dynamical
models for infrared regularization of perturbative QCD and non-perturbative QCD
vacuum and only can be answered confronting such models to the
experiment. We recall that in the conventional DGLAP analysis the effect
of soft gluons is reabsorbed into the input gluon distributions.

The $\vec{\kappa}$-factorization formulas (\ref{sigmat}) and (\ref{sigmal})
correspond to the full-phase space extension of the LO DGLAP approach at
small $x$. For this reason our Ans\"atze for ${\cal F}_{hard}(x,Q^{2})$
will be based on LO DGLAP fits to the gluon structure function of the
proton $G_{pt}(x,Q^2)$. We consider the GRV98LO \cite{GRV}, CTEQ4L,
version 4.6 \cite{CTEQ} and MRS LO 1998 \cite{MRS} parameterizations.
We take the liberty of referring to our Anz\"atze for DGSF based on those
LO DGLAP input as D-GRV, D-CTEQ and D-MRS parameterizations, respectively.

Our formulas (\ref{sigmat}), (\ref{sigmal}) describe the sea component
of the proton structure function. Arguably these LL${1\over x}$
formulas are applicable at $x\lsim x_{0}= 1\div 3 \cdot 10^{-2} $.
At large $Q^{2}$ the experimentally attainable values of $x$ are not so
small. In order to give a crude idea on finite-energy effects at moderately
small $x$, we stretch our fits to $x\gsim x_{0}$ multiplying the above
Ansatz for DGSF by the purely phenomenological factor $(1-x)^{5}$ motivated
by the familiar large-$x$ behaviour of DGLAP parameterizations of the gluon
structure function of the proton. We also add to the sea components
(\ref{sigmat}), (\ref{sigmal}) the contribution from DIS on valence quarks
borrowing the parameterizations from the respective GRV, CTEQ and MRS fits.
The latter are only available for $Q^2 \geq Q_{c}^2$.
At $x\lsim 10^{-2}$ this valence contribution is small and fades away rapidly
with decreasing $x$, for instance see \cite{BFKLRegge}.

\section{The parameters of DGSF for different DGLAP inputs}

Our goal is a determination of small-$x$ DGSF in the whole range of
$\vec{\kappa}^{2}$ by adjusting the relevant parameters to the experimental
data on small-$x$ $F_{2p}(x,Q^2)$ in the whole available region of $Q^2$
as well as the real photoabsorption cross section. The theoretical
calculation of these observables is based on Eqs.~(\ref{sigmat}),
~(\ref{sigmal}),~(\ref{eq:4.7}).

The parameters which we did not try adjusting but
borrowed from early work in the color dipole picture are $R_{c}=0.27$ fm,
i.e., $\mu_{pt}=0.75$ GeV and the frozen value of the LO QCD coupling
with $\Lambda_{QCD}=0.2$ GeV:
\be
\alpha_{S}(Q^{2})=
{\rm min}\left\{0.82,{4\pi\over \beta_{0}\log{Q^{2}\over
\Lambda_{QCD}^{2}}}\right\}\, .
\label{eq:5.1}
\ee
We recall that the GRV, MRS and CTEQ fits
to GSF start the DGLAP evolution at quite a different soft-to-hard interface
$Q_{c}^{2}$ and diverge quite a lot, especially at moderate and small
$\vec{\kappa}^{2}$. The value of $Q_{c}^{2}$ is borrowed from these fits and
is not a free parameter.

The adjustable parameters are $\mu_{soft}$, $a_{soft}$, $m_{u,d}$,
$\vec{\kappa}_s^2$ and $\vec{\kappa}_h^2$ (for the heavier quark masses we take
$m_s=m_{u,d}+0.15 $GeV and $m_{c}=1.5$ GeV). The both $m_{u,d}$
and $\mu_{soft}$ have clear physical meaning and we have certain
insight into their variation range form the early work on color dipole
phenomenology of DIS.  The r\^ole of these parameters is as follows.
The quark mass $m_{u,d}$ defines the transverse size of the $q\bar{q}
=u\bar{u},d\bar{d}$ Fock state of the real photon, whose natural scale
is the size of the $\rho$-meson. Evidently, roughly equal values
of $F_{2p}(x,Q^2)$ can be obtained for somewhat smaller ${\cal F}(x,Q^{2})$
at the expense of taking smaller $m_{u,d}$, 
i.e. larger size of the photon, and vise versa.
Therefore, though the quark mass does not explicitly
enter the parameterization  for ${\cal F}(x,\vec{\kappa}^2)$,
the preferred value of $m_{u,d}$ could have been correlated with the
DGLAP input. We find that it is sufficient to take the universal
$m_{u,d} = 0.22$ GeV.

The $\mu_{soft}^{-2}$ defines the soft scale in which the non-perturbative
glue is confined, and controls the $r$-dependence of, and in conjunction
with $a_{soft}$ sets the scale for, the dipole cross section for large
size $q\bar{q}$ dipoles in the photon. We find that it is sufficient to
take the universal $a_{soft} = 2$ and $\kappa_{s}^2=3$ GeV$^{2}$
for the parameter of suppression of the hard tale of non-perturturbative
soft glue.

The magnitude of the dipole cross section at large and moderately small
dipole size depends also on the soft-to-hard interpolation of DGSF,
which is sensitive to DGLAP inputs for perturbative component
$G_{pt}(x,Q^{2})$. This difference of DGLAP inputs can be corrected
for by adjusting $\mu_{soft}^{2}$ and the hard-to-soft interface
parameter $\vec{\kappa}_h^2$. The slight rise of $\vec{\kappa}_h^2$ helps
to suppress somewhat too strong $x$-dependence of the soft tale of the
perturbative glue. The specific parameterizations for $\vec{\kappa}_h^2$
depend on the DGLAP input and are presented in table 1. Only
$\vec{\kappa}_h^2$ and $\mu_{soft}$ varied from one DGLAP input
to another. The soft components of the D-GRV and D-CTEQ
parameterizations turn out identical.
The eye-ball fits are sufficient for the
purposes of the present exploratory study. The parameters found are
similar to those used in \cite{NPZLT,Twist4} where the focus has been
on the description of diffractive DIS.

\begin{center}
{Table 1. The parameters of differential gluon structure function
for different DGLAP inputs.\vspace{0.3cm}\\}

 \begin{tabular}{|r|c|c|c|}
\hline
& D-GRV & D-MRS & D-CTEQ  \\ \hline\hline
 LO DGLAP input & GRV98LO \cite{GRV}&  MRS-LO-1998 \cite{MRS} &
CTEQ4L(v.4.6) \cite{CTEQ}\\
$Q_c^2$, GeV$^2$            & 0.895 & 1.37 & 3.26  \\
$\kappa_h^2$, GeV$^2$ & $\left(1 + 0.0018\log^4{1\over x}\right)^{1/2}$
& $\left(1 + 0.038\log^2{1\over x}\right)^{1/2}$ &
$\left(1 + 0.047\log^2{1\over x} \right)^{1/2}$  \\
$\mu_{soft}$, GeV     & 0.1 & 0.07 & 0.1\\ \hline
 \end{tabular}
\end{center}

One minor problem encountered in numerical differentiation of all
three parameterizations for $G_{pt}(x,Q^2)$
was the seesaw $\vec{\kappa}^2$-behavior of the resulting DGSF (\ref{eq:4.1}),
which was an artifact of the grid interpolation routines. Although
this seesaw behavior of DGSF would be smoothed out in integral observables
like $G(x,Q^2)$ or $F_{2p}(x,Q^2)$, we still preferred to remove
the unphysical seesaw cusps and have smooth DGSF.
This was achieved by calculating DGSF from (\ref{eq:4.1})
at the center of each interval of the $Q^2$-grid and further interpolating
the results between these points. By integration of the so-smoothed
${\cal F}_{pt}(x,Q^2)$ one recovers the input $G_{pt}(x,Q^2)$.
The values of $Q_c^2$ cited in Table 1 corresponds to centers of the
first bin of the corresponding $Q^{2}$-grid.

\begin{figure}[!htb]
   \centering
   \epsfig{file=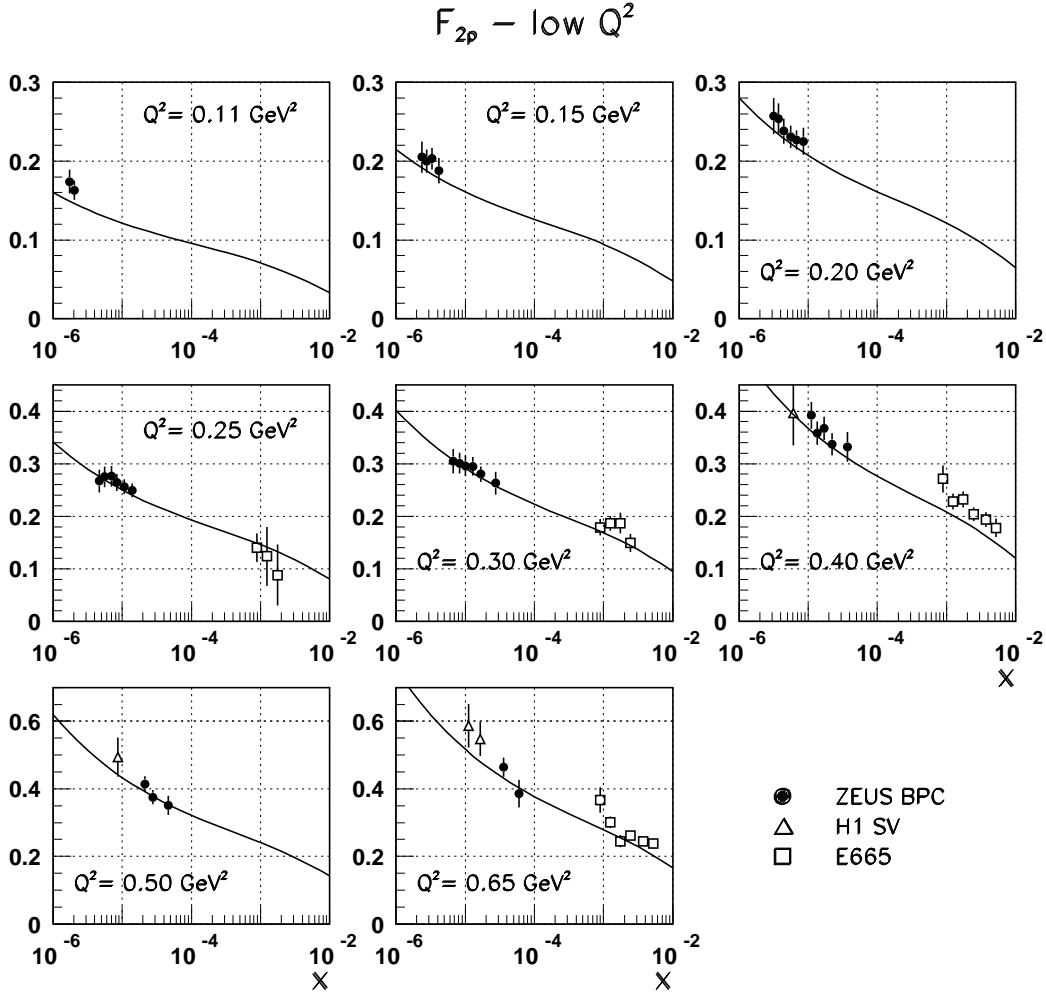,width=16cm}
   \caption{\em The $\vec{\kappa}$-factorization description
of the experimental data on $F_{2p}(x,Q^2)$ in the low $Q^2$ region;
black circles are ZEUS BPC data \cite{ZEUSBPC},
open triangles denote H1 shifted vertex (SV) data \cite{H1shifted},
open squares are E665 data \cite{E665}.
Solid line represents $\vec{\kappa}$-factorization results for the D-GRV parameterization
of the differential gluon structure function ${\cal F}(x,\vec{\kappa}^2)$.}
   \label{F2protonLowQ2}
\end{figure}

\section{The description of the proton structure function $F_{2p}(x,Q^2)$ }

We focus on the sea dominated leading log${1\over x}$ region of $x< 10^{-2}$.
The practical calculation of the proton structure function involves the
two running arguments of DGSF: $x_{g}$ and $\vec{\kappa}^{2}$. We recall that in
the standard collinear DGLAP approximation one has $\vec{\kappa}^{2} \ll \vec{k}^{2}
\ll Q^{2}$ and $x_{g}\approx 2x$, see eq.~(\ref{eq:3.2.1}).
Within the $\vec{\kappa}$-factorization one finds that the
dominant contribution to $F_{2p}(x,Q^{2})$ comes from $M_{t}^{2}\sim Q^{2}$
with little contribution from $M_{t}^{2}\gsim Q^{2}$.
Because at small $x_g$ the $x_g$ dependence of ${\cal F}(x_g,Q^{2})$ is
rather steep, we take into account the $x_{g}-x$ relationship
(\ref{eq:3.1.11}). Anticipating the results on effective intercepts
to be reported in section 7, we
notice that for all practical purposes one can neglect the impact of $\vec{\kappa}$
on the relationship (\ref{eq:3.1.11}), which simplifies greatly the numerical
analysis. Indeed, the $x_{g}$ dependence of
${\cal F}(x_g,\vec{\kappa}^{2})$ is important only at large $\vec{\kappa}^{2}$, which
contribute to $F_{2p}(x,Q^{2})$ only at large $Q^{2}$; but the larger
$Q^{2}$, the better holds the DGLAP ordering $\vec{\kappa}^{2} \ll k^{2}, Q^{2}$.
Although at small to moderate $Q^{2}$ the DGLAP the ordering breaks down,
the $x_{g}$ dependence of ${\cal F}(x_g,\vec{\kappa}^{2})$ is weak here.

\begin{figure}[!htb]
   \centering
   \epsfig{file=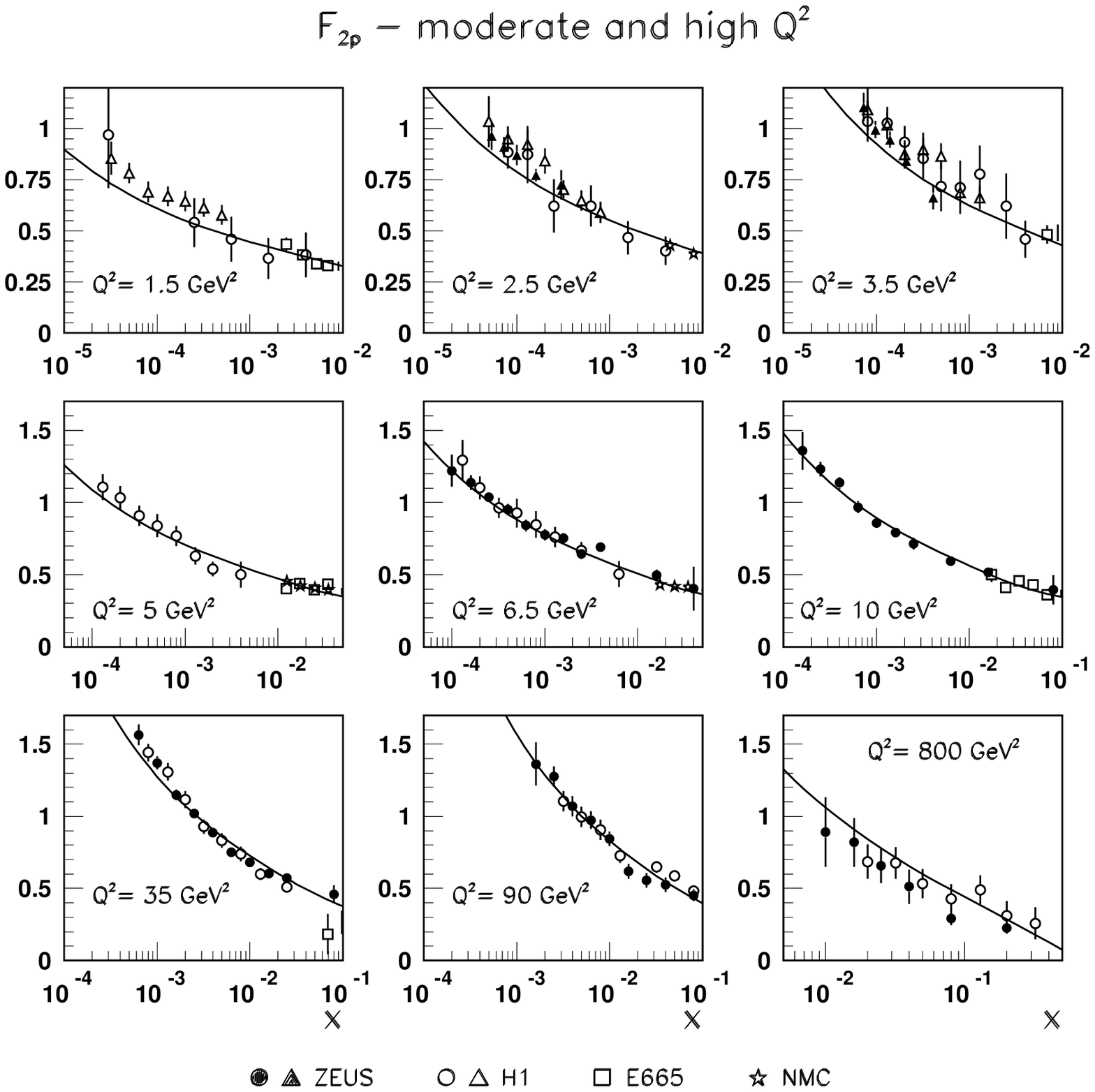,width=16cm}
   \caption{\em The $\vec{\kappa}$-factorization description
of the experimental data on $F_{2p}(x,Q^2)$ in the
moderate and high $Q^2$ region;
black circles and triangles are ZEUS data \cite{ZEUSlarge},
\cite{ZEUSshifted}, open circles and triangles
show H1 data \cite{H1large}, \cite{H1shifted},
open squares are E665 data \cite{E665},
stars refer to NMC results \cite{NMC}.
Solid line represents $\vec{\kappa}$-factorization results for the D-GRV parameterization
of the differential gluon structure function ${\cal F}(x,\vec{\kappa}^2)$.}

\label{F2protonLargeQ2}
\end{figure}

\begin{figure}[!htb]
   \centering
   \epsfig{file=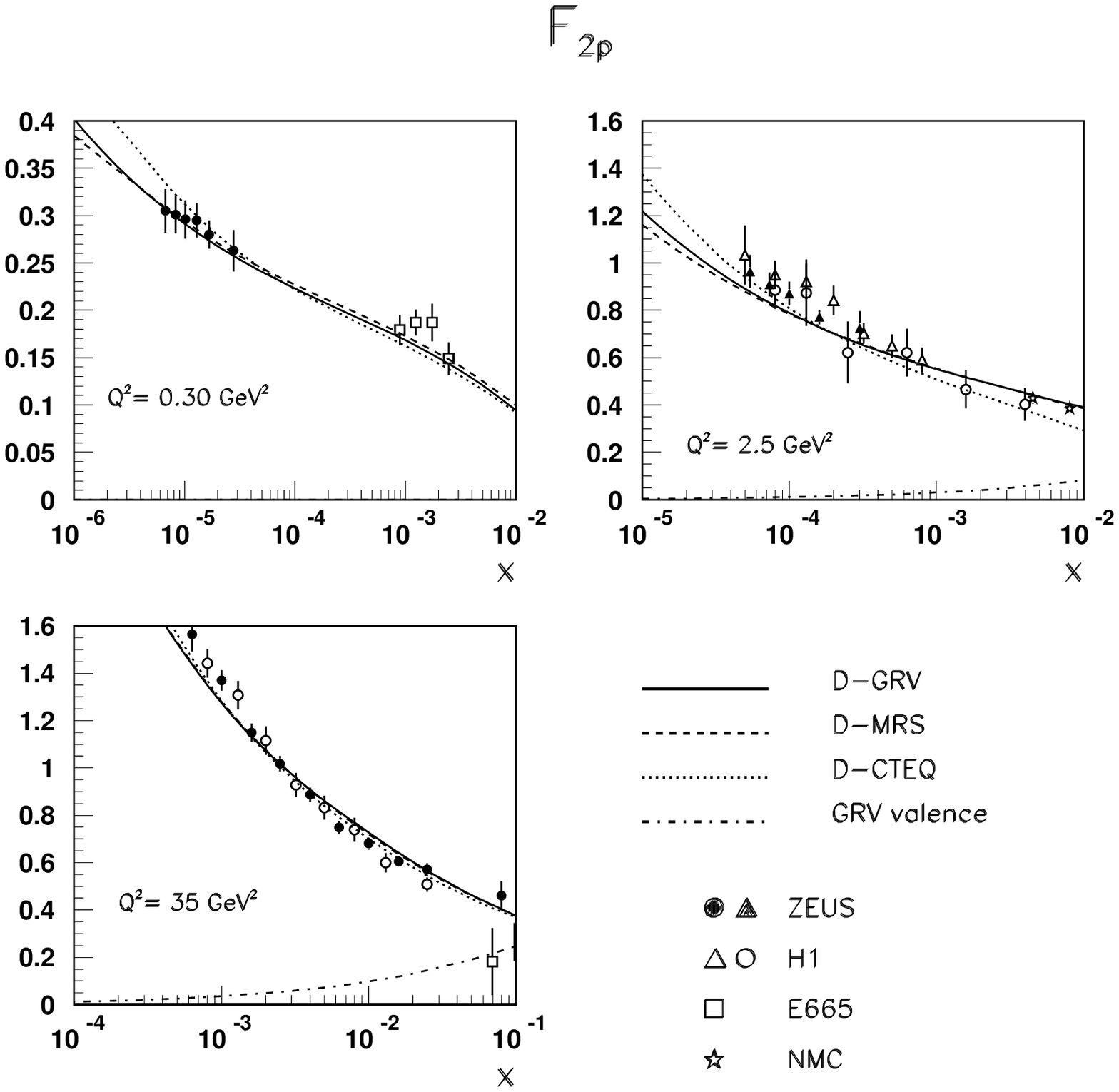,width=14cm}
   \caption{\em A comparison of the $\vec{\kappa}$-factorization description
of the experimental data on $F_{2p}(x,Q^2)$ for several values of
$Q^{2}$ with the D-GRV, D-CTEQ and D-MRS
parameterizations of the differential gluon structure
function ${\cal F}(x,\vec{\kappa}^2)$. The contribution to $F_{2p}(x,Q^2)$
from DIS off valence quarks is shown separately for larger $Q^2$.}
   \label{F2pcompare}
\end{figure}

Obviously, achieving a good agreement
from small to moderate to large $Q^{2}$ is a highly nontrivial task,
because strong modification of the soft contribution to ${\cal F}(x,\vec{\kappa}^2)$
unavoidably echos in the integrated gluon SF
throughout the whole range of $Q^2$ and shall affect the
calculated structure function from small to moderate to large $Q^{2}$.

The quality of achieved description of the experimental data
on the small-$x$ proton structure function is illustrated by
figs.~\ref{F2protonLowQ2},~\ref{F2protonLargeQ2}.
The data shown include recent HERA data
(ZEUS \cite{ZEUSlarge}, ZEUS shifted vertex \cite{ZEUSshifted},
ZEUS BPC \cite{ZEUSBPC}, H1 \cite{H1large},
H1 shifted vertex \cite{H1shifted}),
FNAL E665 experiment \cite{E665} and CERN NMC experiment \cite{NMC}.
When plotting the E665 and NMC data, we took the liberty of shifting
the data points from the reported
values of $Q^2$ to the closest $Q^2$ boxes for which
the HERA data were available. For $Q^{2} < Q_{c}^2 =0.9$ GeV$^2$
the parameterizations for valence distributions are not available
and our curves show only the sea component of $F_{2p}(x,Q^2)$,
at larger $Q^{2}$ the valence component is included.

\begin{figure}[!htb]
   \centering
   \epsfig{file=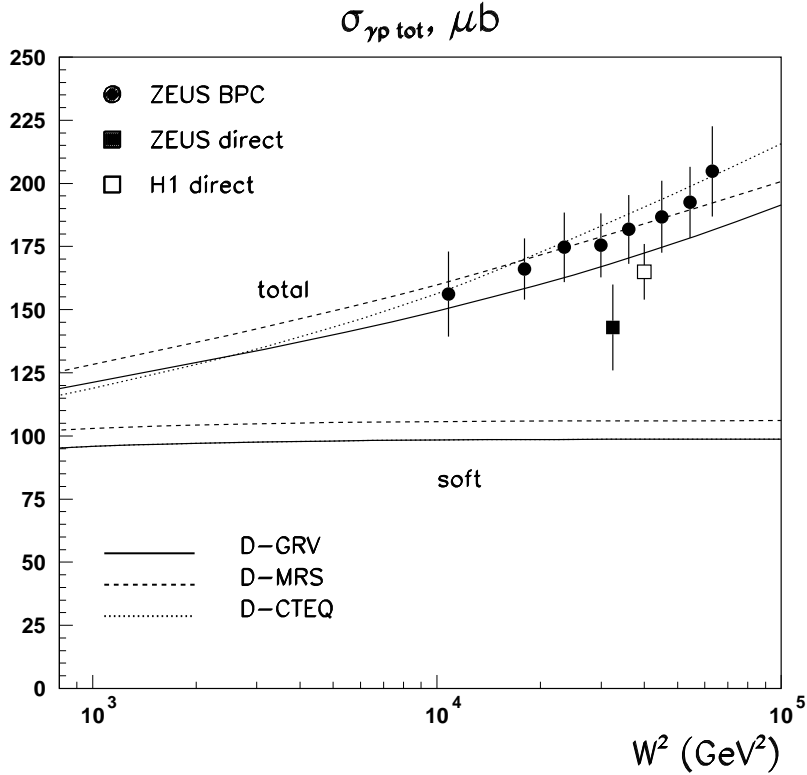,width=12cm}
   \caption{\em A comparison of the $\vec{\kappa}$-factorization description
of the experimental data on real photoabsorption cross section
based on the D-GRV, D-CTEQ and D-MRS
parameterizations  of the differential gluon structure
function ${\cal F}(x,\vec{\kappa}^2)$. The squares
show the experimental data from 1992-93 direct measurements, the
bullets are the results of extrapolation of virtual
photoabsorption to $Q^{2}=0$ (\cite{ZEUSBPC} and references therein).
The soft component of
photoabsorption cross section is shown separately. }
   \label{RealPhoton}
\end{figure}

At $x< 10^{-2}$ the accuracy of our D-GRV description
of the proton structure function is commensurate to
that of the accuracy of standard LO GRV fits. In order not to
cram the figures with nearly overlapping curves, we show the results
for D-GRV parameterization. The situation with
D-CTEQ and D-MRS is very similar, which is seen in Fig.~\ref{F2pcompare},
where we show on a larger scale simultaneously the results
based on the D-GRV, D-CTEQ and D-MRS DGSFs for several selected values
of $Q^{2}$. Here at large $Q^2$ we show separately the contribution from
valence quarks.
The difference between the results for $F_{2p}(x,Q^2)$
for different DGLAP inputs is marginal for all practical purposes,
see also a comparison of the results for $\sigma^{\gamma p}$
for different DGLAP inputs in Fig.~\ref{RealPhoton}.


\section{Real photoabsorption cross section $\sigma^{\gamma p}$}

In the limiting case of $Q^2=0$ the relevant observable is the real
photoabsorption cross section $\sigma^{\gamma p}$. Although the
Bjorken variable is meaningless at very small $Q^{2}$, the
gluon variable $x_{g}$ remains well defined at $Q^{2}=0$,
see eq.~(\ref{eq:3.1.11}). In Fig.~\ref{RealPhoton} we present
our results alongside the results of the direct measurements
of $\sigma^{\gamma p}$ and the results of extrapolation of virtual
photoabsorption cross sections to $Q^{2}=0$,
for the summary of the experimental data see \cite{ZEUSBPC}.
The soft contribution to the cross section is shown separately.
We recall that our parameterizations for
${\cal F}(x,\vec{\kappa}^2)$ give identical soft cross sections for the
GRV and CTEQ inputs (see table 1).
The barely visible decrease of $\sigma^{\gamma p}_{soft}$
towards small $W$ is a manifestation of $\propto (1-x)^5$ large-$x$
behaviour of gluon densities. The extension to lower energies requires
introduction of the secondary reggeon exchanges which goes beyond
the subject of this study.

We emphasize that we reproduce well the observed magnitude
and pattern of the energy dependence of $\sigma^{\gamma p}$
in an approach with manifestly energy-independent
soft contribution to the total cross section.
In this scenario the energy dependence of
$\sigma^{\gamma p}$ is entirely due to the $x_g$-dependent hard component
${\cal F}_{hard}(x_g,Q^2)$ and as such this rise of the total cross
section for soft reaction can be regarded as driven entirely by
very substantial hard-to-soft diffusion. Such a scenario has repeatedly
been discussed earlier \cite{NZHERA,BFKLRegge,VMscan}. Time and time again
we shall see similar effects of hard-to-soft diffusion and vise versa.
Notice that hard-to-soft diffusion
is a straightforward consequence of full phase space calculation of
partonic cross sections and we do not see any possibility for decoupling of
hard gluon contribution from the total cross sections of
any soft interaction, whose generic example is the real photoabsorption.


\chapter{Properties of differential gluon structure
function}

\section{DGSF in the momentum space}

\subsection{Soft/hard decomposition of DGSF}

\begin{figure}[!htb]
   \centering
   \epsfig{file=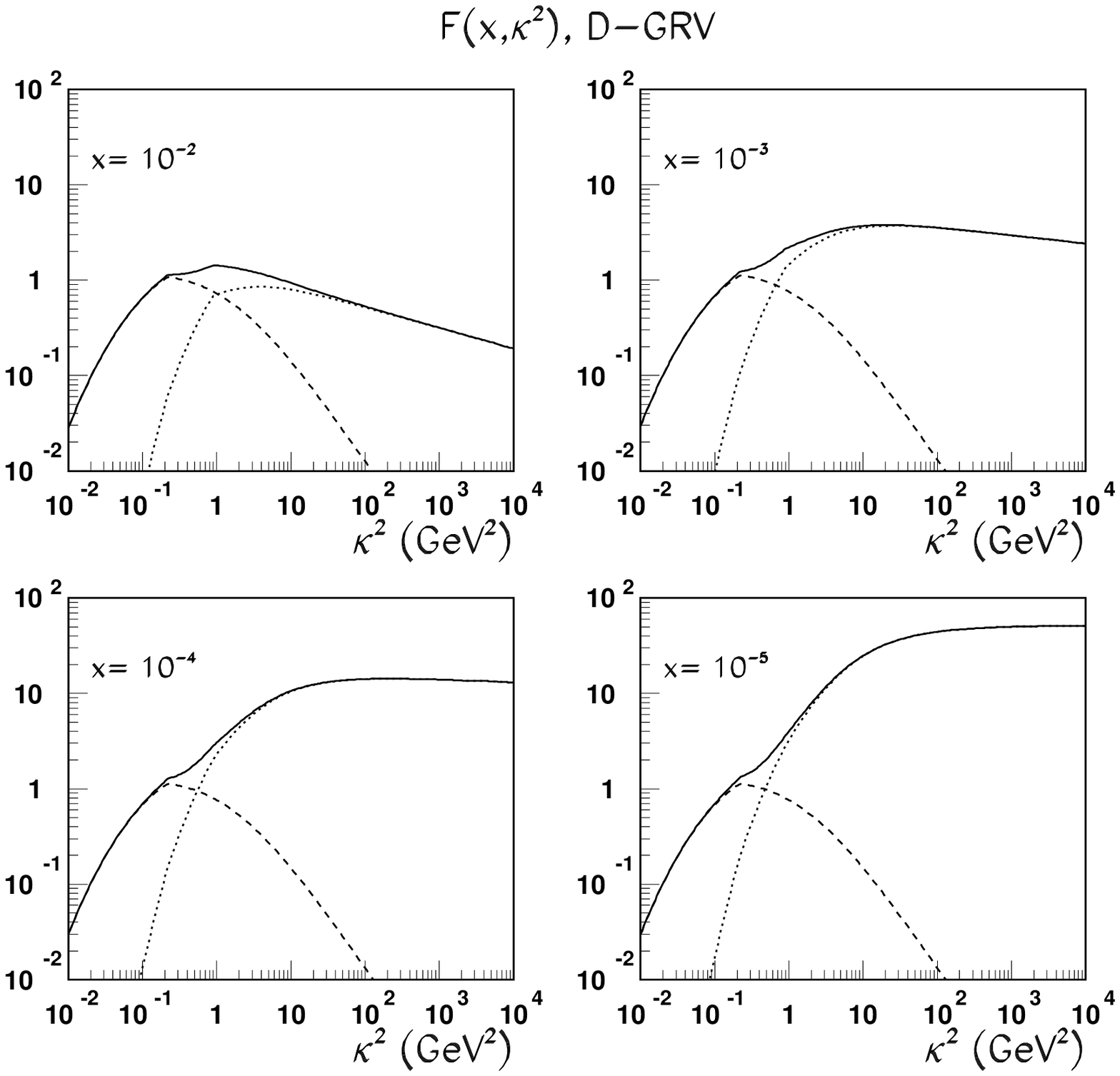,width=14cm}
   \caption{\em D-GRV differential gluon structure function
${\cal F}(x,\vec{\kappa}^2)$ as a function of
$\vec{\kappa}^2$ at several values of $x$. Dashed and dotted
lines represent the soft and hard components; the total
unintegrated gluon density is shown by the solid line}
   \label{DGSF}
\end{figure}

\begin{figure}[!htb]
   \centering
   \epsfig{file=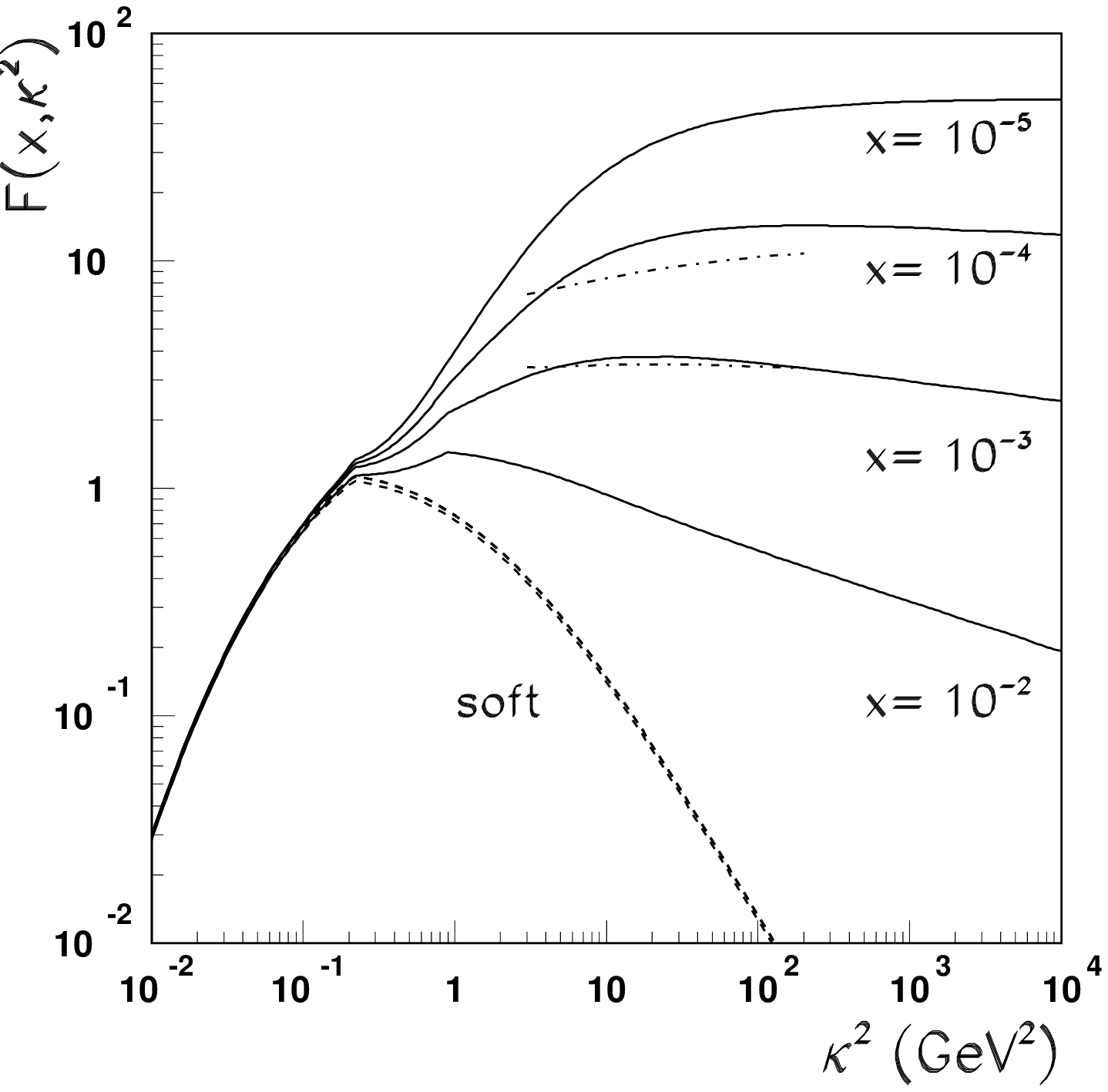,width=10cm}
   \caption{\em The same as in Fig.~\ref{DGSF} but overlaid onto one graph
for illustration of the $x$-dependence of ${\cal F}(x,\vec{\kappa}^2)$. The dashed
lines shows the soft component ${\cal F}_{soft}(x,\vec{\kappa}^2)$ and its slight
variation with $x$ due to the finite-$x$ factor $(1-x)^5$. The dot-dashed
curves show the Kwiecinski et al. \protect{\cite{KMS}} results for DGSF
from a $\vec{\kappa}$-factorization phenomenology of $F_{2p}(x,Q^2)$ based
on the solution of the unified  BFKL/DGLAP equation.}
   \label{DGSFoverlaid}
\end{figure}

\begin{figure}[!htb]
   \centering
   \epsfig{file=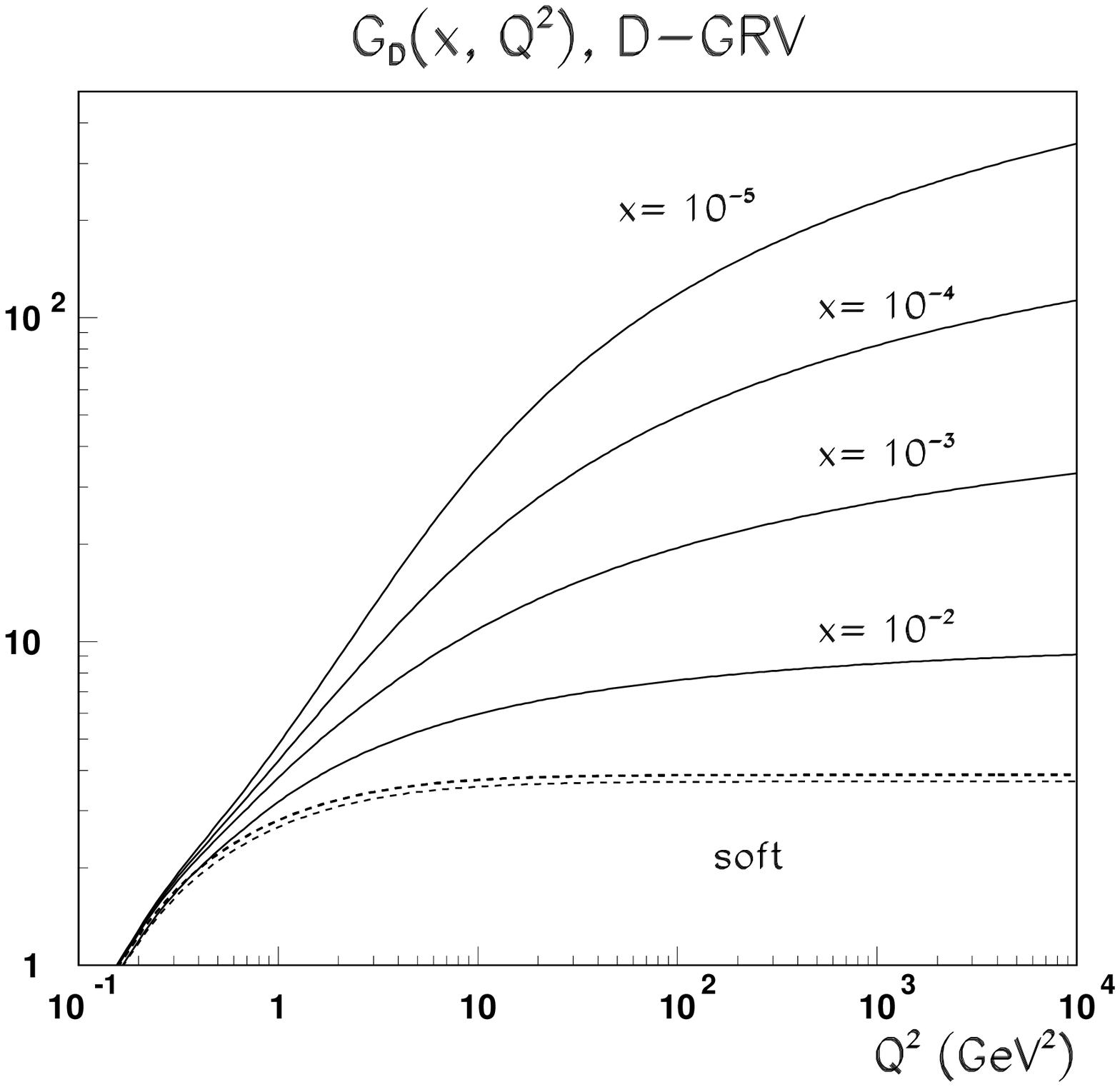,width=10cm}
   \caption{\em The same as Fig.~\ref{DGSF} but for integrated gluon
structure function $G_{D}(x,Q^2)$ as given by the D-GRV
parameterization of the differential gluon structure
function ${\cal F}(x,\vec{\kappa}^2)$, for the discussion see Section 6.2.}
   \label{GSF}
\end{figure}

Now we focus on the $x$ and $\vec{\kappa}^2$ behavior of the
so-determined DGSF starting for the reference with the D-GRV
parameterization. The same pattern holds for DGSF
based on CTEQ and MRS DGLAP inputs, see below.
In figs.~\ref{DGSF} and \ref{DGSFoverlaid}
we plot the differential gluon density ${\cal F}(x_g,Q^2)$,
while in Fig.~\ref{GSF} we show the integrated gluon density
\be
G_{D}(x,Q^{2})=\int^{Q^{2}}_{0} {d\vec{\kappa}^{2} \over \vec{\kappa}^{2}}
{\cal F}(x,\vec{\kappa}^{2})\, .
\label{eq:5.2.1}
\ee
Here the subscript {\sl D} is a reminder that the integrated
$G_{D}(x,Q^{2})$ is derived from DGSF. As such, it
must not be confused with  the DGLAP parameterizations
$G_{pt}(x,Q^{2})$ supplied with the subscript $pt$.

Figs.~\ref{DGSF} and \ref{DGSFoverlaid} illustrate
the interplay at various $x$ of the nonperturbative soft
component of DGSF and perturbative hard contribution supplemented
with the above described continuation into $\vec{\kappa}^{2}\leq Q_{c}^2$.
The soft and hard contributions are shown by dashed and dotted lines
respectively; their sum is given by solid line.

Apart from the large-$x$ suppression factor
$(1-x)^5$ our non-perturbative soft component does not depend on $x$.
At not so small $x=10^{-2}$ it dominates the soft region of
$\vec{\kappa}^2 \lsim 1\div 2 $ GeV$^2$,
the hard component takes over at higher $\vec{\kappa}^2$. The  soft-hard
crossover point is close to $\mu_{pt}^{2}$, but because of the
hard-to-soft diffusion it moves with decreasing $x$
to a gradually smaller $\vec{\kappa}^{2}$.

In this determination of DGSF we focus on the ready-to-use parameterizations,
the dynamical evolution properties of the so-found DGSF will be addressed
elsewhere. In what concerns the relationship between DGSF and the observable
proton structure function, the early work by Kwiecinski et al. \cite{KMS} is
close in spirit to ours, the difference being in a treatment of the
nonperturbative soft component and subjecting DGSF to unified BFKL/DGLAP
evolution. In Fig.~\ref{DGSFoverlaid} 
we present DGSF read off the plots in \cite{KMS};
the agreement with our results is good, which indicates a consistency of
our purely phenomenological parameterizations of DGSF with general
expectations from the BFKL dynamics.

\subsection{Soft/hard decomposition of the integrated gluon structure function}

The r{\^o}le of the soft component if further illustrated by Fig.~\ref{GSF},
where we show the integrated gluon density (\ref{eq:5.2.1})
and its soft and hard components $G_{soft}(x, Q^2)$ and $G_{hard}(x, Q^2)$,
respectively. The soft contribution $G_{soft}(x, Q^2)$
is a dominant feature of the integrated
gluon density $G_{D}(x, Q^2)$ for  $Q^{2}\lsim 1$ GeV$^2$.
It builds up rapidly with $Q^{2}$ and receives the major
contribution from the region $\vec{\kappa}^2 \sim 0.3 \div 0.5 $ GeV$^2$.
Our Ansatz for ${\cal F}_{soft}(x,\vec{\kappa}^2)$ is such that it starts decreasing
already at $\vec{\kappa}^2 \sim 0.2$ GeV$^2$ and  vanishes rapidly
beyond $\vec{\kappa}^{2} \gsim \kappa_{soft}^2$, see figs. 10,11. Still the residual
rise of the soft gluon density beyond $Q^2 \sim  0.5 $ GeV$^2$
is substantial: $G_{soft}(x, Q^2)$ rises by about the factor two
before it flattens at large $Q^{2}$. We emphasize that
$G_{soft}(Q^2)$ being finite at large $Q^{2}$ is quite natural
--- a decrease of $G_{soft}(Q^2)$ at large $Q^{2}$ only is possible
if ${\cal F}_{soft}(Q^{2})$ becomes negative valued at large $Q^2$, which
does not seem to be a viable option.

At moderately small $x=10^{-2}$ the scaling violations are still weak
and the soft contribution $G_{soft}(x, Q^2)$ remains a substantial
part, about one half, of integrated GSF $G_{D}(x, Q^2)$ {\em at all} $Q^{2}$.
At very small $x\lsim 10^{-3}$ the scaling violations in the gluon
structure function are strong and $G_{hard}(x,Q^{2}) \gg G_{soft}(x,Q^{2})$
starting from $Q^{2} \sim$ 1-2 GeV$^2$.

\subsection{Soft/hard decomposition of the proton structure
function $F_{2}(x,Q^{2})$ }

\begin{figure}[!htb]
   \centering
   \epsfig{file=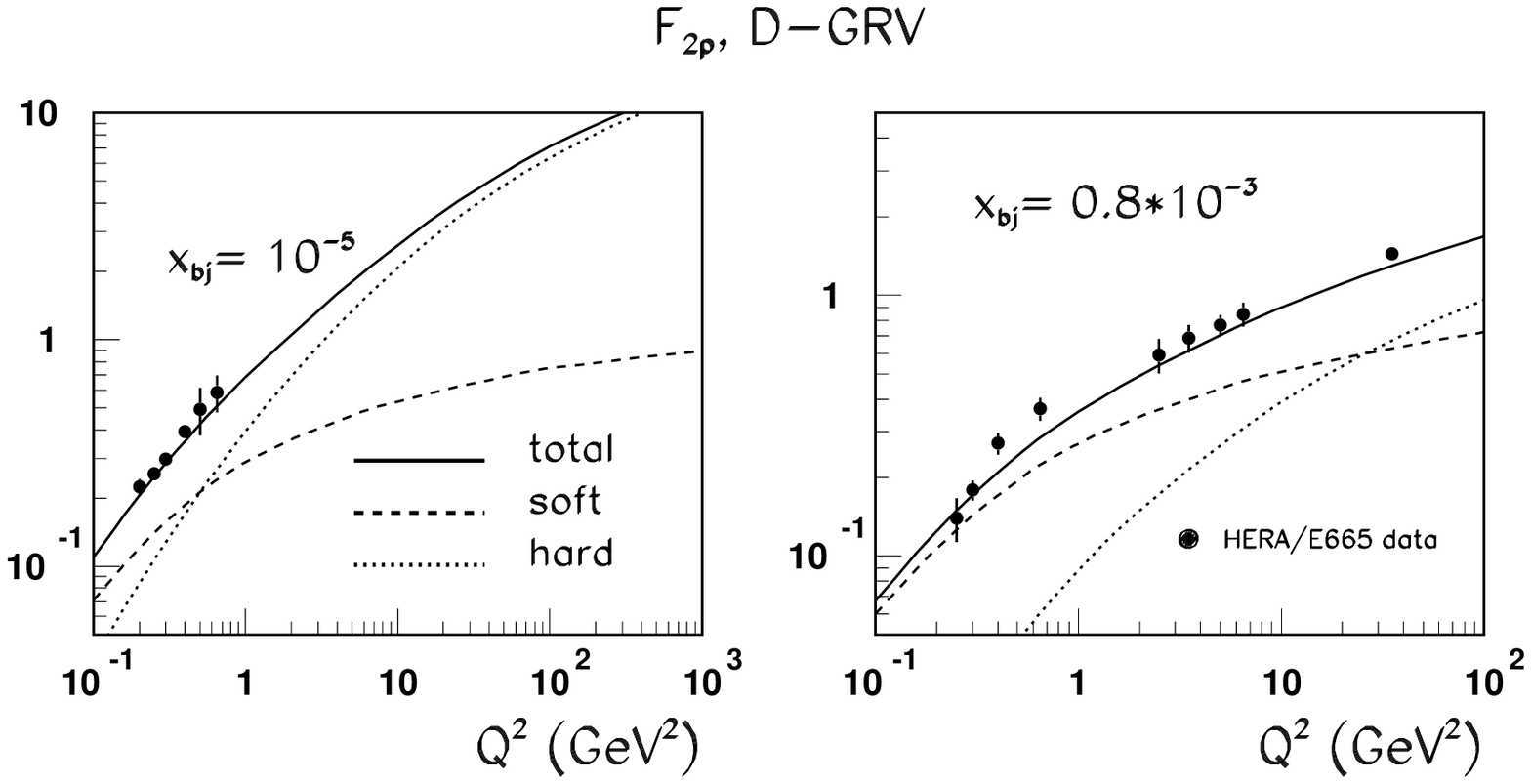,width=16cm}
   \caption{\em The soft-hard decomposition of $\vec{\kappa}$-factorization
results for the
proton structure function $F_{2p}(x,Q^{2})$
evaluated with the
D-GRV parameterization of the differential gluon structure
function ${\cal F}(x,\vec{\kappa}^2)$.}
   \label{Soft-Hard.F2p}
\end{figure}

Eqs. (\ref{sigmat}), (\ref{sigmal}) define
the soft/hard decomposition of the proton structure
function. In Fig.~\ref{Soft-Hard.F2p} we show $F_{2p}^{hard}(x,Q^{2})$ and
$F_{2p}^{soft}(x,Q^{2})$ as a function of $Q^{2}$ for two
representative values of $x$.
Notice how significance of soft component as a function
of $Q^{2}$ rises from fully
differential ${\cal F}(x,Q^{2})$ to integrated $G_{D}(x,Q^2)$ to
doubly integrated $F_{2p}^{soft}(x,Q^{2})$. At a moderately small
$x\sim 10^{-3}$, the soft contribution is a dominant part of
$ F_{2p}(x,Q^{2})$, although the rapidly rising hard component
$F_{2p}^{hard}(x,Q^{2})$ gradually takes over at smaller $x$.

Notice that not only does $F_{2p}^{soft}(x,Q^2)$ not vanish at large $Q^{2}$,
but also it rises slowly with $Q^{2}$ as
\be
F_{2p}^{soft}(x,Q^{2})\sim \sum e_{f}^2{4 G_{soft}(Q^{2}) \over 3 \beta_{0}}
 \log{1 \over \alpha_S(Q^2)}\,.
\label{eq:6.3.1}
\ee
Again, the decrease of $F_{2p}^{soft}(x,Q^{2})$ with $Q^{2}$
would only be possible at the expense of unphysical negative
valued $G_{soft}(Q^2)$ at large $Q^2$.

\section{DGSF in the $x$-space: effective
intercepts and hard-to-soft diffusion}

\begin{figure}[!htb]
   \centering
   \epsfig{file=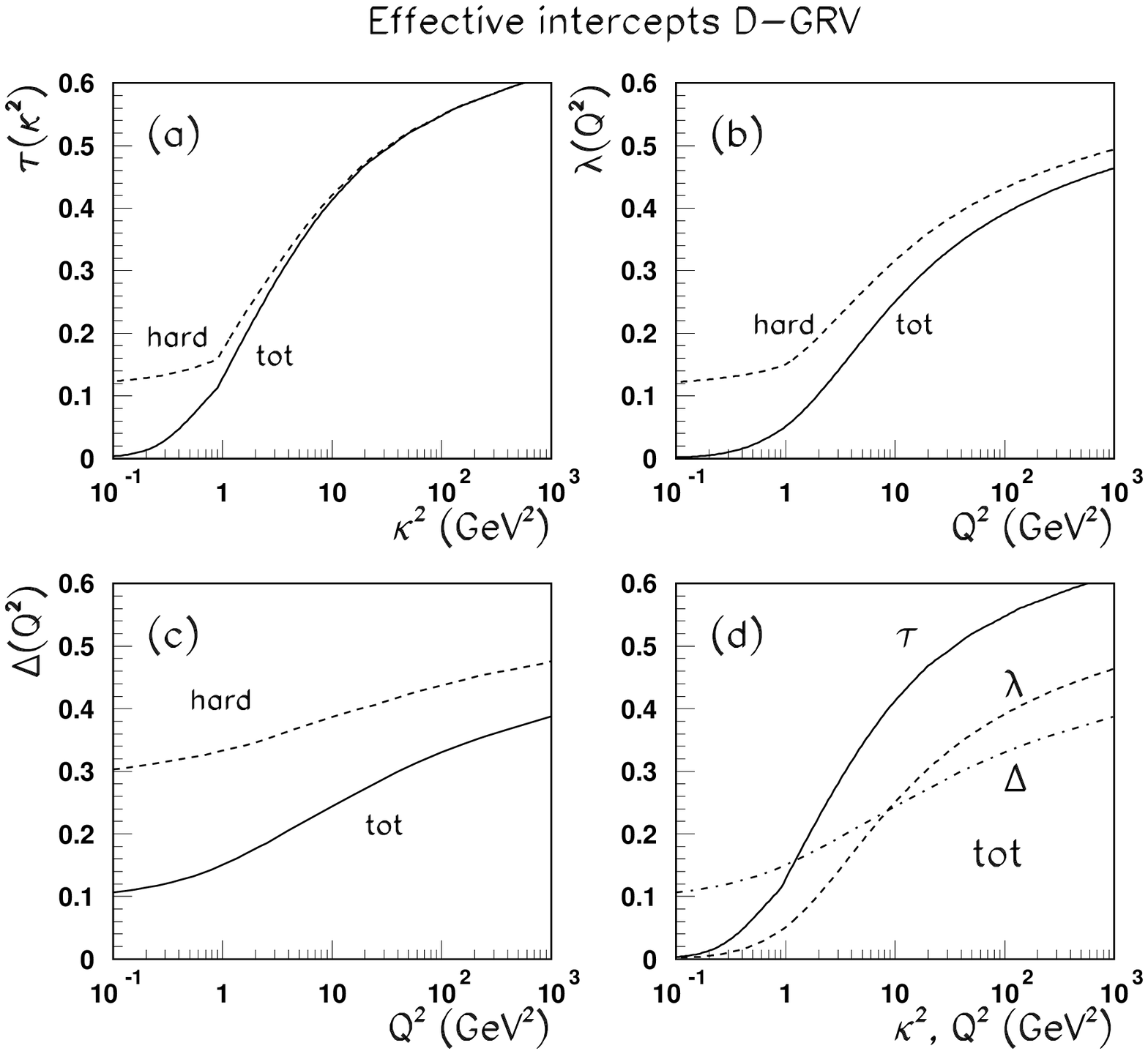,width=10cm}
   \caption{\em Effective intercepts for total, and
hard components of, (a) the differential
gluon structure function ${\cal F}(x,Q^{2})$; (b)
integrated gluon  structure function $G_{D}(x,Q^2)$ and
(c) proton structure function $F_{2p}(x,Q^{2})$ evaluated with the
D-GRV parameterization of the differential gluon structure
function ${\cal F}(x,\vec{\kappa}^2)$. In the box (d) we compare
the effective intercepts $\tau_{eff}(Q^2), \lambda_{eff}(Q^2)$
and $\Delta_{eff}(Q^2)$ for ${\cal F}(x,Q^{2})$, $G_{D}(x,Q^2)$
and $F_{2p}(x,Q^{2})$, respectively. }
   \label{InterceptsGRV}
\end{figure}

\begin{figure}[!htb]
   \centering
   \epsfig{file=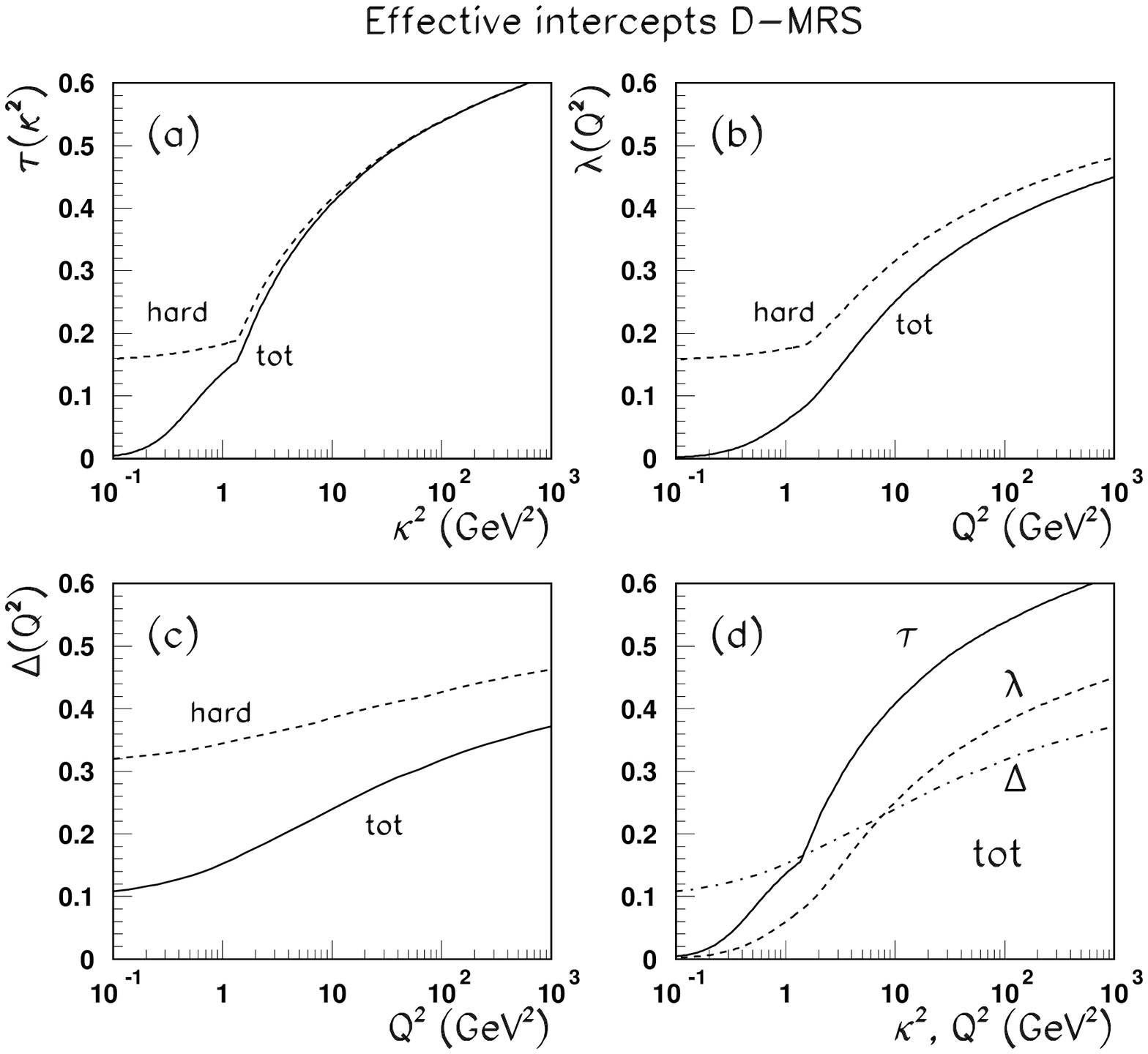,width=10cm}
   \caption{\em Effective intercepts for total, and
hard components of, (a)the differential
gluon structure function ${\cal F}(x,Q^{2})$; (b)
integrated gluon  structure function $G_{D}(x,Q^2)$ and (c)
proton structure function $F_{2p}(x,Q^{2})$ evaluated with the
D-MRS parameterization of the differential gluon structure
function ${\cal F}(x,\vec{\kappa}^2)$. In the box (d) we compare
the effective intercepts $\tau_{eff}(Q^2), \lambda_{eff}(Q^2)$
and $\Delta_{eff}(Q^2)$ for ${\cal F}(x,Q^{2})$, $G_{D}(x,Q^2)$
and $F_{2p}(x,Q^{2})$, respectively.}
   \label{InterceptsMRS}
\end{figure}
\begin{figure}[!htb]
   \centering
   \epsfig{file=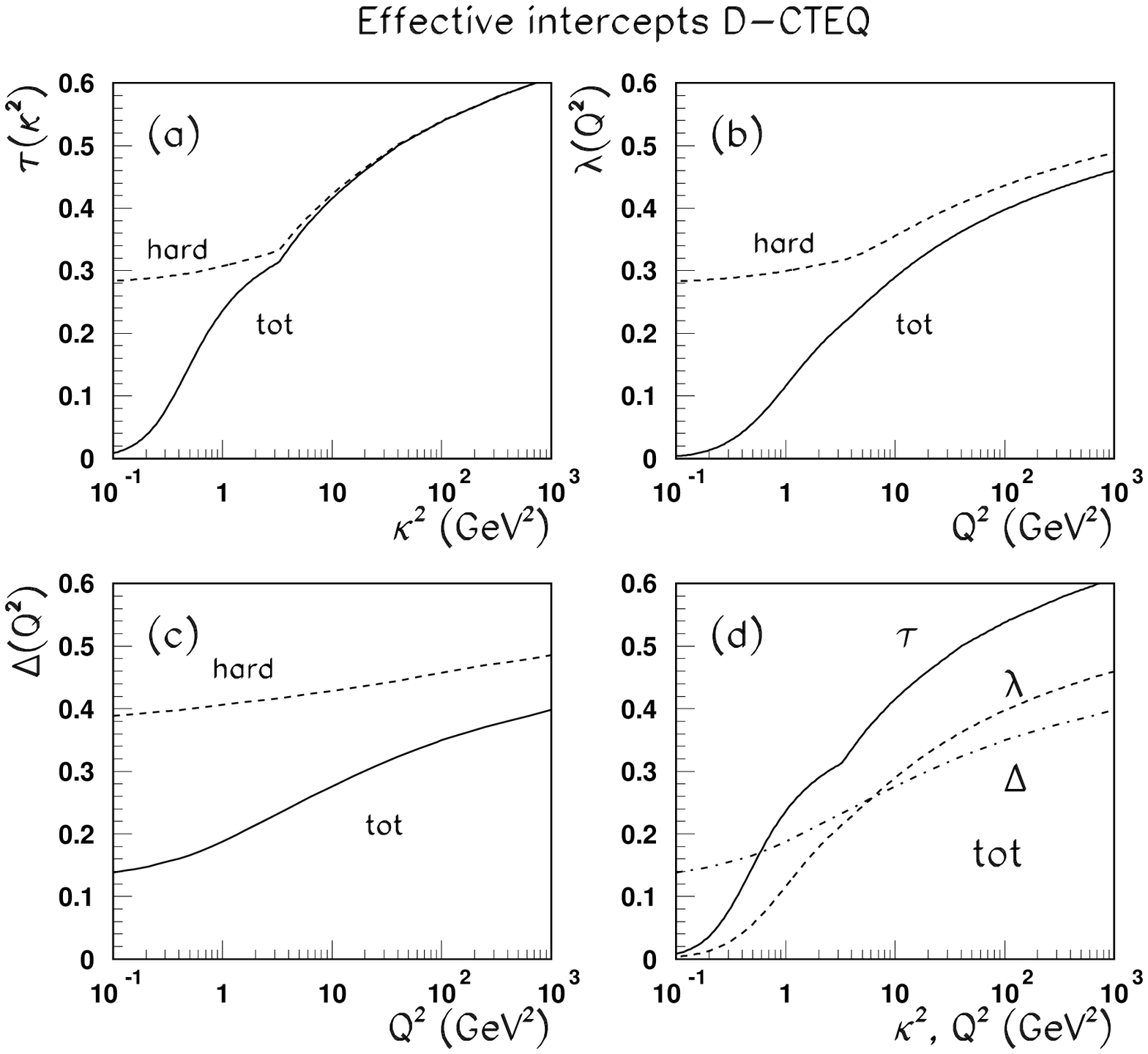,width=10cm}
   \caption{\em Effective intercepts for total, and
hard components of, (a) the differential
gluon structure function ${\cal F}(x,Q^{2})$; (b)
integrated gluon  structure function $G_{D}(x,Q^2)$ and (c)
proton structure function $F_{2p}(x,Q^{2})$ evaluated with the
D-CTEQ parameterization of the differential gluon structure
function ${\cal F}(x,\vec{\kappa}^2)$. In the box (d) we compare
the effective intercepts $\tau_{eff}(Q^2), \lambda_{eff}(Q^2)$
and $\Delta_{eff}(Q^2)$ for ${\cal F}(x,Q^{2})$, $G_{D}(x,Q^2)$
and $F_{2p}(x,Q^{2})$, respectively.}
   \label{InterceptsCTEQ}
\end{figure}

It is instructive to look at the change of the $x$-dependence
from the differential gluon structure function ${\cal F}(x,\vec{\kappa}^2)$
to integrated gluon structure function $G_{D}(x,Q^2)$
and further to proton structure function $F_{2p}(x,Q^2)$.
It is customary to parameterize the $x$ dependence of various
structure functions by the effective intercept. For instance,
the effective intercept $\tau_{eff}$ for differential gluon
structure function is defined by the parameterization
\be
{\cal F}(x,\vec{\kappa}^2) \propto \left({1 \over x}
\right)^{\tau_{eff}(\vec{\kappa}^2)}\,.
\label{eq:6.1}
\ee
One can define the related intercepts $\tau_{hard}$ for the
hard component  ${\cal F}_{hard}(x,\vec{\kappa}^2)$. Notice, that in
our Ansatz $\tau_{soft}\equiv 0$.

The power law (\ref{eq:6.1}) is only a crude approximation to the
actual $x$ dependence of DGSF and the effective intercept $\tau_{eff}$
will evidently depend on the range of fitted $x$. To be more
definitive, for the purposes of the present discussion we define
the effective intercept as
\be
\tau_{eff}(\vec{\kappa}^2) = {\log [{\cal F}( x_{2},\vec{\kappa}^2)/
{\cal F}( x_{1},\vec{\kappa}^2)] \over
\log(x_{1}/x_{2}) }
\label{eq:6.2}
\ee
taking ${x}_{2}=10^{-5}$ and ${x}_{1}=10^{-3}$. The effective intercept
$\tau_{hard}(\vec{\kappa}^2)$ is defined by (\ref{eq:6.2}) in terms of
${\cal F}_{hard}(x,\vec{\kappa}^2)$.

\begin{figure}[!htb]
   \centering
   \epsfig{file=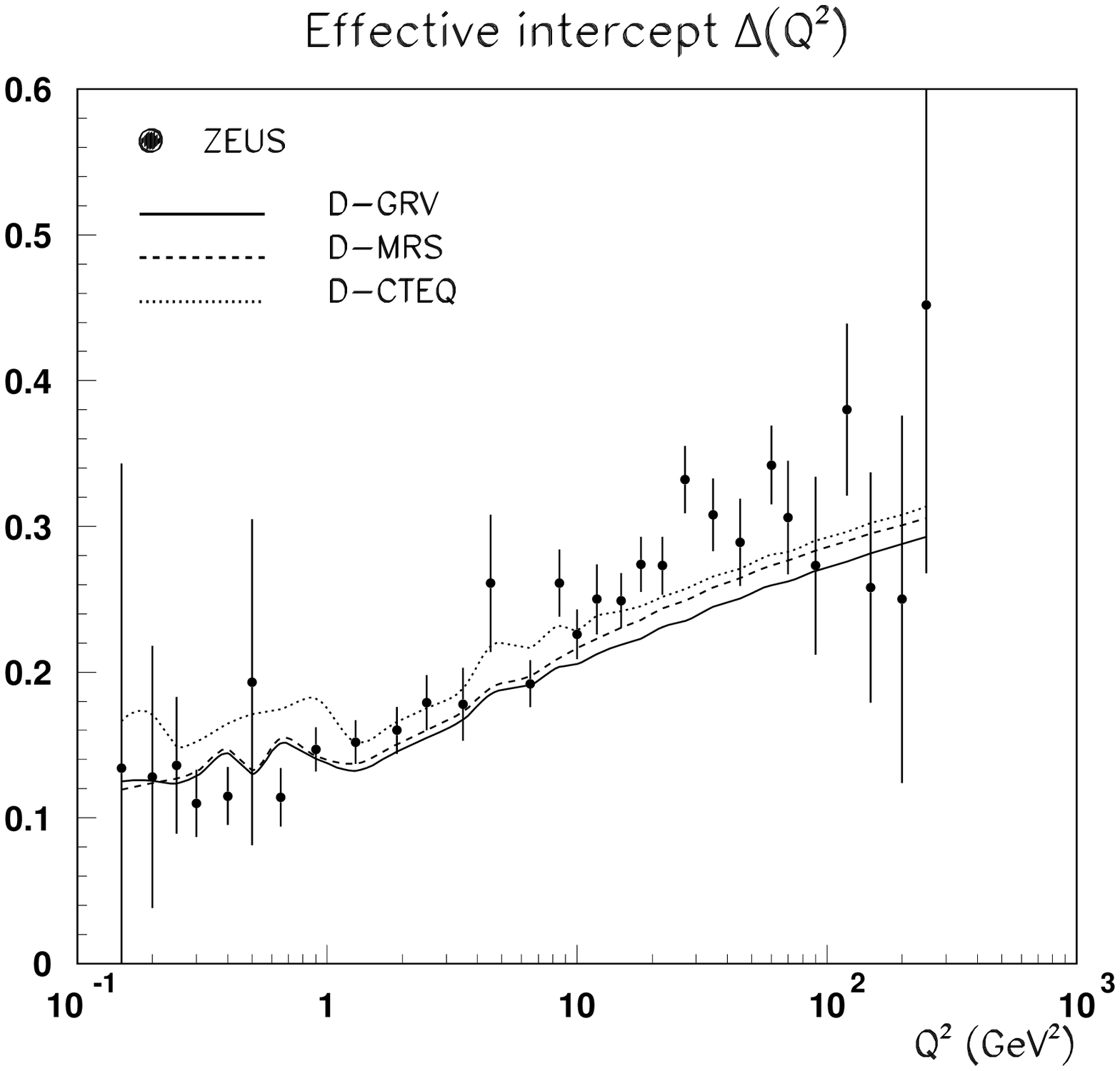,width=10cm}
   \caption{\em Effective intercepts $\Delta(Q^2)$ of the
proton structure function $F_{2p}(x,Q^{2})$ in the HERA domain
evaluated for the D-GRV, D-MRS and D-CTEQ parameterizations for
the differential gluon structure
function ${\cal F}(x,\vec{\kappa}^2)$;
the experimental data points are from ZEUS \cite{ZEUSshifted}}
   \label{Intercepts-HERA}
\end{figure}

One can define the related intercepts $\lambda_{eff},\lambda_{hard}$
for the integrated gluon structure function $G_{D}(x,Q^2)$:
\be
G_{D}(x,Q^2) \propto \left({1 \over x}
\right)^{\lambda_{eff}(Q^2)}\,.
\label{eq:6.3}
\ee

In the case of $F_{2p}(x,Q^2)$ we define the intercept $\Delta(Q^2)$
in terms of the variable ${\overline x}$ defined as
\be
{\bar x} = {Q^2 +M_{V}^2 \over W^2 +Q^2}\sim x_g\, ,
\label{eq:5.2.0}
\ee
where $M_{V}$ is the mass of the ground state vector meson
in the considered flavor channel. Such a replacement allows one
to treat on equal footing $Q^{2} \lsim 1$ GeV$^2$,
where the formally defined Bjorken variable $x$
can no longer be interpreted as a lightcone momentum
carried by charged partons.  For the purposes of the direct comparison with
$\tau(Q^2),\lambda(Q^2)$ and in
order to avoid biases caused by the valence structure function,
here we focus on intercepts  $\Delta_{eff},\Delta_{hard}$  for the
sea component of the proton structure function $F_{2p}^{sea}(x,Q^2)$:
\be
F_{2p}^{sea}(x,Q^2) \propto \left({1 \over {\overline x}}
\right)^{\Delta_{eff}(Q^2)}\,.
\label{eq:6.4}
\ee
The results for the effective intercepts are shown in
figs.~\ref{InterceptsGRV}, \ref{InterceptsMRS} and \ref{InterceptsCTEQ}.

In our simplified hard-to-soft extrapolation of ${\cal F}_{hard}(x,Q^2)$
we attribute to ${\cal F}_{hard}(x,Q^2)$ at $Q^2 \leq Q_{c}^{2}$ the same
$x$-dependence as at $Q^{2}=Q_c^2$ modulo to slight modifications for
the $x$-dependence of $\vec{\kappa}_h^2$. This
gives the cusp in $\tau_{hard}(Q^{2})$ at $Q^2=Q_{c}^2$, i.e., the
first derivative of $\tau_{hard}(Q^{2})$ is discontinuous at $Q^2=Q_{c}^2$.

A comparison of Fig.~\ref{DGSFoverlaid} with
Fig.~\ref{GSF} and further with Fig.~\ref{Soft-Hard.F2p} shows clearly
that only in DGSF ${\cal F}(x,Q^2)$ the effect of the soft component
is concentrated at small $Q^{2}$. In integrated $G_{D}(x,Q^2)$ and especially
in the proton structure function $F_{2p}(x,Q^2)$ the impact of
the soft component extends to much larger $Q^{2}$. The larger
the soft contribution, the stronger is the reduction of $\tau_{eff}$
from $\tau_{hard}$ and so forth, the pattern which is evident from
Fig.~\ref{InterceptsGRV}a to \ref{InterceptsGRV}b to \ref{InterceptsGRV}c,
see also figs.~\ref{InterceptsMRS} and \ref{InterceptsCTEQ}.

The change of effective intercepts from differential ${\cal F}(x,Q^2)$
to integrated $G_{D}(x,Q^2)$ is straightforward, the principal effect is
that $\lambda_{hard}(Q^{2}) < \tau_{hard}(Q^{2})$ and
$\lambda_{eff}(Q^{2}) < \tau_{eff}(Q^{2})$ which reflects the growing
importance of soft component in $G_{D}(x,Q^2)$.
The change of effective intercepts from ${\cal F}(x,Q^2)$ and $G_{D}(x,Q^2)$
to  $ F_{2p} (x,Q^2)$ is less trivial and exhibits two dramatic consequences
of the hard-to-soft and soft-to-hard diffusion. If the standard DGLAP
contribution (\ref{eq:3.2.2}) were all,
then the change from the intercept $\lambda(Q^2)$ for integrated
gluon density to the intercept $\Delta(Q^{2})$ for the proton structure
function $ F_{2p} (x,Q^2)$ would be similar to the
change from $\tau(Q^2)$ to $\lambda(Q^2)$, i.e., the effective intercept
$\Delta_{eff}(Q^2)$ would have been close to zero for $Q^2 \lsim 1$ GeV$^2$.
However, by virtue of the hard-to-soft diffusion phenomenon inherent to the
$\vec{\kappa}$-factorization, $ F_{2p}(x,Q^2)$ receives a contribution
from gluons with $\vec{\kappa}^2 > Q^2$, which enhances substantially
$\Delta_{hard}(Q^{2})$ and $\Delta_{eff}(Q^{2})$.
The net result is that at small to moderately large
$Q^{2}$ we find $\Delta_{hard}(Q^{2})> \lambda_{hard}(Q^{2})$
and $\Delta_{eff}(Q^{2})> \lambda_{eff}(Q^{2})$.  As we emphasized
above in sections 5.3, the rise of real photoabsorption cross
section is precisely of the same origin.

The second effect is a dramatic flattening of effective hard intercept,
$\Delta_{hard}(Q^{2})$, over the whole range of $Q^{2}$. For all
three DGLAP inputs $\Delta_{hard}(Q^{2})$ flattens at approximately
the same $\Delta_{hard} \approx 0.4$.

The whole set of figs.~\ref{InterceptsGRV}--\ref{InterceptsCTEQ}
also shows that the systematics of intercepts
in the hard region of $Q^{2} > Q_{c}^{2}$
is nearly identical for all the three DGLAP inputs.
In the soft region we have a slight inequality $\left.
\tau_{hard}(\vec{\kappa}^{2})\right|_{D-MRS} >  \left.\tau_{hard}(\vec{\kappa}^{2})
\right|_{D-GRV}$, which can be readily attributed to a slight inequality
$Q_{c}^2(MRS)  > Q_c^2(GRV)$. In the case of  CTEQ4L(v.4.6) input
the value of $Q_{c}^{2}(CTEQ)$ is substantially larger than
$Q_{c}^2(MRS), Q_c^2(GRV)$. In the
range  $Q_{c}^2(MRS), Q_c^2(GRV) < \vec{\kappa}^{2}
< Q_{c}^{2}(CTEQ)$ the effective intercept $\tau_{hard}(\vec{\kappa}^2)$
rises steeply with $\vec{\kappa}^2$. This explains a
why in the soft region $\left. \tau_{hard}(\vec{\kappa}^{2})\right|_{CTEQ}$
is significantly larger than for the D-GRV and D-MRS parameterizations. The
difference among intercepts for the three parameterizations decreases
gradually from differential ${\cal F}(x,\vec{\kappa}^2)$ to integrated
$G_{D}(x,Q^2)$ gluon density to the proton structure function $F_{2p}(x,Q^2)$.

Finally, in Fig.~\ref{Intercepts-HERA} we compare our results
for $\Delta_{eff}(Q^2)$ with the recent experimental data
from ZEUS collaboration \cite{ZEUSshifted}.
Since in the experimental fit the range of
$x=[x_{max},x_{min}]$  varies from point
to point, we mimicked the experimental
procedure in our evaluation of $\Delta_{eff}$ from eq.~(\ref{eq:7.4})
by taking ${\overline x}_{2}=x_{max}$ and ${\overline x}_{1}=x_{min}$.
This explains the somewhat irregular $Q^{2}$ dependence.
The experimental data include both sea and valence components.
At $Q^2 > Q_{c}^{2}(GRV) =0.9$ GeV$^2$ we included the valence
component of the structure function taking the GRV98LO parameterization.
For CTEQ4L(v.4.6) and MRS-LO-1998
the values of $Q_{c}^2$ are substantially larger. However, the valence
component is a small correction and we took a liberty of extracting
the valence contribution $F_{2p}^{val}(x,Q^2)$
from GRV fits for $ Q_{c}^{2}(GRV)<Q^2<
Q_{c}^{2}(MRS), Q_{c}^{2}(CTEQ)$. The overall agreement
with experiment is good.
Difference among the three parameterization is marginal and
can of course be traced back to
figs.~\ref{InterceptsGRV}--\ref{InterceptsCTEQ}.

\section{How the gluon densities of $\vec{\kappa}$-factorization
differ from DGLAP gluon densities}

\begin{figure}[!htb]
   \centering
   \epsfig{file=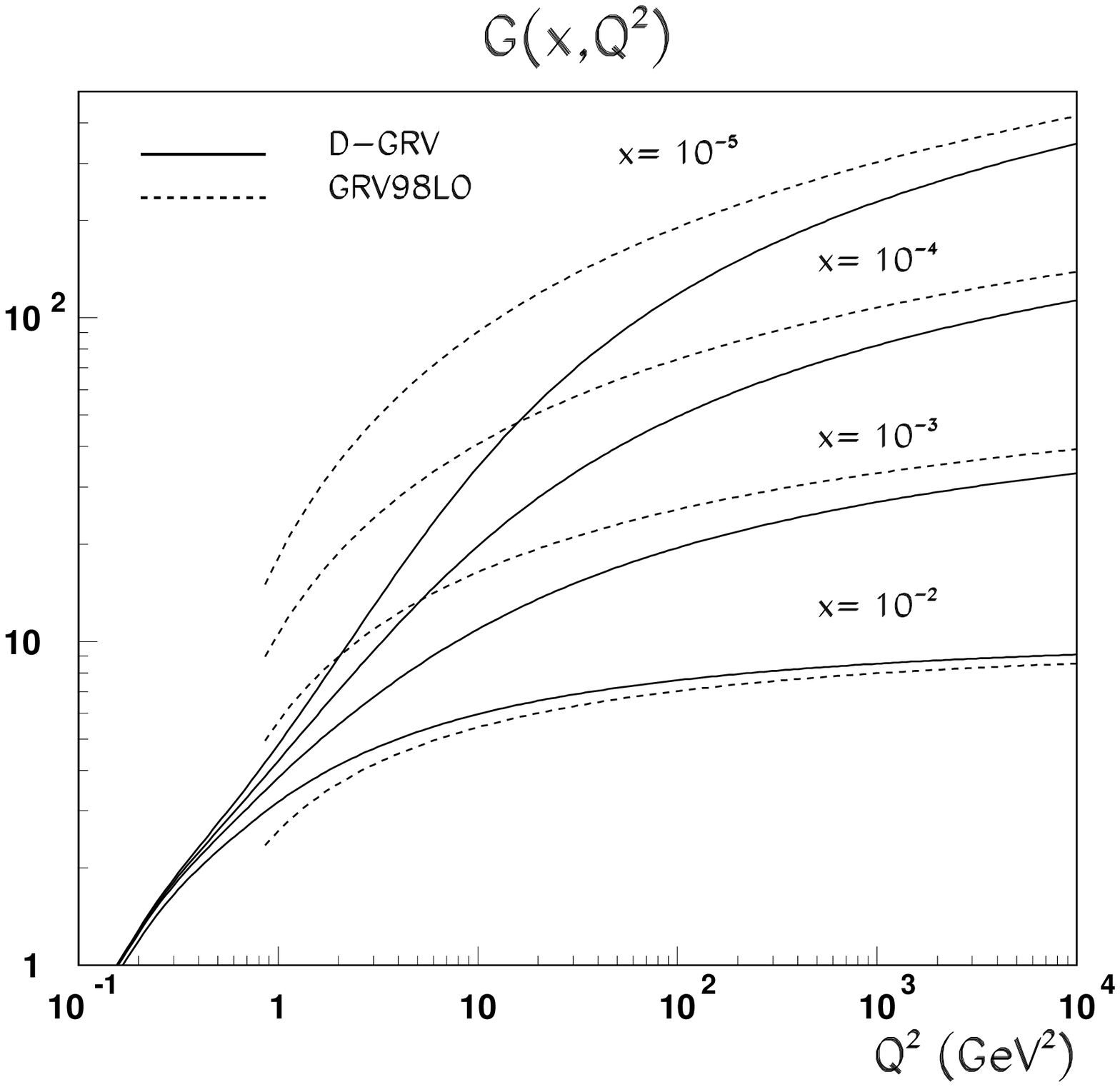,width=10cm}
   \caption{\em Comparison of our results for integrated gluon density
$G_{D}(x,Q^{2})$ evaluated with the
D-GRV parameterization of the differential gluon structure
function ${\cal F}(x,\vec{\kappa}^2)$  with the GRV98L0 DGLAP input parameterization
$G_{pt}(x,Q^2)$. }
   \label{D-GRV.vs.GRV}
\end{figure}

\begin{figure}[!htb]
   \centering
   \epsfig{file=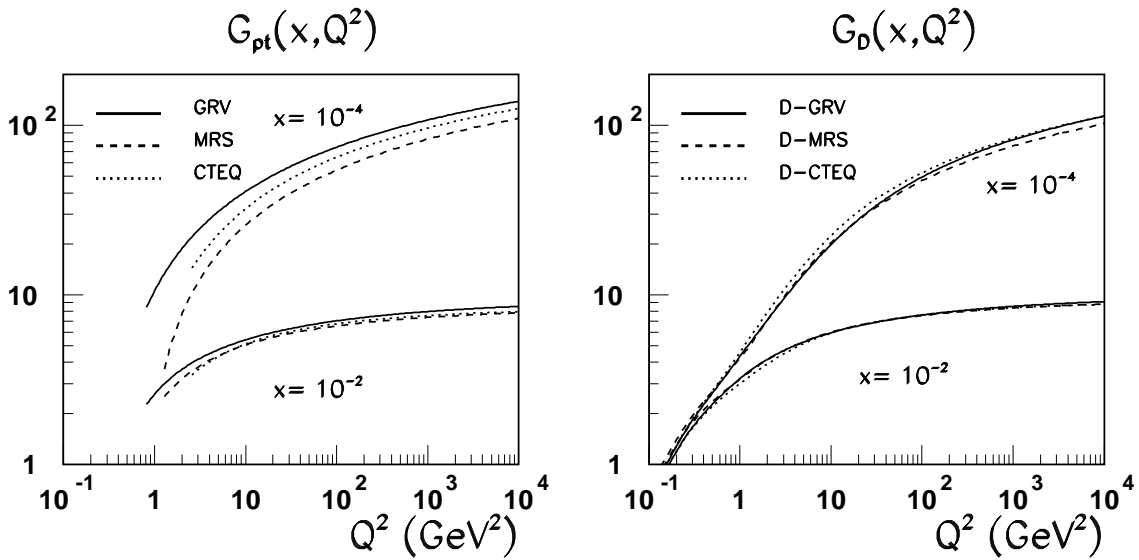,width=15cm}
   \caption{\em A comparison of the divergence of
GRV98L0, CTEQ4L(v.4.6) and MRS-LO-1998 gluon
structure functions $G_{pt}(x,Q^2)$ in the left box
with the divergence of our integrated gluon structure
functions $G_{D}(x,Q^2)$ evaluated for the
D-GRV, D-CTEQ and D-MRS parameterizations for
differential gluon structure function ${\cal F}(x,Q^2)$
at two typical values of $x$}
   \label{G_input.vs.G}
\end{figure}

\begin{figure}[!htb]
   \centering
   \epsfig{file=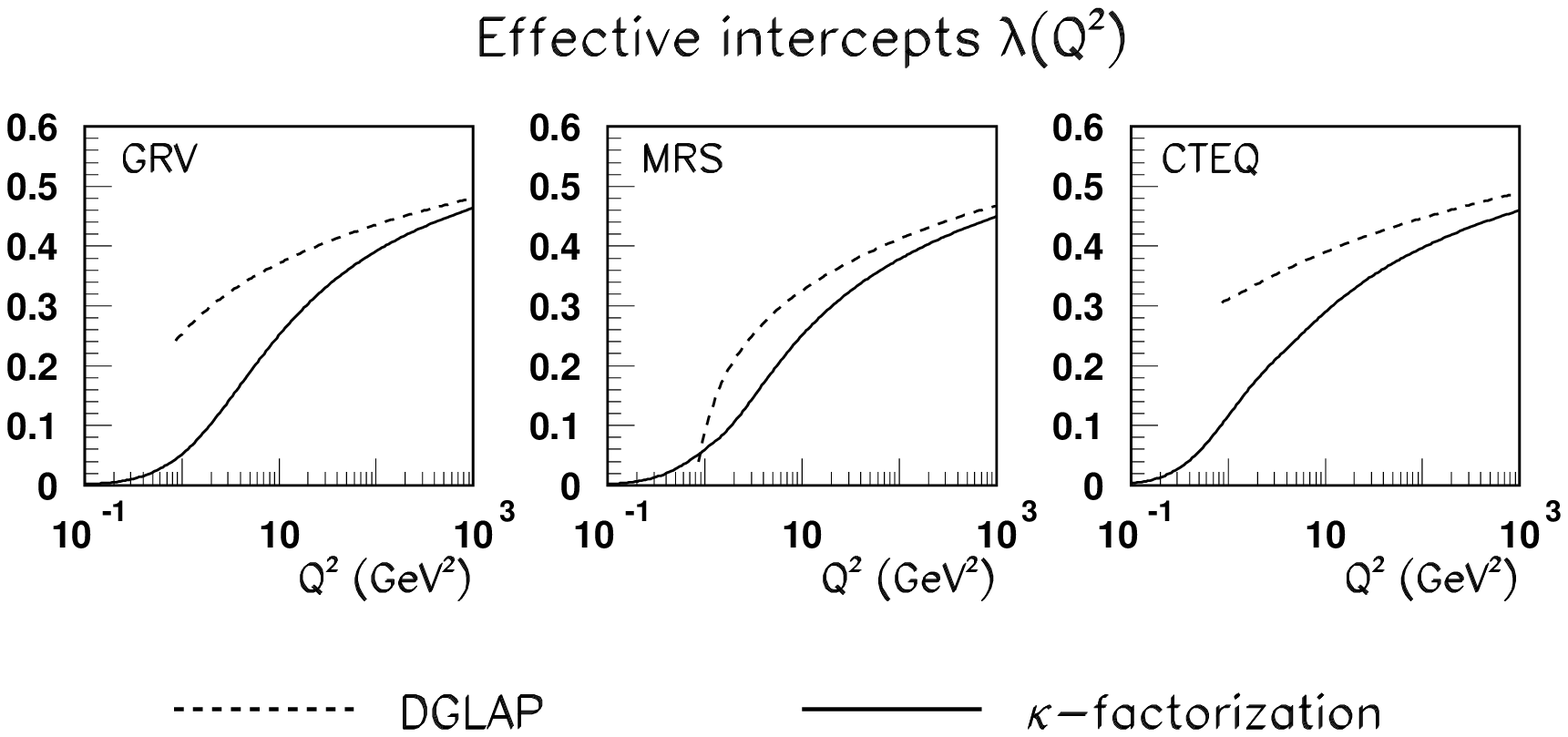,width=16cm}
   \caption{\em A comparison of the intercept $\lambda_{eff}^{(pt)}(Q^2)$
of the $x$-dependence of the GRV98L0, CTEQ4L(v.4.6) and MRS-LO-1998 gluon
structure functions $G_{pt}(x,Q^2)$  with their  counterpart
$\lambda_{eff}(Q^2)$ for integrated $G_{D}(x,Q^2)$ evaluated
with D-GRV, D-CTEQ and D-MRS parameterizations for
differential gluon structure function ${\cal F}(x,Q^2)$.}
   \label{Intercepts-DGLAP.vs.Differ}
\end{figure}

It is instructive also to compare our results for integrated
GSF (\ref{eq:5.2.1}) with the conventional DGLAP fit $G_{pt}(x,Q^2)$.
In Fig.~\ref{D-GRV.vs.GRV} we present such a comparison
 between our integrated D-GRV distribution
(the solid curves) and the GRV98LO distribution
(the dashed curves). As was anticipated in section 3.2,
at very large $Q^{2}$ the two gluon distributions converge.
We also anticipated that at small $x$ and moderate $Q^{2}$ the DGLAP
gluon structure functions $G_{pt}(x,Q^2)$ are substantially larger
than the result of integration of DGSF, see eq.~(\ref{eq:5.2.1}).
At $x=10^{-5}$ they differ by as much as
the factor two-three over a broad range of $Q^{2} \lsim 100$ GeV$^{2}$.
The difference between integrated DGSF and the DGLAP fit decreases
gradually at large $x$, and is only marginal at $x=10^{-2}$.

Recall the substantial divergence of the GRV, MRS and CTEQ gluons
structure functions of DGLAP approximation
$G_{pt}(x,Q^2)$ at small and moderate $Q^{2}$. Contrary to
that, the   $\vec{\kappa}$-factorization D-GRV, D-CTEQ and D-MRS gluon
structure functions $G_D(x,Q^2)$ are
nearly identical. We demonstrate this property
in Fig.~\ref{G_input.vs.G} where we
show integrated $G_D(x,Q^2)$ and their DGLAP counterparts $G_{pt}(x,Q^2)$
for the three parameterizations at two typical
values of $x$. Because of an essentially unified treatment of
the region of $\vec{\kappa}^{2} \leq Q_{c}^{2}$ and strong constraint
on DGSF in this region from the experimental data at small $Q^{2}$,
such a convergence of D-GRV, D-CTEQ and D-MRS DGSFs is not
unexpected.

One can also compare the effective intercepts for our integrated GSF
$G_{D}(x,Q^2)$ with those obtained from DGLAP gluon distributions
$G_{pt}(x,Q^2)$. Fig.\ref{Intercepts-DGLAP.vs.Differ} shows large
scattering of $\lambda^{(pt)}_{eff}(Q^2)$ from one DGLAP input to another.
At the same time, this divergence of different DGLAP input
parameterizations
is washed out to a large extent
in the $\vec{\kappa}$-factorization description of
physical observables (see also (\ref{Intercepts-HERA})).

\section{How different observables probe the DGSF}

The issue we address in this section
is how different observables map the $\vec{\kappa}^2$ dependence of
${\cal F}(x_{g},\vec{\kappa}^{2})$.
We expand on the qualitative discussion in section 3.2 and
corroborate  it with numerical analysis following the discussion
in \cite{NZglue}. We start
with the two closely related quantities ---  longitudinal
structure function $F_{L}(x,Q^{2})$ and scaling violations
$\partial F_{2}(x,Q^{2})/\partial \log Q^{2}$ --- and proceed to
$F_{2p}(x,Q^{2})$ and the charm structure function of the proton
$F_{2p}^{c\bar{c}}(x,Q^{2})$.  This mapping is best studied if in
(\ref{sigmat}) and (\ref{sigmal}) we integrate first over
$\vec{k}$ and $z$. In order to focus on the $\vec{\kappa}^2$
dependence we prefer presenting different observables in terms of
${\cal F}(2x,\vec{\kappa}^{2})$ and $G_{D}(2x,\vec{\kappa}^{2})$
\bea
F_{L}(x,Q^{2}) =
{\alpha_{S}(Q^{2}) \over 3\pi} \sum e_{f}^{2}
\int {d\vec{\kappa}^{2}\over \vec{\kappa}^{2}}\Theta_{L}^{(f\bar{f})}(Q^{2},\vec{\kappa}^{2})
{\cal F}(2x,\vec{\kappa}^{2})\, ,
\label{eq:7.1}
\eea
\bea
{\partial F_{2}(x,Q^{2})\over \partial \log Q^{2}} =
{\alpha_{S}(Q^{2}) \over 3\pi} \sum e_{f}^{2}
\int {d\vec{\kappa}^{2}\over \vec{\kappa}^{2}}\Theta_{2}^{(f\bar{f})}(Q^{2},\vec{\kappa}^{2})
{\cal F}(2x,\vec{\kappa}^{2})\, .
\label{eq:7.2}
\eea
\begin{figure}[!htb]
   \centering
   \epsfig{file=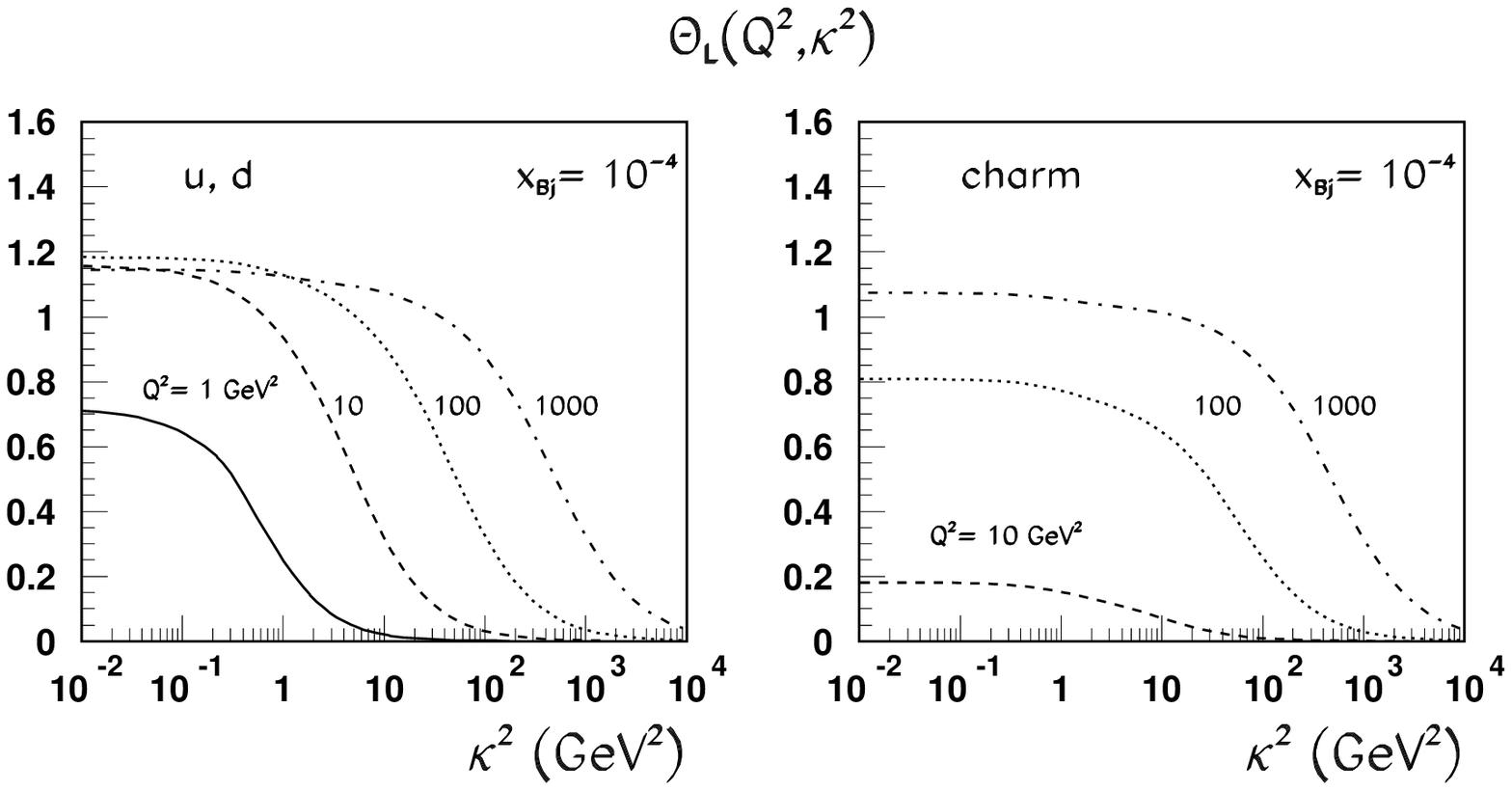,width=16cm}
   \caption{\em The weight function $\Theta_{L}$ for mapping
of the differential gluon structure function ${\cal F}(x,\vec{\kappa}^{2})$ as a function of
$\vec{\kappa}^{2}$ for several values of $Q^{2}$. We show separately the results
for
light flavours, $u,d$, and charm. }
   \label{ThetaL}
\end{figure}

\begin{figure}[!htb]
   \centering
   \epsfig{file=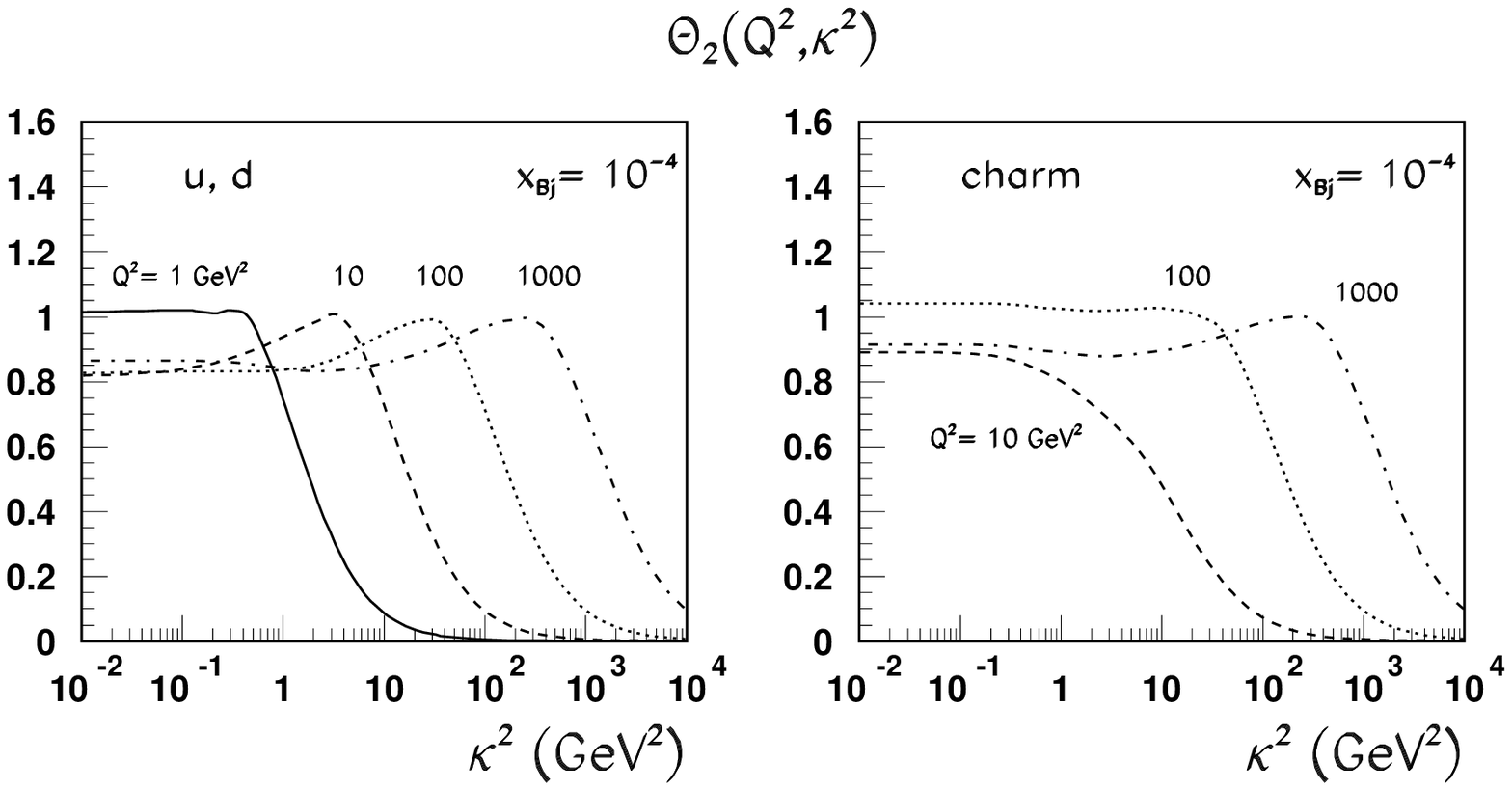,width=16cm}
   \caption{\em The weight function $\Theta_{2}$ for mapping
of the differential gluons structure function ${\cal F}(x,\vec{\kappa}^{2})$
as a function of
$\vec{\kappa}^{2}$ for several values of $Q^{2}$. We show separately the results
for
light flavours, $u,d$, and charm.}
   \label{Theta2}
\end{figure}

In the numerical calculation of $F_{L}(x,Q^{2})$ starting from
eq.~(\ref{sigmal}) we have $x_{g}$ and $\vec{\kappa}^{2}$
as the two running arguments of  ${\cal F}(x_{g},\vec{\kappa}^{2})$.
As discussed above, the mean value of $x_{g}$ is close to $2x$, but the exact
relationship depends on $\vec{\kappa}^2$.  The $\vec{k},z$ integration
amounts to averaging of ${\cal F}(x_{g},\vec{\kappa}^{2})$ over certain
range of $x_{g}$. The result of this averaging is for the most part
controlled by the effective intercept $\tau_{eff}(\vec{\kappa}^2)$:
\be
\langle {\cal F}(x_{g},\vec{\kappa}^{2})\rangle =
\left\langle {\cal F}(2x,\vec{\kappa}^{2})
\left({2x \over x_{g}}\right)^{\tau_{eff}(\vec{\kappa}^{2})}\right\rangle
=r(\vec{\kappa}^2) {\cal F}(2x,\vec{\kappa}^{2})\, .
\label{eq:7.3}
\ee
Because the derivative of $\tau_{eff}(\vec{\kappa}^2)$ changes rapidly around
$\vec{\kappa}^2=Q_{c}^2$, the rescaling factor $r(\vec{\kappa}^2)$ also has
a rapid variation of the  derivative at $\vec{\kappa}^2=Q_{c}^2$,
which in the due turn generates the rapid change of derivatives of
$\Theta_{L,2}^{(f\bar{f})}(Q^{2},\vec{\kappa}^{2})$ around $\vec{\kappa}^2=Q_{c}^2$.
As far as the mapping of differential ${\cal F}(2x,\vec{\kappa}^{2})$
is concerned, this is an entirely marginal effect.
However, if we look at the mapping of integrated gluon structure
function $G_{D}(x,Q^{2})$, which is derived from (\ref{eq:7.1}),
(\ref{eq:7.2}) by integration by parts:
\bea
F_{L}(x,Q^{2}) =
-{\alpha_{S}(Q^{2}) \over 3\pi} \sum e_{f}^{2}
\int {d\vec{\kappa}^{2}\over \vec{\kappa}^{2}}{\partial \Theta_{L}^{(f\bar{f})}(Q^{2},\vec{\kappa}^{2})
\over \partial \log\vec{\kappa}^2}
G_{D}(2x,\vec{\kappa}^{2})
\, ,
\label{eq:7.4}
\eea
\bea
{\partial F_{2}(x,Q^{2})\over \partial \log Q^{2}} =
-{\alpha_{S}(Q^{2}) \over 3\pi} \sum e_{f}^{2}
\int {d\vec{\kappa}^{2}\over \vec{\kappa}^{2}}{\partial \Theta_{2}^{(f\bar{f})}(Q^{2},\vec{\kappa}^{2})
\over \partial \log\vec{\kappa}^2}
G_{D}(2x,\vec{\kappa}^{2})\, ,
\label{eq:7.5}
\eea
then the weight functions $\partial \Theta_{2,L}^{(f\bar{f})}
(Q^{2},\vec{\kappa}^{2}) / \partial \log\vec{\kappa}^2$
will exhibit a slightly irregular behaviour around
$\vec{\kappa}^2=Q_{c}^2$. Evidently, such an irregularity appears in any region of
fast variation of $\tau_{eff}(\vec{\kappa}^2)$;
in our simplified model it is somewhat amplified
by the cusp-like $\vec{\kappa}^2$ dependence of $\tau_{eff}(\vec{\kappa}^2)$.

\begin{figure}[!htb]
   \centering
   \epsfig{file=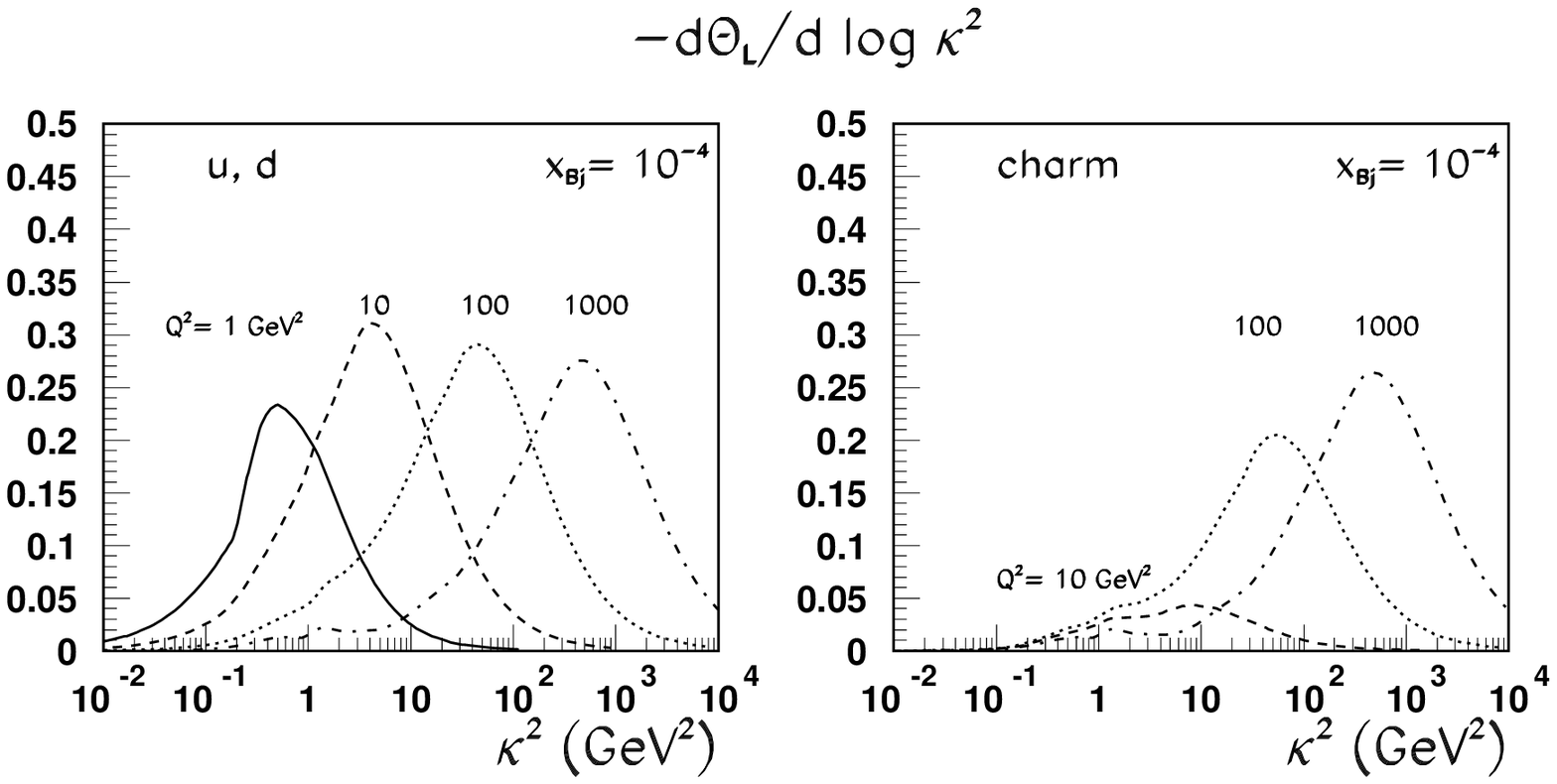,width=16cm}
   \caption{\em The same as Fig.~\ref{ThetaL} but for mapping
of the integrated gluon structure function $G_{D}(x,\vec{\kappa}^{2})$
as a function of $\vec{\kappa}^{2}$ for several values of $Q^{2}$.
We show separately the results for light flavours and charm. }
   \label{PeakL}
\end{figure}

Finally, starting from (\ref{eq:7.5}) one obtains a useful representation
for how the proton structure function $F_{2p}(x,Q^2)$ maps the integrated
gluon structure function:
\bea
F_{2p}(x,Q^{2})=-\int_{0}^{Q^{2}}{dq^2 \over q^{2}}
{\alpha_{S}(q^{2}) \over 3\pi} \sum e_{f}^{2}
\int {d\vec{\kappa}^{2}\over \vec{\kappa}^{2}}{\partial \Theta_{2}^{(f\bar{f})}(q^{2},\vec{\kappa}^{2})
\over \partial \log\vec{\kappa}^2}
G_{D}(2x,\vec{\kappa}^{2}) \nonumber\\
={1\over 3\pi} \sum e_{f}^{2} \int {d\vec{\kappa}^{2}\over \vec{\kappa}^{2}}W^{(f\bar f)}_{2}(Q^{2},\vec{\kappa}^{2})\alpha_{S}(\vec{\kappa}^2)
G_{D}(2x,\vec{\kappa}^{2})
\label{eq:7.6}
\eea

\begin{figure}[!htb]
   \centering
   \epsfig{file=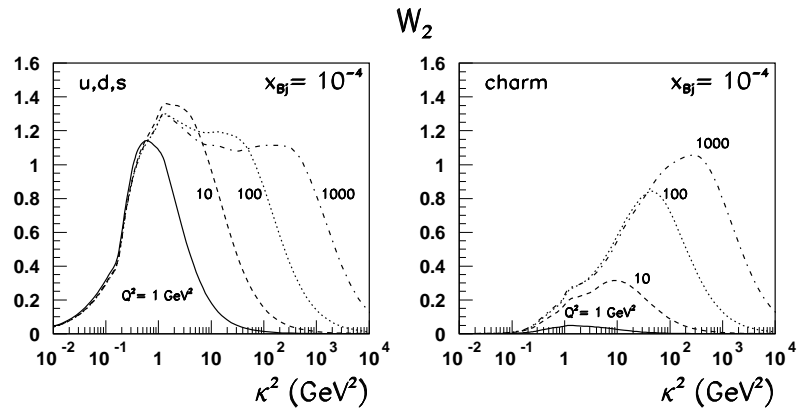,width=12cm}
   \caption{\em The weight function $W_{2}$ for mapping
of the integrated gluons structure function $G_{D}(x,\vec{\kappa}^{2})$
as a function of $\vec{\kappa}^{2}$ for several values of $Q^{2}$.
We show separately the results for light flavours and charm}
   \label{W2}
\end{figure}

\begin{figure}[!htb]
   \centering
   \epsfig{file=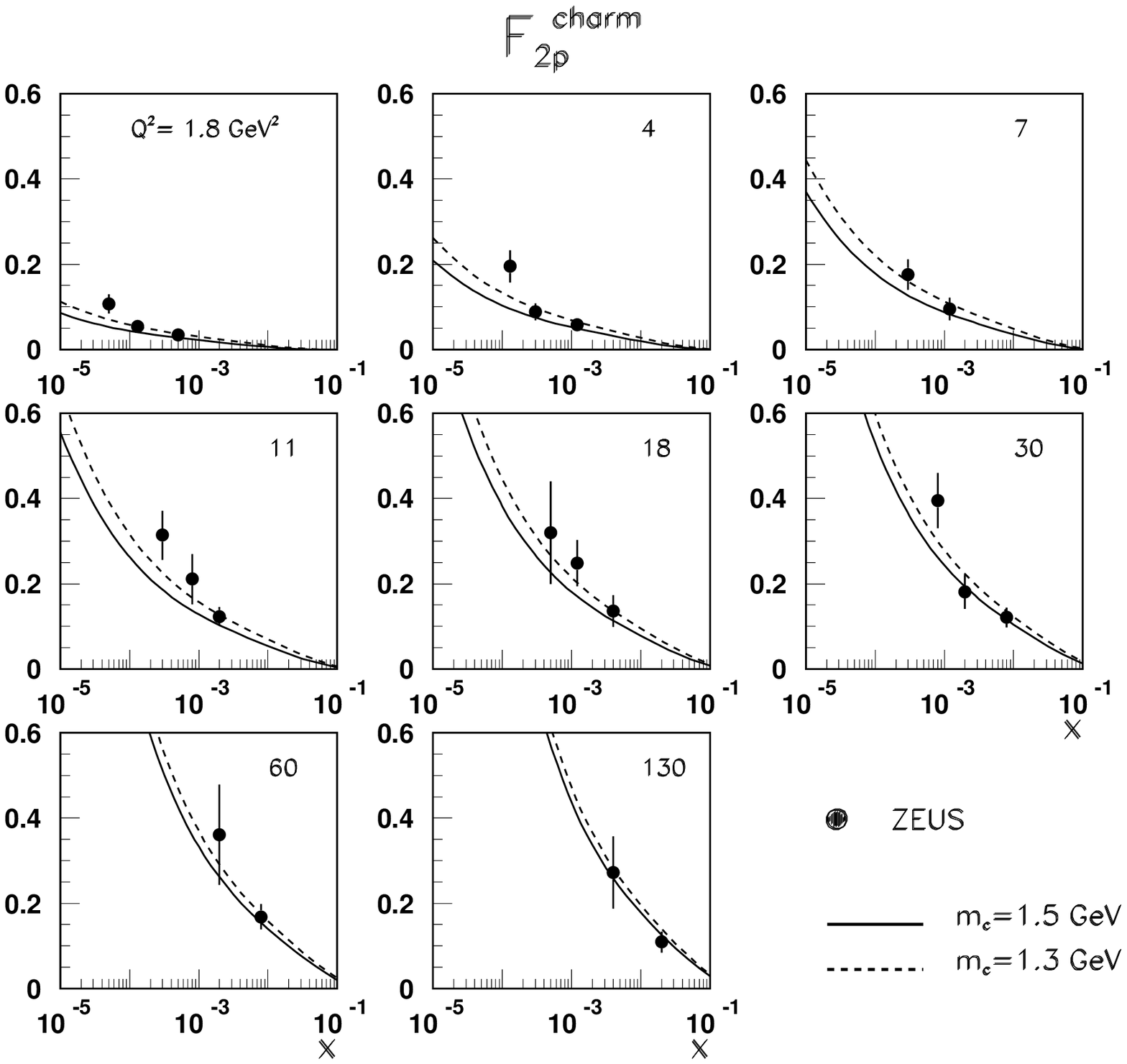,width=12cm}
   \caption{\em A comparison of the  experimental data from ZEUS \cite{ZEUScharm}
on the charm structure function of the proton with $\vec{\kappa}$-factorization
results for $F_{2}^{c\bar{c}}(x,Q^2)$ based
on the D-GRV parameterization of the differential gluon
structure function ${\cal F}(x,Q^2)$.}
   \label{F2charm}
\end{figure}

In figs.~\ref{ThetaL} and \ref{Theta2} we show the weight functions
$\Theta_{L}$ and $\Theta_{2}$.
Evidently, for light flavours and very large $Q^{2}$ they can be approximated
by step-functions
\be
\Theta_{L,2}^{(f\bar{f})}(Q^{2},k^{2})\sim \theta(C_{L,2}Q^{2}-\vec{\kappa}^2)\,,
\label{eq:7.7}
\ee
where the scale factors $C_{L} \sim {1\over 2}$ and $C_{2} \sim 2$ can
be readily read from figures, for the related discussion see \cite{NZglue}.
Note that the value $C_2\sim 2$ corresponds to $\bar C_2 \sim 8$
introduced in Section 3.2.
Recall that the development of the plateau-like behaviour of $\Theta_{L}$ and
$\Theta_{2}$ which extends to $\vec{\kappa}^{2}\sim Q^{2}$ signals the onset of
the leading log$Q^{2}$ approximation. For large $Q^{2}$ in the approximation
(\ref{eq:7.7}) the $\vec{\kappa}^2$ integration can be carried out explicitly and
$F_{L}(x,Q^{2}) \propto  G_{D}(2x,C_{L}Q^2)$. Similarly, $
\partial F_{2}(x,Q^{2})/\partial \log Q^{2} \propto  G_{D}(2x,C_{2}Q^2)$, cf.
eq.~(\ref{eq:3.2.6}).

Still better idea on how $F_{L}$ and scaling violations map the integrated
GSF is given by figs.~\ref{PeakL},\ref{W2}, where we show results for
$-\partial \Theta{(f\bar f)}_L/\partial \log\vec{\kappa}^2$ and $W^{(f\bar f)}_2$.
The first quantity is sharply peaked at $\vec{\kappa}^2 \sim C_{L}Q^{2}$.
The second quantity visibly develops a plateau at large $Q^2$.
As can be easily seen, scaling violations do receive a substantial contribution
from the beyond-DGLAP region of $\vec{\kappa}^{2}> Q^{2}$.

Because of the heavy mass, the case of the charm structure function
$F_{2p}^{c\bar{c}}(x,Q^{2})$ is somewhat special.
Figs.~\ref{PeakL} and \ref{W2} show weak sensitivity of
$F_{2p}^{c\bar{c}}(x,Q^{2})$ to the soft component of
${\cal F}(x,\vec{\kappa}^{2})$, which has an obvious
origin: long-wavelength soft gluons with $\kappa \lsim m_{c}$ decouple from
the color neutral $c\bar{c}$ Fock state  of the
photon, which has a small transverse size $\lsim {1\over m_{c}}$. Our results
for $F_{2p}^{c\bar{c}}(x,Q^{2})$ are shown in Fig.~\ref{F2charm},
the agreement with the recent precision experimental data
from ZEUS \cite{ZEUScharm} is good.

\chapter{Improved determination of the differential glue in proton:
DGD2002 analysis}

During the last two years new high-accuracy measurements of structure function
$F_{2p}$ in the expanded parameter space were presented by 
 ZEUS \cite{ZEUSF2newlow}, \cite{ZEUSF2newhigh} and H1 \cite{H1F2new}. 
In the light of our analysis, the most important improvement was further 
plunging into low-$x$ region (down to $x=6.7\cdot 10^{-7}$, 
\cite{ZEUSF2newlow}) and $2\div 3\%$ accurate determination
of $F_{2p}$ in the small-$Q^2$ ($0.25$ GeV$^2 < Q^2 < 0.65$ GeV$^2$) and 
relatively small-$x$ ($10^{-5} < x < 10^{-3}$) region.
Even a brief comparison of $k_t$-factorization predictions for $F_{2p}$ in this
region based on the differential glue from the previous section 
showed a systematic several-sigma deviation from experimental points.
The second problem with old differential glue was that we did not
quite match the  experimentally measured exponent of $F_{2p}$ 
rise towards high $1/x$. As our analysis showed 
(see Fig.~\ref{Intercepts-HERA}), the predicted intercept 
was $1\div 2 \sigma$ lower at $Q^2 > 10$ GeV$^2$.

Thus, not being able to claim that we reproduce with old
differential glue the $F_{2p}$ data well enough,
we re-extracted the differential gluon structure function.
This time, the eye-ball fits were not acceptable, 
and therefore a $\chi^2$-minimization procedure was carried out.
This communication presents the results of this reevaluation.

\section{Fitting procedure and parameters of DGSF}

Our goal is a determination of small-$x$ DGSF in the whole range of 
$\vec{\kappa}^{2}$ by adjusting the relevant parameters to the experimental 
data on small-$x$ $F_{2p}(x,Q^2)$ in the whole available region of $Q^2$. 
The $k_t$-factorization predictions for $F_{2p}$ 
were calculated at $N=191$ points in $(x, Q^2)$ space and 
compared to the experimentally values  measured at HERA.
These points include all data available at the moment
for low-$Q^2$ region ($Q^2<1$ GeV$^2$) as well as
all data points for several higher values of $Q^2$,
namely, at $Q^2 = 1.5\,\ 2.5\,\ 2.7\,\ 3.5\,\ 4.5\,\ 5.0\,\ 
6.5\,\ 10\,\ 35$ GeV$^2$.The $\chi^2$ was calculated according to
\be
\chi^2 = \sum_{i=1}^N {(F_{2p}^{theor.}-F_{2p}^{exp.})^2 \over 
\sigma^2_{stat} + \sigma^2_{syst}}\,.
\ee

The parameters which we did not try adjusting but
borrowed from early work in the color dipole picture are $R_{c}=0.27$ fm, 
i.e., $\mu_{pt}=0.75$ GeV and the frozen value of the LO QCD coupling
with $\Lambda_{QCD}=0.2$ GeV (\ref{eq:5.1}).
Our previous analysis showed that the parametrizations
D-GRV, D-MRS, and D-CTEQ (that is, DGSF based on GRV, MRS, and
CTEQ fits to integrated gluons) run very close to one another
throughout the whole $x$, $\vec{\kappa}^2$ space. This can be verbalized 
as no matter from what
particular DGLAP parametrization we start, we will arrive
at virtually the same shape of the differential gluon structure function.
To this end, we used only GRV98LO fits, since they are available
for the widest range of $x$ and $Q^2$ values. 

The adjustable parameters are $\mu_{soft}$, $a_{soft}$, $m_{u,d}$, 
$\vec{\kappa}_s^2$ and $\vec{\kappa}_h^2$ (for the heavier quark masses we take
$m_s=m_{u,d}+0.15 $GeV and $m_{c}=1.5$ GeV). The both $m_{u,d}$ 
and $\mu_{soft}$ have clear physical meaning and we have certain 
insight into their variation range form the early work on color dipole
phenomenology of DIS.  The r\^ole of these parameters is as follows.
The quark mass $m_{u,d}$ defines the transverse size of the $q\bar{q}
=u\bar{u},d\bar{d}$ Fock state of the real photon, whose natural scale
is the size of the $\rho$-meson. Evidently, roughly equal values 
of $F_{2p}(x,Q^2)$ can be obtained for somewhat smaller ${\cal F}(x,Q^{2})$ 
at the expense of taking smaller $m_{u,d}$ and vise versa. 
Therefore, though the quark mass does not explicitly 
enter the parameterization  for ${\cal F}(x,\vec{\kappa}^2)$,
the preferred value of $m_{u,d}$ could have been correlated with the 
DGLAP input. We find that it is sufficient to take the universal 
$m_{u,d} = 0.21$ GeV, which is slightly lower than 0.22 GeV, used
before. 

Parameter $\mu_{soft}^{-2}$ defines the soft scale in which the 
non-perturbative 
glue is confined, and controls the $r$-dependence of, and in conjunction 
with $a_{soft}$ sets the scale for, the dipole cross section for large 
size $q\bar{q}$ dipoles in the photon. We find that it is sufficient to 
take the universal $\mu_{soft} = 0.1$ GeV and $\kappa_{s}^2=1$ GeV$^{2}$. 

The magnitude of the dipole cross section at large and moderately small 
dipole size depends also on the soft-to-hard interpolation of DGSF, 
which is sensitive to DGLAP inputs for perturbative component 
$G_{pt}(x,Q^{2})$. This difference of DGLAP inputs can be corrected 
for by adjusting the hard-to-soft interface 
parameter $\vec{\kappa}_h^2$. The slight rise of $\vec{\kappa}_h^2$ helps
to suppress somewhat too strong $x$-dependence of the soft tale of the
perturbative glue. The specific parameterizations for $\vec{\kappa}_h^2$ 
depend on the DGLAP input and are presented in Table 2. Only 
$\vec{\kappa}_h^2$ and $\mu_{soft}$ varied from one DGLAP input 
to another.

In order to be able to assess the uncertainty in determination
of differential glue, we performed several $\chi^2$ minimization procedures
using slightly different sets of free parameters.
The resulting parameters of the fits are shown in Table 2;
below we comment on each fit in detail.

\subsubsection{Fit 1}

When obtaining Fit 1, we adjusted values of $a_{soft}$, transition
point $Q_c^2$ in (\ref{eq:4.5}),
functional form of $\kappa_h^2=\kappa_h^2(x)$, which we took
as a first order polynomial $a+b\log(1/x)$, and the power $\zeta$ of 
the high-$x$ suppressing factor $(1-x)^\zeta$.

The resulting value of $\zeta$ turned our uncomfortably large
$\zeta \approx 11$. Numerically, such a strong suppression
serves as a remedy against somewhat too slow $x$ behavior of $F_{2p}$
as we approach $x\sim 10^{-2}$ from the low-$x$ side.
Although rather artificial, this suppressing factor
does not invalidate our approach since the exact behavior
of the glue in the limit $x\to 1$ lies far beyond the scope of
the present approach. Still, we would like to note the alarming tendency
that even with this factor large part of the overall $\chi^2$
comes precisely from the region we wanted to correct
($10^{-3} < x <10^{-2}$). This might be an indication that
our understanding of this region is far from perfect.
More analysis is needed to settle up this issue.

\subsubsection{Fit 2}

In Fit 2 we opened up another degree of freedom, namely,
we allowed for shifting in the second argument of $G(x,Q^2)$.
As early analyses showed, the best $k_t$-factorization vs. DGLAP
correspondence would be
\be
{\cal F}(x,\vec{\kappa}^2) \leftrightarrow {\partial G(x,Q^2) \over
\partial \log Q^2} \Bigg|_{Q^2 = C \kappa^2}\,.\label{shiftkappa}
\ee
The early analysis gave $C\approx 2$, but in our approach
we treated $C$ as free parameter.

We started from Fit 1 and let the $\chi^2$ slide to its minimum
as we freed parameter $C$. We expected $C>1$ at minimum point.
Indeed, our predictions with Fit 1 for high-$Q^2$ region 
suffered from too high values at high $x$ and 
too low values at very small $x$. 
As early analysis indicated, when we shift the $\vec{\kappa}^2$ scale 
according to (\ref{shiftkappa}), we make the unintegrated gluon
density increase at $x<10^{-3}$ and decrease at higher values of $x$.
We expect this tendency to survive the multiple integration
procedure and to echo in the proton structure $F_{2p}$.

The minimization procedure gave $C \approx 1.1$
with slight adjustment of other parameters, see Table 1 for details.

\subsubsection{Fit 3}

In the region of very hard gluons both Fit 1 and Fit 2 
rely on the leading order DGLAP parametrizations of $G(x,Q^2)$.
Although it is desirable that integrated gluon structure function
$G_D(x,Q^2)$ based on our parametrizations approaches 
in the double logarithmic limit the 
conventional gluon density obtained from DGLAP evolution,
there is certainly no requirement that our fits be {\em built}
on these DGLAP fits. One should only make sure that
at $\log(1/x)\gg 1$ and $\log(Q^2/\Lambda^2_{QCD})\gg 1$
behavior of our fits is compatible with the corresponding
behavior of the DGLAP fits.

The properties of the DGLAP-evolved gluon density in the limit
is well understood. Since the anomalous dimension of gluons
is higher than that of the sea quarks, the secondary gluons
in this limit tend to be radiated off gluons as well. The evolution
of the integrated gluon density $G(x,Q^2) = x g(x,Q^2)$
separates out and is governed by
\be
{\partial G(x,Q^2) \over \partial \log Q^2} =
{\alpha_s(Q^2) \over 2\pi} \int_x^1 {dz}
G\left({x\over z},Q^2\right)P_{GG}(z)\,,
\ee
with splitting function
\be
P_{GG}(z)=2N_c\left[z(1-z) + {z \over(1-z)_+} + {1-z \over z}\right]
+ {11 \over 6}N_c \delta(1-z)\,.
\ee
The Regge-type behavior
\be
G(x,Q^2) = f(Q^2)\cdot \left({1\over x}\right)^\Delta
\ee
with constant $\Delta$ is compatible with the DGLAP equations
and leads to
\bea
{d \log f(Q^2) \over ds} &=& 4N_c \int_x^1 dz\ z^\Delta 
\left[z(1-z) + {z \over(1-z)_+} + {1-z \over z} + 
\delta(1-z){11 \over 12}\right]\nonumber\\
&=&4N_c\left[{1 \over \Delta} - {1 \over \Delta+1}
+ {1 \over \Delta+2} - {1 \over \Delta+3} + {11\over 12 }
+ {\cal C} + \psi(\Delta + 2)\right] \nonumber\\
&\equiv& \delta\,.
\eea
Here 
\be
s = \log\left[{\log\left(Q^2/ \Lambda^2_{QCD}\right)
\over \log\left(Q_0^2/ \Lambda^2_{QCD}\right)}\right]
\ee
and ${\cal C}$  and $\psi$ are the Euler constant and the digamma function
respectively.
One immediately gets the solution for the integrated gluon structure function
\be
G(x,Q^2) \propto \left[\log\left({Q^2 \over \Lambda^2_{QCD}}\right)\right]
^{\delta}\cdot\left({1 \over x}\right)^\Delta\,,
\ee
which leads to the differential glue of the form
\be
{\cal F}(x,\vec{\kappa}^2) \propto 
\left[\log\left({\vec{\kappa}^2 \over \Lambda^2_{QCD}}\right)\right]^{\delta-1}
\cdot\left({1 \over x}\right)^\Delta\,.
\ee
This analysis inspires us search for a parametrization of
${\cal F}_{pt}$ that would be power-like both in 
$\log\left(\vec{\kappa}^2/\Lambda^2_{QCD}\right)$ and $\log(1/x)$
in the double logarithmic regime.

The functional form of  ${\cal F}_{pt}$ Fit 3 is based on is
\be
{\cal F}_{pt}(x,\vec{\kappa}^2)|_{Fit 3} = 0.245
\cdot\left[\log\left({\vec{\kappa}^2 \over \Lambda_{QCD}}\right)\right]
^{0.34 -6\sqrt{x}}
\cdot\left({1 \over x}\right)^{0.4}\,.\label{fit3pt}
\ee

\begin{center}
{Table 2. The parameters of DGD2002 fits to differential gluon 
densities.
\vspace{0.3cm}\\}

 \begin{tabular}{|r|c|c|c|}
\hline
 & Fit 1 & Fit 2 & Fit 3  \\ \hline\hline
 hard input & GRV98LO &  GRV98LO & Eq.(\ref{fit3pt})  \\
 $\vec{\kappa}^2$-shift&& $C=1.1$ & \\
$Q_c^2$, GeV$^2$      & 1.45 & 1.45 & 1.4   \\
$a_{soft}$            & 2.66 & 2.63  & 2.6      \\
$\kappa_h^2$, GeV$^2$ & $0.4 + 0.245\log{1\over x}$
& $0.38 + 0.245\log{1\over x}$ & $0.31\log{1\over x}$ \\ \hline\hline
total $\chi^2$ & 257 & 245 & 226 \\ \hline
 \end{tabular}
\end{center}

\section{The properties of the gluon structure function}

\begin{figure}[!htb]
   \centering
   \epsfig{file=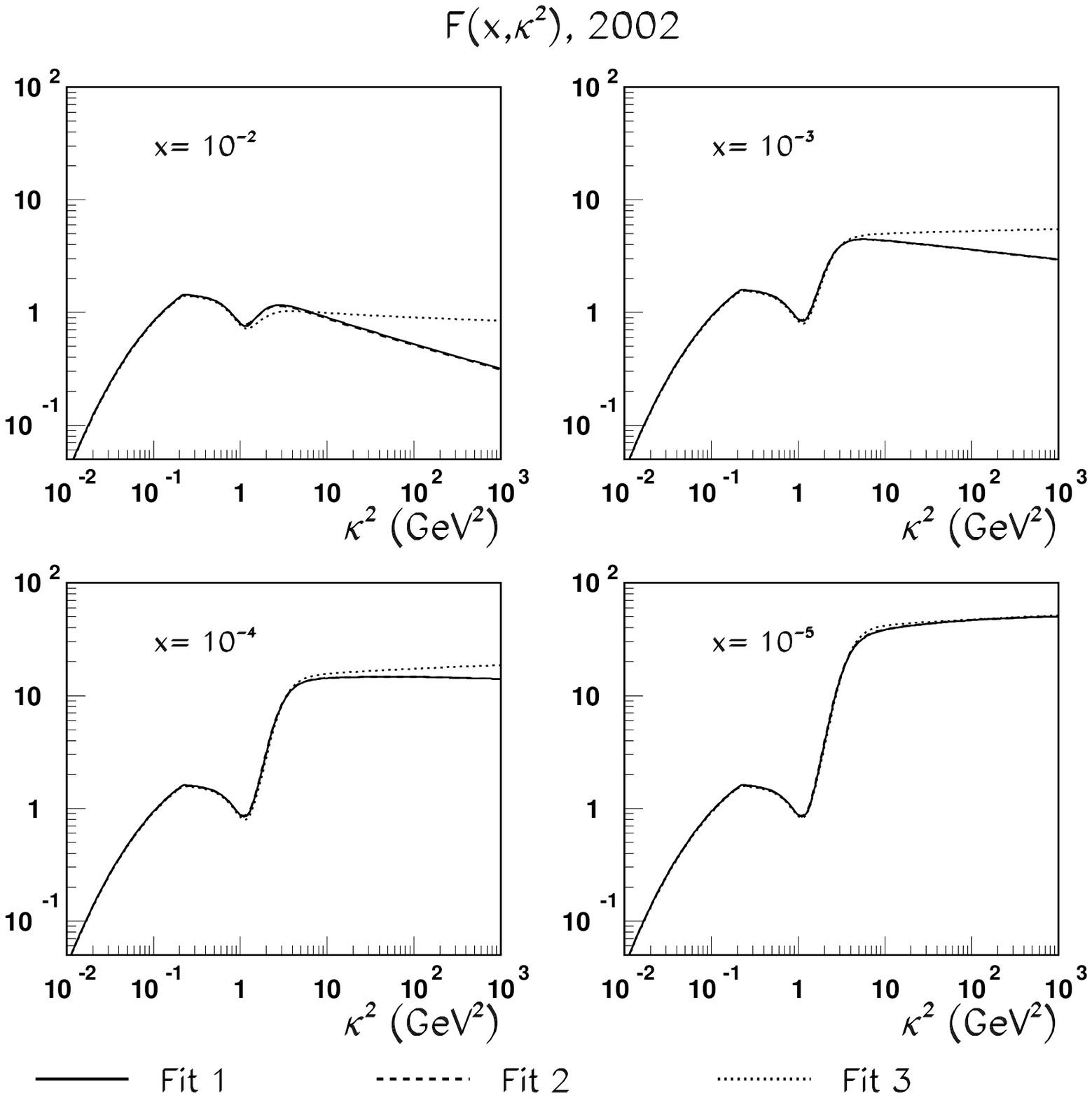,width=16cm}
   \caption{\em Differential gluon structure function as function
of $\kappa^2$: DGD2002 analysis}
   \label{fig_dgd2002}
\end{figure}

In Fig.~\ref{fig_dgd2002} we plotted the three fits to the 
differential gluon structure functions obtained
above vs. $\vec{\kappa}^2$ at several values of $x$. One observes here
a number of interesting features.
First, the three curves display rather similar behavior at very small $x$,
but at $x$ as high as $10^{-2}$ the difference among them throughout 
the region $\vec{\kappa}^2>1$ GeV$^2$ is quite sizable. Still, all of these
parametrizations of DGSF do provide a reasonably accurate description
of $F_{2p}$. Thus, we conclude that modern experimental data on $F_{2p}$
do not place severe restrictions on the shape of differential glue
at not very small $x$ (say, $x\gsim 10^{-3}$).

The second feature of the curves presented is their salient
two-peak shape. The technical origin of this clear separation
of the soft and hard exchange mechanisms is, of course,
the very abrupt extrapolation of soft gluons into hard region
and vica versa, generated by high powers $\zeta$. 
This might seem artificial, but as we described, 
such abrupt extrapolation seemed necessary in order to obtain
the correct behavior of effective intercept of $F_{2p}$
and therefore it was quite essential for getting better $\chi^2$.
Thus, we inclined to think that such clear separation of the soft and hard
mechanisms is indeed preferred by experiment.

\begin{figure}[!htb]
   \centering
   \epsfig{file=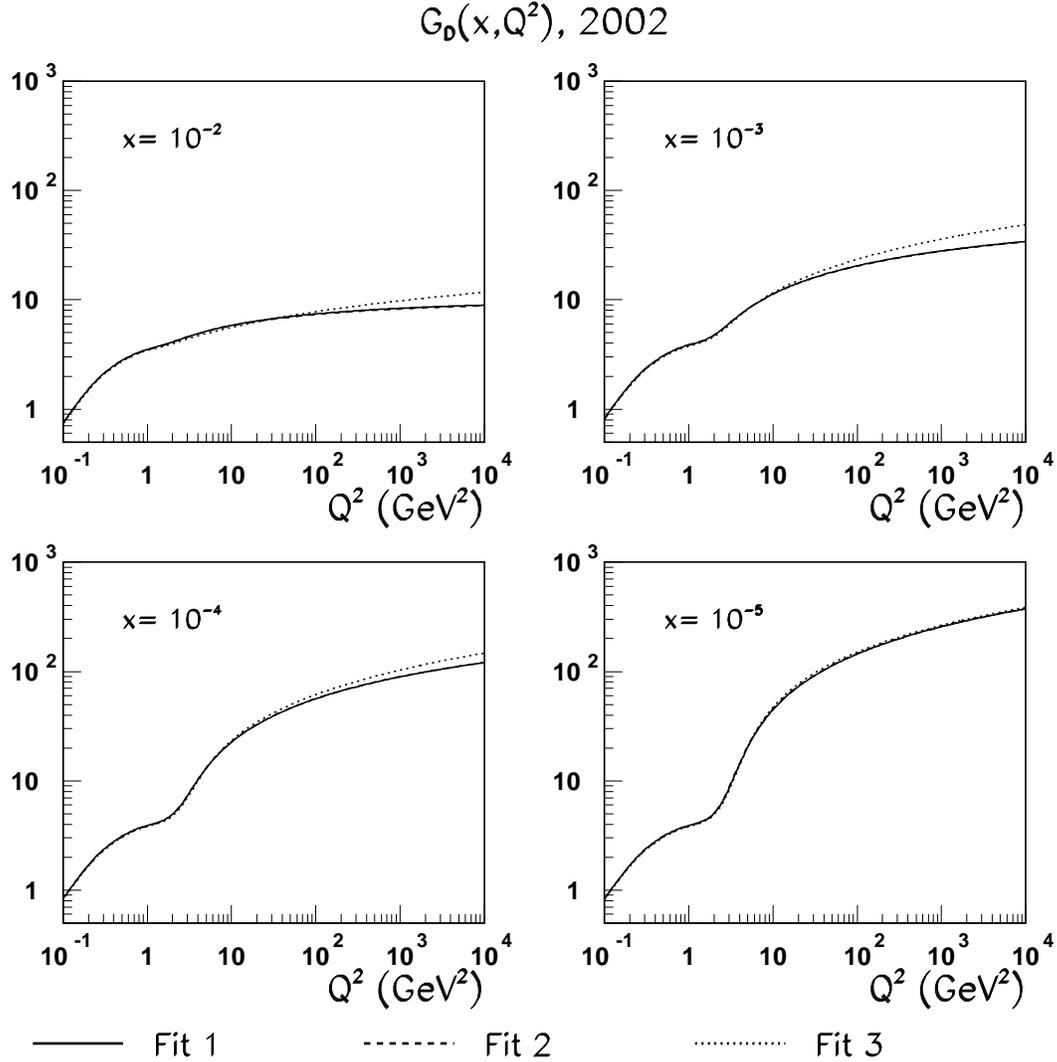,width=16cm}
   \caption{\em Integrated gluon structure function as function
of $Q^2$: DGD2002 analysis}
   \label{fig_intgd2002}
\end{figure}

The integrated gluon structure function is shown in Figs.~\ref{fig_intgd2002}.
A similar observation, though in a subdued form, can be made.
At $x = 10^{-2}\div 10^{-3}$ one can notice some departure among the curves,
which die out as we shift towards lower and lower $x$.
As we go to smaller $x$,  we observe plateau around 
$\vec{\kappa}^2 \sim 1\div 3$ GeV$^2$, which originates from the two-peak
shape of ${\cal F}(x,\vec{\kappa}^2)$, become more and more prominent.

Note also that at small $x$ 
and very large $Q^2$ all three $G_D(x,Q^2)$ curves, including
Fit 3 with simple-formula parameterization of ${\cal F}_{hard}$, 
approach each other.

\section{The observables}

\subsection{Structure function $F_{2p}$ and its derivatives}

\begin{figure}[]
   \centering
   \epsfig{file=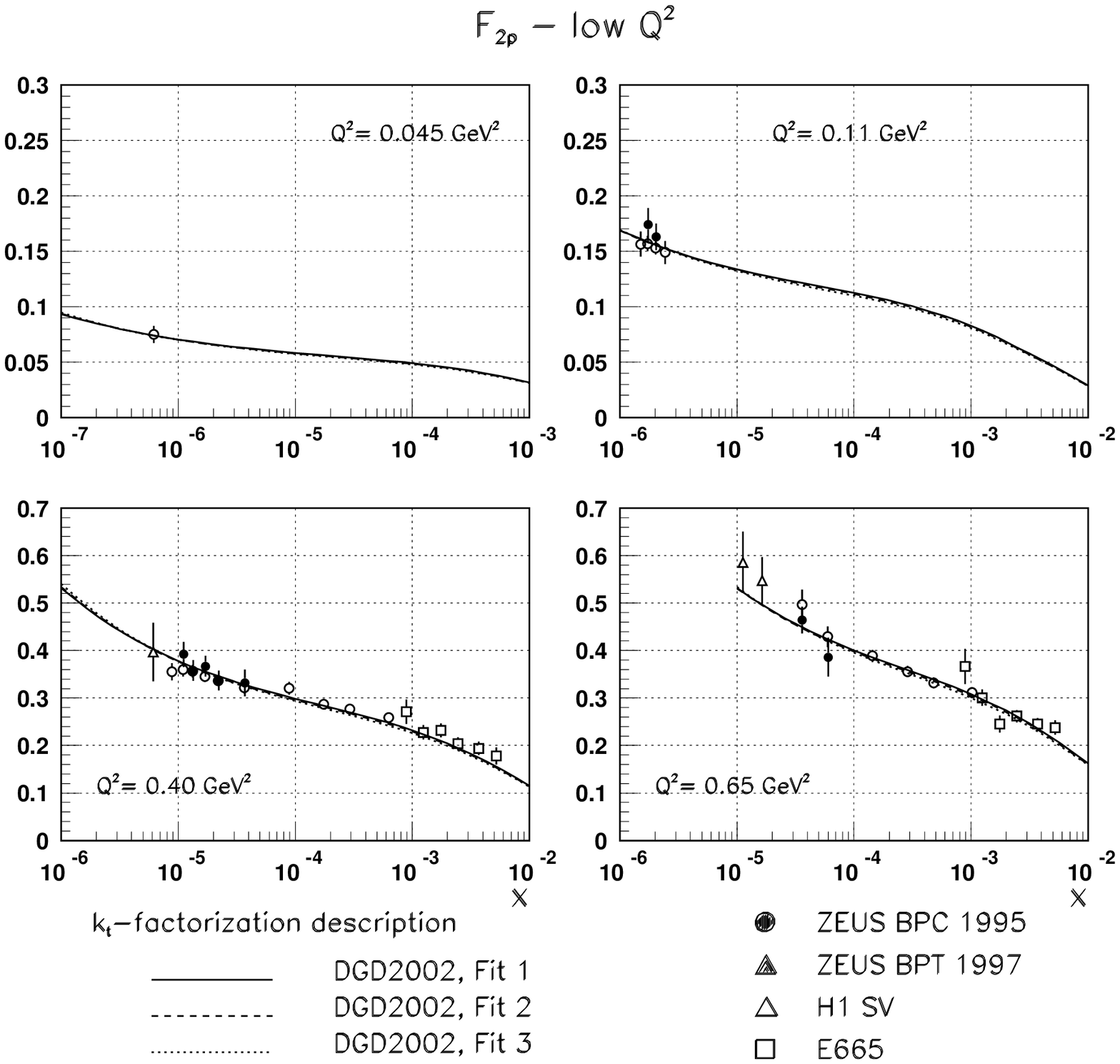,width=18cm}
   \caption{\em The $k_t$-factorization results for the 
structure function $F_{2p}(x,Q^2)$ in the small $Q^2$ region}
   \label{fig_lowq2_2002}
\end{figure}

\begin{figure}[]
   \centering
   \epsfig{file=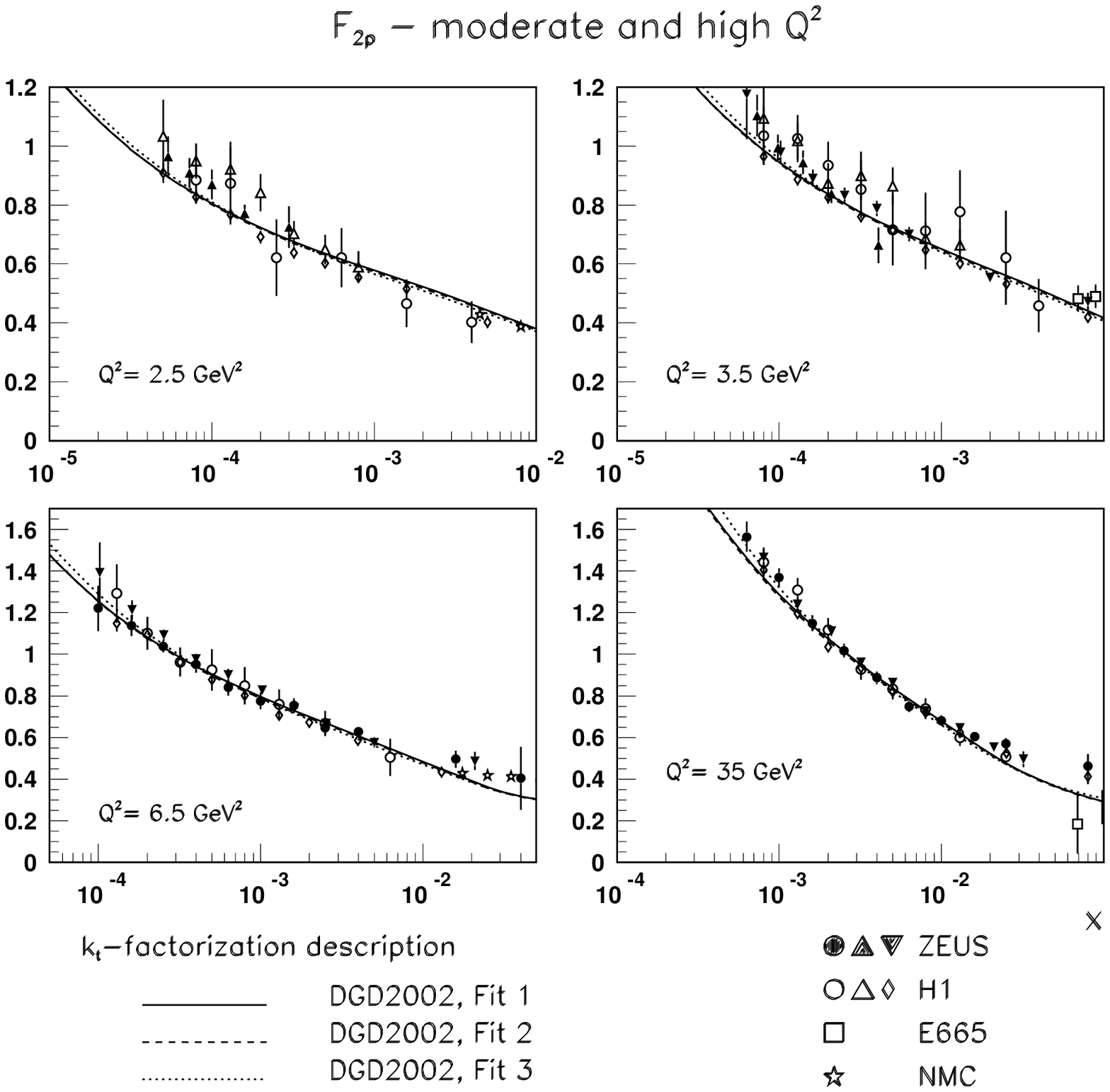,width=18cm}
   \caption{\em The $k_t$-factorization results for the 
structure function $F_{2p}(x,Q^2)$ in the moderate and high $Q^2$ region}
   \label{fig_highq2_2002}
\end{figure}

Since the overall behavior of structure function $F_{2p}(x,Q^2)$ 
was the quantity we tried to fit, one can expect a very good
description of the data. Indeed, as Figs.~\ref{fig_lowq2_2002} 
and \ref{fig_highq2_2002} show,
our calculations for $F_{2p}$ based on all three DGSF fits go almost directly
through the experimental points. This trend is somewhat spoiled
beyond the region of fitting, namely, at $x>0.01$, but still
we do not run into any severe discrepancy even here.
Note also that throughout the fitting region all three curves
differ less than 5\%.

\begin{figure}[!htb]
   \centering
   \epsfig{file=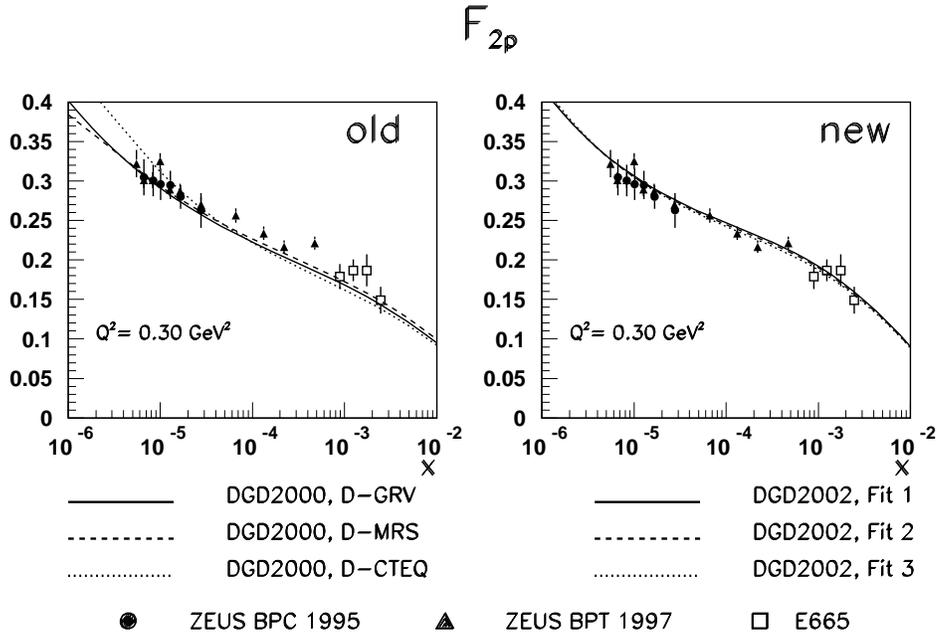,width=14cm}
   \caption{\em The improvement of the $k_t$-factorization predictions
for the structure function $F_{2p}$ based on DGD2000 (left pane) 
and DGD2002 (right pane) parametrizations of the differential 
gluon structure function}
   \label{fig_f2compare2002}
\end{figure}

Fig.~\ref{fig_f2compare2002} illustrates the improvement
in the description of the structure function $F_{2p}$
by the $k_t$-factorization calculations as we switch from
old DGD2000 parametrizations of DGSF to the new DGD2002 fits.
The curves now go directly through very constraining new data points
and therefore have less tendency to depart from each other outside
the fitting range.

\begin{figure}[!htb]
   \centering
   \epsfig{file=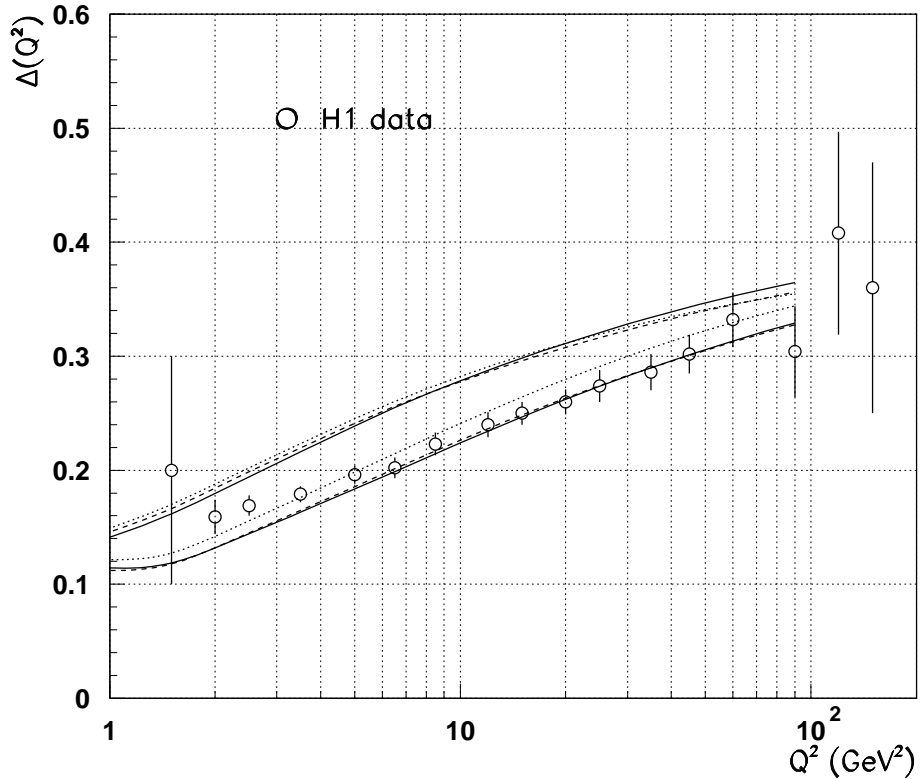,width=13cm}
   \caption{\em The $k_t$-factorization predictions
for the effective intercepts of the structure function
$F_{2p}$ in the moderate $Q^2$ region confronted with H1 data 
(solid circles)}
   \label{fig_intercepts_h1}
\end{figure}

Fig.~\ref{fig_intercepts_h1} 
shows the effective intercept of structure function $F_{2p}$ in the 
moderate $Q^2$ domain together with recent H1 data \cite{H1F2newrise}.
The intercepts were calculated according to
\be
\Delta(Q^2) = {\log[F_{2p}(x_1,Q^2)/F_{2p}(x_2,Q^2)] \over
\log(x_2/x_1)}\,.
\ee
For each differential gluon density fit we plotted here two curves:
the upper one corresponds to the effective intercept taken between
$x_1 =10^{-5}$ and $x_2=10^{-4}$, while the lower one
is for $x_1 =10^{-4}$ and $x_2=10^{-3}$. Significant deviation
between the two curves indicates the fact that the powerlike law
\be
F_{2p}(x,Q^2) \propto \left({1 \over x}\right)^{\lambda(Q^2)}
\ee
is only a very crude approximation.

The agreement with the data is reasonable, especially
when one takes into account that at lower $Q^2$ one should compare
the data with the upper curves and at higher $Q^2$
the data should be compared with the lower curves.
This is due to the experimental procedure used by H1 to
determine the intercepts: at smaller $Q^2$ the value of the intercept
comes from data points in the range $x\in (10^{-5},10^{-3})$, while
at higher $Q^2$ only $x\sim 10^{-3}\div 10^{-2}$ were available.

\subsection{Structure function $F_L$}

\begin{figure}[!htb]
   \centering
   \epsfig{file=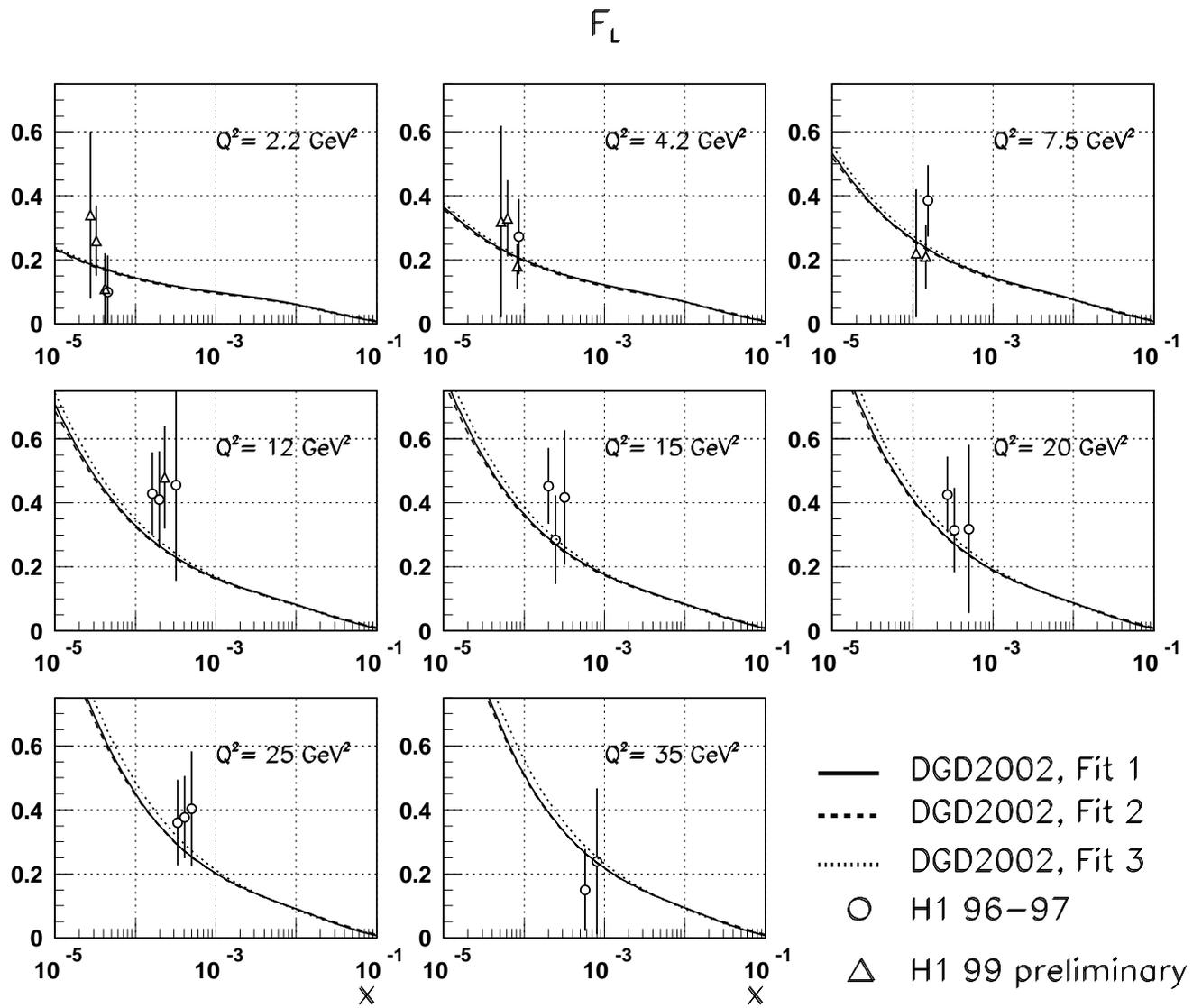,width=18cm}
   \caption{\em The longitudinal structure function
$F_{L}$ of the proton in the moderate $Q^2$ region}
   \label{fig_fl2002}
\end{figure}

In 1999 H1 published improved data on determination of structure 
function $F_L(x,Q^2)$. We show them in Fig.~\ref{fig_fl2002} together with
our predictions. The general agreement can be seen, 
however, since the data are not very constraining,
little futher information can be extracted from this plot.

\subsection{Real photoabsorption cross section}

\begin{figure}[!htb]
   \centering
   \epsfig{file=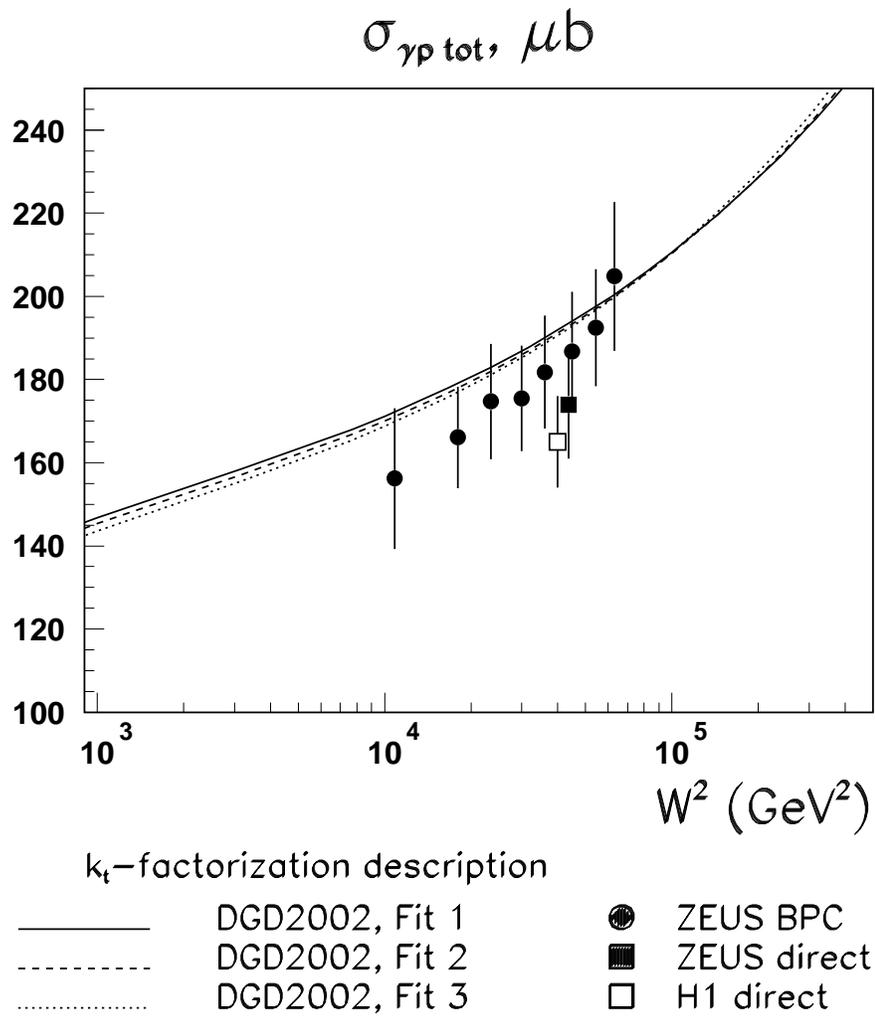,width=14cm}
   \caption{\em The predictions of the real photoabsorption cross section
as function of total energy based on DGD2002 parametrization}
   \label{fig_real2002}
\end{figure}

Finally, in Fig.~\ref{fig_real2002} we show experimental data for real photoabsorption
cross section $\sigma_{\gamma p}$ together with our predictions.

\chapter{Numerical analysis of the vector meson producton}

\section{$1S$ states: a brief look}

\subsection{Absolute values of cross sections and scaling phenomenon}

\begin{figure}[!htb]
   \centering
   \epsfig{file=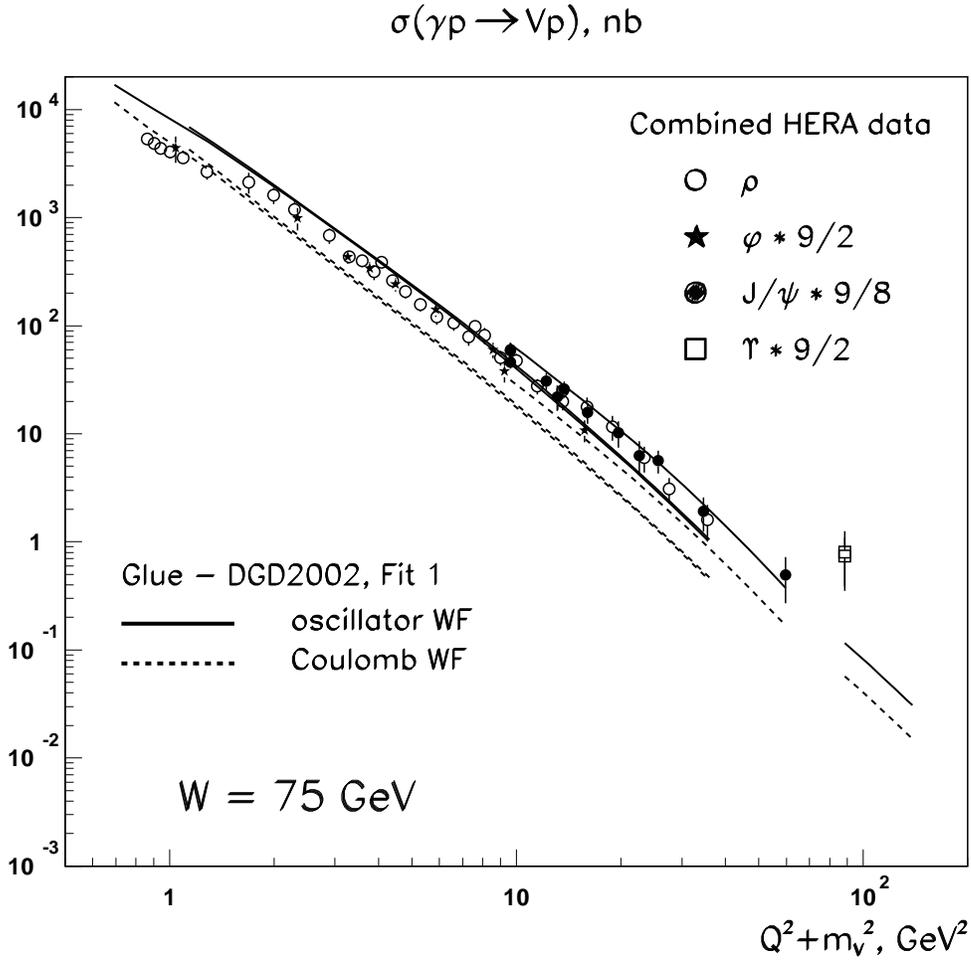,width=15cm}
   \caption{\em The total cross section of diffraction electroproduction
 of various vector mesons against scaling variable $Q^2+m_V^2$. 
Results are scaled according to flavor content 
to equilibrate the electric charges of different flavours.
The $k_t$-factorization predictions based on oscillator (solid lines) 
and suppressed Coulomb (dashed lines) are also shown.
All calculation are performed for $W=75$ GeV; the experimental points
are either taken in this energy range or consistently extrapolated
to this energy.}
   \label{num_tot_all}
\end{figure}

An overview of the experimental results on vector meson
production together with our predictions is given by
Fig.~(\ref{num_tot_all}). Here we plotted the experimentally measured 
values of $\rho$ meson cross sections  for photoproduction
(H1 \cite{H1rhophoto} and ZEUS \cite{ZEUSrhophoto})
and electroproduction (H1 \cite{H1rho} and ZEUS \cite{ZEUSrho}) cases,
$\phi$ mesons electroproduction cross sections
(H1 \cite{H1phi} and ZEUS \cite{ZEUSphi}, \cite{ZEUSphi2}), $J/\psi$
photoproduction (H1 \cite{H1jpsiphoto}) and electroproduction
cross sections (H1 \cite{H1jpsi} and ZEUS \cite{ZEUSrho}, \cite{ZEUSjpsi2}), 
and $\Upsilon(1S)$ photoproduction cross sections
(H1 \cite{H1jpsiphoto} and ZEUS \cite{ZEUSupsilon}).
All experimental data points are either taken at $W=75$ GeV
or are extrapolated to this energy. Whenever possible, we
used the energy dependence measured experimentally 
in the corresponding papers. In the case of $\Upsilon$
mesons photoproduction no reliable data on energy are available,
and we used power law $\sigma(\Upsilon) \propto W^\delta$
with $\delta = 1.75$, which comes from our calculations.

In the case of $\phi$, $J/\psi$, and $\Upsilon$ mesons
the cross sections were multiplied by appropriate
flavor factors in order to remove trivial sensitivity of the cross 
sections to the mean-square charge $\langle e_i^2\rangle$ 
of the quark content in a vector meson.

As suggested by the twist expansion analysis, the total production
cross sections should exhibit an approximate scaling in
variable $\Qb^2$. Indeed, as Fig.~(\ref{num_tot_all}) shows, 
the experimental data for various vector mesons do possess 
such a scaling property: data points for $\rho$, $\phi$, and 
$J/\psi$ taken at the same values of $\Qb2$ almost coincide.
It is worth noting that the scaling phenomenon takes place
even at small $\Qb^2$.
Note that the $k_t$-factorization predictions also exhibit 
approximate scaling phenomenon.

\subsection{The energy and $|t|$-dependence}

The remarkable scaling in variable $\Qb^2$ is observed
not only in the magnitude of the production cross sections,
but also in the patterns of energy dependence and $|t|$-dependence.

The energy dependence of the vector meson production
cross sections is compatible within experimental errors
with the power-like Regge-type growth:
\be
\sigma \propto W^\delta\,;\quad \delta = 4[\alpha_\Pom-1]\,.
\ee
The latter equality reflects the assumption that the
energy behavior comes from the gluon content of the proton,
which is usually linked to the Pomeron intercept $\alpha_\Pom$ at
$|t|=0$. 

The $t$-dependence of the differential cross section $d\sigma/d|t|$
can be approximated at $|t|\lsim 0.5$ GeV$^2$ by a simple exponential
law
\be
{d\sigma\over d|t|} \propto e^{-B|t|}\,.
\ee
The magnitude of the slope parameter $B$ shows how "fragile"
the proton and the produced meson are.

\begin{figure}[!htb]
   \centering
   \epsfig{file=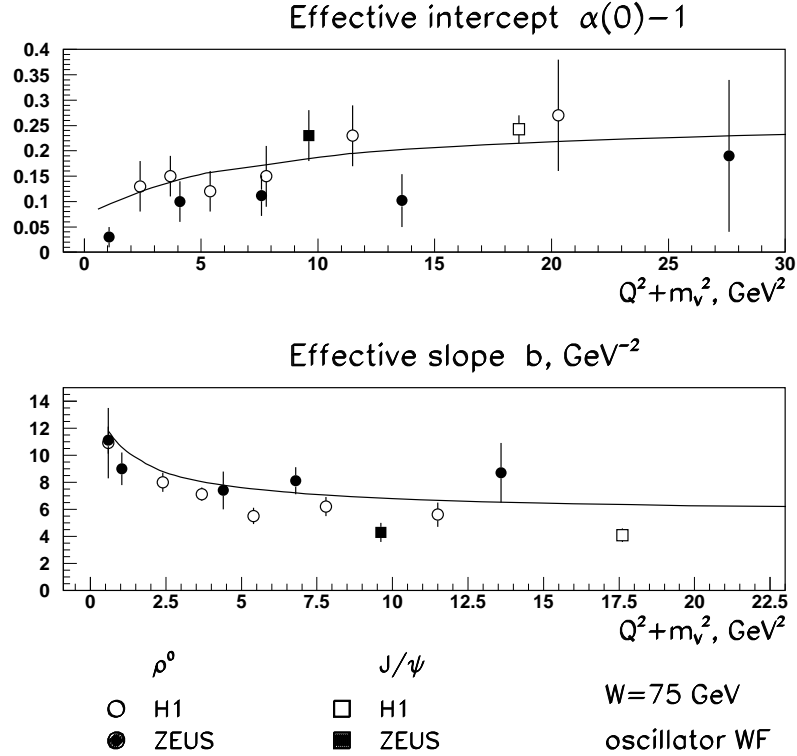,width=12cm}
   \caption{\em The effective intercepts (upper pane) 
and effective slopes(lower pane) of the vector meson production cross 
sections vs. scaling variable $Q^2 + m_V^2$. The data points
represent $\rho$ meson and $J/\psi$ meson results. The curves 
are the $k_t$0-factorization calculations for the $\rho$ meson
with the oscillator wave function. }
   \label{num_int_slope}
\end{figure}

Both quantities --- the intercept and the slope --- do exhibit
$Q^2$ dependence, but again via scaling variable $\Qb^2$.
Fig.~\ref{num_int_slope} shows a plot from \cite{DIS2000},
where the $k_t$-factorization predictions for these quantities 
are compared with the experimental results then available.
Although the agreement is not perfect, the tendency
is caught by the $k_t$-factorization calculations.
Below we will give more detailed investigation
of both quantities.

\subsection{The level of accuracy}

Before proceeding further, we should have a clear idea
of what level of accuracy one can expect from the 
$k_t$-factorization predictions.

There is a number of quantities that can be a source of uncertainty
in the final answers. Below we discuss them.

\subsubsection{The gluon density}

The {\bf gluon content of the proton} is an intrinsically
non-perturbative quantity, and therefore it is not calculable
within the $k_t$-factorization approach. Our extensive analysis
of the structure function $F_{2p}$ lead us to an accurate determination
of the differential gluon density ${\cal F}(x,\vec{\kappa}^2)$ in the proton.
The $k_t$-factorization results for various physical observables
based on distinct fits to ${\cal F}(x,\vec{\kappa}^2)$ differ from each
other at the level of several percent.

As we discussed above, the vector meson production amplitudes
are related not to the diagonal quantity ${\cal F}(x,\vec{\kappa}^2)$,
but to the off-diagonal (skewed) gluon distribution 
${\cal F}(x_1,x_2,\vec{\kappa}, \vec{\Delta})$. 
The situation can be partially cured by linking the
skewed gluon distribution to the diagonal one by means of Shuvaev
formula (\ref{shuvaev}) or its simplified version (\ref{shuvaevsimple}). 
Since such a linking is strictly valid
only when $x_1$ or $x_2$ vanishes, and therefore it is
a good approximation in the cases $Q^2 \gg m_V^2$ and $Q^2 \ll m_V^2$.

If $Q^2\approx m_V^2$, this linking becomes not well
justified, and the whole procedure introduces an uncertainty. 
In order to test the magnitude of the uncertainty associated with the
skewness of the gluon density, we performed an additional check, namely
we calculated the $\rho$ meson cross sections at low $Q^2$ 
using strictly forward and simplified off-forward (\ref{shuvaevsimple})
Ans\"atze for the gluon structure function. We found that at $Q^2 = 0$
using forward instead of non-forward (the later is default option
for all calculations here) reduced the cross section by a factor
of $1.07$. Obviously, smallness of this factor originates from low average
value of energy growth exponent $\lambda$.

Thus, the inaccuracy introduced at low $Q^2$ by forward/off-forward
Ansatz for gluon density is no more than 10\%.

\subsubsection{The wave function}

With the gluon density being brought under reasonable control,
the only major uncalculable piece of the pie is vector meson wave function.
As described above, we focus on the soft wave function,
in particular, we used the oscillator and "suppressed Coulomb" 
wave function Ans\"atze. Being virtually the two limiting cases
of how the radial part of the wave function can look like,
they represent fairly well the region of uncertainty introduced
by a specific choice of the wave function. 

As can be seen from the above Figures, the calculations based
on the oscillator wave function Ansatz are roughtly twice higher
than those obtained with the suppressed Coulomb WF.

\subsubsection{The width of vector meson}

In our calculations we treated the produced vector mesons
as particles with negligible width. This is not the case for the 
$\rho$ meson, whose width is about $1/5$ of its mass.
Usually, the incorporation of a finite width of a produced particle
is conducted via effective 'smearing' of the results (which
depend on the mass of the particle produced) over a certain
mass interval. If the cross sections calculated for a given mass 
happen to have convex dependence on mass (which is precisely
the case in the $\rho$ meson production), then such a smearing will
lower the values of the cross sections.
Thus, from a very general arguments we can expect such a smearing
in our case as well. Roughly, it should amount to decreasing
of the cross sections by factor of $~ (1+\Gamma_\rho/m_\rho) 
\approx 1.2$.

A more accurate calculation of this effect is a non-trivial task.
The problem is that in our treatment of the vector meson
production we {\em never} refer to the vector meson mass.
We deal only with the effective invariant mass of the $q\bar q$ pair.
Therefore accurate calculation of the smearing effect 
requires solution of  conceptually non-trivial problems.

\subsubsection{The limits of $k_t$-factorization approach}

Finally, the very approach we use has limited applicability domain.
In particular, it would be fallacious to extend our calculations
to high $Q^2$ region. A rough criterion to border of the applicability
domain can be given by $Q^2_{max} \sim \sqrt{W^2 \mu_{soft}^2}$,
which is about 50 GeV$^2$ for the HERA energy range.
Above this values, the logarithms $\log(Q^2/\mu_{soft}^2)$
will be more important than $\log(1/x)$, and one can expect
that  $k_t$-factorization will underestimate observed  cross sections.

\section{The $\rho$ meson production}

In this section we provide a detailed description
of the $k_t$-factorization predictions for the $\rho$ meson
production.  Note that throughout this chapter
we treat the physical $\rho$ meson as pure $1S$ state.
Whether this is indeed realized in Nature and
what changes if $S/D$ wave mixing occurs, will be discussed 
in Chapter 11.

Whenever experimental data available, 
we compare them with our results.

\subsection{$Q^2$ dependence}

\begin{figure}[!htb]
   \centering
   \epsfig{file=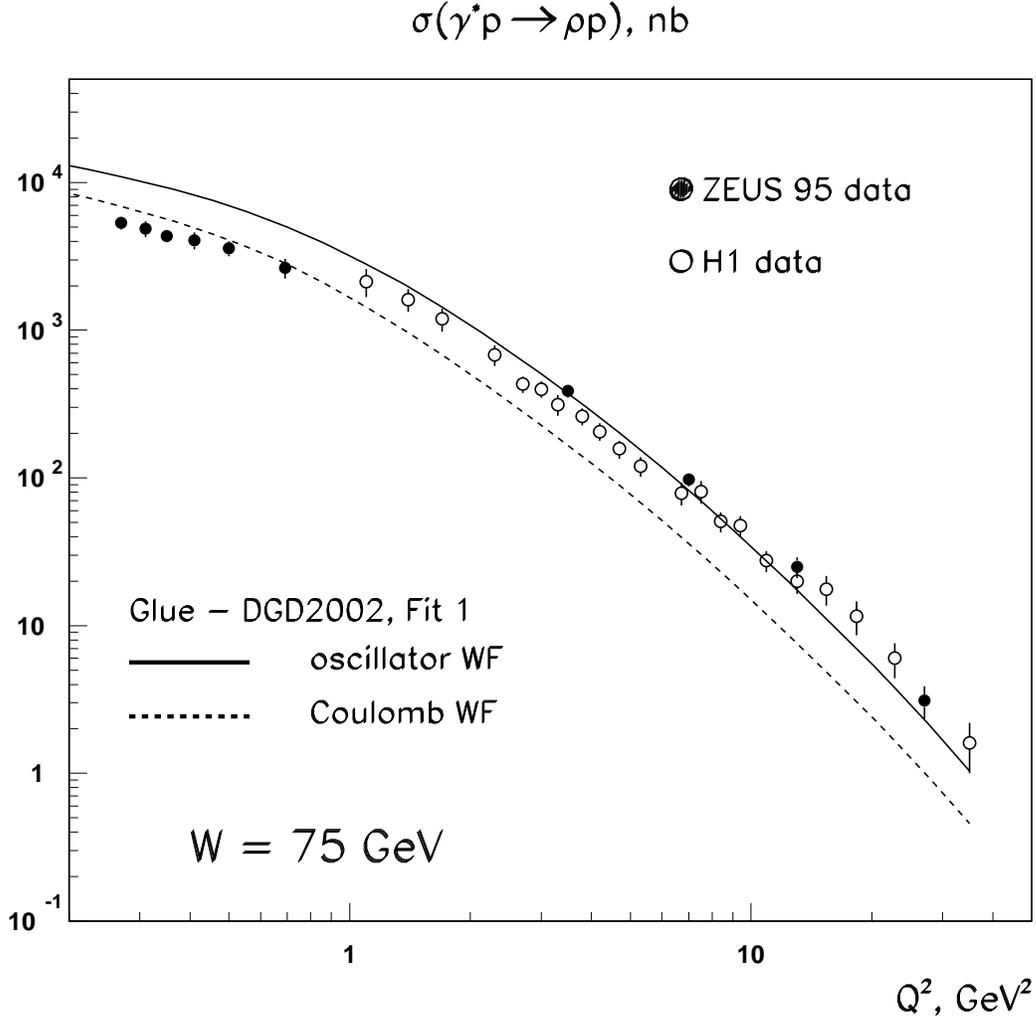,width=16cm}
   \caption{\em Total cross section of the diffractive $\rho$ meson
production as a function of $Q^2$. 
The $k_t$-factorization predictions based on oscillator (solid lines) 
and suppressed Coulomb (dashed lines) are also shown.
All calculation are performed for $W=75$ GeV using DGD2002, Fit 1.}
   \label{num_tot_rho}
\end{figure}

The $Q^2$-dependence of the $\rho$ meson production cross section
is shown in Fig.~\ref{num_tot_rho}. 
One sees that as we slide to higher values of $Q^2$,
the cross section drops sharply. 
Although for the major part of the $Q^2$ region shown
the experimental data points fall roughly 
between the oscillator and Coulomb wave function predictions,
two separate places of discrepancy are easily visible.

\subsubsection{Low $Q^2$ region}
The first problematic point is low-$Q^2$ region.
As we come from moderate $Q^2$ down to $Q^2=1$ GeV$^2$ and below,
our predictions, if compared to the experimental points, 
tend to climb too high.

\begin{figure}[!htb]
   \centering
   \epsfig{file=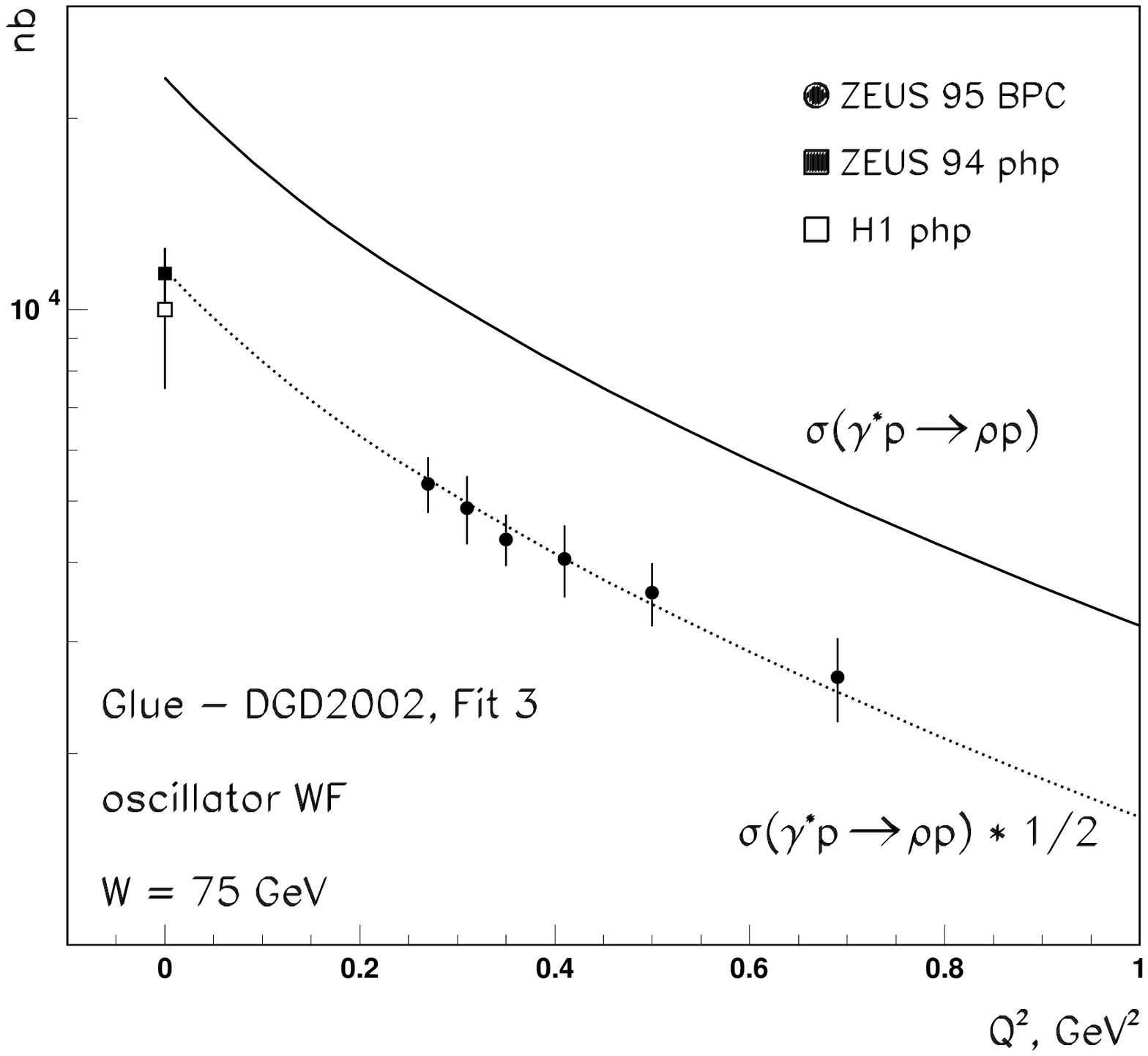,width=16cm}
   \caption{\em  Total cross section of the diffractive $\rho$ meson
production as a function of $Q^2$ in the small-$Q^2$ region.
The solid curve represents the $k_t$-factorization predictions;
dashed line shows the same prediction divided by two.}
   \label{num_lowq2}
\end{figure}

\begin{figure}[!htb]
   \centering
   \epsfig{file=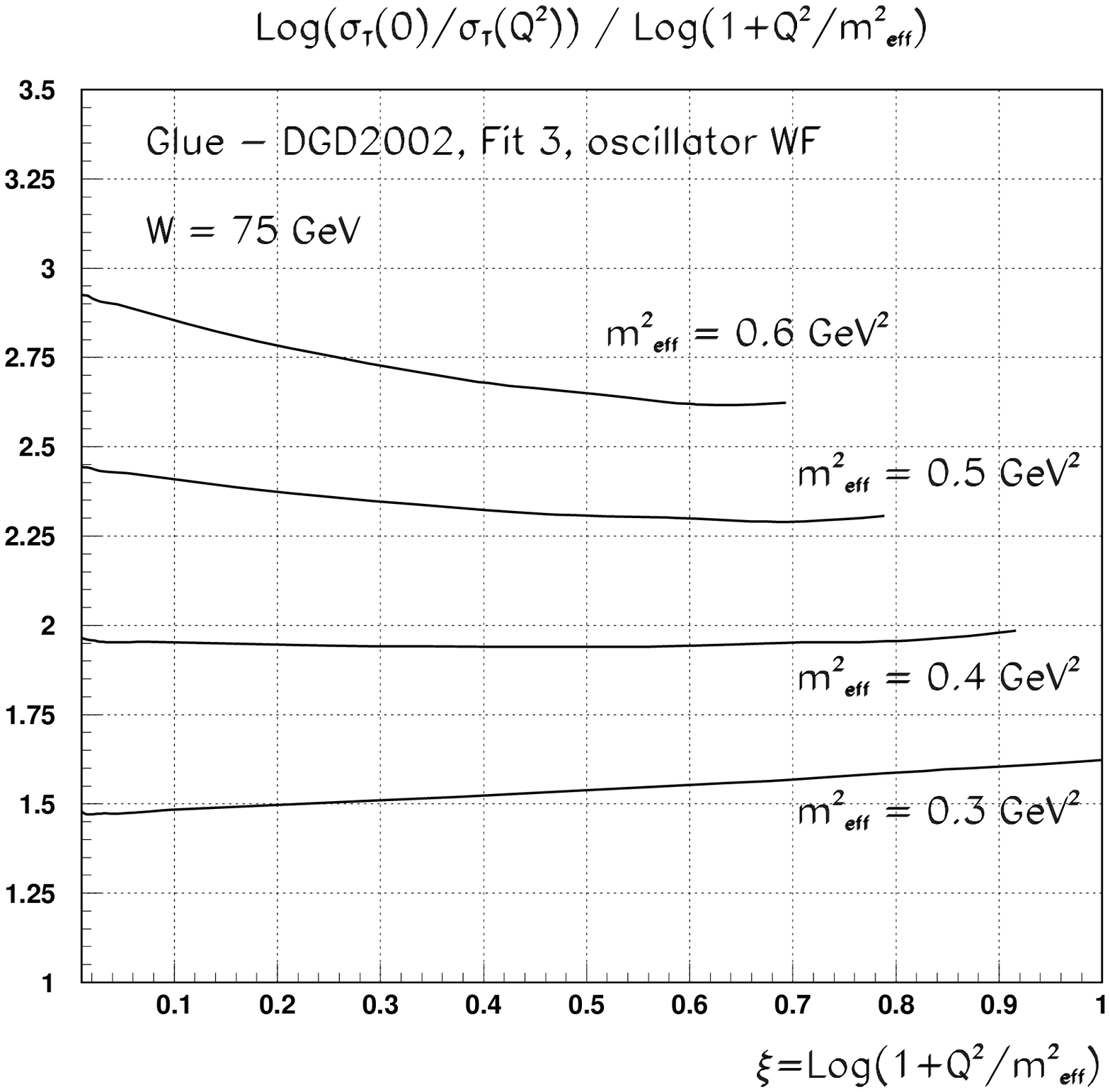,width=16cm}
   \caption{\em Effective $Q^2+m_{eff}^2$ exponent of the transverse
$\rho$ meson production cross section in the low-$Q^2$ region as a function
of $\xi = 1+ Q^2/m_{eff}^2$ at various values of $m_{eff}$}
   \label{num_lowq2power}
\end{figure}

The first thought would be to suspect that our predictions
rise too steeply as $Q^2 \to 0$. An accurate analysis shows, however,
that this suspicion is off the point. In fact,
the $Q^2 \to 0$ behavior of our predictions is perfectly compatible
with experimentally observed tendencies.
This ather unexpected fact is illustrated by
Figs.(\ref{num_lowq2}) and (\ref{num_lowq2power}).

In Fig.(\ref{num_lowq2}) we zoom in on the region $Q^2<1$ GeV$^2$,
where the available experimental data include ZEUS 95 BPC \cite{ZEUSrho},
ZEUS 94 photoproducton \cite{ZEUSrhophoto} 
and H1 photoproduction \cite{H1rhophoto} measurements. 
The $k_t$-factorization prediction based on the oscillator
wave function (which seems to be a more reasonable choice for the $\rho$
mesons than the Coulomb WF) and on DGD2002, Fit 3 are shown with solid line.
The $k_t$ predictions overshoot the data throught the whole region shown here.
However, when simply divided by 2, the predictions pass exactly through
all the data points (dotted curve in Fig.(\ref{num_lowq2})),
including the photoproduction point.

Alternatively, one can try parametrizing the cross sections at
low $Q^2$ by a simple formula:
\be
\sigma(Q^2) \propto {1 + R(Q^2)\over (Q^2 + m^2_{eff})^n}\,;\quad
R(Q^2) \equiv {\sigma_L(Q^2) \over \sigma_T(Q^2)}\,,
\ee
which is equivalent to
\be
\sigma_T(Q^2) \propto {1\over (Q^2 + m^2_{eff})^n}\,.\label{lowq2fit}
\ee
In order to find parameters $m^2_{eff},\,n$ that reproduce
the low-$Q^2$ behavior of our predictions,
we plotted in Fig.~(\ref{num_lowq2power}) the quantity
\be
{\cal R} = {\log[\sigma_T(Q^2=0)/\sigma_T(Q^2)] \over \log(1+Q^2/m^2_{eff})}\,
\ee
against $\xi \equiv 1+Q^2/m^2_{eff}$. If Eq.(\ref{lowq2fit}) holds,
this quantity should be equal to $n$ for all values of $\xi$.
We see that when $m_{eff}^2 \approx 0.4$ GeV$^2$, ${\cal R}$
has the most flat shape vs. $\xi$ and is $\approx 2$. 

This result is in perfect agreement with low-$Q^2$ analysis
of the ZEUS BPC data \cite{ZEUSrho}: when fitted by the formula
\be
\sigma(Q^2) \propto {1 + R(Q^2)\over (Q^2 + m^2_{eff}(exp))^2}\,,
\ee
the data yield 
$$
m_{eff}(exp) = 0.66 \pm 0.11\mbox{ GeV}\,,
$$
in perfect agreement with our $m_{eff} \approx 0.6\div 0.65$ GeV.
Note that although in our analysis power $n$ was a free parameter,
it turned our close to 2.

Three conclusions can be drawn from this analysis.

First, we showed that although there is a sizable (by factor of 2)
departure between {\em the magnitude} of the experimentally 
measured cross sections and the $k_t$ factorization predictions,
{\em the shape of low-$Q^2$ behavior} is perfectly
reproduced by the calculations based on the oscillator wave function.

Second, this analysis proves that the value $m_{eff} \approx 0.65$ GeV
is dynamically generated. Indeed, the light quark mass used
in our approach was pre-fixed by the gluon structure function
analysis at the level $m_q = 0.21$ GeV. 
Still $m_{eff}^2 > 4m_q^2$, which means that approximately
one half of $m_{eff}^2$ comes from quark momentum.
The exact proportion depends of course on the particular choice
of the vector meson wave function. It is still to be investigated
whether the oscillator wave function was an unexpectedly lucky guess,
or whether the perfect reproduction of low-$Q^2$ behavior
is accidental.

Third, the above means that the actual region where our predictions
and the experimental data really mismatch is
not the low-$Q^2$, but rather moderate $Q^2$ region 
($Q^2 \approx 1$ GeV$^2$).

\subsubsection{High $Q^2$ region}

The second region where our predictions tend to depart from
the data is high-$Q^2$ region, $Q^2 \gsim 5\div 10$ GeV$^2$.
If the cross section is (locally) fitted by the powerlike
fall-off
\be
\sigma(Q^2) \propto {1 \over (Q^2 + m_\rho^2)^n}\,,\label{highq2fit}
\ee
then this discrepancy can be expressed numerically in terms of 
effective $Q^2$ exponent $n$.
The experimental determination of this exponent resulted in the
following values.
The ZEUS 95 data \cite{ZEUSrho} with $Q^2>5$ GeV$^2$ 
are consistent with fit (\ref{highq2fit}) with 
energy independent $n_{exp}$, 
whose avegare value is found to be $2.32\pm 0.10$.
More copious H1 data sample \cite{H1rho} taken at $W=75$ GeV results
in $n_{exp} = 2.24\pm 0.09$, which is in agreement with ZEUS fit.

\begin{figure}[!htb]
   \centering
   \epsfig{file=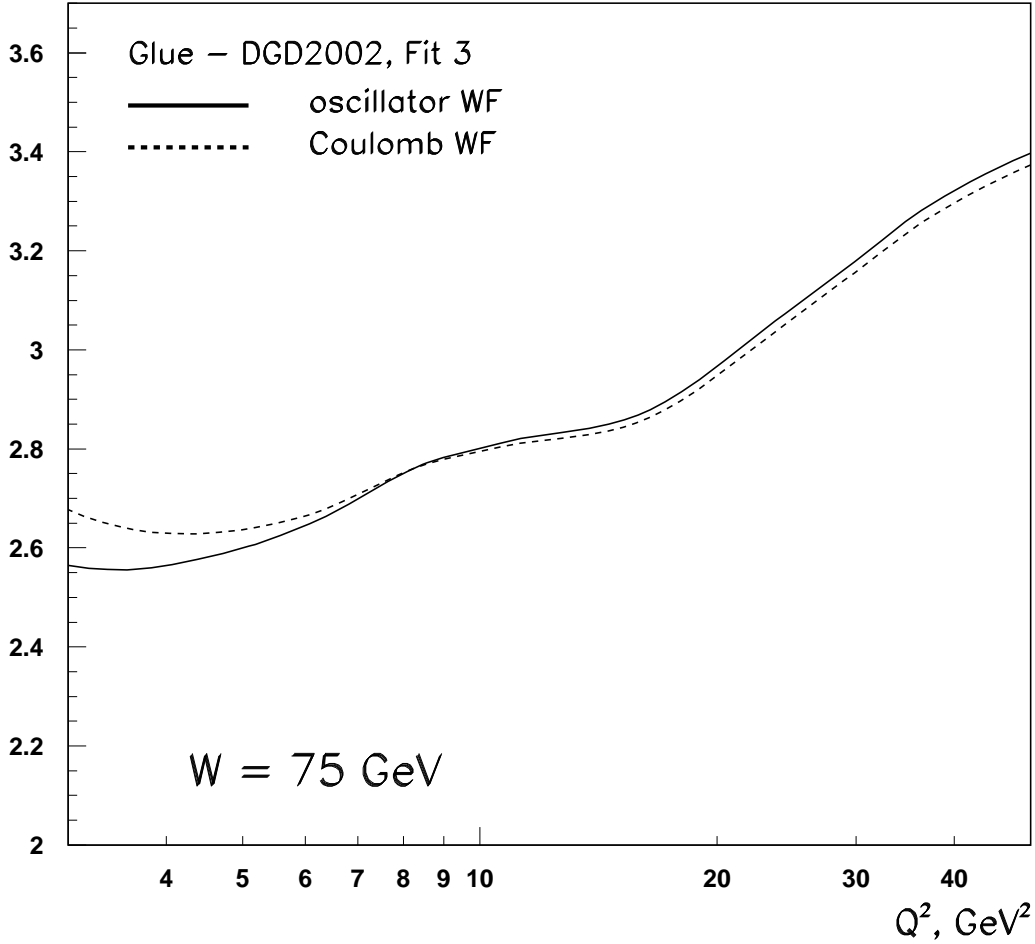,width=16cm}
   \caption{\em Effective $Q^2+m_{eff}^2$ exponent of the total
$\rho$ meson production cross section in the high-$Q^2$ region 
as a function of $Q^2$. The solid and dashed curves
represent calculations for oscillator and suppressed Coulomb
wave functions respectively}
   \label{num_q2power}
\end{figure}

The $k_t$ factorization predictions for this exponent is shown 
in Fig.~(\ref{num_q2power}). Here we plotted local analogue of the
exponent $n_{exp}$, i.e. quantity
\be
n\left(Q^2=\sqrt{Q_1^2\,Q^2_2}\right) = {\log[\sigma(Q_1^2)/\sigma(Q^2_2)]
\over \log[(Q_2^2+m_\rho^2)/(Q_1^2+m_\rho^2)]}\,.\label{localq2exp}
\ee
Again, if fit (\ref{highq2fit}) holds, $n$ should be independent of $Q^2$.

Graphs in Fig.~\ref{num_q2power} show that this is not the case.
At intermediate $Q^2$ $n$ starts already from about $2.5$, then grows
s $Q^2$ increases, and at $Q^2>20$ GeV$^2$ it is even higher than $3.0$.

Although being in constrast with fits to experimental data,
such a $Q^2$ growth of $n(Q^2)$ is still firmly grounded theoretically.
Qualitatively, this can be understood from the analysis of 
the leading $\log Q^2$ result. At fixed $W^2$ and high enough $Q^2$, 
the $Q^2$ dependence of the cross section comes from
\be
\sigma(Q^2) \propto {1 + R(Q^2) \over (Q^2 + m_\rho^2)^4}\cdot
\left[G\left(c{Q^2 \over W^2},Q^2/4\right)\right]^2\,,
\ee
where $c\approx 0.41$ comes from the approximate representation of the
off-forward gluon distribution (\ref{shuvaevsimple}). 
The $Q^2$ dependence of slope $B$ and of running
coupling $\alpha_{s}$ is inessential for our point.

\begin{figure}[!htb]
   \centering
   \epsfig{file=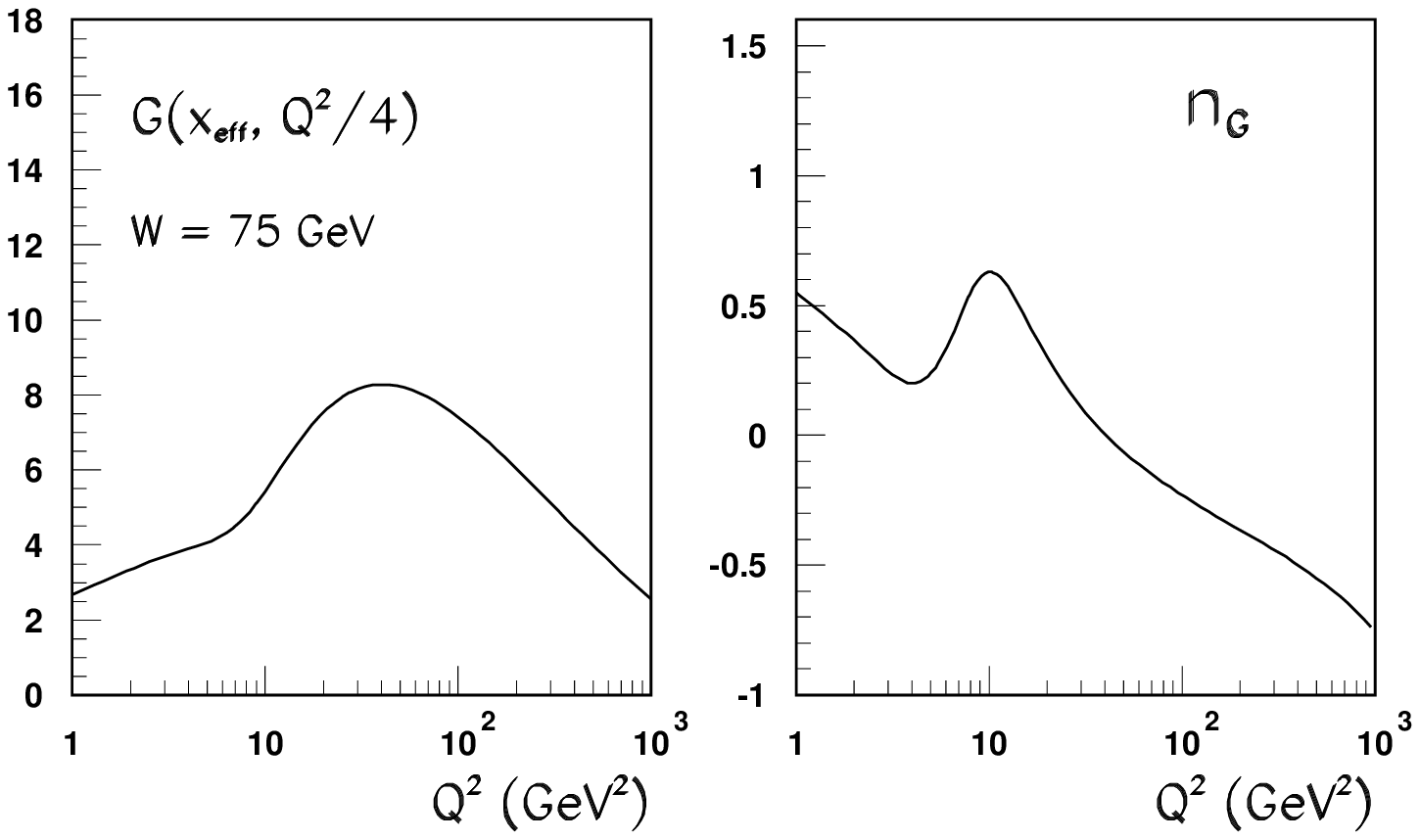,width=14cm}
   \caption{\em (Right pane) The $Q^2$ behavior of the integrated gluon density
$G(0.41Q^2/W^2,Q^2/4)$ at fixed value of $W=75$ GeV obtained 
by integration of the DGD2002 Fit 1; 
(left pane) The effective $Q^2$ exponent of this integrated gluon density}
   \label{num_intgd_w75}
\end{figure}

The non-trivial $Q^2$ behavior arises from the integrated gluon
structure function. In Fig.~\ref{num_intgd_w75}, left pane, we show
its $Q^2$ behavior at $W=75$ GeV. The origin of the peaked shape
is obvious: at moderate $Q^2$ the integrated glue grows due to
the sharp explicit $Q^2$ dependence (that is, due to large values
of unintegrated gluon density ${\cal} F(x,\vec{\kappa}^2)$), while for 
larger $Q^2$ the effect of decreasing $x_{eff}$ overpowers
and leads to decreasing of $G$ as $Q^2$ grows further.

On the right pane of Fig.~\ref{num_intgd_w75} we show the 
local $Q^2$ exponent $n_G$ of the gluon density
(the gluon density contribution to the local $Q^2$ exponent $n$
is equal to $2n_G$)
\be
n_G\left(Q^2=\sqrt{Q_1^2\,Q^2_2}\right) = {\log[G(Q_1^2)/G(Q^2_2)]
\over \log[(Q_2^2+m_\rho^2)/(Q_1^2+m_\rho^2)]}\,.
\ee
One sees that at moderate $Q^2$, when $G$ is stil rising, 
it tampers the $Q^2$ fall, but when $Q^2 \gsim 10$ GeV$^2$,
gluon density starts decreasing on its own.
This is precisely the reason why at higher $Q^2$ the $Q^2$ exponent
$n$ (Fig.~\ref{num_q2power}) boosts up.

The arguments that justify such a behavior seem to be universal,
and to this end, it is surprising why the experimental data
do not possess such a behavior.

\subsection{$\sigma_L-\sigma_T$ decomposition}

An important insight of the $Q^2$ behavior of the $\rho$ meson
production cross sections comes from the separate analysis
of $\sigma_L(Q^2)$ and $\sigma_T(Q^2)$, that is proportion of the
$\rho$ meson production rates cause by transverse and longitudinal photons.

\begin{figure}[!htb]
   \centering
   \epsfig{file=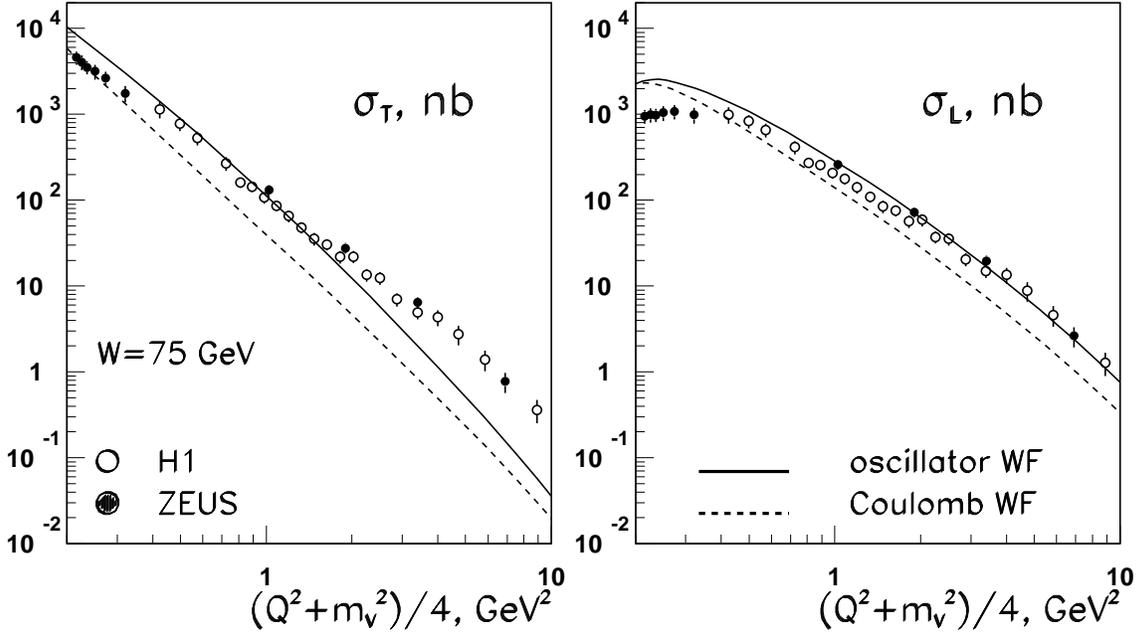,width=16cm}
   \caption{\em Experimental data on transverse $\sigma_T$ 
and longitudinal $\sigma_L$ cross sections of $\rho$ meson production
compared with the $k_t$-factorization approach.}
   \label{num_rho_lt}
\end{figure}

Fig.~\ref{num_rho_lt} represents the results for these cross sections
within the $k_t$-factorization approach confronted with experimental data
\cite{sigmalt}.

One sees that at high $Q^2$ we do provide a reasonably good description
of the $\sigma_L$, but our $\sigma_T$ curves sink significantly deeper
as $Q^2$ grows. Thus, it is mostly a way too steep $Q^2$-behavior
of $\sigma_T$ that causes departure of our curves from the data.

If analyzed in terms of power-like fits
\be
\sigma_T(Q^2)\propto (Q^2+m_\rho^2)^{-n_T}\,;\quad
\sigma_L(Q^2)\propto (Q^2+m_\rho^2)^{-n_L}\,,
\ee
the experimental data yield \cite{sigmalt}
\be
n_T(exp) = 2.47 \pm 0.03\,.
\ee
We found no direct expermental results for $n_L$, but clearly
it should be even less than $n_T$.

\begin{figure}[!htb]
   \centering
   \epsfig{file=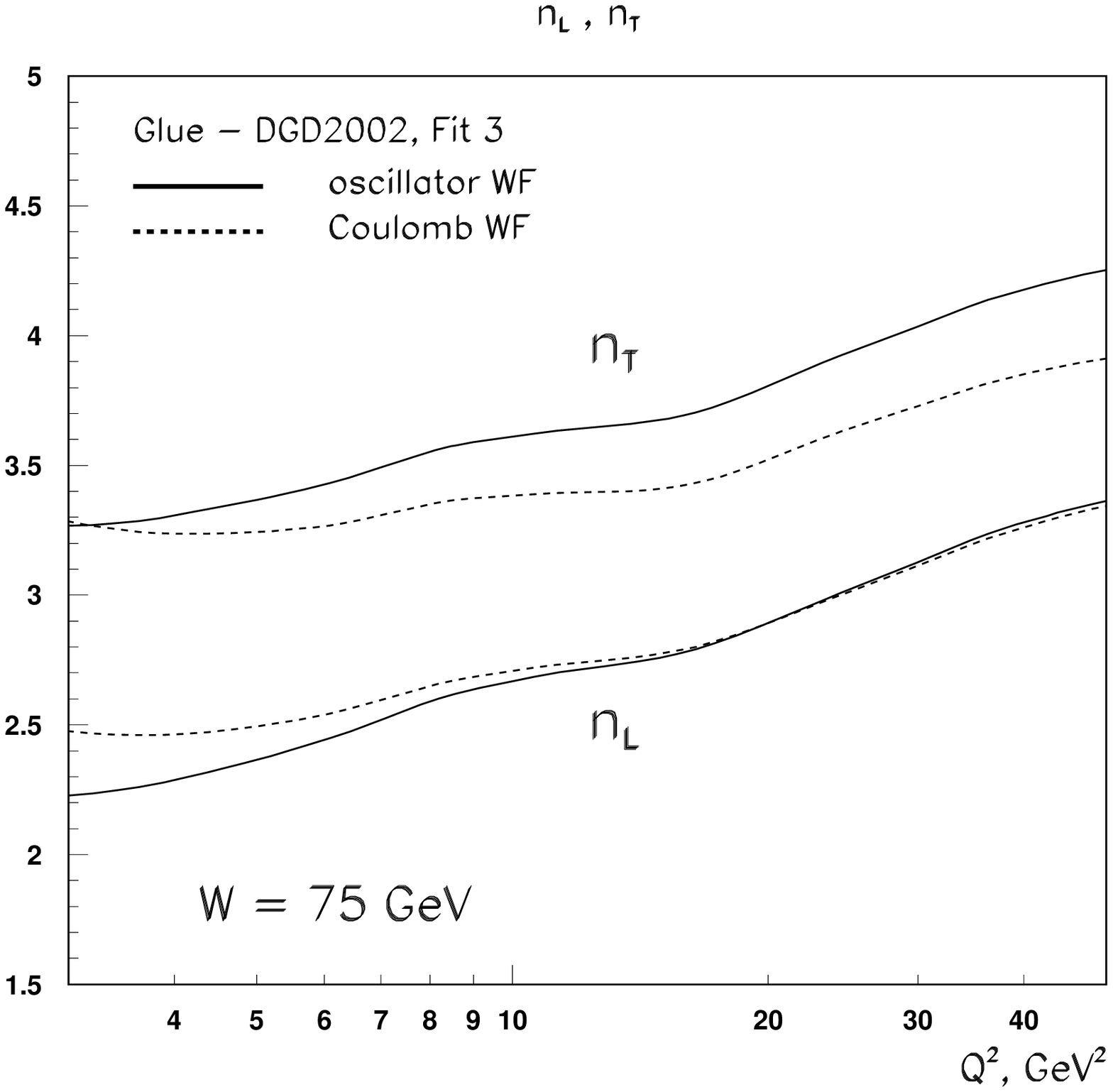,width=16cm}
   \caption{\em The effective $Q^2$ exponents shown separately 
for $\sigma_L$ and $\sigma_T$ as functions of $Q^2$. 
Solid and dashed lines correspond to oscillator and suppressed
Coulomb wave functions respectively}
   \label{num_q2power_lt}
\end{figure}

The $k_t$-factorization predictions for the local values 
of $n_L$ and $n_T$, defined similarly to (\ref{localq2exp}),
are shown in Fig.~\ref{num_q2power_lt}. At higher $Q^2$,
$n_L$ and $n_T$ grow up to 3 and 4 respectively, 
the latter being in stark contast to the data.

\begin{figure}[!htb]
   \centering
   \epsfig{file=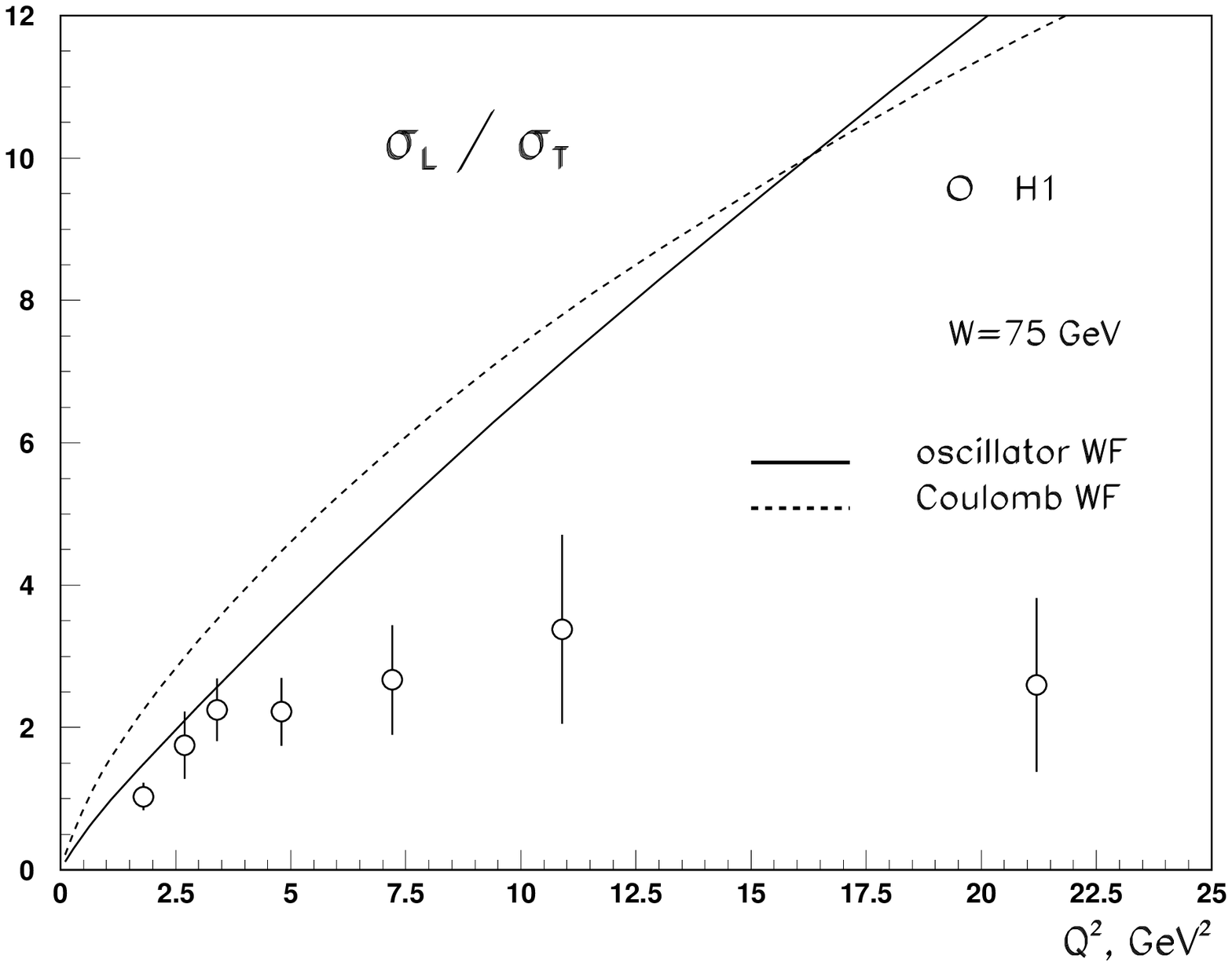,width=16cm}
   \caption{\em Ratio $R = \sigma_L/\sigma_T$ as a function of $Q^2$
for $\rho$ meson as a function of $Q^2$.
Solid and dashed lines correspond to oscillator and suppressed
Coulomb wave functions respectively}
   \label{num_ratiolt}
\end{figure}

Fig.~\ref{num_ratiolt} depicts the ratio
\be
R(Q^2) = {\sigma_L(\gamma^* p \to \rho p) \over 
\sigma_T(\gamma^* p \to \rho p)}\,,
\ee
experimental points taken from \cite{H1rho}.
The glaring diagreement at higher $Q^2$ is, of course, caused
by too much suppressed $\sigma_T$ out calculations predict.
Evidently, if we found a way to increase $\sigma_T$, the ratio
$R(Q^2)$ would be authomatically cured.

\subsection{Energy dependence}

The growth of the vector meson production cross sections
is a well-established fact. It is linked basically to the energy growth
of the Pomeron exchange, and therefore fitting the cross sections
to the energy power law
\be
\sigma(W) \propto W^\delta\label{w2depfit}
\ee
seems a natural way to quantitize the energy growth.

\begin{figure}[!htb]
   \centering
   \epsfig{file=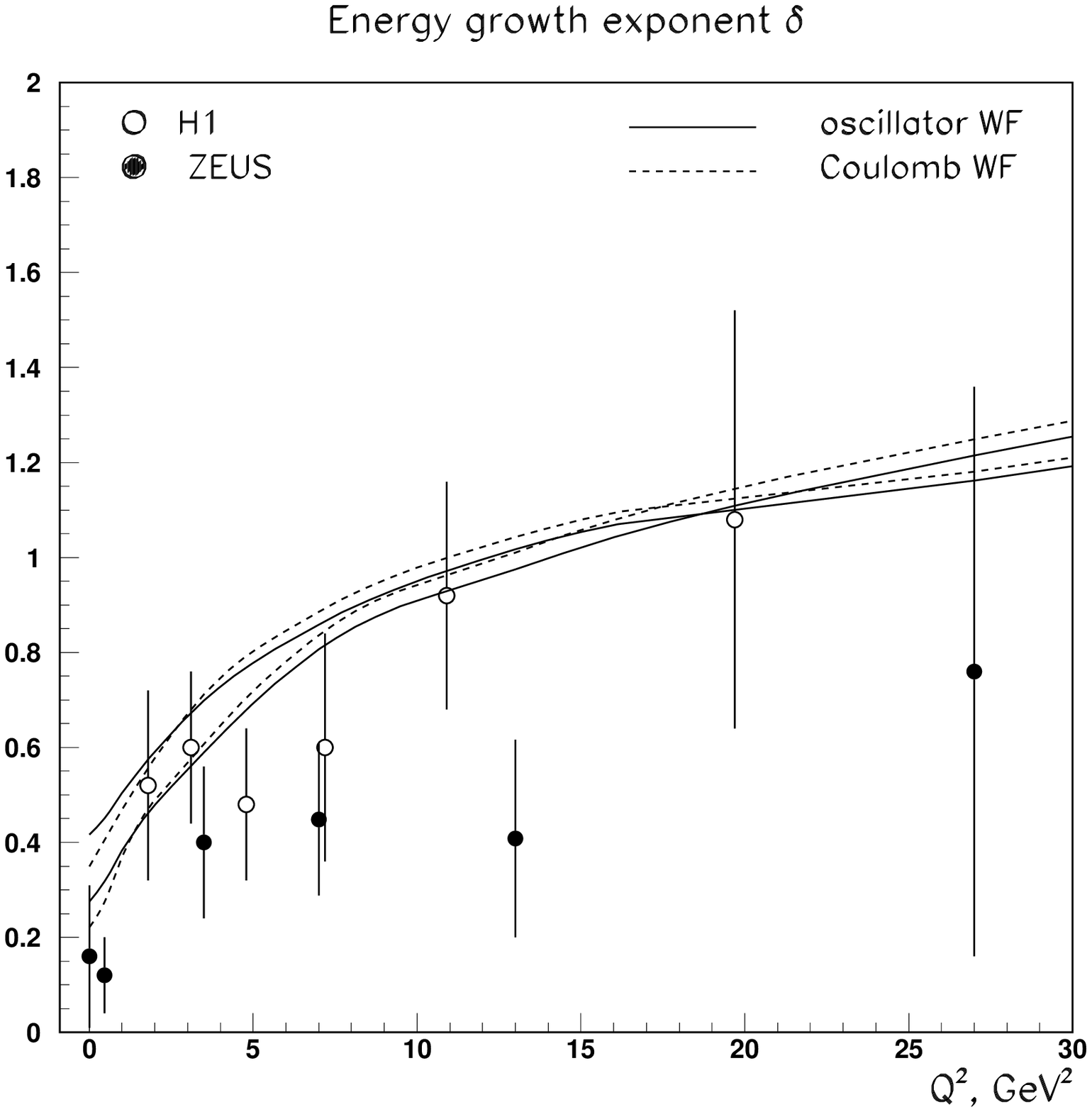,width=16cm}
   \caption{\em Effective exponents $\delta$ of energy growth
for the $\rho$ meson production cross sections.
Solid and dashed lines correspond to oscillator and suppressed
Coulomb wave functions respectively. The upper pair of curves (at low $Q^2$)
corresponds to $\delta$ calculated between 50 and 75 GeV, while
the lower pair is the result for $\delta$ calculated between
110 and 150 GeV.}
   \label{num_w-exponents}
\end{figure}

Fig.~\ref{num_w-exponents} shows the experimental data
on $\delta$ from ZEUS \cite{ZEUSrho} and H1 \cite{H1rho}
together with the $k_t$-factorization
predictions based on oscillator and Coulomb wave function. 
Since the true $W$ dependence of the cross sections
can deviate from simple power law (\ref{w2depfit}),
the exponent $\delta$ can depend on $W$ as well.
At lower $Q^2$ the upper pair of curves corresponds to $\delta$
calculated between 50 and 75 GeV, while the upper pair corresponds
to energy range between 110 and 150 GeV.

One sees that the agreement is rather good,
although a tendency that our curves go slightly higher than the 
(ZEUS) data is noticeable. However due to still significant experimental
errors, it is too early to draw any more definite conclusions.

\subsection{$t$-dependence}

The analysis of $t$-dependence of the differential cross sections
within a perturbative framework has an ambiguous status.
On the one side, if we deal with proton in final state,
this dependence is governed largely by the intrinsically 
nonperturbative (multiparticle) formfactor of the proton.
Therefore, in order to have a plausible $t$-dependence,
we have to introduce a certain "educated guess".
On the other hand, our gluon density analysis already
uses this formfactor, and therefore places some contraints on it.
Finally, as the analysis suggests, the at low $t$, 
within diffraction cone, several other mechanisms
modify the $t$ dependence of the proton formfactor.
If one parametrizes the differential cross sections
at low $t$ ($t<0.5$ GeV$^2$) by an exponential law with slope parameter $b$,
\be
{d\sigma\over d|t|} \propto e^{-b|t|}\,,
\ee
one finds that various sources of $t$-dependence
can be treated in terms of cotributions
to the overall slope $b$ approximately additively.

In principle, there can be three sources of the non-zero 
contributions to the slope: the proton transition,
the $\gamma^*\to V$ transition, and the exchange 
(the Pomeron propagation):
\be
b = b_{p\to p}  + b_{exch.} + b_{\gamma^*\to V}\,.
\ee
First term appears in our calculations explicitly as an effective slope
of the energy- and $\vec{\kappa}^2$-independent dipole formfactor 
\be
F(\vec\Delta^2) = {1 \over \left(1+\vec\Delta^2/\Lambda^2\right)^2}\,;
\quad \Lambda = 1\mbox{ GeV}\,; \quad \to \quad 
b_{p\to p} = 4 \mbox{ GeV}^{-2}\,.
\ee
The second term is also introduced explicitly.
It is responsible for the shrinkage of the diffractive cone 
with energy growth.
The third term appears from the accurate QCD treatment of the 
$\gamma^*\to V$ transition. It possesses a characteristic 
$1/(Q^2 + m_V^2)$ shape, which leads to the scaling phenomenon
mentioned earlier.

Due to the above blend of non-calculable soft and perturbative hard
contributions, the $k_t$-factorization approach should not be expected
to yield first-principles predictions for the absolute value
of the slope parameter. However, several kinematical dependencies
observed in the experiment must be confronted with the predictions.

In principle, one can invoke several definitions of the effective
slope parameter. However there is no siginficant difference
among them. Below in Table~3 we compare results for three
definitions of the effective slope:
\be
b(\mbox{def.1}) = - {d \log(d\sigma/d|t|) \over d|t|}\Bigg|_{|t|=0}\,;\quad 
b(\mbox{def.2}) = {1\over \sigma}{d\sigma\over d|t|}\Bigg|_{|t|=0}\,;\quad
b(\mbox{def.3}) = {1 \over \langle|t|\rangle} = {\int d\sigma
\over \int|t|d\sigma}\,,
\ee
calculated at several $Q^2$ points.

\begin{center}
{Table 3. Various possible definitions of the effective slope
and their values obtained from $k_t$-factorization calculations
at three characteristic values of $Q^2$.\vspace{0.3cm}\\}

 \begin{tabular}{|r|c|c|c|}
\hline
$Q^2$, GeV$^2$ & $b$(def.1) & $b$(def.2) & $b$(def.3)  \\ \hline
0    & 14.0 & 13.1  & 11.4  \\
2.2  & 10.2 &  9.6  & 8.7  \\
27   &  6.7 &  6.3  & 6.3  \\ \hline
 \end{tabular}
\end{center}

\begin{figure}[!htb]
   \centering
   \epsfig{file=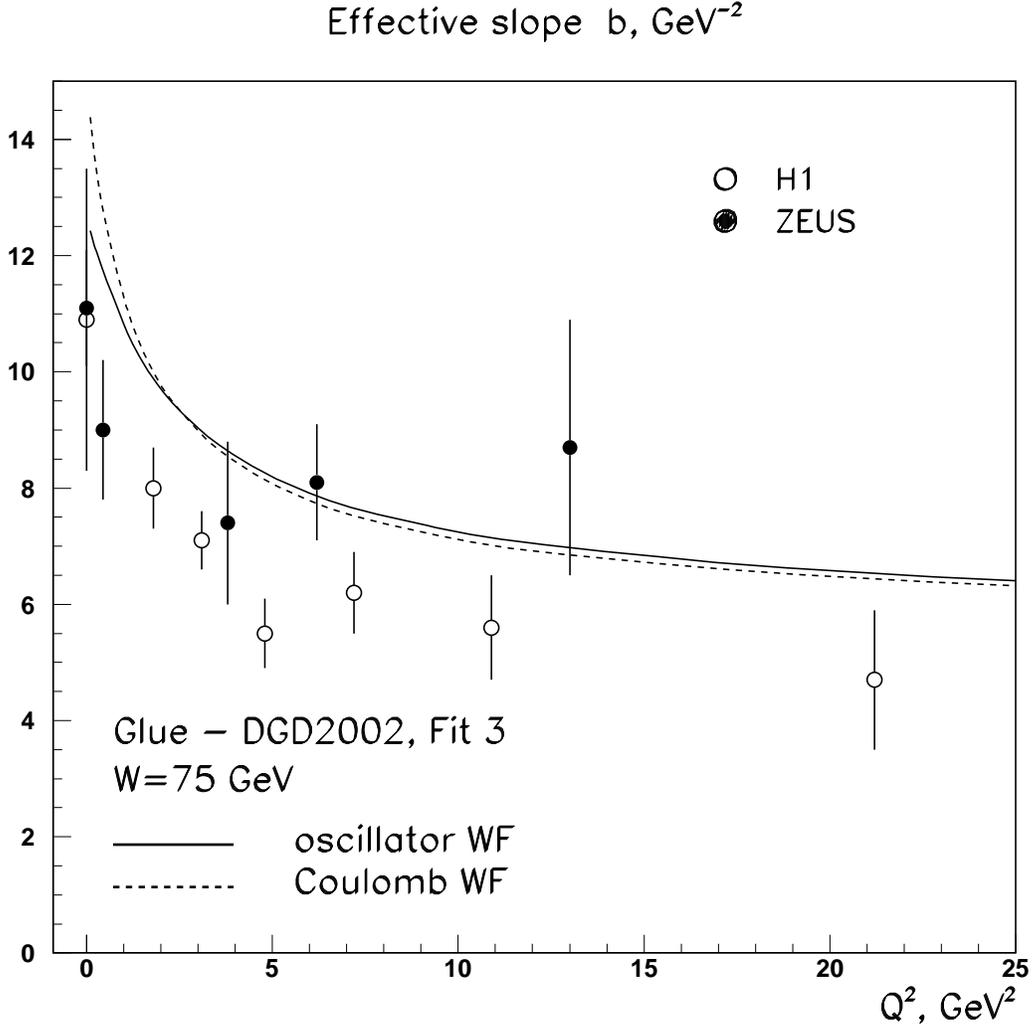,width=16cm}
   \caption{\em Effective slopes $b$ of the $\rho$ meson 
differential cross sections within diffraction cone as functions of $Q^2$.
Solid and dashed lines correspond to oscillator and suppressed
Coulomb wave functions respectively.}
   \label{num_rho_slope}
\end{figure}

In Fig.~\ref{num_rho_slope} we present our results for the slope
parameter of the $\rho$ meson production cross sections.
The results for our calculations give somewhat too high values of the 
slope than the experimentally measured numbers 
(ZEUS \cite{ZEUSrho} and H1 \cite{H1rho}), 
but nevertheless they exhibit the right $Q^2$ dependence.

\begin{figure}[!htb]
   \centering
   \epsfig{file=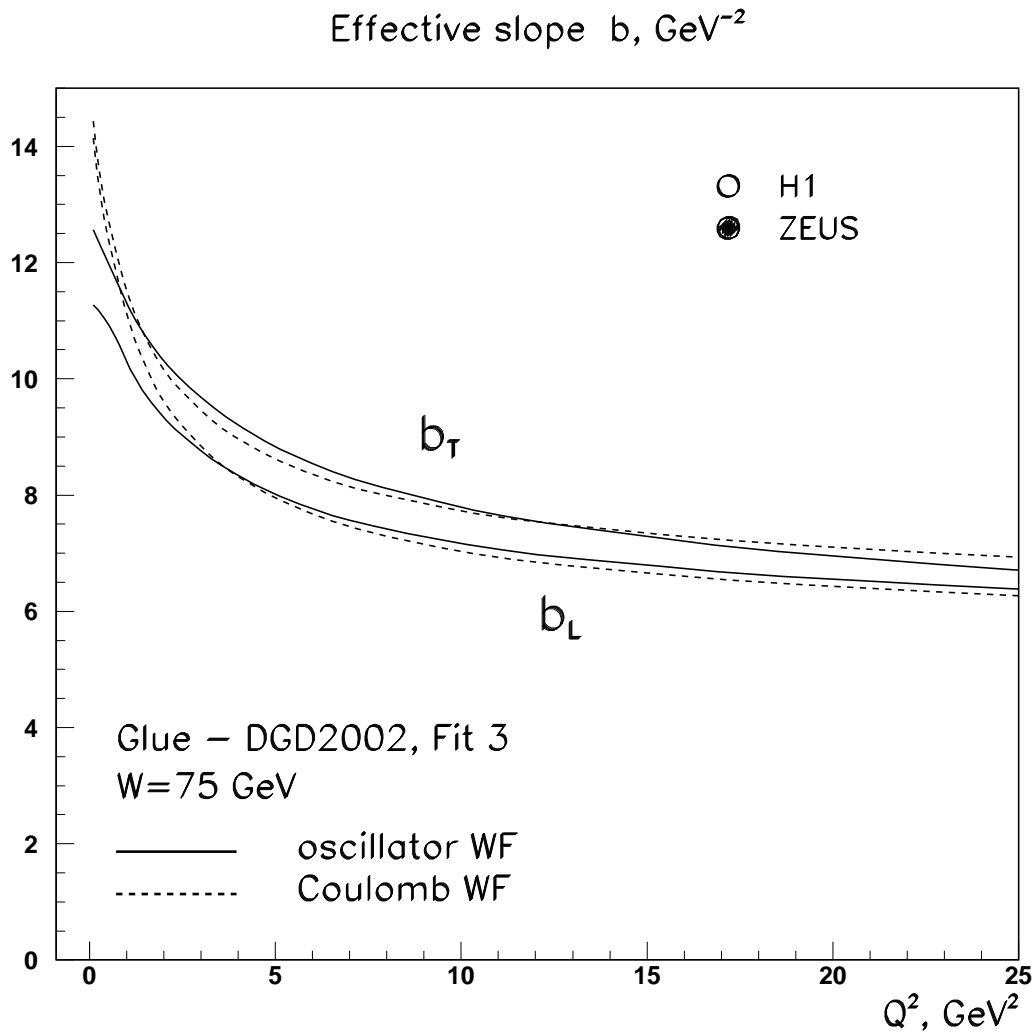,width=16cm}
   \caption{\em Effective slopes $b_L$ and $b_t$ of the $\rho$ meson 
longitudinal and transverse differential cross sections 
within diffraction cone as functions of $Q^2$.
Solid and dashed lines correspond to oscillator and suppressed
Coulomb wave functions respectively.}
   \label{num_rho_slope_lt}
\end{figure}

It is interesting to note that the slopes for $\sigma_L$ and $\sigma_T$
slightly differ, see Fig.~\ref{num_rho_slope_lt}. Note also that
at small $Q^2$ the calculations based on the Coulomb wave function
give higher results for slopes.

\begin{figure}[!htb]
   \centering
   \epsfig{file=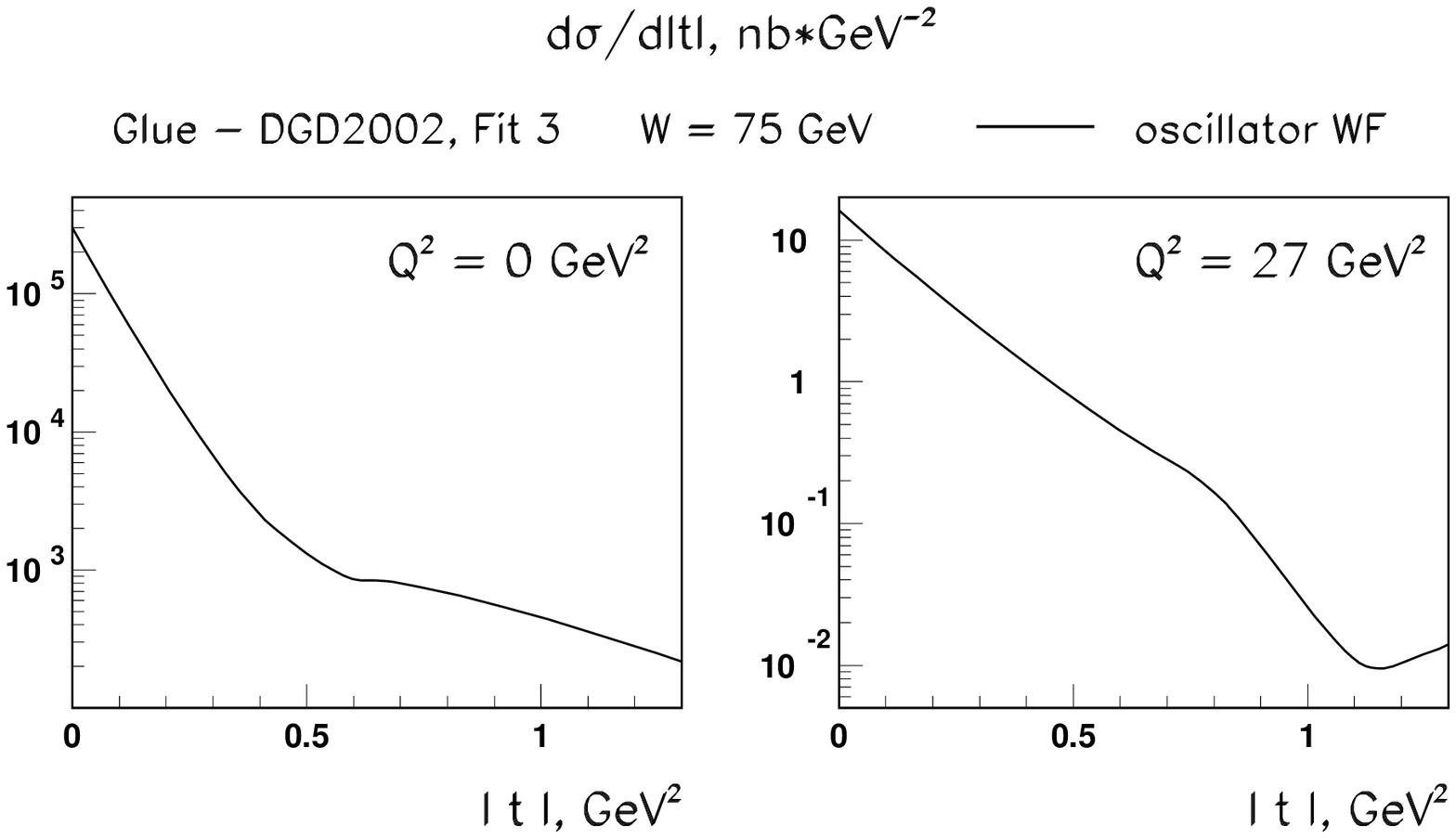,width=14cm}
   \caption{\em The transverse momentum squared dependence of
the differential $\rho$ meson production cross sections at 
two characteristic values of $Q^2$.}
   \label{num_rho_tdep}
\end{figure}

Fig.~\ref{num_rho_tdep} shows a typical pattern of
the $|t|$-dependence of the differential cross sections
in the region of small to moderate $t$ ($0<|t|<1.5$ GeV$^2$).
Oscillator wave function was used everywhere in this Figure.
One sees that initial approximately exponentially decreasing
of the differential cross section flattens at higher $|t|$
as the process leaves the diffractive peak.

\subsection{Helicity amplitudes}

Our analysis explicitly takes into account all possible helicity
amplitudes $\gamma^*(\lambda_\gamma)\to V(\lambda_V)$, with 
$\lambda_\gamma, \lambda_V = 0, \pm 1$. Since the Pomeron exchange
does not distiguish left from right, only five independent
helicity amplitudes survive.

\begin{figure}[!htb]
   \centering
   \epsfig{file=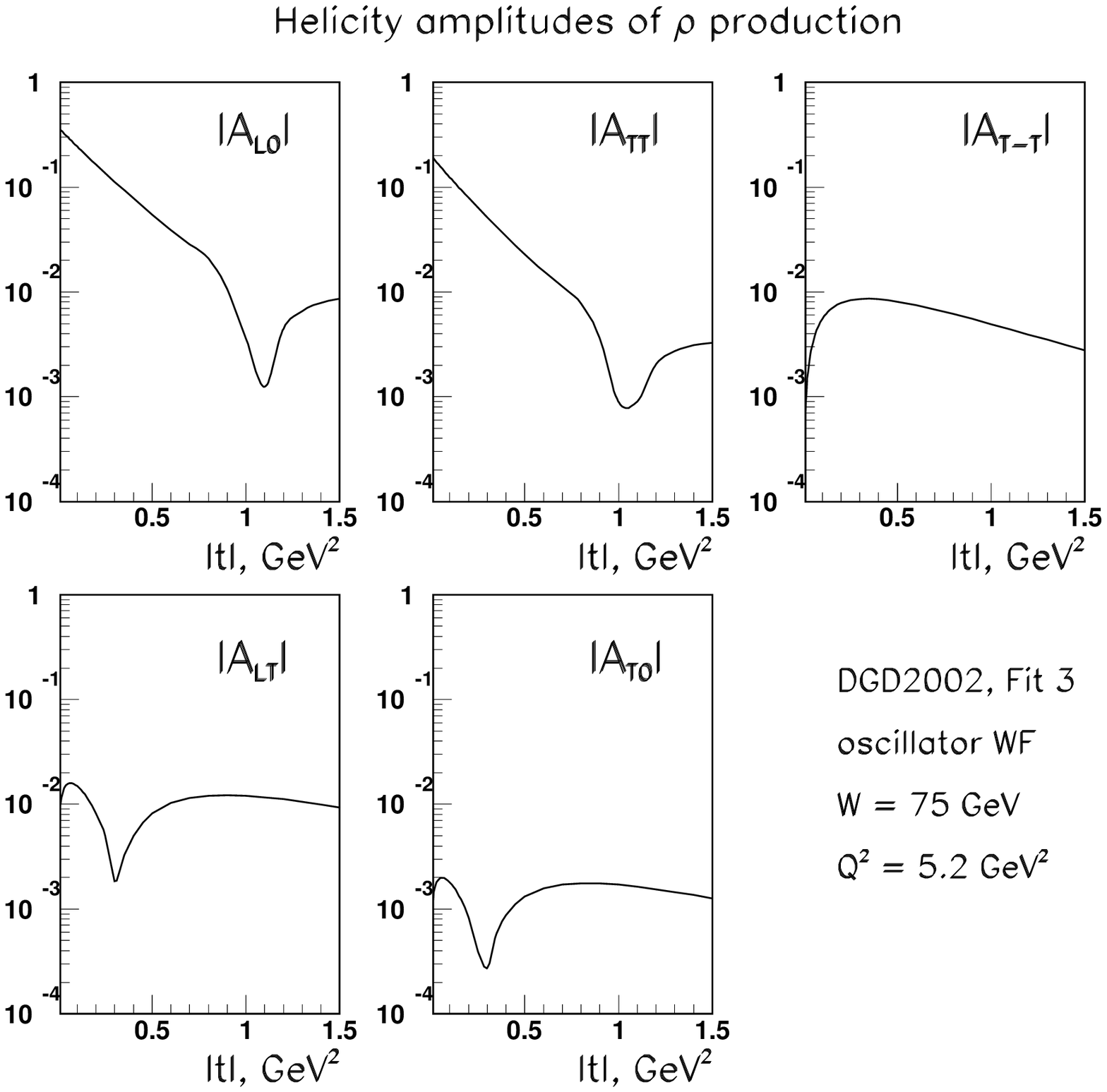,width=14cm}
   \caption{\em  The transverse momentum squared dependence of
the five helicity amplitudes of $\rho$ meson production at
$Q^2 = 5.2$ GeV$^2$.}
   \label{num_rho_helicity}
\end{figure}

\begin{figure}[!htb]
   \centering
   \epsfig{file=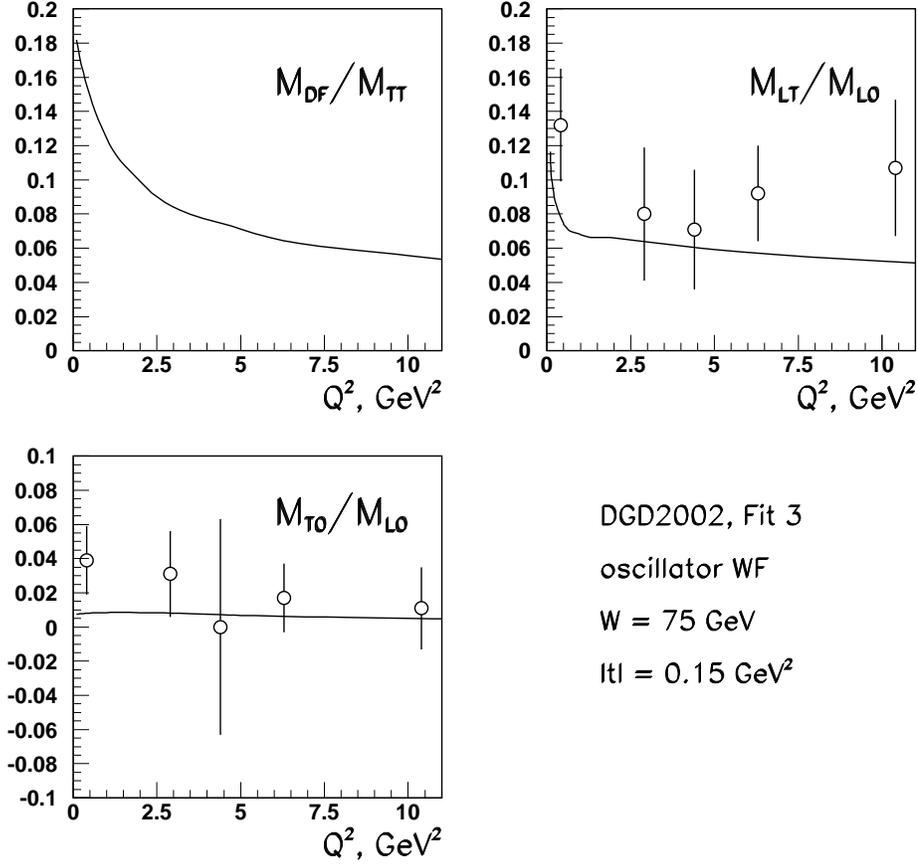,width=14cm}
   \caption{\em  The $Q^2$ dependence of the spin-flip to non-spin-flip
amplitudes in the $\rho$ meson production compared with combined HERA
data.}
   \label{num_rho_flips}
\end{figure}

\begin{figure}[!htb]
   \centering
   \epsfig{file=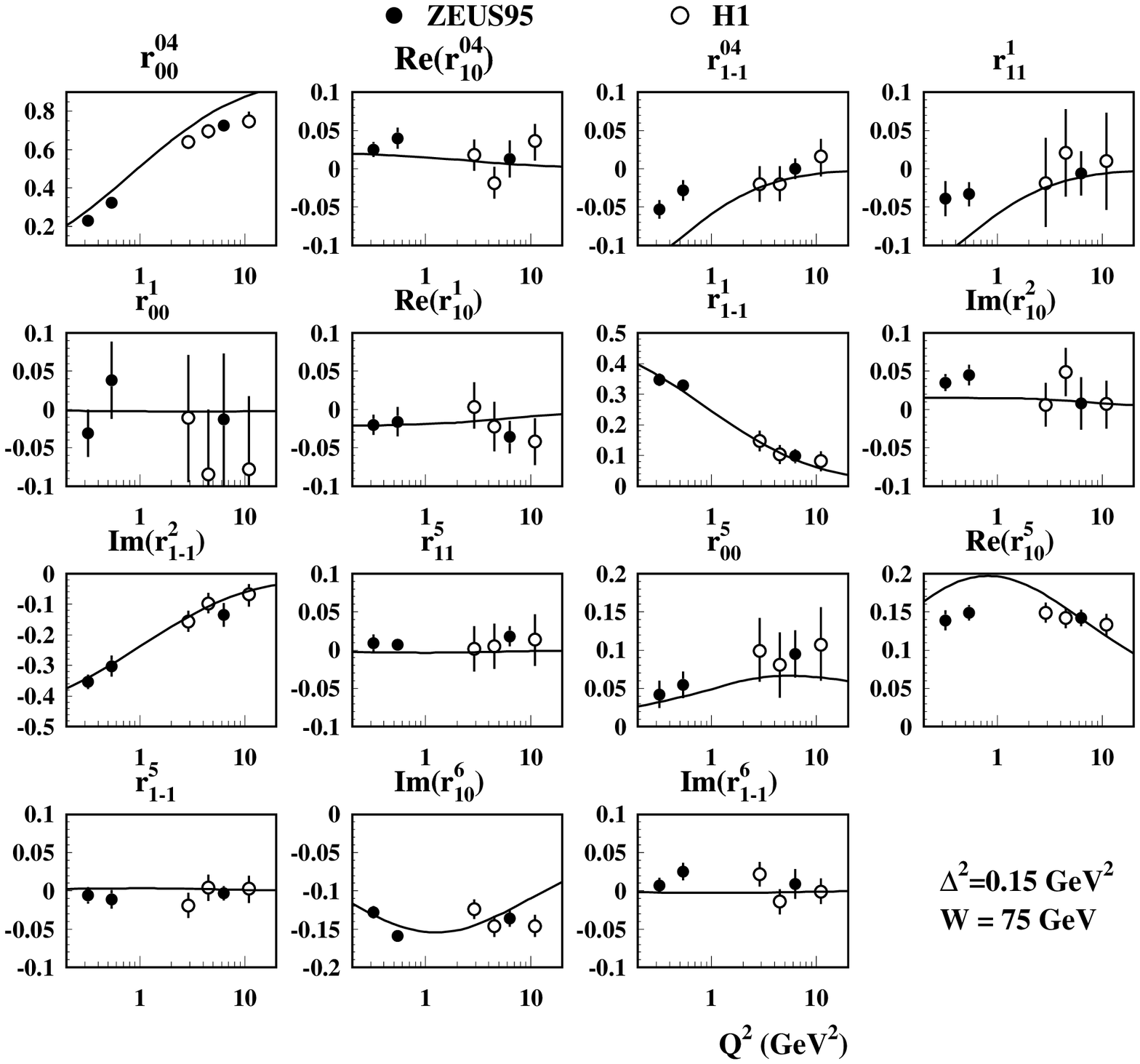,width=17cm}
   \caption{\em Experimental data on spin density matrix for 
the $\rho$ meson production as function of $Q^2$ compared with
$k_t$-factorization predictions based on oscillator wave function.}
   \label{num_rho_density_hera}
\end{figure}

Fig.~\ref{num_rho_helicity} shows the absolute values of the 
five helicity amplitudes against the momentum transfer squared $|t|$.
Within diffraction cone one sees the charachterictic behavior
of all the amplitudes. In the region of moderate $|t|$ one can observe
diffractive dips, whose location changes from one amplitude to the other.
Certainly, since we did not focus on large $t$, we cannot be sure that
the dips are located exactly where we predicted. Still this picture shows
a general pattern of the $t$ behavior of the helicity amplitudes.

It is clear that the presence of helicity flip amplitudes leads
to the breaking of the $s$-channel helicity conservation.
It is therefore interesting to check what is the magnitude
of the helicity flip amplitudes. 

Fig.~\ref{num_rho_flips} gives the answer to this question. 
Here we show ratios of the helicity flip to helicity non-flip amplitudes
\be
{|A(\gamma_{T}\to V_{-T})| \over 
|A(\gamma_{T}\to V_{T})|}\,;\quad
{|A(\gamma_{T}\to V_{L})| \over 
|A(\gamma_{0}\to V_{L})|}\,;\quad
{|A(\gamma_{0}\to V_{T})| \over 
|A(\gamma_{0}\to V_{L})|}\,,
\ee
and whenever possible compare $k_t$-factorization predictions
to the avaliable experimental data, taken from \cite{sigmalt}.

Finally, we made predictions to the full number of spin density
matrix elements for the $\rho$ meson production and compared them
with H1 \cite{H1rho} and ZEUS \cite{ZEUSrhospin} data. Results
are shown in Fig.~\ref{num_rho_density_hera}.

\section{$\phi$ mesons}

Production of $\phi$ mesons has much similarity with $\rho$ meson
production. Therefore we will not provide as detailed discussion 
of the predictions as we did for $\rho$ meson, but will rather
show our direct predictions for the quantities that have been
measured experimentally for the $\phi$ mesons.

\begin{figure}[!htb]
   \centering
   \epsfig{file=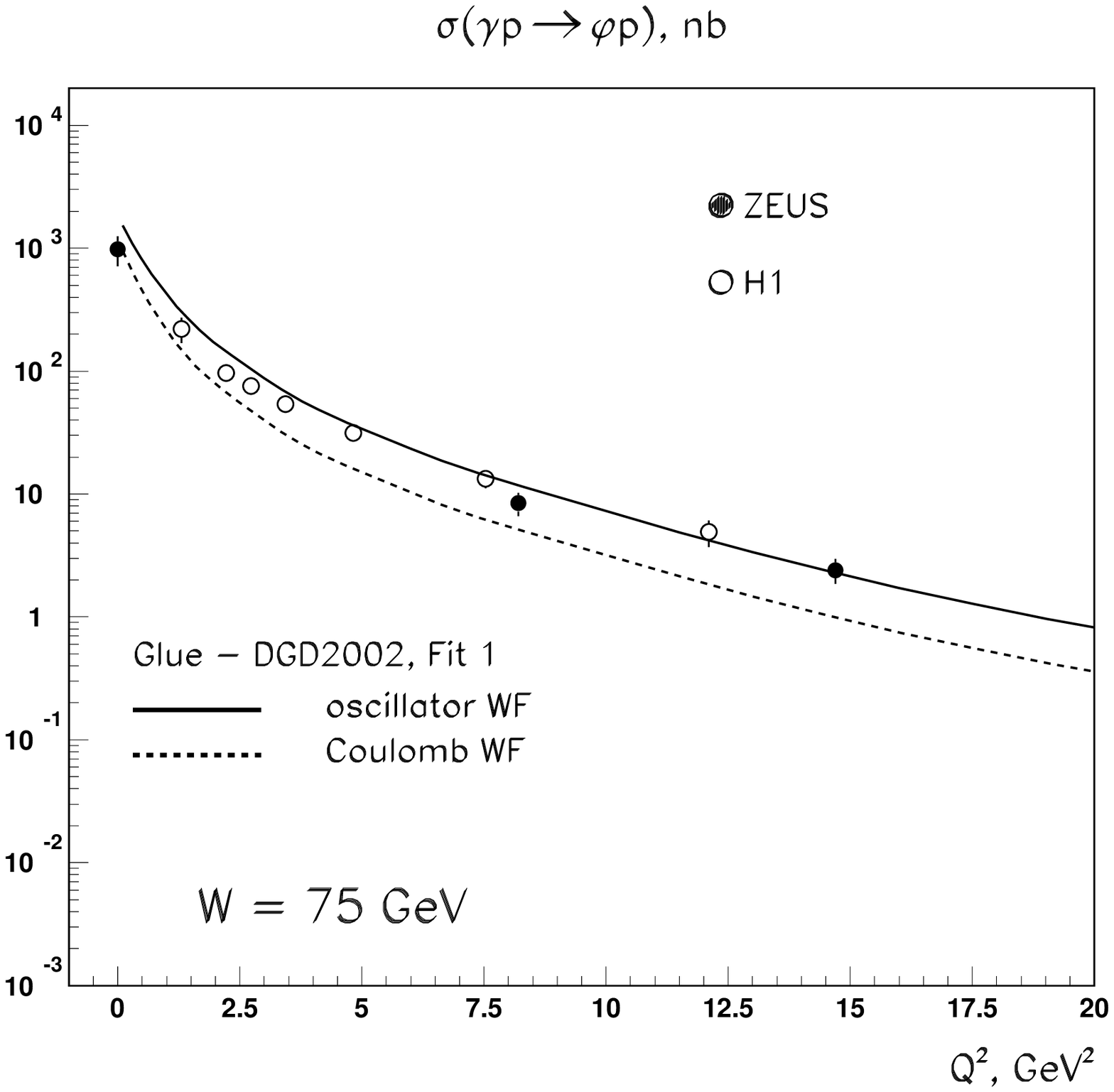,width=16cm}
   \caption{\em Total cross section of the diffractive $\phi$ meson
production as a function of $Q^2$. 
The $k_t$-factorization predictions based on oscillator (solid lines) 
and suppressed Coulomb (dashed lines) are also shown.
All calculation are performed for $W=75$ GeV using DGD2002, Fit 1.}
   \label{num_phi_total}
\end{figure}

\begin{figure}[!htb]
   \centering
   \epsfig{file=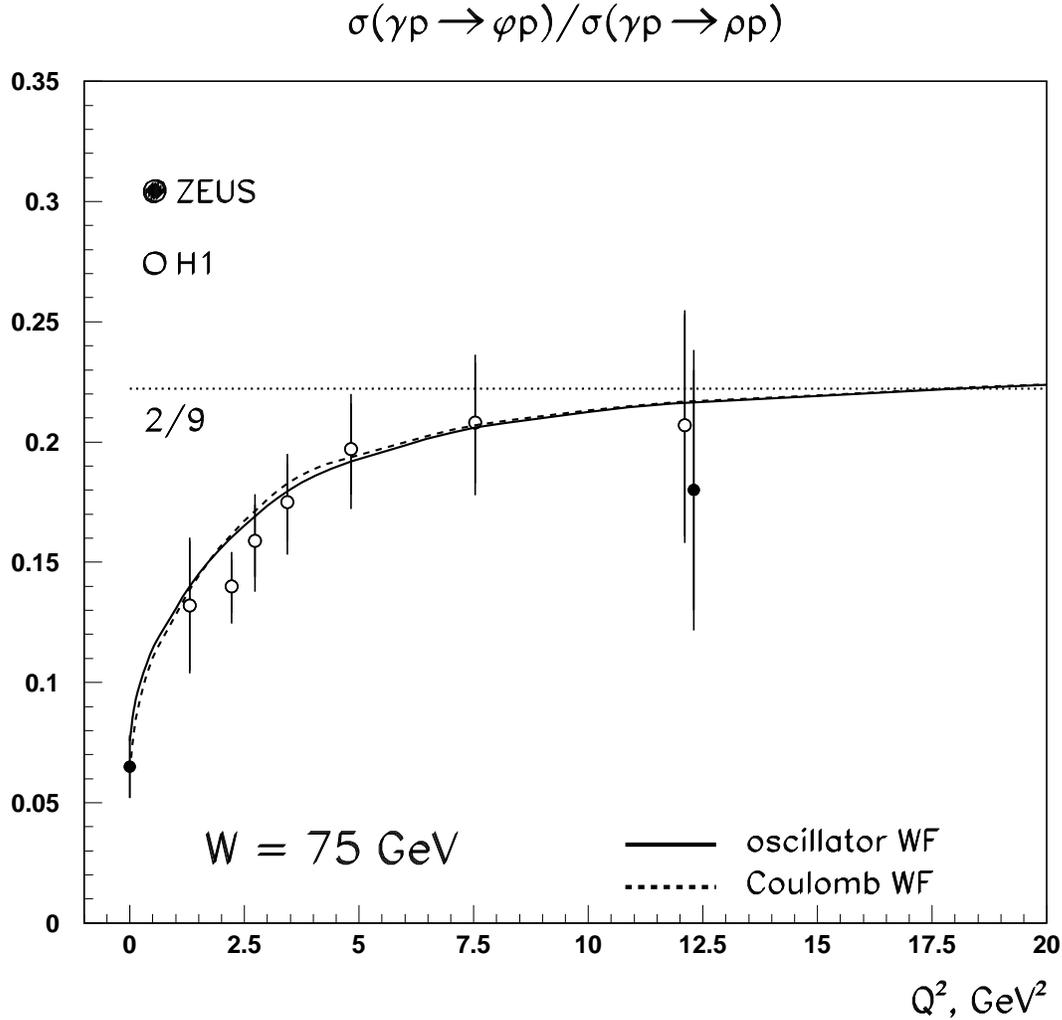,width=16cm}
   \caption{\em Ratio of $\phi$ meson to $\rho$ meson
total production cross sections as function $Q^2$. 
Solid and dashed lines correspond to oscillator and suppressed
Coulomb wave functions respectively.}
   \label{num_phi_rho_ratio}
\end{figure}

\begin{figure}[!htb]
   \centering
   \epsfig{file=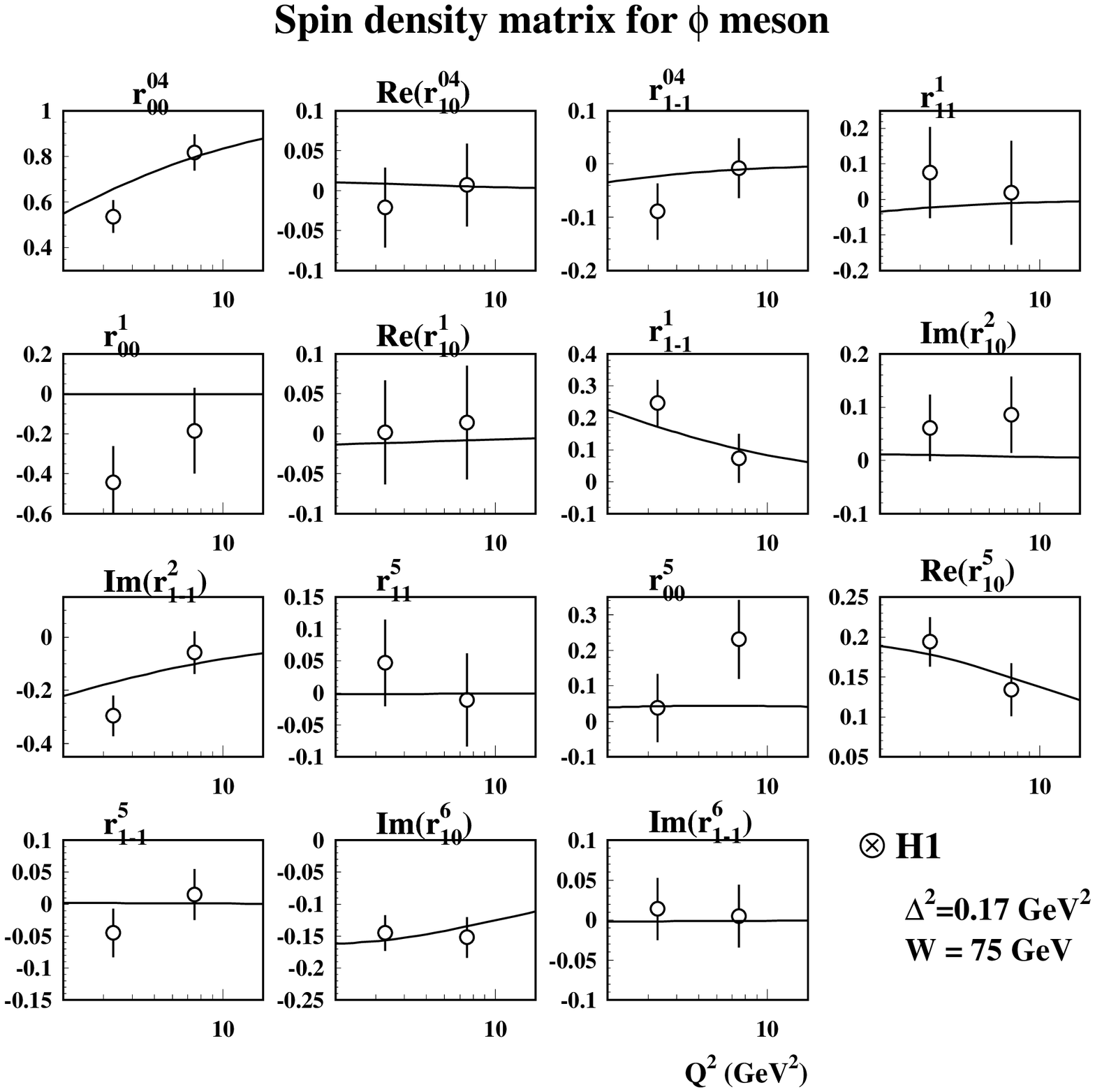,width=16cm}
   \caption{\em  Experimental data on spin density matrix for 
the $\phi$ meson production as function of $Q^2$ compared with
$k_t$-factorization predictions based on oscillator wave function.}
   \label{num_phi_density}
\end{figure}

Fig.~\ref{num_phi_total} shows the total cross sections
of diffractive $\phi$ meson production as a function of $Q^2$
data taken from H1 \cite{H1phi} and ZEUS \cite{ZEUSphi}, \cite{ZEUSphi2}.
Remembering the scaling phenomenon of the vector meson production
cross sections, we should expect the picture similar to what happens in
the $\rho$ meson case. Indeed, we see at this figure
a reasonably good description of the data for moderate $Q^2$
and a slight overshooting of our predictions as we shift towards small
$Q^2$.

It is interesting to directly compare $\rho$ meson and $\phi$
meson production cross sections taken at equal $Q^2$.
Fig.~\ref{num_phi_rho_ratio} shows our predictions for the ratio 
$\sigma(\gamma^*p\to\phi p)/\sigma(\gamma^*p\to\rho p)$
toether with experimental data.
If the scaling phenomenon holds, at higher $Q^2$ 
the ratio is expected to approach constant value of $2/9$,
a tendency to do so is indeed visible at the figure.
However, at smaller $Q^2$ the ratio goes down. This behavior should
be expected, for ta smaller $Q^2$ the ratio can be approximated by
\be
{\sigma(\gamma^*p\to\phi p) \over \sigma(\gamma^*p\to\rho p)}
\propto \left({Q^2+m^2_{\rho eff} \over Q^2+m^2_{\phi eff}}\right)^2\,,
\ee
with $m_{\rho eff} < m_{\phi eff}$.

One sees that our predictions for this ration agree with 
the data very well. Moreover, note a remarkable coincidence
of the results based on the oscillator and Coulomb wave functions.
The reason for that is of course the fact that we study here
not absolute values of cross sections, but their ratios.
This removes a significant part of ambiguity present
in this or that specific choice of the wave function
and reveals the features of the $k_t$ factorization approach
in its pure form.

Finally, Fig.~\ref{num_phi_density} shows the experimental results
for the $\phi$ meson density matrix measurement, 
published by H1 \cite{H1phi}. Our predictions agree with the data well.

\section{$J/\psi$ and $\Upsilon$ mesons}

\begin{figure}[!htb]
   \centering
   \epsfig{file=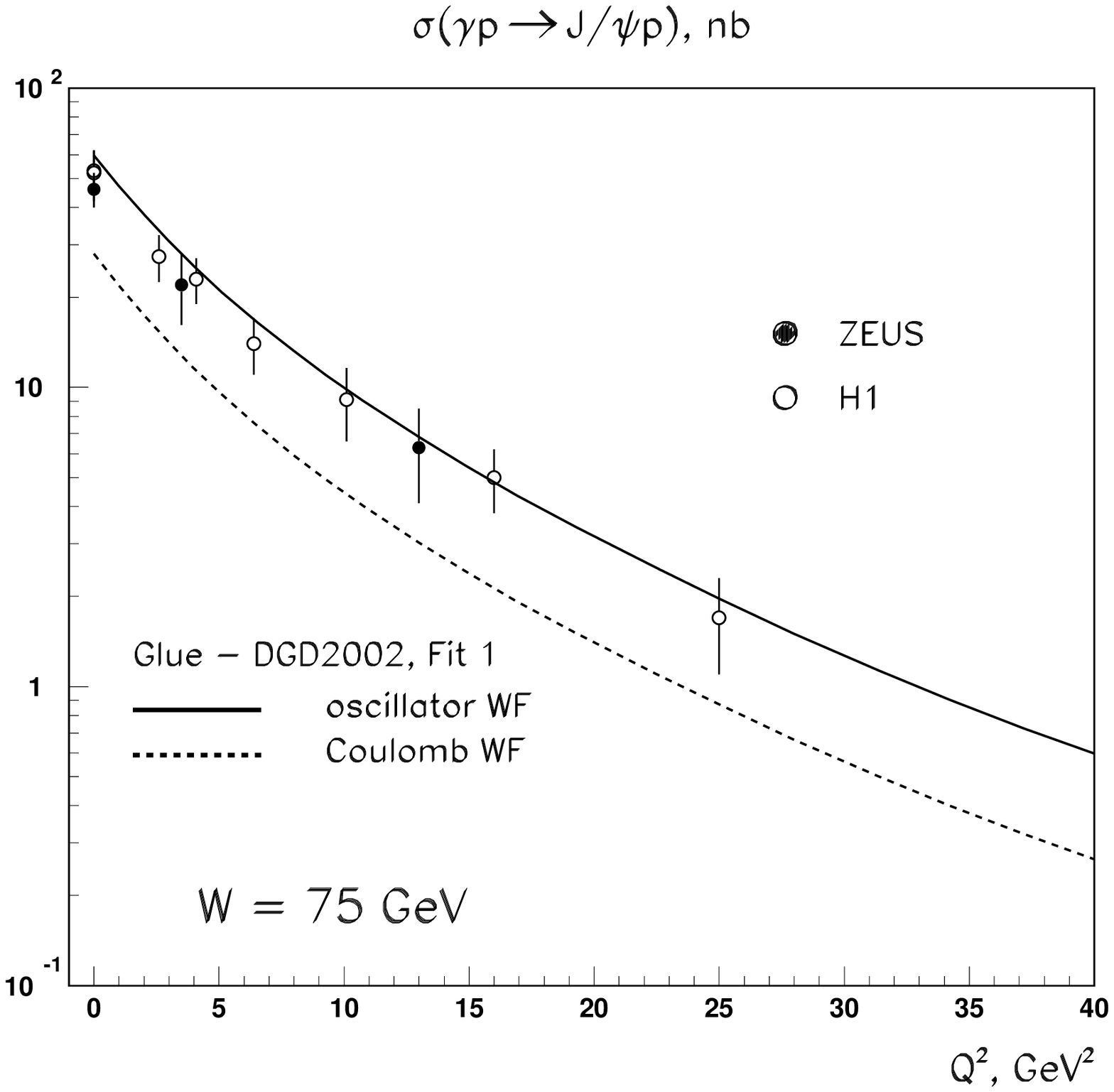,width=16cm}
   \caption{\em  Total cross section of the diffractive $J/\psi$ meson
production as a function of $Q^2$. 
The $k_t$-factorization predictions based on oscillator (solid lines) 
and suppressed Coulomb (dashed lines) are also shown.
All calculation are performed for $W=75$ GeV using DGD2002, Fit 1.}
   \label{num_jpsi_total}
\end{figure}

Fig.~\ref{num_jpsi_total} shows the $k_t$-factorization predictions
for the total cross sections of the $J/\psi$ meson electroproduction 
as a function of $Q^2$ compared with available data from
H1 \cite{H1jpsiphoto}, \cite{H1jpsi} and ZEUS \cite{ZEUSrho}, 
\cite{ZEUSjpsi2}. A reasonable agreement throughout the whole
$Q^2$ range is seen.

\begin{figure}[!htb]
   \centering
   \epsfig{file=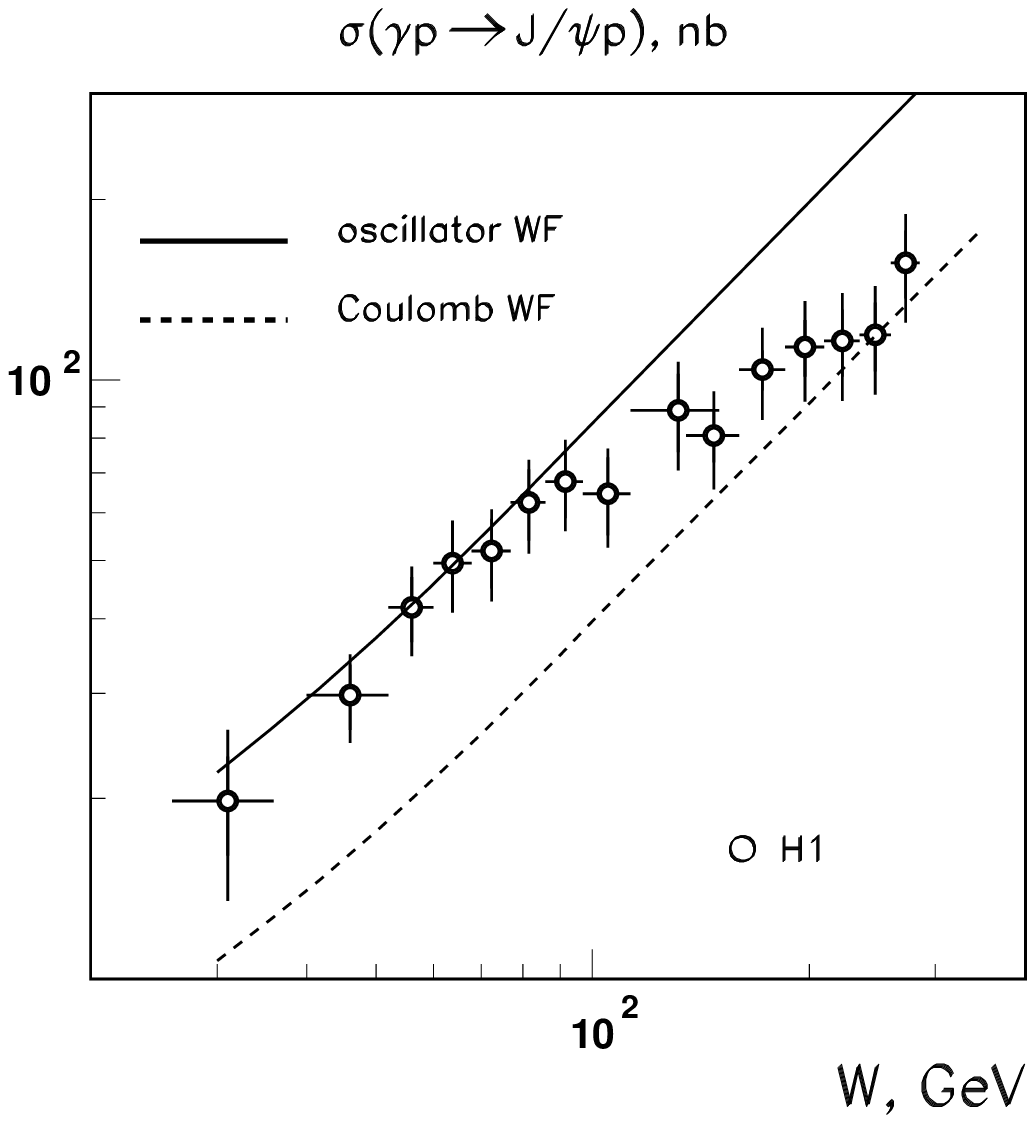,width=16cm}
   \caption{\em The energy dependence of the total $J/\psi$ meson
photoproduction cross section. 
The $k_t$-factorization predictions based on oscillator (solid lines) 
and suppressed Coulomb (dashed lines) are also shown.
All calculation are performed for $W=75$ GeV using DGD2002, Fit 1.}
   \label{num_jpsi_w2}
\end{figure}

\begin{figure}[!htb]
   \centering
   \epsfig{file=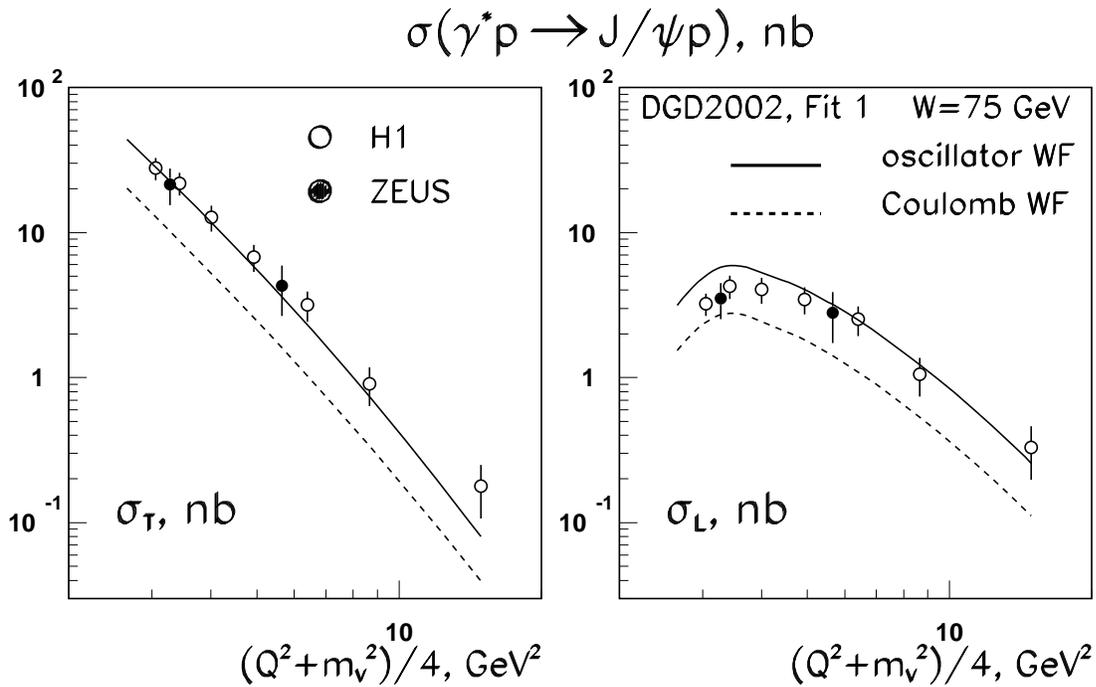,width=16cm}
   \caption{\em Decomposition of the $J/\psi$ production 
cross sections in $\sigma_L$ and $\sigma_T$}.
   \label{num_jpsi_lt}
\end{figure}

\begin{figure}[!htb]
   \centering
   \epsfig{file=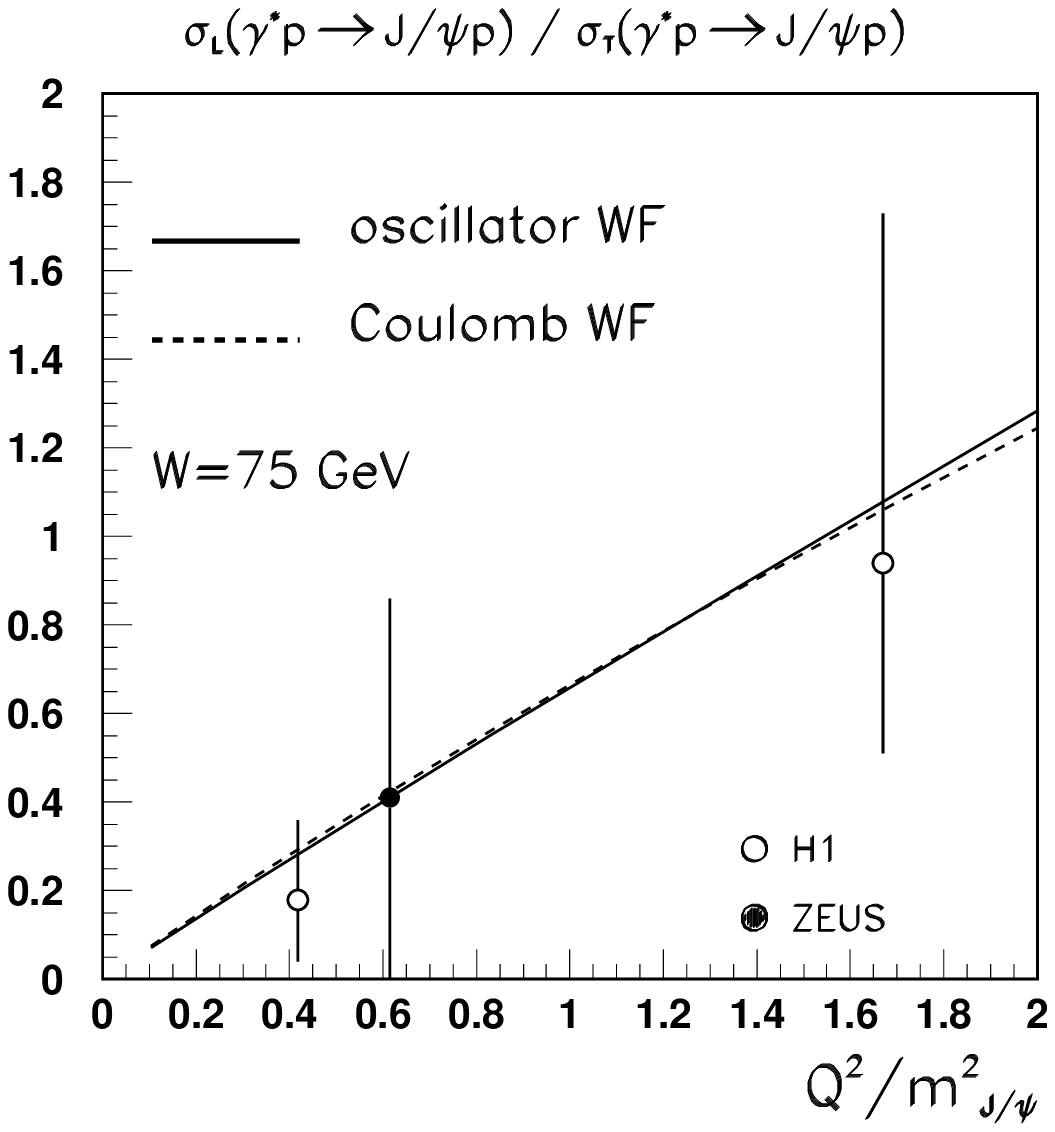,width=16cm}
   \caption{\em Ratio $R = \sigma_L/\sigma_T$ as a function of $Q^2$
for $J/\psi$ meson as a function of $Q^2$.
Solid and dashed lines correspond to oscillator and suppressed
Coulomb wave functions respectively}
   \label{num_jpsi_ratiolt}
\end{figure}

The photoproduction cross sections are shown at Fig.~\ref{num_jpsi_w2}
versus total energy $W$ of the $\gamma p$ collision.
One can see again that the $k_t$-factorization calculations
give prediction with accuracy roughly of factor 2,
uncertainty coming from the meson wave function. Still,
the experimental data fit between the two curves.

Fig.~\ref{num_jpsi_lt} shows $\sigma_L - \sigma_T$ decomposition
of the $J/\psi$ meson production cross sections, taken from
\cite{sigmalt}. An agreement at a similar level of accuracy is
observed as well. The ratio $R(Q^2) = \sigma_L/\sigma_T$ is shown 
at Fig.~\ref{num_jpsi_ratiolt}. The data are taken from
ZEUS and H1.

\begin{figure}[!htb]
   \centering
   \epsfig{file=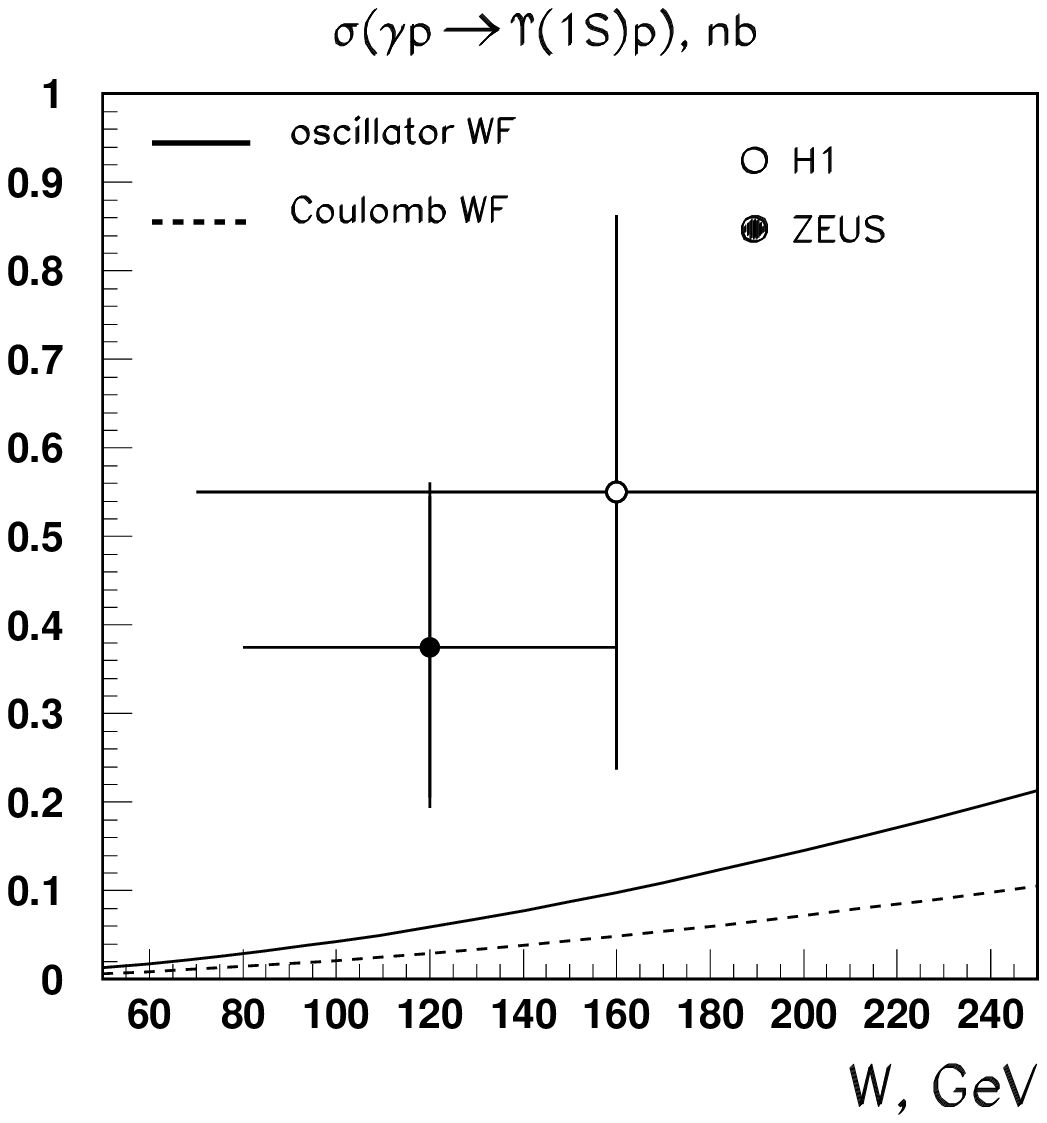,width=16cm}
   \caption{\em  The energy dependence of the total $\Upsilon(1S)$ meson
photoproduction cross section. 
The $k_t$-factorization predictions based on oscillator (solid lines) 
and suppressed Coulomb (dashed lines) are also shown.
All calculation are performed for $W=75$ GeV using DGD2002, Fit 1.}
   \label{num_upsilon_w2}
\end{figure}

The energy dependence of the $\Upsilon(1s)$ meson 
photoproduction is shown at Fig.~\ref{num_upsilon_w2}.
One observes that the experimental points soar about 10 times
higher than the both curves. The origin of this discrepancy
is not clear. At the one hand, the $k_t$ factorization
predictions for the $\Upsilon$ production cross sections
follow the scaling behavior in Fig.~\ref{num_tot_all}, while
the $\Upsilon$ photoproduction data do not comply
with this tendency. It should be noted also that 
a similar behavior is observed for other processes:
that it, the production rates for $\Upsilon$ seems to be higher
than expected.

\section{Production of excited states}

In this section we will give some of the most prominent
features in the reaction of diffractive production of $2S$ and
$D$-wave vector mesons.

\begin{figure}[!htb]
   \centering
   \epsfig{file=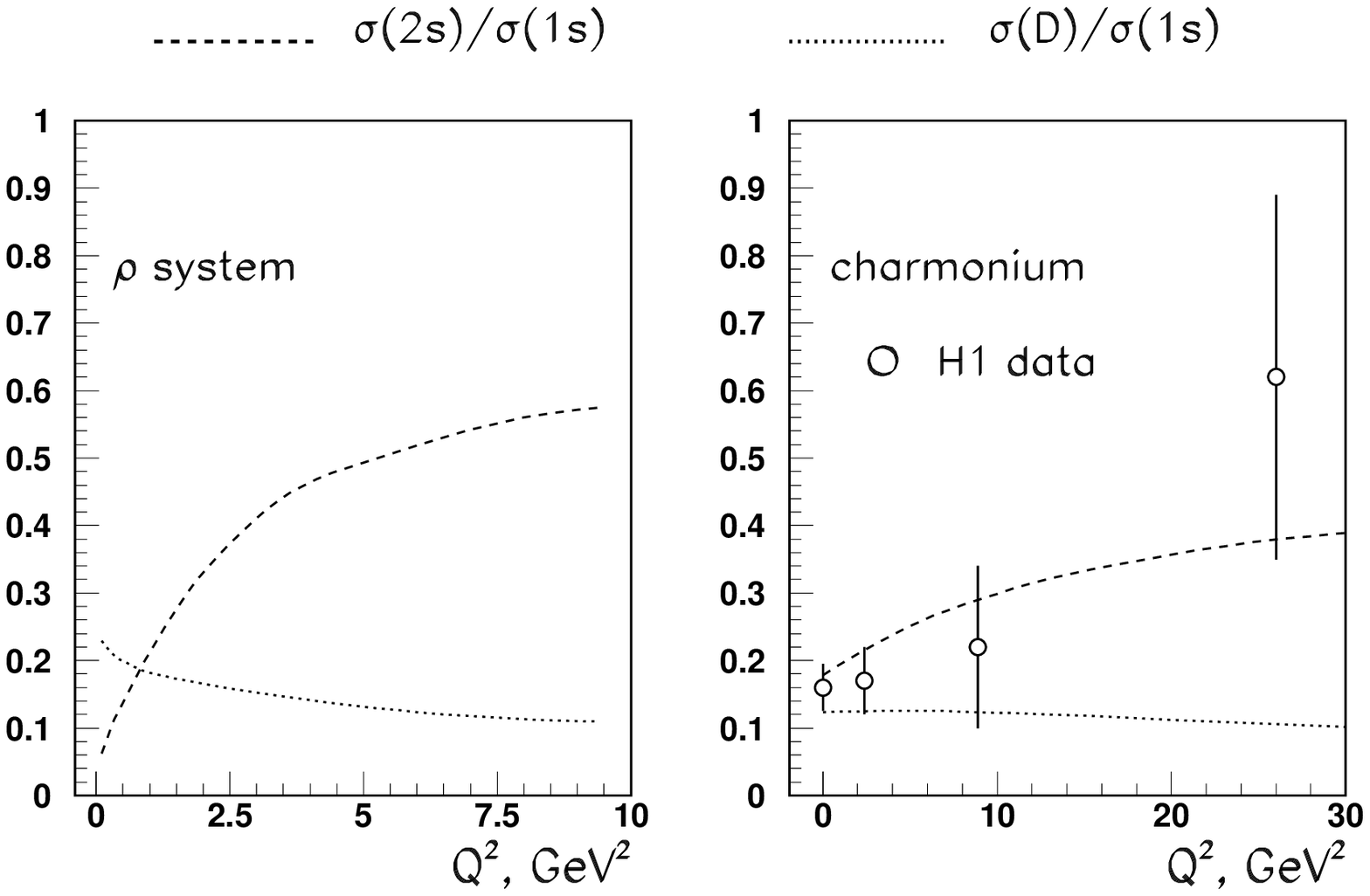,width=16cm}
   \caption{\em The $Q^2$ behavior of the ratio of excited
to ground state vector meson production cross sections 
in the case of $\rho$ system and charmonium. Dashed and dotted lines
represent $\sigma(2S)/\sigma(1S)$ and $\sigma(D)/\sigma(1S)$ ratios
respectively. In all calculations, oscillator wave function was used.}
   \label{num_excited_tot}
\end{figure}

In Fig.~\ref{num_excited_tot} we show the ratios of the excited-to-ground
state production cross sections 
\be
r(2S/1S) = {\sigma(\gamma p\to V(2S) p) \over \sigma(\gamma p\to V(1S) p)}
\,;\quad r(D/1S) = 
{\sigma(\gamma p\to V(D) p) \over \sigma(\gamma p\to V(1S) p)}\,.
\ee
for $\rho$ system and for charmonium. 

A brief look reveals that although the production
rate both of $2S$ and of $D$-wave states are suppressedin respect to
ground state, a radically different $Q^2$ behavior of the suppressing
factors $r(2S/1S)$ and $r(D/1S)$ is observed.
This difference is due to distinct nature of suppression
in these two cases.

The suppression for $2S$ state production comes from
much discussed node effect of $2S$ state wave function.
Indeed, if one looks at the $2S$ production amplitude in the 
impact parameter space, then one product of two factors:
the photon wave function multiplied by dipole cross section 
$\Psi_\gamma\cdot \sigma_{dip}$ and
the radial wave function of the produced meson.
At low $Q^2$, $\Psi_\gamma\cdot \sigma_{dip}$ is a wide function,
peaked at a hadronic scale, that is precisely where the
node of the vector meson wave function is located. This  
results in significant calcellation between contributions
from impact parameter regions smaller and higher than the node position.
At higher $Q^2$, the photon wave function shrinks, and the peak of 
$\Psi_\gamma\cdot \sigma_{dip}$ shifts toward smaller $r$, that is away
from the node. Thus, the effect of calcellation vanished.

As for the $D$ wave meson suppression,
it arises from the angular part of the integrals.
In fact, if the initial photon were built only of the $S$ wave
$q\bar q$ pair, then, due to ortogonality between pure $S$ and $D$ waves,
there would be no $D$ wave vector mesons at all.
However, the spinorial structure of the photon coupling to the quark 
line does not correspond to the pure $S$ wave, but contains 
an admixture of $D$ wave as well. This $D$ wave part leads to the $D$ wave
meson producton, and since it is rather small, the cross sections
turn out suppressed as well.

The situation is basically the same both in the $\rho$ system and in the
charmonium. The major difference is the energy scale of the node effect
suppression, which can be directly related to the mass of the corresponding
meson. 

In the case of charmonium the $k_t$-factorization predictions 
are compared with the available H1 data on $\psi(2s)$ 
production \cite{H1psi2sphoto}, \cite{H1psi2s}. A good agreement is seen.

\begin{figure}[!htb]
   \centering
   \epsfig{file=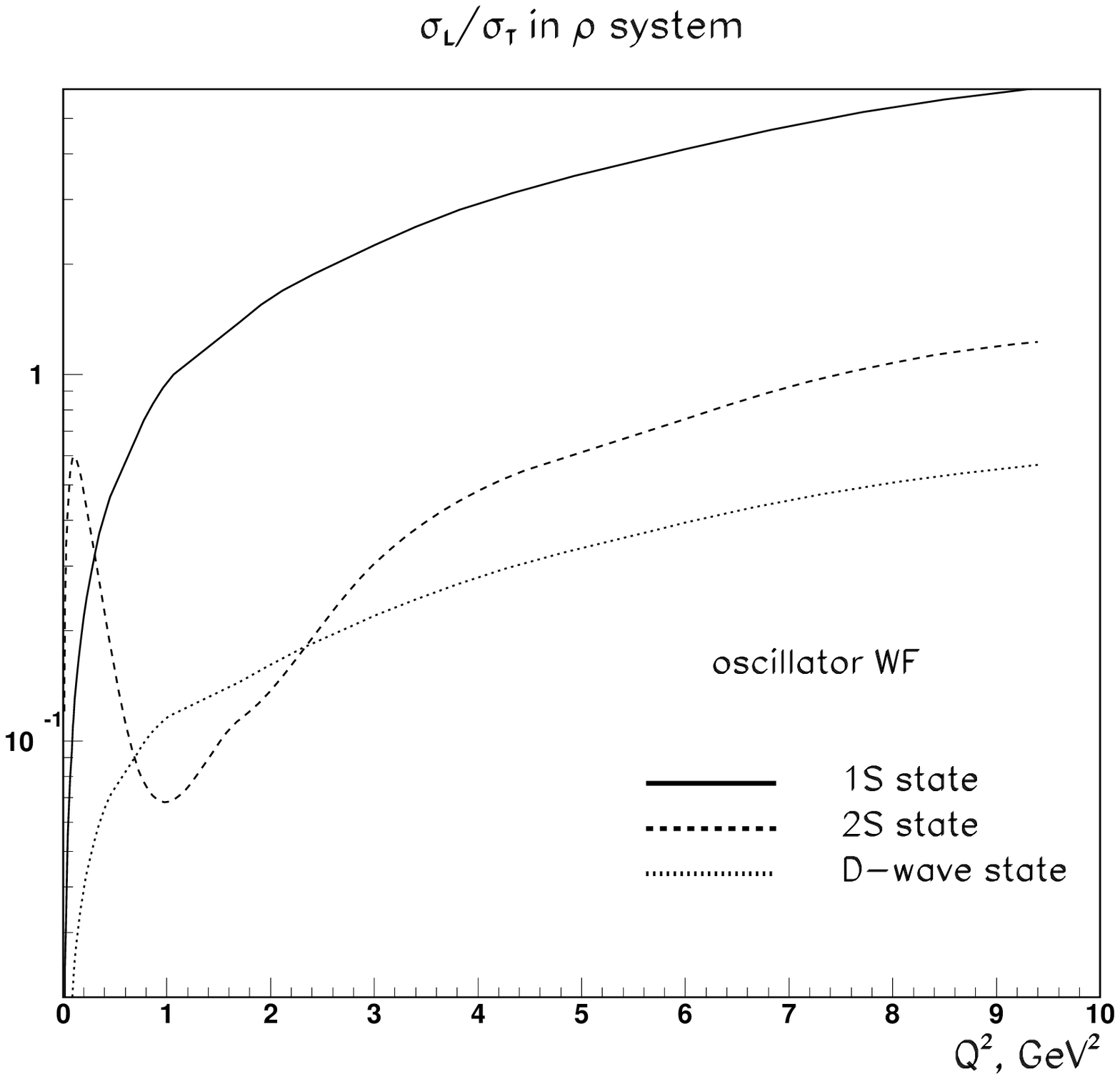,width=16cm}
   \caption{\em  The $Q^2$ behavior of the $\sigma_L/\sigma_T$
ratio for $1S$ (solid line), $2S$ (dashed line) and $D$ wave
(dotted line) states in the $\rho$ system. 
In all calculations, oscillator wave function was used.}
   \label{num_excited_lt}
\end{figure}

As vividly illustrated by Fig.~\ref{num_excited_tot},
a mere measurement of the production cross section for
high-lying states in the $\rho$ spectrum (such as $\rho'(1450)$ and
$\rho'(1700)$) at high $Q^2$ might be enough to get
some insight into the spin-angular structure of these states.
However, even more dramatic distinction between $S$ and $D$ wave states
is offered by the $R_V = \sigma_L(V)/\sigma_T(V)$ measurements.

Fig.~\ref{num_excited_lt} shows rations $R(1S)$, $R(2S)$, and $R(D)$
for the $\rho$ system. We see that at higher $Q^2$ the following
hierarchy takes place
\be
R(1S) \gg R(2S) \gg R(D)\,,
\ee
with about one order of magnitude difference between the $1S$ and 
$D$-wave states production rates.
At smaller $Q^2$, $R(2S)$ exhibits very characteristic 
spectacular wiggles. Starting from zero at $Q^2=0$, 
it springs to a local maximum of .,.. at $Q^2 \approx $ GeV$^2$,
then rapidly fall down, and then rises again.
This behavior is the manifestation of the node effect as well.
Indeed, since the structure of the transverse and longitudinal 
photons is not identical, the strongest cancellation 
takes places at different $Q^2$ values in these two cases.
In particular, the curve suggests that as start from $Q^2 = 0$ 
and we slide along $Q^2$ scale, the strongest cancellation
occurs in $\sigma_T$ earlier than in $\sigma_L$.

A word of caution shoud be said now.
The exact position of the node in the $2S$ state wave function
cannot be predicted accurately and depends on the particular
Ansatz for the wave function chosen. This means that
the wiggly shape of $R(2S)$ in Fig.~\ref{num_excited_lt}
--- with the $Q^2$ positions and the values of local extrema ---
should not be regarded as an accurate prediction.
In fact, one cannot even be sure that the non-monotonous
$Q^2$ behavior of $R(2S)$ actually takes place.
This curve simply illustrates what kind of effect can be expected.
The specific profile of this curve is not of course predictable,
as long as we rely on simple vector meson wave functions.

An immediate conclusion from here is that we found an observable
that is extremely sensitive to the minute details of the vector meson
wave function --- a powerful tool that would help distunguish
among various models of the vector meson structure.

\begin{figure}[!htb]
   \centering
   \epsfig{file=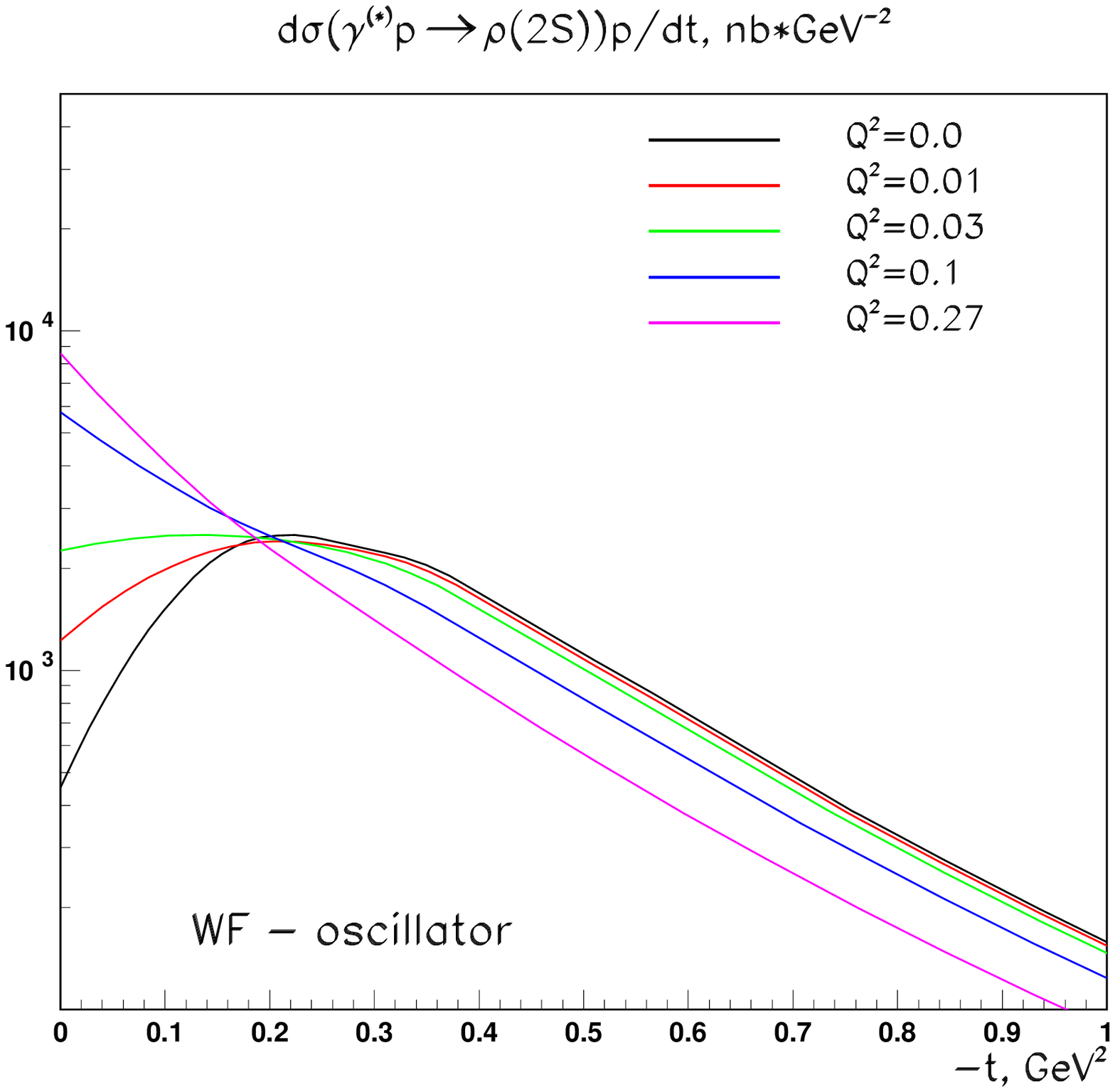,width=16cm}
   \caption{\em Strong forward dip for $2S$ state in $\rho$ system:
the change in $t$ profile of the differential cross section
wihtin narrow $Q^2$ region.}
   \label{num_2s_tdep}
\end{figure}

The $t$-dependence of the $2S$ state production also shows
remarkable features originating from the node effect.
Fig.~\ref{num_2s_tdep} shows how the $t$-profile of the differential
cross section changes over a narrow $Q^2$ region.
The strongest cancellation due to the node effect takes place
at $Q^2 =0\,, |t|=0$. When we increase $|t|$ or $Q^2$,
the strength of the cancellation diminishes, and the differential
cross section grows.

\begin{figure}[!htb]
   \centering
   \epsfig{file=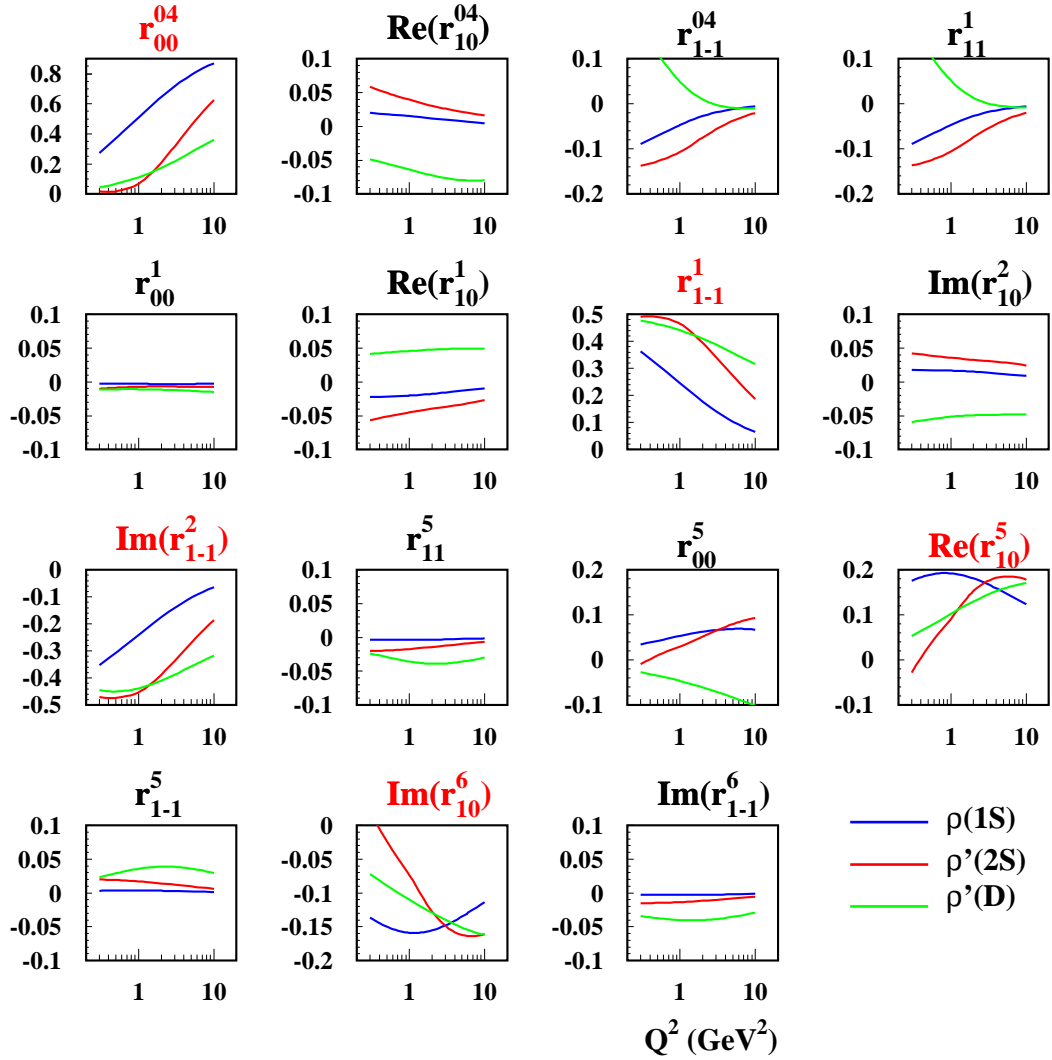,width=16cm}
   \caption{\em The $Q^2$ behavior of the spin ensity matrix 
for $1S$, $2S$, and $D$ wave states in the $\rho$ system.}
   \label{num_excited_spin}
\end{figure}

Finally, in Fig.~\ref{num_excited_spin} we show the spin-angular 
density matrix for the $1S$, $2S$, and $D$ states  in the $\rho$ system.
A dramatic difference among the $Q^2$ profiles both SCHC and SCHNC
matrix elements is seen. Note that some of the matrix elements
are even of opposite sign ofr $S$ and $D$ wave states.

\chapter{The $\sigma_T$ puzzle}

The detailed comparison of the numerical results 
for $\rho$ meson production obtained
in the $k_t$-factorization approach with the experimental data
shows that the scheme used fails to reproduce the
correct $Q^2$ behavior of the transverse cross section $\sigma_T$,
which was dubbed by us as $\sigma_T$ puzzle.
In this chapter we explore two possible causes
of this mismatch. Namely, we explore the effect
of color Coulomb interaction of $q\bar q$ pair and show
that it leads to $\sigma_T$ increase in respect to $\sigma_L$.
Then we analyze the issue of possible $S/D$ wave mixing in the 
$\rho$ system and try to obtain a better description
of the $\sigma_L/\sigma_T$. 

As we will see,
the first method might turn our the remedy for the $\sigma_T$
puzzle, but even an accurate formulation of this method
forces us to go beyond the lowest Fock state, thus making only rough
estimates of the effect possible. The second mechanism
will be shown indeed to decrease the $\sigma_L/\sigma_T$ ratio
at higher $Q^2$ down to an acceptable values, but this 
happens at the expense of too low values of $\sigma_L$,
not increased values of $\sigma_T$. Therefore, this mechanisms
does not provide the solution to the $\sigma_T$ puzzle,
at least in its pure form.

\section{The Coulomb tail of the wave function}

As we discussed before, the major source of ambiguity in our
approach is the vector meson wave function. Without accurate
knowledge of the radial wave function, one is bound only to an
educated guess.

In our treatment we stuck everywhere to the {\em soft wave
function} Ansatz, that is, we assumed that any integrated
involving the wave function would be saturated by small-$p$ region
($|\bp|\sim 1/R_V$). This implies that at large-$p$ the wave
function must vanish fast enough, or to be precise, faster than
any $\bp^2$ power present in numerators of the above formulas.
Since the highest momentum power is $\vec k^2$, which appears, for
instance, in the transverse amplitudes, we conclude that the
"softness" of the wave function means that
\be
\psi(z,\vec k)|_{\vec k^2 \to \infty} =  o\left({1 \over
\vec k^4 }\right)\,.
\ee
The Gaussian wave function Ansatz used
everywhere above certainly satisfies this requirement.

It turns out, however, that the above requirement is violated when
the {\em true short-range Coulomb-like interquark potential} is
taken into account. As well known, if the color Coulomb potential
were the only source of the quark-antiquark interaction, the
resulting Coulomb-type wave function of the vector meson would
look like
\be
\psi(\bp) \propto {1 \over (\bp^2 + a^2)^2}.
\ee

In reality, the interquark forces are much more complicated.
However, without any need to know the precise form of interquark
forces, one can assert that at short distances there must be some
resemblance of the Coulomb-like wave high-$\bp^2$ tail.

\subsection{The strategy}

Certainly, there is a multitude of approaches that claim to
account for the color Coulomb interaction at short distances. One
of them would consist in obtaining an accurate numeric solution of
a given potential model. If one prefers to used a fully QFT-based
approach, one must deal then with a set of diagrams. Ideally, one
should start with free quarks and then, by taking their
interaction into account, arrive at the physical vector meson with
a (presumably) uniquely defined $q\bar q$ wave function.

Realizing such a program is still unresolved task and attempts to
accomplish it would certainly go far beyond this study. Moreover,
even if we were capable of doing it, we would still run into
interpretational difficulties, at least in the framework of our
scheme. Indeed, when calculating the diagrams, we will see the
higher Fock states intrude into our $q\bar q$ description of the
vector meson. So, even if we still wanted to follow only
quark-antiquark distribution in a meson, we will be forced to
switch from the wave function to the density matrix description.
Another problem would be the presence of higher angular momenta of
the $q\bar q$ pair due to gluon radiation. This will leave no room
for our simple $S$ wave/$D$wave description of the vector meson.
Finally, the $q\bar q$ wave function (or density matrix) will have
no unique, process independent definition. Indeed, in order to
preserve the gauge invariance at any given order of perturbation
theory, we will have to include corrections to the rest of a
diagram (the kernel) as well as corrections that entangle the
kernel with the wave function. As a result, the two particle
irreducibility of the process will be lost.

The conclusion is that we cannot expect a reasonable answer to the
question "how to account for short range Coulomb interaction" and
still stay within the framework of our approach.

Nevertheless, the mere formulation of this question does not force
us to go beyond our lowest Fock state, two-particle irreducible
approach. So, it can be reiterated as "The Coulomb tail of the
wave function must be there; given that, can we develop a
reasonable understanding of its impact on dynamics of the vector
meson production?" In this formulation, we now ask for a
QFT-inspired model that would produce an estimate of the hard tail
without forcing us to run into problems just mentioned.

We suggest the following procedure that would satisfy this need.
We start with the soft wave function Ansatz and perform a sort of
evolution procedure, that will generate the hard tail. This
evolution will basically consist in allowing for gluon exchange
between the quark and the antiquark. The result will be
interpreted as the hard piece of the vertex function $\Gamma$, and
eventually as the hard piece of the wave function itself.

\subsection{The quantum mechanics of the Coulomb tail}

Suppose that our potential is a sum of an oscillator and Coulomb
potential, the latter being a "perturbation" and bearing an
intrinsic small parameter:
\bea
&&V(p) = V_{osc}(p) + V_{coul}(p)\,;\nonumber\\ &&V_{osc}(p) = -
{\mu\omega^2 \over 2} \cdot{12 \pi^2 \over p^2}\delta''(p)\,,
\quad V_{coul}(p) = - {4\pi \over p^2 + \mu_G^2} \alpha_s(p^2)
C_F\,.
\eea
Here coupling constant is supposed to be the small parameter.
Effective parameter $\mu_G$ accounts for confinement.

The total wave function is represented as sum of the soft part
$\psi_s = c\cdot \exp(- k^2 a^2/2)$, which is the solution of
Schrodinger equation with oscillator potential only, and the hard
part $\psi_h$, determined via:
\be
\left({p^2 \over 2\mu} - E\right)\psi_h(p) = - {1 \over (2\pi)^3}
\int dk V_{coul}(k)\psi_s(p-k)\label{coul1}
\ee
Let's for a moment neglect running of the coupling constant. Then,
\be
\psi_h(p) = {1 \over {p^2 \over 2\mu} - E} \cdot {4\pi \alpha_s
C_F \over (2\pi)^3}\cdot I(p)\,, \label{psihp}
\ee
so that
\be
\psi(p) = \exp\left(- {k^2a^2 \over 2}\right) + {1 \over {p^2
\over 2\mu} - E} \cdot {4\pi \alpha_s C_F \over (2\pi)^3}\cdot
I(p)\,. \label{psitot}
\ee
Here
\bea
I(p)& =& \int {d^3\vec{k} \over (\bp-\vec{k})^2 + \mu_G^2} \exp\left(-
{k^2a^2 \over 2}\right) = 2\pi \int_0^\infty k^2 dk {1 \over 2pk}
\log\left[{(p+k)^2+\mu_G^2 \over (p-k)^2+\mu_G^2}\right]
\exp\left(- {k^2a^2 \over 2}\right)\nonumber\\ &=& 2\pi
\int_0^\infty {k^2 dk \over k^2 + \mu_G^2} {e^{-{(k-p)^2a^2 \over
2}} - e^{-{(k+p)^2a^2 \over 2}} \over kpa^2} \,. \label{ip}
\eea
This function cannot be evaluated exactly. One can, however ask
for its the asymptotic large-$p$ behavior, which can be
evaluated directly from (\ref{coul1}) by replacing $k\to p-k$
and taking $V_{coul}(p-k) \approx V_{coul}(p)$ out of the integral:
\be
\psi_h(p \to \infty) = 4\pi C_F\alpha_s{2\mu \over p^4}\cdot \psi(r=0)
\label{psihasym}\,.
\ee
Note, however, that one should not use
this simple analytic form because it spoils the large distance
behavior of the wave function. The wave function (\ref{psihasym}),
even when regularized as $p^2 \to p^2 + \mu_G^2$, leads to
exponential decrease of the wave function as $\propto
\exp(-r\mu_G)$, while the honest wave function exhibits gaussian
decrease $\propto \exp(-r^2a^2/2)$.

\subsection{Derivation of $\Delta\psi_{Coul}$}

\begin{figure}[!htb]
   \centering
   \epsfig{file=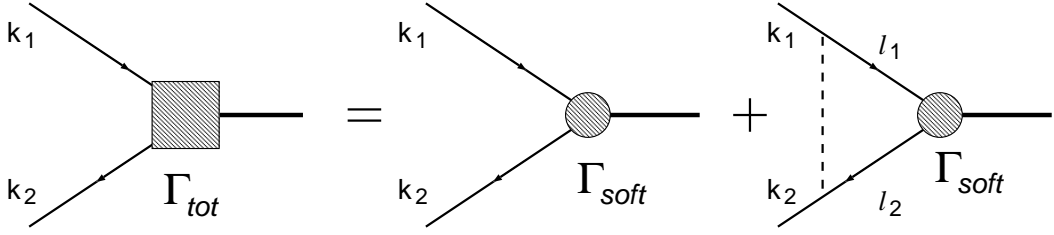,width=14cm}
   \caption{\em Diagrmmatic representation of
the effect of Coulomb tail at small distances}
   \label{coulombtail}
\end{figure}

The diagrammatical representation of this procedure is show in
Fig.~\ref{coulombtail}. Instead of implementing soft vertex function
$\Gamma^\alpha_{soft}$, we use properly normalized
\be
\Gamma^\alpha_{total} = {1 \over \sqrt{N}} (\Gamma^\alpha_{soft} +
\Delta\Gamma^\alpha)\,,
\ee
where $1/\sqrt{N}$ factor accounts for the proper normalization.
 The Coulomb correction $\Delta\Gamma^\alpha$ reads then
\be
\Delta\Gamma(z,\vec k)^\alpha = -i \int {d^4 l \over (2\pi)^4}
\cdot g^2 C_F \cdot {\gamma^\mu (- \hat
l_2+m)\Gamma_{soft}^\alpha(\hat l_1 + m) \gamma_\mu \over (l_1^2 -
m^2 + i\varepsilon)(l_2^2 - m^2 + i\varepsilon) [(k_1-l_1)^2 -
\mu_G^2 + i\varepsilon]}\,.\label{coulomb-dg1}
\ee
The Sudakov decomposition of the momenta of the initial and intermediate
particles' reads
\bea
k_1^\mu &=& y p'^\mu + z q'^{\mu} + k^\mu_\perp\,;\nonumber\\
k_2^\mu &=& (1-y) p'^\mu + (1-z) q'^{\mu} - k^\mu_\perp\,;\nonumber\\
l_1^\mu &=& y' p'^\mu + z' q'^{\mu} + l^\mu_\perp\,;\nonumber\\
l_2^\mu &=& (1-y') p'^\mu + (1-z') q'^{\mu} - l^\mu_\perp\,;\nonumber\\
(k_1-l_1)^\mu &=& (y-y') p'^\mu + (z-z') q'^{\mu} + (k_\perp-l_\perp)^\mu\,.
\eea
We already know from the above analysis that one of the initial
particles is on-mass shell. Therefore, one of the two
conditions
$$
y = {\vec k^2 + m^2 \over z m_V^2} \quad \mbox{or} \quad
y = 1-{\vec k^2 + m^2 \over (1-z) m_V^2}
$$
is fulfilled.

The numerator constitutes the most challenging part. Certainly,
there is absolutely no problem to perform gluon index summation:
\be
- \gamma^\mu (-\hat l_2+m)\gamma^\alpha(\hat l_1+m) \gamma_\mu =
2m^2\gamma^\alpha -2\hat l_1 \gamma^\alpha\hat l_2 - 4 (l_1 - l_2
)^\alpha\,.\label{coulomb-num1}
\ee
The problem however is that this spinorial structure cannot be
reduced to the scalar multiplicative renormalization of the
original structure $\gamma^\alpha$. Moreover, after integration
the spinorial structure cannot be expressed in terms of $S$ and
$D$ waves only. Finally, the resulting spinorial structure will
lead to not gauge invariant correction to the wave function.

In order to go as far as we can in trying to obtain a reasonable
estimate for the Coulomb tail and avoid trouble, we propose the
following procedure. First, we look at the transverse vector meson
production. Since the vector meson polarization vector $V^\alpha$
possesses only transverse components ($\vec V^\alpha$), only
transverse terms
\be
2m^2\vec \gamma^\alpha +2\hat{\vec l}\vec\gamma^\alpha\hat{\vec l}
-2[z'(1-y')\hat p' \vec\gamma^\alpha \hat q' + y'(1-z')\hat q'
\vec\gamma^\alpha \hat p'] - 8 {\vec l}^\alpha\,.
\ee
will survive from the whole expression (\ref{coulomb-num1}). Then,
in order to simplify life, we will proceed for the $\rho$ meson as
if it were a heavy quarkonium. That is, we will neglect ${\vec
l}^2$ in comparison with $m^2$ and put $z'={1\over 2}$. Finally,
the last stroke will be to put $m_V^2=4m^2$ in numerator, so that
$y'$ can be replaced by $1/2$ as well. The result of this
procedure will give
\be
- \gamma^\mu (-\hat l_2+m)\vec \gamma^\alpha(\hat l_1+m)
\gamma_\mu \to 4m^2 \vec\gamma^\alpha,.
\ee
Finally, we {\em postulate} that the same scalar renormalization
holds for longitudinal vector mesons as well.

The denominator is calculated in the same way as before. We note
that the numerator in our calculation does not depend on $y'$, and
therefore, the pole structure on the $y'$ plane comes from the
propagators only. The position of poles, however, changes with
$z'$:  at $z'<0$ or $z'>1$ all poles lie in the same half-plane,
and thus by closing the contour we can nullify the contribution of
these $z'$ regions to the integral. Then, the interval $0<z'<1$
naturally breaks into two parts: $z'<z$ and $z'>z$ intervals. In
the former region we prefer to close the contour from below and
take the residue at the position of pole 1. The two other
propagators become
\bea
\mbox{pole 2} &\to &(1-z')(m_V^2 - M'^2) +
i\varepsilon\,;\nonumber\\ \mbox{pole 3} &\to &(z-z')\left(ym_V^2
- {\vec l^2 + m^2 \over z'} \right) - (\vec k - \vec l)^2 -
\mu_G^2 + i\varepsilon\,.
\eea
In the second $z'$ region we close the contour from above and take
residue at $y'=1-(\vec l^2+m^2)/[(1-z')m_V^2]$, so that the other
poles become
\bea
\mbox{pole 1} &\to& z'(m_V^2 - M'^2) + i\varepsilon\;;\nonumber\\
\mbox{pole 3} &\to& -(z'-z)(M'^2 - m_V^2) -(z'-z)\left(ym_V^2 -
{\vec l^2 + m^2 \over z'} \right) - (\vec k - \vec l)^2 - \mu_G^2
+ i\varepsilon\,.
\eea
Note that in both cases the familiar $M'^2 - m_V^2$ appears in the
denominator, which fuses with $\Gamma_{soft}(M'^2)$ to produce
$\psi_{soft}(M'^2)$.
Thus, the result of the $y'$ integration is
\bea
&&\int dy' {\Gamma_{soft}(M'^2) \over (l_1^2 - m^2 + i\varepsilon)
(l_2^2 - m^2 + i\varepsilon)[(k_1-l_1)^2 - \mu_G^2 +
i\varepsilon]} \nonumber\\[3mm] &&=\theta(z-z'){1 \over \Delta_1 +
(\vec k - \vec l)^2 + \mu^2} +\theta(z'-z){1 \over \Delta_2 +
(\vec k - \vec l)^2 + \mu^2}
\eea
with
\bea
\left.\begin{array}{l}
\Delta_1 = -(z-z')\left({\vec k^2 + m^2 \over z} -
{\vec l^2 + m^2 \over z'}\right) \\[2mm]
\Delta_2 = (z'-z)\left({\vec k^2 + m^2 \over z} - {\vec l^2 + m^2 \over z'}\right)
+ (z'-z)(M'^2 -m_V^2)
\end{array}
\right\}&& \mbox{$k_1$ on mass shell;}\nonumber\\[5mm]
\left.\begin{array}{l}
\Delta_1 = (z-z')\left({\vec k^2 + m^2 \over 1-z} - {\vec l^2 + m^2 \over 1-z'}\right)
+ (z'-z)(M'^2 -m_V^2) \\[2mm]
\Delta_2 = -(z'-z)\left({\vec k^2 + m^2 \over 1-z} - {\vec l^2 + m^2 \over 1-z'}\right)
\end{array}
\right\}&& \mbox{$k_2$ on mass shell.}\nonumber
\eea

The analysis of the positive definiteness of the gluon
propagator reveals that in one case the positiveness indeed holds due to
inequality
$$
-{z-z' \over z} \vec k^2 + {z-z' \over z'}\vec l^2 + (\vec k - \vec
l)^2 = \left(\sqrt{z' \over z} \vec k - \sqrt{z \over z'} \vec
l\right)^2 > 0\,,
$$
while in the other case the gluon pole can in principle arise in
the allowed kinematical region. Such a pole would correspond to
the situation when both quark and gluon are simultaneously on mass
shell. However, in the case of $\rho$ mesons such configurations
are avoided due to small enough mass of the vector meson ($m_V^2 <
2\mu^2_G$).

So, the expression for $\Delta\Gamma(z,\vec k)$ takes form
\bea
\Delta\Gamma(z,\vec k) &=& C_F {\alpha_s(\vec k^2) \over 4 \pi^2}
\ 4m^2\ \int {dz'\over z'(1-z')}d^2\vec l \
\psi_{soft}(M'^2)\nonumber\\ &\times&\left[\theta(z-z'){1 \over
\Delta_1 + (\vec k - \vec l)^2 + \mu^2} +\theta(z'-z){1 \over
\Delta_2 + (\vec k - \vec l)^2 + \mu^2}\right]\,.
\label{coulomb-dg2}
\eea

With $\Delta\Gamma(z,\vec k)$ calculated according to
(\ref{coulomb-dg1}), we can now construct the hard part of the wave
function $\psi(z,\vec k)$. However, the straightforward answer
$\delta\psi = \Delta\Gamma/(M^2-m_V^2)$ will not be a satisfactory
option due to unphysical pole at $M^2=m_V^2$. We remember that
when constructing the soft wave function, we forced the
corresponding vertex function $\Gamma_{soft}$ to have zero at
$M^2=m_V^2$ in order to cancel the unphysical pole. Here we do not
have such a freedom in manipulation with $\Delta\Gamma$, which is,
by the way, always positive. However, we can again tune the {\em
soft vertex function} $\Gamma_{soft}$ so that the entire
expression $\Gamma_{soft} + \Delta\Gamma$ does have zero at the
required point. This can be achieved by shifting the node position
of the soft vertex function
\be
\Gamma_{soft}(z,\vec k^2) \to \psi_{soft}(M^2)\left[M^2 - m_V^2
- {\Delta\Gamma(z,\vec k^2_0) \over \psi_{soft}(m_V^2)}\right],
\ee
where ${k_0^2+m^2 \over z(1-z)} = m_V^2$.
The total wave function will then turn into
\bea
\psi_{total}(z,\vec k^2) &=& { \psi_{soft}(M^2)\left[M^2 - m_V^2 -
{\Delta\Gamma(z,\vec k^2_0) \over \psi_{soft}(m_V^2)}\right] +
\Delta\Gamma(z,\vec k^2)\over M^2 - m_V^2}\nonumber\\ &=&
\psi_{soft}(M^2) + {1 \over M^2 - m_V^2} \left[
\Delta\Gamma(z,\vec k^2) - \Delta\Gamma(z,\vec
k^2_0){\psi_{soft}(M^2) \over \psi_{soft}(m_V^2)}
\right]\,.\label{coulomb-wf1}
\eea
The wave function constructed in this manner is regular at $M^2 =
m_V^2$ and does have the expected large $M^2$ behavior.

\subsection{The large $Q^2$ asymptotics of $\sigma_T$: analytical result}
The honest integration in (\ref{coulomb-dg2}) cannot be done
analytically, but in order to grasp the asymptotics, a simple
estimate can be performed.

For $\vec k^2 \gg \vec l^2,\ \mu_G^2,\ m^2$ the denominators turn
into $\vec k^2 z'/z$ and $\vec k^2 (1-z')/(1-z)$ for on mass shell
initial quarks 1 or 2 respectively. If we then suppose that the
soft wave function is given by the Gaussian Ansatz with $R_V^2 m^2
\gg 1$, we can put $z' \to 1/2$ and do the integration completely.
The answer reads
\be
\Delta\Gamma(z,\vec k^2) = C_F {\alpha_s(\vec k^2) \over \pi^2}
\cdot m^2 {2 \over m} {2 z \over \vec k^2}\cdot \int d^3\bl
\psi_{soft}(M'^2) = {z \over \vec k^2} \ C_F {\alpha_s(\vec k^2)
\over \pi^2 }\cdot 4m  a_\psi \cdot \left({2 \pi \over R_V^2
}\right)^{{3 \over 2}}\,.
\ee
We see that the hard tail of the wave function $\Delta\Gamma$
falls off as $1/\vec k^4$ at large $\vec k^2$, in accordance with
our expectations.

The Coulomb tail of the vector meson wave function being finally
obtained, we are now ready to estimate the expected change in the
$\sigma_L/\sigma_T$ ratio for high-$Q^2$ production of $\rho$
mesons. Before we take a look at the numerical values, let us make
an analytical estimate of the $Q^2$ asymptotics. We plug the hard
tail of the wave function $\Delta\psi(z,\vec k^2)$ into the
leading twist expression for $A_{T\to T}$, which we write as
\bea
Im A_T &=& s {c_V\sqrt{4\pi\alpha_{em}}\over 4\pi^2} \int {dz
\over z(1-z) } d^2\vec k \psi_{total}(z,\vec k^2)\cdot \int
{d^2\vec\kappa\over \vec \kappa^4} \alpha_s(\vec\kappa^2) {\cal
F}(x_g,\vec\kappa^2)\nonumber\\&&\times  {2\vec\kappa^2\over
[z(1-z)Q^2+\vec\kappa^2+\vec k^2]^2} \left\{m^2 +
2[z^2+(1-z)^2]\vec k^2\right\}\nonumber\\[4mm] &\approx & s
{c_V\sqrt{4\pi\alpha_{em}}\over 4\pi^2} \cdot
2\pi\alpha(\Qb^2)G(\Qb^2)\cdot {\cal J}_{total}(Q^2)\,.
\eea
The integral
$$
{\cal J}_{total}(Q^2) \equiv \int_0^1 {dz \over z(1-z)}d^2\vec k
\psi_{total}(z,\vec k^2){m^2 + 2[z^2+(1-z)^2]\vec k^2 \over
[z(1-z)Q^2+\vec\kappa^2+\vec k^2]^2}
$$
is naturally split into ${\cal J}_{soft}(Q^2)$ and ${\cal
J}_{hard}(Q^2)$. The former is calculated in the heavy-quarkonium
approximation with Gaussian wave function and yields
\be
{\cal J}_{soft}(Q^2)=\int {4 \over m_V}d^3\vec{k}\psi_{soft} {4m_V^2
\over Q^4} = {16 m_V \over Q^4} a_\psi\cdot \left({2 \pi \over
R_V^2 }\right)^{{3 \over 2}}\,.
\ee
The hard piece is calculated as
\bea
{\cal J}_{hard}(Q^2) &=&\int_0^1 {dz \over z(1-z)}\
\pi\int_{\mu^2}^{z(1-z)Q^2} d\vec k^2 {32 [z^2 + (1-z)^2]\vec k^2
\over Q^4 } \cdot {z(1-z)\over \vec k^2}\nonumber\\
&&\times {1 \over 2\vec k^2} \
C_F {\alpha_s(\vec k^2) \over \pi^2 }\cdot 2m_V a_\psi \cdot
\left({2 \pi \over R_V^2 }\right)^{{3 \over 2}}\nonumber\\ &=&
2m_V a_\psi \cdot \left({2 \pi \over R_V^2 }\right)^{{3 \over 2}}
\cdot {C_F \over \pi} {16 \over Q^4} \cdot \int_0^1dz [z^2 +
(1-z)^2] \cdot\int_{\mu^2}^{Q^2/4}{d\vec k^2 \over \vec
k^2}\alpha_s(\vec k^2)\,.
\eea
The ratio $A_{hard}/A_{soft}$ is then
\bea
{A_{hard}(Q^2)\over A_{soft}(Q^2)} = {4C_F \over 3\pi}
\cdot\int_{\mu^2}^{Q^2/4}{d\vec k^2 \over \vec k^2}\alpha_s(\vec
k^2) = {64 \over 81} \log\left[{\alpha_s(\mu^2)\over \alpha_s({1
\over 4 }Q^2)}\right]\,.
\eea

This expression represents the asymptotic large-$Q^2$ behavior of
the $A_{hard}/A_{soft}$ ratio. As it could be expected, it is
governed by $\log\left[{\alpha_s(\mu^2)\over \alpha_s({1 \over 4
}Q^2)}\right]$, which rises with $Q^2$ growth, but extremely
slowly. Within the experimentally studied region of $Q^2$ (not
higher than 100 GeV$^2$), this logarithm, which is supposed to be
large, is still less than unity (basically $0.5\div 0.7$). This
indicates that the asymptotic regime is not reached at HERA, and
the estimate just derived seems to be irrelevant to the real
magnitude of the hard tail effects. Nevertheless, this estimate
tells us that no abnormally suppressing factors appear in our
problem.

The numerical analysis of the Coulomb tail impact on the
production amplitudes will be given elsewhere.

\section{The $S/D$ wave mixing}

The presence of tensor forces in two-fermion bound states
is always a natural consequence of relativistic corrections.
They lead mixing of states with definite angular momentum,
which in the case of a vector particle translates into
$S$ wave -- $D$ wave mixing.

The most famous example is provided by the Breit potential
in a hydrogen atom.  Such forces are
present in a deuteron, where they lead to about 5\% admixture of
$D$ wave state, and they can be present in vector meosns as well.
Review \cite{tensor} provides further examples.

How are we going to mix $S$ and $D$ wave? When calculating the
normalization or decay constant for a vector meson, we observed a
two-fold dependence on the vector meson state: in the spinorial
structure and in the particular form of the radial wave function.
In all cases, {\it the rest} was absolutely insensitive to the
vector meson state. Therefore, the two quantities we should mix,
are ${\cal S}_\mu \psi_S({\bf p}^2)$ and ${\cal D}_\mu \psi_D({\bf
p}^2)$:
\begin{equation}
|\Gamma_\mu  \psi({\bf p}^2) \rangle= |{\cal S}_\mu \psi_S({\bf
p}^2)\rangle \cos\theta + |{\cal D}_\mu \psi_D({\bf p}^2)
\sin\theta\rangle\,.\label{mix1}
\end{equation}
If mixing angle $\theta$ is constant, then this vector meson state
satisfies automatically the normalization condition:
\be
\langle\Gamma_\mu\psi|\Gamma_\nu\psi\rangle =\langle{\cal
S}_\mu\psi_S|{\cal S}_\nu\psi_S\rangle \cos^2\theta + \langle{\cal
D}_\mu\psi_D|{\cal D}_\nu\psi_D\rangle \sin^2\theta =
g_{\mu\nu}\,.
\ee

There can be several types of $S/D$ mixing. First, we can mix $D$
state with $1S$, $2S$ or even higher $S$ states. It is not clear
{\it a priori} which mixing should be stronger. Therefore in our
analysis below we accounted for possible $1S/D$ and $2S/D$ mixing
on the equal foot. Second, the mixing angle $\theta$ can be an
explicit function of ${\bf p}^2$. This case is a bit more
complicated but nothing seems to rule out such possibility.

\subsection{Cooking up $\gamma_\mu$ vertex}

The first issue we want to elaborate is how to construct the
$\gamma_\mu$ vertex from accurate ${\cal S}_\mu$ and ${\cal
D}_\mu$ spinorial structures. This question arises when one tries
to check how the naive $\bar u \gamma_\mu u$ treatment of the
quark-antiquark-vector meson vertex differs from the accurate one.
To do so, in the expression
\begin{equation}
\Gamma_\mu  \psi({\bf p}^2) = \left(\gamma_\mu - { 2 p_\mu \over M
+ 2m}\right) \psi_S({\bf p}^2) \cos\theta({\bf p}^2) + \left({\bf
p}^2\gamma_\mu + (M+m)p_\mu\right) \psi_D({\bf p}^2)
\sin\theta({\bf p}^2)\,.\label{mix1a}
\end{equation}
all terms with $p_\mu$ must be banished by properly adjusting
mixing angle $\theta = \theta({\bf p}^2)$. Assuming for
simplicity, that we use oscillator ansatz wave functions with all
$a_i$ equal, one obtains for $1S/D$ mixing
\begin{equation}
\tan\theta({\bf p}^2) = - {2 \over (M+2m)(M+m)}{c_1 \over c_D} = -
{\sqrt{30} \over (M+2m)(M+m) a^2}\,.\label{mix2}
\end{equation}
A very rough estimate for $\rho$ system gives $\tan\theta \sim
0.2$. For $2S/D$ mixing one has
\begin{equation}
\tan\theta({\bf p}^2) = - {\sqrt{45} \left(1 - 2 {\bf p}^2
a^2/3\right) \over (M+2m)(M+m) a^2}\,.\label{mix3}
\end{equation}

This time the mixing angle even swings with ${\bf p}^2$ growth.
After $p_\mu$ terms are canceled in (\ref{mix1a}), the residual
expression is proportional only to $\gamma_\mu$ and has slightly
modified ${\bf p}^2$ dependence via explicitly ${\bf p}^2$
dependent mixing angle. However, this modification is completely
inessential. Moreover, for $2S/D$ mixing this modification does
not shift the position of the node. Therefore, it seems that the
impact of this specific mixing on the radial wave function is
minimal. The magnitude of this mixing can be also described with
only one number --- the integrated mixing angle:
\begin{equation}
\tan^2\theta^* = {\langle\sin^2\theta({\bf p}^2)\rangle_D \over
\langle\cos^2\theta({\bf p}^2)\rangle_S} \label{mix4}
\end{equation}
where $\langle...\rangle_{S,D}$ mean the normalization integral
for $S/D$ state. For $1S/D$ and $2S/D$ types of mixing,
$\tan^2\theta^*$ is equal to 0.027 ($\theta^* = 9.3^\circ$) and
0.062 ($\theta^* = 14^\circ$) respectively.

\begin{figure}[!htb]
   \centering
   \epsfig{file=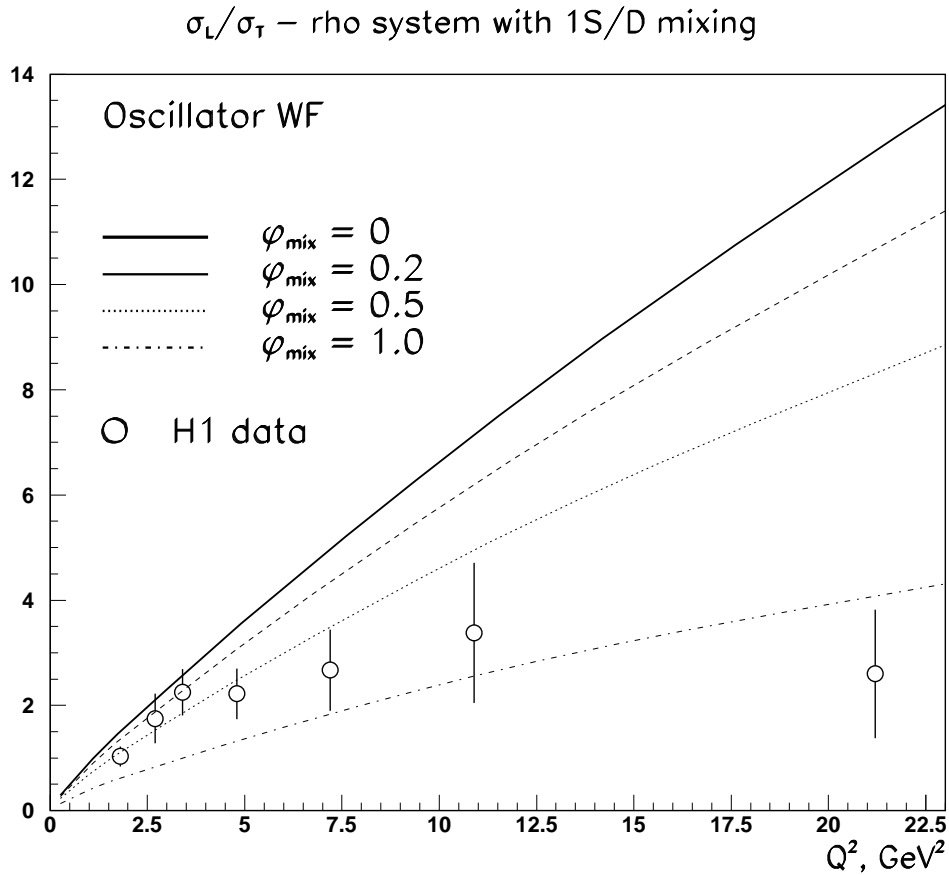,width=14cm}
   \caption{\em The impact of $1S/D$ mixing in the $\rho$
system on the $\sigma_L/\sigma_T$ ratio for the lowest energy state
production.}
   \label{num_mix_ltratio}
\end{figure}

\subsection{Impact of $S/D$ wave mixing on $\rho$ meson production}

For the simplicity, we will restrict ourselves to the
constant-angle variant of $S/D$ wave mixing. Since our
primary motivation here is to check how mixing alters the
$\sigma_L/\sigma_T$ ratio in $\rho$ production, we will focus only
on $\rho$ meson.

Since a generic $\rho$ meson production amplitude is linear in the
meson wave function (understood in its complete sense, i.e.
spinorial structure times the scalar wave function), one
immediately has
\be
{\cal A}_\rho = {\cal A}_S \cos\theta + {\cal A}_D \sin\theta\,.
\ee
With this expression in mind, we can now predict the effect of
$S/D$ wave mixing by simply looking at the pictures for pure $S$
and $D$ wave states production.

\begin{figure}[!htb]
   \centering
   \epsfig{file=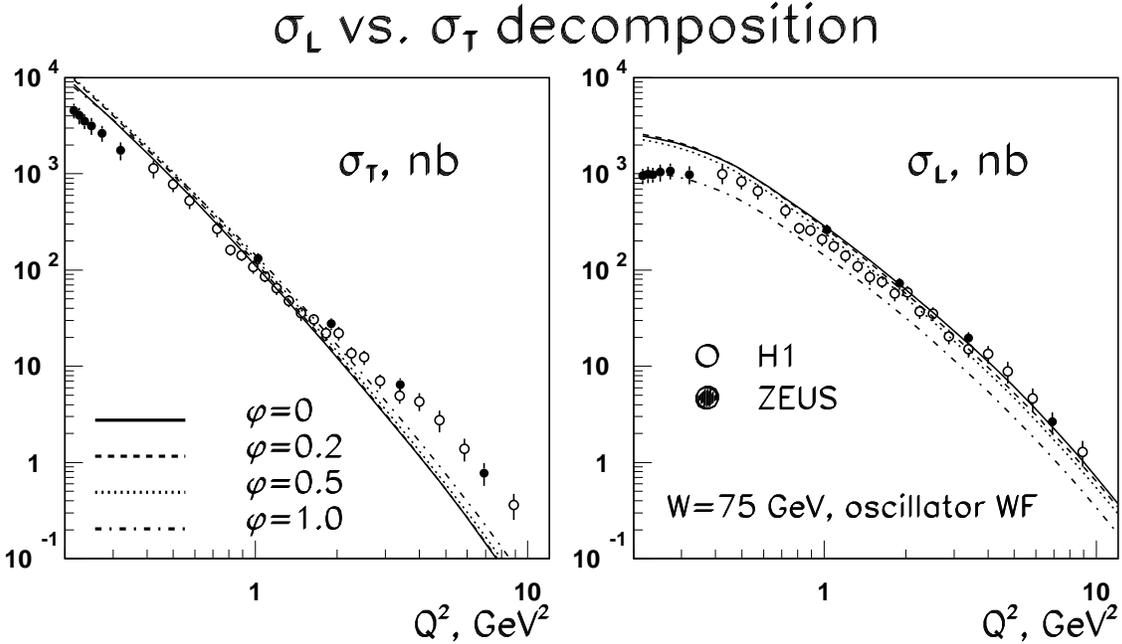,width=16cm}
   \caption{\em Changes in $Q^2$ profiles of $\sigma_T$ and $\sigma_L$
cross sections for $\rho$ system ground states caused by
$1S/D$ mixing.}
   \label{num_mix_lt}
\end{figure}

Fig.~\ref{num_mix_ltratio} and Fig.~\ref{num_mix_lt} show the changes
in the longitudinal and transverse $\rho$ meson cross sections
caused by the $1S/D$ mixing. One sees that at $\phi_{mix}>0$ grows,
the ratio $\sigma_L/\sigma_T$ decreases. 
Fig.~\ref{num_mix_ltratio} implies that $\phi \sim 0.7$ would
do the best job in describing the data points. Note however that the shape
of $\sigma_L/\sigma_T$ ratio remains the same and does not significantly
flatten, which would be needed for a better description of the experimental
points. Even more problems come from separate analysis of 
 $\sigma_L$ and $\sigma_T$ cross sections.
One sees that decrease of $\sigma_L/\sigma_T$ arises 
not from making $\sigma_T$ higher, but at the expense of making $\sigma_L$
significantly lower than the data. 
Thus, although the ratio $\sigma_L/\sigma_T$
can indeed be corrected in a simple mixing scenario, the cross sections
themselves still deviate at large $Q^2$ from the measured values.
This analysis lead us to a conclusion that 
the $\sigma_T$ puzzle in $\rho$ production still persists
in our approach. 

Although the $S/D$ wave mixing failed to completely resolve
the $\sigma_T$ puzzle, it is still an interesting issue on its own.
In constrast to all previous calculations of the vector meson
production amplitudes, our approach allows for a clear formulation
and detailed analysis of this phenomenon.

\chapter{Conclusion}

In the present work we formulated the $k_t$-factorization approach
to the calculation of the diffractive production of vector mesons
in DIS. Since this approach is organically linked to the BFKL
dynamics, no requirements on the value of $Q^2+m_V^2$ was placed,
which allowed us to investigate the production cross sections
from the photoproduction limit up to highest $Q^2$ values
attainable in the experiment.

When describing the vector meson, we limited ourselves
to the lowest Fock state only, that is we treated
a vector meson as a bound state of quark-antiquark pair only.
Being of course an approximation, this limitation
allowed for the strict construction of the pure $S$-wave and $D$-wave
states of the $q\bar q$ pair sitting inside the meson,
which was then applied to the corresponding production amplitudes.

On the other side of the reaction, we related
the production amplitudes to the unintegrated 
gluon structure function of the proton.
We undertook an extensive study of the differential gluon density, 
which included its first ever extraction from the experiment
and detailed investigation of its properties.
This was then used in the vector meson production calculations
and yielded reliable numerical results.

Here we give a detailed list of main results of the present work.

\begin{enumerate}
\item  Differential gluon structure function of the proton
\begin{itemize}
  \item First ever extraction of the unintegrated gluon structure 
   function and casting it into a form of simple 
   and ready-to-use parametrizations;
  \item detailed analysis of the soft-hard decomposition of 
   various observables and the impact of soft-to-hard diffusion 
   phenomenon on them;
  \item observation of self-generated two-reggeon-like structure
   of $F_{2p}$ predicted by $k_t$-factorization;
  \item showing by means of $\chi^2$ analysis that 
   the high-precision experimental data on $F_{2p}$ suggests
   rather strong separation of soft and hard parts
   of the unintegrated gluon density.
\end{itemize} 
\item Diffractive vector meson production: analytical study
\begin{itemize}
  \item Developing for the first time a theory for vector meson 
  spin-angular coupling and using it consistently in the 
  derivation of the meson production  amplitudes;
  \item observing remarkably different $Q^2$-dependence of $S$ and $D$ wave
  type amplitudes, providing thus a way to discern $S$ and
  $D$ wave states that are indistinguishable at $e^+e^-$
  colliders;
  \item at large $Q^2$, observing a dramatic role 
  of higher twist contributions to the $D$
  wave vector mesons, which even forced sign change for $L\to L$
  amplitude and led to non-monotonous $\sigma_L/\sigma_T$ ratio;
  \item observation of very large helicity violating effects 
  for $D$ wave vector states, which do not get suppressed 
  even in the case of heavy  quarkonia;
  \item confirmation of the soft dominance of the double spin flip
  amplitude in the case of both  $S$ and $D$ wave states;
  \item establishing the borders of our approach during discussing
   the hard Coulomb tail of the wave function.
\end{itemize}
\item  Diffractive vector meson production: numerical study
\begin{itemize}
  \item Showing that $k_t$-factorization approach leads to a good
  overall agreement with availble experimental data on all types
  of vector mesons. Namely, we showed that the overall shape 
  of $Q^2$ dependence, energy growth, $t$-dependence, 
  picture of $s$-channel helicity violation observed in $\rho$,
  $\phi$, and $J/\psi$ mesons production are in good agreement
  with $k_t$-factorization prediction.
  \item Recognizing that two particular issues --- the transverse 
  cross section $\sigma_T$ for $\rho$ mesons at higher $Q^2$ 
  and the magnitude of $\Upsilon(1s)$ state --- still remain unresolved.
  The causes of this discrepancy and the resolution possibilities 
  are discussed.
  \item Predicting many previously unknown features of the excited vector meson
  production reactions, including opposite signs for the 
  largest spin flip amplitude
  $T \to L$ for $S$ and $D$ wave vector mesons and dramatically different
  $\sigma_L/\sigma_T$ ratio;
  \item investigating $S/D$-mixing induced phenomena; 
  indicating that large $S/D$ wave mixing in $\rho$ system
  can be the origin of $\sigma_T$ puzzle.
\end{itemize}
\end{enumerate}

Since the work contains a large number of  predictions on the observables
that have not yet been investigated in experiment,
I hope that the thesis will serve as a guide to the directions of
future experimental research. The dramatic differences
between $S$ and $D$ wave vector meson production predicted
in this work demand confirmation and offer a novel way to study
the structure of hadrons.

Finally, in this work I intended  not merely to list 
the results, but also give a detailed and pedagogical presentation
of all steps. In particular, Part I, where I introduced the $k_t$-factorization
scheme, Chapter 4, where I construct the theory of the vector meson
spin-angular coupling, and Appendices 
contain very detailed intermediate calculations,
so that everyone can follow the entire line of derivation. 
To this end, I hope that the present text
can be used as a means of learning as well.

\newpage

\appendix
\chapter{Denominator evaluation: details} \label{apa}

Below we give a detailed denominator calculus that is
used in all $k_t$-factorizatoin processes. The major
guideline in this derivation will be the analysis of pole positions and
setting some of the propagators on mass shell by taking
appropriate residues.\\

\section{Forward Compton scattering}

\subsection{The $s$-channel diagram: all details}

We first start with the case of forward virtual Compton scattering
amplitude. The integral to be calculated is
\begin{eqnarray}
&&\int dy\ dz\ d\alpha\ d\beta\ {1 \over [(q-k)^2 -m^2 +i\epsilon]
[k^2 -m^2 +i\epsilon] [k^2 -m^2 +i\epsilon] [(k+\kappa)^2 -m^2
+i\epsilon]}\nonumber\\ &&{1 \over [\kappa^2 -\mu^2
+i\epsilon]^2[(p-\kappa)^2 -m^2 +i\epsilon]} \label{denom-initial1}
\end{eqnarray}
With Sudakov's decomposition
\be
q_\mu = q_\mu' - {Q^2 \over s}p_\mu'\,; \quad p_\mu = p_\mu' +{m^2
\over s}q_\mu'\,; \quad  k_\mu= y{p_\mu}' + z{q_\mu}' + \vec
k_\mu\,;\quad  \kappa_\mu = \alpha {p_\mu}' + \beta {q_\mu}'
+ \vec \kappa_\mu\,;
\ee
one can rewrite all 7 propagators as:
\begin{eqnarray}
&\langle 1\rangle&\quad (q-k)^2 - m^2 +i\epsilon =
\left(-y-{Q^2\over s}\right)(1-z)s -(\vec{k}^2 +m^2) +i\epsilon\,,\nonumber\\
&\langle 2\rangle, \langle 4\rangle&\quad k^2 - m^2 +i\epsilon =
yzs -(\vec k^2 +m^2) +i\epsilon \,,\nonumber\\[2mm]
&\langle 3\rangle&\quad (k+\kappa)^2- m^2 +i\epsilon =
(y+\alpha)(z+\beta)s -[(\vec{k}+\vec\kappa)^2 +m^2] + i\epsilon\,,
\nonumber\\[2mm]
&\langle 5\rangle, \langle 6\rangle &\quad \kappa^2 - \mu^2 +i\epsilon =
\alpha\beta s -(\vec\kappa^2 +\mu^2) + i\epsilon \,,\nonumber\\[2mm]
&\langle 7\rangle&\quad (p-\kappa)^2 - m^2 +i\epsilon =
(1-\alpha)\left({m^2 \over s}-\beta\right) s -(\vec\kappa^2 +m^2) + i\epsilon
\,.\nonumber\label{denom-poles1}
\end{eqnarray}
Let us analyze the position of propagator poles (namely, the sign
of $i\epsilon)$ in complex $y$ plane as functions of $z$ and
$\beta$. One gets
\bea
&\langle 1\rangle& y = -{Q^2 \over s}-{\vec k^2 + m^2 \over (1-z)s}
+ {i\epsilon \over 1-z}\,;\nonumber\\
&\langle 2\rangle, \langle 4\rangle& y = {\vec k^2 + m^2 \over zs}
- {i\epsilon \over z}\,;\nonumber\\
&\langle 3\rangle& y = -\alpha + {(\vec k + \vec \kappa)^2 + m^2 \over (\beta + z)s}
- {i\epsilon \over z+\beta}\,.
\eea
For the poles on $\alpha$ plane we get
\bea
&\langle 3\rangle& \alpha = -y + {(\vec k + \vec \kappa)^2 + m^2 \over (\beta + z)s}
- {i\epsilon \over z+\beta}\,;\nonumber\\
&\langle 5\rangle, \langle 6\rangle& \alpha = {\vec \kappa^2 + \mu^2 \over \beta s}
- {i\epsilon \over \beta}\,;\nonumber\\
&\langle 7\rangle& \alpha = 1 + {\vec \kappa^2 + m^2 \over \beta s - m^2}
+ {i\epsilon \over \beta - m^2/s}\,.\label{denom-poles2}
\eea

Since the function we integrate (namely, the product of
propagators) is an analytic function, which vanishes at large $|y|$ and
large $|\alpha|$ fast enough, we can switch from $(-\infty, +\infty)$
integration to the complex plane contour integration both on $y$
and $\alpha$ planes. This is done by adding half-circle of infinite radius
either on the upper or the lower half planes. What we now have to
done is just to trace how many poles we have inside the contours.

As clearly seen from (\ref{denom-poles1}) and
(\ref{denom-poles2}), the position of the poles on $y$ and
$\alpha$ planes depends on the values of $z$ and $\beta$. If
values of of $z$ and $\beta$ are such that no poles happen to fall
inside the contours, the integral turns zero, and corresponding
$(z,\beta)$ region does not contribute to the total integral. So,
our task now transforms into searching for such $(z,\beta)$
regions that both $y$ and $\alpha$ contour integrations result in
non-zero values.

A convenient way to perform this analysis is to do it graphically.
Figure illustrates the sign of $i\epsilon$ in $y = \dots \pm
i\epsilon$ (left pane) and $\alpha = \dots \pm i\epsilon$ (right
pane) for all values of $(z,\beta)$. Each shaded line here
corresponds to a propagator from (\ref{denom-poles1}) and
(\ref{denom-poles2}), the shaded side indicating the half plane
with positive value of $i\epsilon$. For example, propagator
$\langle 1\rangle$ from (\ref{denom-poles1}) has positive
$i\epsilon$ when $z<1$.

If we now take a closer look at each of possible regions
$(z,\beta)$, we see that the only $(z,\beta)$ pairs that result in
a non-zero expression lie inside a triangle. For
convenience, we break up the whole area into three sub-regions A,
B, and C. For each of these regions, we get only one pole inside
$y$ and one pole inside $\alpha$ contours:
\bea
\mbox{Region A:} & \left\{\begin{array}{c}
0<z<1 \\[2mm] -z < \beta < 0 \end{array}\right\} & \quad
\mbox{pole $\langle 1\rangle$ for $y$; pole $\langle 3\rangle$ for $\alpha$.}
\nonumber\\
\mbox{Region B:} & \left\{\begin{array}{c}
0<z<1 \\[2mm] 0 < \beta < m^2/s \end{array}\right\} & \quad
\mbox{pole $\langle 1\rangle$ for $y$; pole $\langle 7\rangle$ for $\alpha$.}
\nonumber\\
\mbox{Region C:} & \left\{\begin{array}{c}
0<\beta <m^2/s \\[2mm] -\beta < z < 0 \end{array}\right\} & \quad
\mbox{pole $\langle 3\rangle$ for $y$; pole $\langle 7\rangle$ for $\alpha$.}
\label{denom-regions1}
\eea

Thus, the integrals over $y$ and $\alpha$ turns into taking the
residues at certain poles.
Let us write the result of this procedure for each of the three
regions.\\

{\bf Region A.}\\
Here we get
\be
y = - {Q^2 \over s} - {\vec k^2 +m^2 \over (1-z)s}\,;\quad \alpha
= {Q^2 \over s} + {\vec k^2 +m^2 \over (1-z)s} + {(\vec k + \vec \kappa)^2 + m^2
\over (z-|\beta|)}\,.
\ee
We know that $\beta <0$, so we explicitly switch for more clear
notation $|\beta|$. The other five propagators turn now into
\bea
\langle 2\rangle, \langle 4\rangle &\to& -{1 \over 1-z}[\vec k^2 + m^2 -
z(1-z)Q^2] + \not{\!\!i\epsilon}\,;\nonumber\\[1mm]
\langle 5\rangle, \langle 6\rangle &\to& -\vec\kappa^2 - \mu^2
-\left[Q^2 + {\vec k^2 + m^2 \over 1-z} + {(\vec k + \vec \kappa)^2 + m^2
\over z-|\beta|}\right] + \not{\!\!i\epsilon}\,;\nonumber\\[2mm]
\langle 7\rangle &\to& -\vec\kappa^2 - m^2
+ \left[s - Q^2 - {\vec k^2 + m^2 \over 1-z} - {(\vec k + \vec \kappa)^2 + m^2
\over z-|\beta|}\right]\left({m^2 \over s} + |\beta|\right) +
i\epsilon\,.
\eea
Here symbol $\not{\!\!i\epsilon}$ means that the rule for pole
passing is not needed since the propagator has definite sign for
the physical values of the parameters.

Thus, the honest result of the integration equals
\bea
&&\int dy\ dz\ d\alpha\ d\beta\ {1 \over \mbox{propagators}} =
\left(-{2\pi i\over s}\right)^2 \int_0^1 dz \int_0^z d|\beta|\ {1-z\over z-|\beta|}
\ {1 \over [\vec k^2 + m^2 + z(1-z)Q^2]^2}\nonumber\\
&&\quad \times {1 \over \left\{ \vec \kappa^2 + \mu^2 + |\beta|\left[Q^2 +
{\vec k^2 + m^2 \over 1-z} + {(\vec k + \vec\kappa)^2 + m^2 \over z-|\beta|}\right]
\right\}^2 }\nonumber\\
&&\quad\times {1 \over \vec\kappa^2 + m^2 - (m^2 + |\beta|s)
\left[1 - {Q^2 \over s} - {\vec k^2 + m^2 \over 1-z} - {(\vec k + \vec\kappa)^2
+ m^2 \over z-|\beta|}\right] - i\epsilon }
\,.\label{denom-res1}
\eea
Now comes the last step. The result (\ref{denom-res1}) has
both real and imaginary part. Since we are hunting for the
imaginary part only, we extract it by using
$$
{1 \over X - i\epsilon} = \mbox{V.P.}\left({1 \over X}\right) +
i\pi\delta(X)
$$
and killing the $\beta$ integral with the means of $\delta$
function. The value of $\beta$ is
\be
|\beta| \approx {\vec \kappa^2 \over s} \ll 1
\ee
and therefore it can be neglected everywhere else in
(\ref{denom-res1}). The result reads:
\be
Im \left\{\int {dy\ dz\ d\alpha\ d\beta \over \mbox{propagators}}\right\} =
{4\pi^2 \over s^3} \int_0^1 dz\ {1-z \over z}
{1 \over [\vec k^2 + m^2 + z(1-z)Q^2]^2}
{1 \over [\vec \kappa^2 + \mu^2]^2}\,.\label{denom-res2}
\ee
Of course, the $z$ integration limits in (\ref{denom-res2}) should
not be understood literally as 0 and 1. In fact, when $z$ is close
enough to 0 or 1 (by a typical value of $\sim m^2/s$), $\beta$ can
no longer be neglected in comparison with $z$. Therefore, the
integrand in (\ref{denom-res2}) does not have the correct small
$z$ and small $1-z$ asymptotics. Nevertheless, since the physical
quantities will have regular $z\to 0$ and $z \to 1$ behavior,
the impact of this difference will be suppressed by $\sim m^2/s$
simply due to small integration measure.

{\bf Regions B and C.}\\
Although region B has much smaller area than region A, one cannot
guarantee a priori that the third pole, which produces the
imaginary part of the integral, does not happen to lie inside it.
However, one can check that it is not the case.

One extracts first the values of $y$ and $\alpha$ according to
(\ref{denom-regions1})
\be
y = - {Q^2 \over s} - {\vec k^2 +m^2 \over (1-z)s}\,;\quad \alpha
= -{\beta s + \vec\kappa^2 \over m^2-\beta s}\,.
\ee
We remind that $0<\beta<m^2/s$, so that no singularity arises
here.
The remaining propagators are
\bea
\langle 2\rangle, \langle 4\rangle &\to& -{1 \over 1-z}[\vec k^2 + m^2 -
z(1-z)Q^2] + \not{\!\!i\epsilon}\,;\nonumber\\[1mm]
\langle 5\rangle, \langle 6\rangle &\to& - \mu^2
-{\beta^2 s^2 + m^2 \vec\kappa^2 \over m^2 - \beta s}
+ \not{\!\!i\epsilon}\,;\nonumber\\[1mm]
\langle 3\rangle &\to& -[(\vec k + \vec\kappa)^2 + m^2]
-\left(Q^2 + {\vec k^2 +m^2 \over 1-z}\right)(z + \beta)
- {\vec \kappa^2 + \beta s \over m^2 - \beta s}(z+\beta)s+ \not{\!\!i\epsilon}\,.
\eea
We see that throughout the whole phase space of the remaining
kinematic parameters all the propagators keep definite sign, which
makes the answer purely real.

The similar picture occurs in region C as well.
Therefore, regions B and C do not
contribute to the imaginary part of our integral.

\subsection{The other three $s$-channel diagrams}

The evaluation scheme just described can be applied to the other
three diagrams. There will be slight modifications in quark
propagators, but the strategy is unchanged.

For example, consider diagram (b) in Fig.~\ref{main2}. 
The only difference is the
expression for propagator $\langle 4 \rangle$:
\bea
\langle 4 \rangle &=& (q-k-\kappa)^2 - m^2 + i\epsilon =
(1-z-\beta)\left(-{Q^2 \over s} - y - \alpha\right) - [(\vec k +
\vec \kappa)^2 + m^2] + -\epsilon\,;\nonumber\\
&\to& y = -\alpha - {Q^2 \over s} - {(\vec k+\vec\kappa)^2+m^2\over (1-z-\beta)s}
+ {i\epsilon \over 1-z-\beta}\,;\nonumber\\
&\to& \alpha = -y - {Q^2 \over s} -
{(\vec k+\vec\kappa)^2+m^2\over (1-z-\beta)s} + {i\epsilon \over 1-z-\beta}\,.
\eea
The graphic representation of the pole position on $y$ and
$\alpha$ planes is shown in Figure. Again, one can check that the
main nonzero contribution comes from the same region A and the
same residues. So, if one again calculates the imaginary part of
the integral, one finds
\bea
&&Im \left\{\int {dy\ dz\ d\alpha\ d\beta \over \mbox{propagators}}\right\}
=\nonumber\\&&\quad ={4\pi^2 \over s^3} \int_0^1 dz
{1 \over [\vec k^2 + m^2 + z(1-z)Q^2][(\vec k+\vec\kappa)^2 + m^2 + z(1-z)Q^2]}
{1 \over [\vec \kappa^2 + \mu^2]^2}\,.\label{denom-res3}
\eea
Again, one can make sure that the small regions B and C do not
contribute to the imaginary part as well, although the proof might
not appear as clean as before. Namely, one can find that regions B
or C will contain a propagator (namely, propagator $\langle 4\rangle$)
that can change its sign, and therefore will contribute to the
real part. However, this can happen at abnormally small
$\vec\kappa^2$
$$
\vec\kappa^2 \sim {m^4 \over s}\,.
$$
Therefore the contribution of this configuration is
$1/s$-suppressed, and we cannot take it into account at our level
of accuracy.

Finally, the answers for the remaining two diagrams are
\bea
{4\pi^2 \over s^3} \int_0^1 dz
{1 \over [(\vec k+\vec\kappa)^2 + m^2 + z(1-z)Q^2][\vec k^2 + m^2 + z(1-z)Q^2]}
{1 \over [\vec \kappa^2 + \mu^2]^2}\quad \mbox{diagram C;}\nonumber\\
{4\pi^2 \over s^3} \int_0^1 dz
{1 \over [(\vec k+\vec\kappa)^2 + m^2 + z(1-z)Q^2]^2}
{1 \over [\vec \kappa^2 + \mu^2]^2}\quad \mbox{diagram D.}\nonumber
\eea

\subsection{The $u$-channel diagrams}

We claimed before that the $u$-channel diagrams
 do not contribute to the imaginary part. Here we
show that it is indeed so.

The only difference between diagrams $t$-channel and $u$-channel diagrams
is that in the $u$-channel case we switch the direction 
of particle $p$. Propagator $\langle 7\rangle$ will be now
\bea
\langle 7 \rangle &=&(p+\kappa)^2 - m^2 +i\epsilon =
(1+\alpha)\left({m^2 \over s}+\beta\right) s -(\vec\kappa^2
+m_p^2) + i\epsilon \,.\nonumber\\
\alpha &=& -1 + {\vec\kappa^2+m^2\over m^2 + \beta s}
- {i\epsilon\over \beta + m^2/s}\,.
\eea
Following the same line as before, we recognize three regions A,
B, and C. In region A ($-1 <\beta <-m^2/s;\ -\beta <z<1$)
the last propagator turns into
\be
\langle 7\rangle \to -\vec\kappa^2 - m^2
- \left[s + Q^2 + {\vec k^2 + m^2 \over 1-z} + {(\vec k + \vec \kappa)^2 + m^2
\over z-|\beta|}\right]\left(|\beta| -{m^2 \over s}\right) +
\not{\!\!i\epsilon}\,,
\ee
which means we do not get any contribution to the imaginary part.
One can check that no contribution to the imaginary part arises
from the other regions on $(z,\beta)$ plane (except for abnormal
cases when $\vec\kappa^2 \sim m^4/s$). The conclusion is that the
$u$-channel diagram gives no leading $1\over s$ contributions to
the imaginary part of the process.

\subsection{The same integral in the $\alpha$-representation technique}

Here, for the purpose of completeness, we show an alternative way
to do the integrations over $y$ and $\alpha$. Sure enough, the
underlying meaning of all manipulations will be absolutely the same as
before. However, in a certain sense this way might appears
simpler, since it does not require one to perform any graphic
analysis or to think what propagator should be put on mass shell.
Everything is done automatically here. It seems that this method
is similar to the so-called $\alpha$ representation of the loop
integrals.

We start with expression (\ref{denom-initial1}) and use the
following representation for each of the seven propagators
\be
{1 \over X - i\epsilon} = {i \over \epsilon + iX} =
i \int_0^\infty dt \exp[-t\epsilon - itX]\,.
\ee
The integral (\ref{denom-initial1}) transforms into
\bea
&&i^7 \int dy\ dz\ d\alpha\ d\beta\ \int_0^\infty dt_1 \cdots dt_7
\ \exp[-t_1\epsilon - it_1(ys+Q^2)(1-z)-it_1(\vec k^2 + m^2)]
\nonumber\\&&\quad\times
\exp[-t_2\epsilon + it_2 yzs-it_2(\vec k^2 + m^2)
-t_3\epsilon + it_3(y+\alpha)(z+\beta)s-it_1[(\vec k+\vec\kappa)^2 + m^2]]
\nonumber\\&&\quad\times
\exp[-t_4\epsilon + it_4 yzs-it_2(\vec k^2 + m^2)
-t_5\epsilon + it_5\alpha\beta s -it_5(\vec \kappa^2 + \mu^2)]
\nonumber\\&&\quad\times
\exp[-t_6\epsilon + it_6\alpha\beta s -it_6(\vec \kappa^2 + \mu^2)
-t_7\epsilon + it_7(m^2-\beta s)(1-\alpha)-it_7(\vec \kappa^2 + m^2)]
\eea
Since the total integral is convergent and thanks to the
factorization of $y$ and $\alpha$ dependencies, we can do the
$y$ and $\alpha$ integration first and obtain
\bea
&&\int dy\ e^{iys\left[-t_1(1-z)+t_2z+t_3(z+\beta)+t_4z\right]}
\int d\alpha\  e^{i\alpha s\left[t_3(z+\beta)+t_5\beta+t_6\beta
+t_7(\beta-m^2/s)\right]} = \left(-{2\pi i\over s}\right)^2\nonumber\\
&&\times\ \delta\left[-t_1(1-z)+t_2z+t_3(z+\beta)+t_4z\right]
\ \delta\left[t_3(z+\beta)+t_5\beta+t_6\beta
+t_7(\beta-m^2/s)\right]\,.
\eea
Since all parameters $t_i$ are positive, the two delta-functions
can be simultaneously non-zero only for certain $(z,\beta)$ pairs.
As should be expected, these pairs lie precisely within the allowed
regions.

Now we can take two of $t_1\dots t_7$ integrations to kill the two
delta-functions. Let them be $t_1$ and $t_3$ (of course,
one can take other pairs as well).
We get
\bea
t_3 &=& -{\beta\over z+\beta}(t_5 + t_6 + t_7) + t_7{m^2 \over
(z+\beta)s}\,;\nonumber\\
t_1 &=& {z \over 1-z}(t_2+t_4)-{\beta\over 1-z}(t_5 + t_6 + t_7 )
+ t_7{m^2 \over (1-z)s}\,.
\eea
Since the integrations in $t_i$ remain factorized, we easily
obtain
\bea
&&\int_0^\infty dt_2 e^{-t_2\epsilon} e^{-it_2(\vec k^2 + m^2)
- {z\over 1-z}(\vec k^2 + m^2) - zQ^2} =
{-i \over {1\over 1-z}[\vec k^2 + m^2 + z(1-z)Q^2] - i\epsilon}\,;
\nonumber\\[2mm]
&&\int_0^\infty dt_4 e^{-t_4\epsilon} e^{-it_4(\vec k^2 + m^2)
- {z\over 1-z}(\vec k^2 + m^2) - zQ^2} =
{-i \over {1\over 1-z}[\vec k^2 + m^2 + z(1-z)Q^2] - i\epsilon}\,;
\nonumber\\[3mm]
&&\int_0^\infty dt_5 e^{-t_5\epsilon} e^{-it_5(\vec \kappa^2 + \mu^2)
+ i t_5 {\beta \over z+\beta}[\vec k^2 + m^2 + Q^2(1-z)
+ (\vec k+\vec\kappa)^2+m^2]} \nonumber\\&&\qquad=
{-i \over \vec \kappa^2 + \mu^2 - i\epsilon - {\beta \over z+\beta}
[Q^2(1-z) + \vec k^2 + m^2 + (\vec k + \vec \kappa)^2 + m^2]}\,;
\nonumber\\[3mm]
&&\int_0^\infty dt_6 e^{-t_6\epsilon} e^{-it_6(\vec \kappa^2 + \mu^2)
+ i t_6 {\beta \over z+\beta}[\vec k^2 + m^2 + Q^2(1-z)
+ (\vec k+\vec\kappa)^2+m^2]} \nonumber\\&&\qquad=
{-i \over \vec \kappa^2 + \mu^2 - i\epsilon - {\beta \over z+\beta}
[Q^2(1-z) + \vec k^2 + m^2 + (\vec k + \vec \kappa)^2 + m^2]}\,;
\nonumber\\[3mm]
&&\int_0^\infty dt_7 e^{-t_7\epsilon} e^{it_y(\mu^2-\beta s)
- i t_7 (\vec \kappa^2 + m^2) - i t_7 {m^2/s - \beta \over z+\beta}
[\vec k^2 + m^2 + Q^2(1-z)+ (\vec k+\vec\kappa)^2+m^2] } \nonumber\\&&\qquad=
{-i \over \vec \kappa^2 + \beta s - i\epsilon + {m^2/s -\beta \over z+\beta}
[Q^2(1-z) + \vec k^2 + m^2 + (\vec k + \vec \kappa)^2 + m^2]}\,.
\eea
If one now brings all pieces together, one will arrive at already
familiar expression.

\section{Vector meson production: the fully off-forward case}

Since this case is done similarly to the Compton scattering, we do
all calculations in less details and pay special attention only to
distinctions.

Strictly speaking, the (virtual) photoproduction of vector mesons
is always an off-forward process, even though the final state might
have no transverse momentum. The reason is that
when the initial and final states have different masses, the
exchange Pomeron must carry a non-zero longitudinal momentum.
In the language of two-gluon exchange it means that the momenta of
the gluons are not identical, in contract to what we had in
previous section.

The integral we deal with is
\bea
&&\int dy\ dz\ d\alpha\ d\beta\ {\Gamma(M^2) \over [(q-k)^2 -m^2 +i\epsilon]
[k^2 -m^2 +i\epsilon] [(k+\Delta)^2 -m^2 +i\epsilon] [(k+\kappa_1)^2 -m^2
+i\epsilon]}\nonumber\\
&&{1 \over [\kappa_1^2 -\mu^2 +i\epsilon][\kappa_2^2 -\mu^2 +i\epsilon]
[(p-\kappa_1)^2 -m^2 +i\epsilon]} \label{denom-initial2}
\eea
With Sudakov decomposition  and notation $\kappa_1 \equiv \kappa + \Delta/2\;;
\kappa_1 \equiv \kappa - \Delta/2$, we have following seven
propagators:

\bea
&\langle 1\rangle&\quad (q-k)^2 - m^2 +i\epsilon =
\left(-y-{Q^2\over s}\right)(1-z)s -(\vec{k}^2 +m^2) +i\epsilon\,,\nonumber\\
&\langle 2\rangle&\quad (k+\Delta)^2 - m^2 +i\epsilon =
(y+\sigma)(z+\delta)s -[(\vec k+\vec\Delta)^2 +m^2] +i\epsilon \,,\nonumber\\[2mm]
&\langle 3\rangle&\quad (k+\kappa_1)^2- m^2 +i\epsilon =
(y+\alpha+{1\over 2}\sigma)(z+\beta+{1\over 2}\delta)s
-[(\vec{k}+\vec\kappa+\Delta/2)^2 +m^2] + i\epsilon\,,\nonumber\\[2mm]
&\langle 4\rangle&\quad k^2 - m^2 +i\epsilon =
y z s -(\vec k^2 +m^2) +i\epsilon \,,\nonumber\\[2mm]
&\langle 5\rangle&\quad \kappa_1^2 - \mu^2 +i\epsilon =
(\alpha+{1\over 2}\sigma)(\beta+{1\over 2}\delta) s
-[(\vec\kappa+{1\over 2}\Delta)^2 +\mu^2] + i\epsilon \,,\nonumber\\[2mm]
&\langle 5\rangle&\quad \kappa_2^2 - \mu^2 +i\epsilon =
(\alpha-{1\over 2}\sigma)(\beta-{1\over 2}\delta) s
-[(\vec\kappa-{1\over 2}\Delta)^2 +\mu^2] + i\epsilon \,,\nonumber\\[2mm]
&\langle 7\rangle&\quad (p-\kappa_1)^2 - m^2 +i\epsilon =
(1-\alpha-{1\over 2}\sigma)\left(-\beta-{1\over 2}\delta\right) s
-[(\vec\kappa+{1\over 2}\vec\Delta)^2 +m^2] + i\epsilon
\,.\nonumber\label{denom-poles3}
\eea
One can write now poles on the $y$ and $\alpha$ plane, draw the
regions of positivity of the corresponding $i\epsilon$'s on the
$(z,\beta)$ plane, and select the regions that leads to non-zero
contributions (there will be five of them this time). One can
again make sure that the leading $1/s$ contribution to the
imaginary part comes from the main region. Another thing that
helps avoid extra poles is the presence of $\mu^2$ (namely, $\mu^2 > m_V^2 - 4m^2$
for all vector mesons).

Setting propagators $\langle 1\rangle$ and $\langle 3\rangle$
on mass shell, we obtain
\be
y = -{Q^2 \over s} - {\vec k^2 + m^2 \over (1-z)s}\;;\quad
\alpha = {Q^2 \over s} - {1\over 2}\sigma + {\vec k^2 + m^2 \over (1-z)s}
+ {(\vec k + \vec \kappa + \vec \Delta/2)^2 + m^2 \over
(z+\beta+\delta/2)s}\,.
\ee
The resulting quark propagators read
\bea
\langle 4\rangle&=& - {1\over 1-z}\left[\vec k^2 + m^2 +
z(1-z)Q^2\right]+\not{\!\!i\epsilon}\,;\nonumber\\
\langle 2\rangle&=&(z+\delta) s \sigma - (z+\delta) s Q^2 -
(z+\delta){\vec k^2 + m^2 \over 1-z} - [(\vec k + \vec
\Delta)^2+m^2]+i\epsilon\,.
\eea
Note that since the invariant mass of the produced $q\bar q$ state
is equal
$$
M^2 = {\vec k^2 + m^2 \over 1-z} + {(\vec k + \vec
\Delta)^2+m^2 \over z+\delta} - \vec\Delta^2\,,
$$
propagator $\langle 2\rangle$ turns into $(z+\delta)(m_V^2 -
M^2)$. Together with the vertex factor $\Gamma(M^2)$, this
propagator gives rise to the wave function.

The gluon propagators are
\bea
\langle 5\rangle&=& (\beta + {1\over 2}\delta)\left(Q^2 +
{\vec k^2 + m^2 \over 1-z} + {(\vec k + \vec \kappa + \vec \Delta/2)^2
+ m^2 \over z+\beta+\delta/2}\right) - [(\vec\kappa+{1\over 2}\Delta)^2
+\mu^2] + \not{\!\!i\epsilon}\,;\nonumber\\
\langle 6\rangle&=& (\beta - {1\over 2}\delta)\left(-m_V^2 - \vec\Delta^2 +
{\vec k^2 + m^2 \over 1-z} + {(\vec k + \vec \kappa + \vec \Delta/2)^2
+ m^2 \over z+\beta+\delta/2}\right) - [(\vec\kappa-{1\over 2}\Delta)^2
+\mu^2] + \not{\!\!i\epsilon}\,.\nonumber
\eea
In the case of gluon $\langle 5\rangle$ the absence of pole was a
trivial thing (due to negativity of both $\beta$ and $\delta$), in
the case of the second gluon the pole can in principle arise. In
our case however this is avoided due to large enough $\mu^2$
(namely, $\mu^2 > m_V^2 - 4m^2$ for all vector mesons).

Finally, the propagator $\langle 7\rangle$ acquires the form
\be
\langle 7\rangle= -(\beta + {1\over 2}\delta)\left[s- Q^2 -
{\vec k^2 + m^2 \over 1-z} - {(\vec k + \vec \kappa + \vec \Delta/2)^2
+ m^2 \over z+\beta+\delta/2}\right] - [(\vec\kappa+{1\over 2}\Delta)^2
+\mu^2] + i\epsilon\,.
\ee
This propagator, as usual, is used to extract the imaginary part
of the integral, which leads to
$$
\beta \approx {(\vec \kappa + {1\over 2} \vec\Delta)^2 \over
s} \ll 1\,.
$$

With all these manipulations done, one finally obtains
\begin{eqnarray}
&&Im \int dy\ dz\ d\alpha\ d\beta\ {\Gamma \over [\mbox{all propagators}]}
\nonumber\\
&&= \left(-{\pi i \over s}\right)\left(-{2\pi i \over s}\right)^2\cdot
\int {dz \over z(1-z)} \psi_V(z,\vec k^2) \cdot{1 \over
[\vec k_1^2 +m^2 +z(1-z)Q^2]} { 1 \over (\vec \kappa^2 + \mu^2)^2}\label{gv14a}
\end{eqnarray}

\chapter{Helicity amplitude technique} \label{apb}

Here we give the derivation of expression for traces
of the following type
\begin{equation}
Sp\left\{
\hat e\  (\hat k_4 + m)\ \hat p'\ (\hat k_3 + m)\ \hat V^*\
(\hat k_2+m)\ \hat p'\ (\hat k_1 + m)
\right\}
\label{sp1}
\end{equation}
in full detail. Though one can calculate this trace covariantly,
a particularly convenient way to do so is given by
light cone helicity amplitude technique \cite{LCQFT}.
We emphasize that both ways are absolutely equivalent.
In the helicity amplitude approach, we recall that all
fermion lines in (\ref{sp1}) can be taken on mass shell
(see detailed derivation of LCWF normalization in
Sect.\ref{sectnorm}) and decomposed into a sum of
light cone helicities
\begin{eqnarray}
&&(\hat k + m) \to \sum_{\lambda = \pm} u_\lambda \bar u_\lambda
\quad \mbox{for quark lines;}\nonumber\\
&&(\hat k + m) = - [ (-\hat k) -m ] \to
- \sum_{\lambda = \pm} v_\lambda \bar v_\lambda
\quad \mbox{for antiquark lines.}\nonumber
\end{eqnarray}
Since the specific choice of this decomposition does not
affect the final result, we are free to take the most
convenient choice of spinors (see \cite{LCQFT} for details), namely,

Here are the spinors we use:
\begin{eqnarray}
u(p,\lambda) &=& {1 \over \sqrt{\sqrt{2}p^+}}\left(\sqrt{2}p^+ +
\beta m + \vec\alpha\vec p\right) \left\{
\begin{array}{cc}
\chi(\uparrow) & \lambda = +1 \\ \chi(\downarrow) & \lambda = -1
\end{array}\right. \nonumber\\
v(p,\lambda) &=& {1 \over \sqrt{\sqrt{2}p^+}}\left(\sqrt{2}p^+ -
\beta m + \vec\alpha\vec p\right) \left\{
\begin{array}{cc}
\chi(\downarrow) & \lambda = +1 \\ \chi(\uparrow) & \lambda = -1
\end{array}\right.
\label{spinors1}
\end{eqnarray}
where
$$
\chi(\uparrow) = {1 \over \sqrt{2}}\left(
\begin{array}{c}
1\\ 0\\ 1\\ 0
\end{array}
\right)\,;\quad \chi(\downarrow) = {1 \over \sqrt{2}}\left(
\begin{array}{c}
0\\ 1\\ 0\\ -1
\end{array}
\right)
$$
We stress that we use here our normal convention for + components
of 4-vectors. For convenience, we also give the explicit
expressions for all these spinors in the Dirac representation of
$\gamma$ matrices:
\begin{eqnarray}
\beta =\gamma^0 = \left(
\begin{array}{cc}
I & 0 \\ 0 & -I
\end{array}\right)\
\gamma^k = \left(
\begin{array}{cc}
0 & \sigma^k \\  -\sigma^k & 0
\end{array}\right)\
\alpha^k = \left(
\begin{array}{cc}
0 & \sigma^k \\  \sigma^k & 0
\end{array}\right)\
\gamma^5 = \left(
\begin{array}{cc}
 0 & I\\  I & 0
\end{array}\right)\
  \label{spinors2}
\end{eqnarray}
The explicit expressions for all spinors $u^+(p),\ u^-(p),\
v^-(p),\ v^+(p)$ read:
\begin{equation}
{1 \over \sqrt{2}} {1 \over \sqrt{\sqrt{2}p^+}}\cdot
\left(\begin{array}{c} \sqrt{2}p^+ + m\\ p_x + ip_y \\ \sqrt{2}p^+
- m\\p_x + ip_y
\end{array}\right)
\left(\begin{array}{c} -p_x + ip_y \\ \sqrt{2}p^+ + m\\ p_x - ip_y
\\ -\sqrt{2}p^+ + m
\end{array}\right)
\left(\begin{array}{c} \sqrt{2}p^+ - m\\ p_x + ip_y \\ \sqrt{2}p^+
+ m\\p_x + ip_y
\end{array}\right)
\left(\begin{array}{c} -p_x + ip_y \\ \sqrt{2}p^+ - m\\ p_x - ip_y
\\ -\sqrt{2}p^+ - m
\end{array}\right)
  \label{spinors3}
\end{equation}

Using these explicit formulas, one can do straightforward
calculations and indeed prove that comply with orthonormality and
completeness rules:
\begin{eqnarray}
  \label{spinors4}
  &&\bar u(p,\lambda)u(p,\lambda') = - \bar v(p,\lambda)v(p,\lambda') =
2m \delta_{\lambda\lambda'}\,;\nonumber\\&& \sum_\lambda
u(p,\lambda)\bar u(p,\lambda) = \hat p +m\,;\quad \sum_\lambda
v(p,\lambda)\bar v(p,\lambda) = \hat p -m\,.
\end{eqnarray}

For $\lambda= \pm 1$ we defined
\begin{equation}
a(\lambda) = - \lambda a_x - i a_y \ .
  \label{eq3}
\end{equation}
Cross product is defined as
\begin{equation}
[\vec{a}\vec{b}] = a_x b_y - b_x a_y \ .
  \label{eq4}
\end{equation}
Moreover {\bf every} matrix element should be multiplied by common
factor $\sqrt{p_+q_+}$.\\ Some useful relations:
\begin{eqnarray}
&&\gamma^5 v_{\lambda}(q) = - \lambda u_{-\lambda}(q)\,;\quad
\gamma^5 u_{\lambda}(q) = \lambda v_{-\lambda}(q)\,;\nonumber\\
&&a(-\lambda)b(\lambda) = - \vec a\vec b + i\lambda [\vec a\vec
b]\,; \quad\sum_\lambda a(-\lambda)b(\lambda) = -2 \vec a\vec
b\,;\nonumber\\ &&a(-\lambda)\left(\vec b\vec a + i \lambda [\vec
b\vec a]\right) = b(-\lambda) \vec a^2\,.
\end{eqnarray}

\begin{figure}[!htb]
   \centering
   \epsfig{file=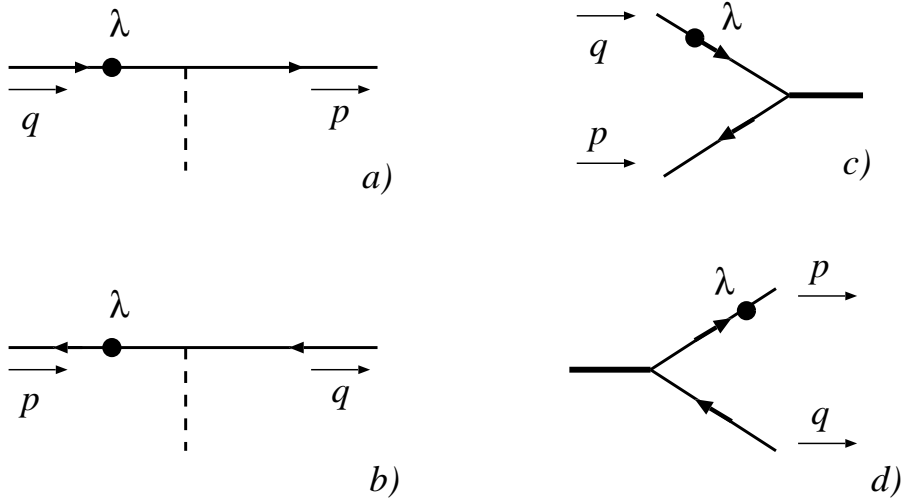,width=120mm}
   \caption{Four types of transitions, for which we give the amplitudes.
Dot with label $\lambda$ indicates the spinor whose helicity is
used as $\lambda$.}
   \label{helicity}
\end{figure}

\newpage

\subsubsection{Straight line elements}
For line in Fig.\ref{helicity}a, one has
\begin{eqnarray}
 & \bar{u}_\lambda(p)...u_\lambda(q) & \bar{u}_{-\lambda}(p)...u_\lambda(q)
\nonumber \\[4mm] \gamma^+ \quad & 2 & 0 \nonumber\\ \gamma^-
\quad& \fr{1}{p_+q_+}\left( m^2 + \vec{p}\vec{q} + i\lambda
[\vec{p}\vec{q}]\right) & {m \over p_+q_+}\left(p(\lambda) -
q(\lambda)\right)  \nonumber\\ \vec{a}\cdot \vec{\gamma} \quad &
\fr{\vec{a}\vec{p}}{p_+} + \fr{\vec{a}\vec{q}}{q_+} - i \lambda
\left( \fr{[\vec{a}\vec{p}]}{p_+} - \fr{[\vec{a}\vec{q}]}{q_+}
\right) & -m a(\lambda) \left({1 \over p_+} - {1 \over q_+}
\right) \nonumber\\ 1  \quad & m \left(\fr{1}{p_+} +
\fr{1}{q_+}\right) & {p(\lambda) \over p_+} - {q(\lambda) \over
q_+} \nonumber\\ \gamma^-\gamma^+\vec{\gamma}\cdot \vec{a} \quad &
\fr{2}{p_+} \left( \vec{a}\vec{p} - i \lambda
[\vec{a}\vec{p}]\right) & - {2 m \over p_+} a(\lambda) \nonumber\\
\vec{a} \cdot\vec{\gamma}\ \gamma^+\gamma^-  \quad & \fr{2}{q_+}
\left( \vec{a}\vec{q} + i \lambda [\vec{a}\vec{q}]\right) & {2 m
\over q_+} a(\lambda) \nonumber\\ \vec{a}\cdot \vec{\gamma}\
\gamma^+ \vec{\gamma}\cdot \vec{b}  \quad & 2 (\vec{a}\vec{b} + i
\lambda [\vec{a}\vec{b}]) & 0 \nonumber\\ \gamma^+ \gamma^5 \quad
& 2 \lambda & 0 \nonumber
\end{eqnarray}

\vspace{4mm} For line in Fig.\ref{helicity}b, one has
\begin{eqnarray}
& \bar{v}_\lambda(p)...v_\lambda(q) &
\bar{v}_{\lambda}(p)...v_{-\lambda}(q)  \nonumber\\[4mm] \gamma^+
\quad & 2 & 0 \nonumber\\ \gamma^- \quad & \fr{1}{p_+q_+}\left(
m^2 + \vec{p}\vec{q} - i\lambda [\vec{p}\vec{q}]\right) & - {m
\over p_+q_+}\left(p(\lambda) - q(\lambda)\right)   \nonumber\\
\vec{a}\cdot \vec{\gamma}  \quad & \fr{\vec{a}\vec{p}}{p_+} +
\fr{\vec{a}\vec{q}}{q_+} + i \lambda \left(
\fr{[\vec{a}\vec{p}]}{p_+} - \fr{[\vec{a}\vec{q}]}{q_+}  \right) &
m a(\lambda) \left({1 \over p_+} - {1 \over q_+} \right)
\nonumber\\ 1  \quad & -  m \left(\fr{1}{p_+} + \fr{1}{q_+}\right)
&  {p(\lambda) \over p_+} - {q(\lambda) \over q_+}\nonumber\\
\gamma^-\gamma^+\vec{\gamma}\cdot \vec{a}  \quad & \fr{2}{p_+}
\left( \vec{a}\vec{p} + i \lambda [\vec{a}\vec{p}]\right) & {2 m
\over p_+} a(\lambda) \nonumber\\ \vec{a} \cdot\vec{\gamma}\
\gamma^+\gamma^-  \quad & \fr{2}{q_+} \left( \vec{a}\vec{q} - i
\lambda [\vec{a}\vec{q}]\right) & - {2 m \over q_+} a(\lambda)
\nonumber\\ \vec{a}\cdot \vec{\gamma}\ \gamma^+ \vec{\gamma}\cdot
\vec{b}  \quad &  2 (\vec{a}\vec{b} - i \lambda [\vec{a}\vec{b}])
&  0 \nonumber\\ \gamma^+ \gamma^5 \quad & - 2 \lambda & 0
\nonumber
\end{eqnarray}

\newpage

\subsubsection{Vertex lines}
For vertex in Fig.\ref{helicity}c, one has
\begin{eqnarray}
 & \bar{v}_\lambda(p)...u_\lambda(q) & \bar{v}_{-\lambda}(p)...u_\lambda(q)
\nonumber \\[4mm] \gamma^+ \quad & 0 & 2 \nonumber\\ \gamma^-
\quad & \fr{m}{p_+q_+}\left(p(\lambda) + q(\lambda)\right) & {1
\over p_+q_+}\left( - m^2 + \vec{p}\vec{q} + i\lambda
[\vec{p}\vec{q}]\right) \nonumber\\ \vec{a}\cdot \vec{\gamma}
\quad & m a(\lambda) \left(\fr{1}{p_+} + \fr{1}{q_+} \right) &
{\vec{a}\vec{p} \over p_+} + {\vec{a}\vec{q} \over q_+} - i
\lambda \left( {[\vec{a}\vec{p}] \over p_+} - {[\vec{a}\vec{q}]
\over q_+}  \right) \nonumber\\ 1  \quad & \fr{p(\lambda)}{p_+} -
\fr{q(\lambda)}{q_+} & - {m \over p_+} + {m \over q_+}
\nonumber\\ \gamma^-\gamma^+\vec{\gamma}\cdot \vec{a} \quad &
\fr{2 m}{p_+} a(\lambda) & {2 \over p_+} \left( \vec{a}\vec{p} - i
\lambda [\vec{a}\vec{p}]\right) \nonumber\\ \vec{a}
\cdot\vec{\gamma}\ \gamma^+\gamma^-  \quad & \fr{2}{\over q_+}
a(\lambda) & {2 \over q_+} \left( \vec{a}\vec{q} + i \lambda
[\vec{a}\vec{q}]\right) \nonumber\\ \vec{a}\cdot \vec{\gamma}\
\gamma^+ \vec{\gamma}\cdot \vec{b}  \quad & 0 & 2 (\vec{a}\vec{b}
+ i \lambda [\vec{a}\vec{b}]) \nonumber\\ \gamma^+ \gamma^5 \quad
& 0 & 2 \lambda \nonumber\\ \gamma^5 \quad &
\lambda\left({p(\lambda) \over p_+} - {q(\lambda) \over
q_+}\right) & - \lambda m \left({1 \over p_+} + {1\over
q_+}\right)\nonumber
\end{eqnarray}

\vspace{4mm} For vertex in Fig.\ref{helicity}d, one has
\begin{eqnarray}
& \bar{u}_\lambda(p)...v_\lambda(q) &
\bar{u}_{\lambda}(p)...v_{-\lambda}(q)  \nonumber\\[4mm] \gamma^+
\quad & 0 & 2 \nonumber\\ \gamma^-  \quad & -
\fr{m}{p_+q_+}\left(p(-\lambda) + q(-\lambda)\right) & {1 \over
p_+q_+}\left( - m^2 + \vec{p}\vec{q} + i\lambda
[\vec{p}\vec{q}]\right) \nonumber\\ \vec{a}\cdot \vec{\gamma}
\quad & - m a(-\lambda) \left(\fr{1}{p_+} + \fr{1}{q_+} \right) &
{\vec{a}\vec{p} \over p_+} + {\vec{a}\vec{q} \over q_+} - i
\lambda \left( {[\vec{a}\vec{p}] \over p_+} - {[\vec{a}\vec{q}]
\over q_+}  \right) \nonumber\\ 1  \quad & \fr{p(-\lambda)}{p_+} -
\fr{q(-\lambda)}{q_+} &  {m \over p_+} - {m \over q_+}
\nonumber\\ \gamma^-\gamma^+\vec{\gamma}\cdot \vec{a} \quad & -
\fr{2 m}{p_+} a(-\lambda) & {2 \over p_+} \left( \vec{a}\vec{p} -
i \lambda [\vec{a}\vec{p}]\right) \nonumber\\ \vec{a}
\cdot\vec{\gamma}\ \gamma^+\gamma^-  \quad & - \fr{2 m}{q_+}
a(-\lambda) & {2 \over q_+} \left( \vec{a}\vec{q} + i \lambda
[\vec{a}\vec{q}]\right) \nonumber\\ \vec{a}\cdot \vec{\gamma}\
\gamma^+ \vec{\gamma}\cdot \vec{b}  \quad & 0 & 2 (\vec{a}\vec{b}
+ i \lambda [\vec{a}\vec{b}]) \nonumber\\ \gamma^+ \gamma^5 \quad
& 0 & 2 \lambda \nonumber\\ \gamma^5 \quad &
-\lambda\left({p(-\lambda) \over p_+} - {q(-\lambda) \over
q_+}\right) & \lambda m \left({1 \over p_+} + {1\over
q_+}\right)\nonumber
\end{eqnarray}

\newpage

\subsubsection{Transverse photon polarization}

As we see from the table, neither of gluon legs can flip quark
helicity. Therefore, there are 4 combinations of all possible
helicity assignments.

In the case both $q\bar{q}$ helicities are the same, one gets

\begin{eqnarray}
&&2zq_+p_-\cdot2zq_+p_-\cdot {\sqrt{z(1-z)} q_+\over
z(1-z)q_+}me^*(\lambda)\cdot {\sqrt{z(1-z)} q_+\over
z(1-z)q_+}(-m)e(-\lambda)\nonumber\\ &&= {z \over
1-z}s^2(-m^2)e^*(\lambda)e(-\lambda)\label{photon23}
\end{eqnarray}
In the case quark and antiquark helicities are opposite, one
obtains
\begin{eqnarray}
&&2zq_+p_-\cdot2zq_+p_-\cdot {\sqrt{z(1-z)} q_+\over z(1-z)q_+}
\left[(-e^*k)z + (e^*k)(1-z) - i\lambda(-[e^*k]z -
[e^*k](1-z))\right]\nonumber\\ &&\cdot{\sqrt{z(1-z)} q_+\over
z(1-z)q_+} \left[(ek)(1-z) - (ek)z - i\lambda([e^*k](1-z) +
[e^*k]z)\right] \nonumber\\ && ={z \over 1-z}s^2\left[(e^*k)(1-2z)
+ i\lambda[e^*k]\right] \left[(ek)(1-2z) -
i\lambda[ek]\right]\label{photon24}
\end{eqnarray}
Thus, performing summation over $q\bar q$ helicities, one obtains
$$
2s^2{z\over 1-z} \left[m^2(ee^*) + (ek)(e^*k)(1-2z)^2 +
[ek][e^*k]\right]
$$
Finally, averaging over azymuthal angle, omitting factor $2s^2$
and including $(-1)$ due to one antiquark propagator reveals
\begin{equation}
{1-z \over z}I^{(a)} = - \left[m^2 + (z^2 +
(1-z)^2)k^2\right]\label{photon25}
\end{equation}

Note that the expression for Diagr.(d) can be instantly obtained
from the above expressions by replacement $\vec{k} \to -\vec{k};\ z
\to 1-z$. In this case 3 antiquark propagators also give factor
$(-1)^3 = -1$. So, the answer for this diagram is
\begin{equation}
{z \over 1-z}I^{(d)} = - \left[m^2 + (z^2 +
(1-z)^2)k^2\right]\label{photon26}
\end{equation}

Finally, one has
\begin{eqnarray}
&&m^2 + (z^2 + (1-z)^2)(k,k+\kappa)\nonumber\\ &&m^2 + (z^2 +
(1-z)^2)(k,k-\kappa)\nonumber
\end{eqnarray}
for Diagrams (b) and (c) respectively.\\

With all these results, Eq.(\ref{photon21}) turns into
\begin{eqnarray}
A^T &=& is {32 \over (2\pi)^2}e_f^2 \cdot \alpha_s^2
\alpha_{em}\cdot2\cdot \int dz {d^2\vec{k} \over\left[\vec{k}^2 +
m^2 + z(1-z)Q^2\right]} \int {d^2 \vec{\kappa} V(\kappa)\over
(\vec{\kappa}^2 + \mu^2)^2} \nonumber\\ &&\times \Biggl[ - {m^2 +
(z^2 + (1-z)^2)\vec{k}^2 \over [\vec{k}^2 + m^2 + z(1-z)Q^2]} +{m^2
+ (z^2 + (1-z)^2)(\vec{k}^2+(\vec{k}\vec{\kappa})) \over
[(\vec{k}+\vec{\kappa})^2 + m^2 + z(1-z)Q^2]} \Biggr]
\,.\label{photon27}
\end{eqnarray}
Note that we changed $\kappa \to -\kappa$ in Diagr.(c), so that it
became identical to Diagr.(b).

\subsubsection{Scalar photon polarization}

The virtual photon scalar polarization is described by
\begin{equation}
  e^\mu_0 = {1 \over Q}({q'}^\mu + x{p'}^\mu) = {1 \over Q}(q_+n^\mu_+
  + x p_-n^\mu_-)\label{photon28}
\end{equation}
One notes that in this case only $q\bar q$ states with opposite
helicities do contribute. For Diagr.(a) an amplitude reads
\begin{eqnarray}
&& 2zq_+p_- \cdot 2zq_+p_- \cdot \left({\sqrt{z(1-z)}q_+ \over
Q}\right)^2 \left[{q_+(-m^2-\vec{k}^2) \over z(1-z)q_+^2} +
2xp_-\right] \left[{q_+(-m^2-\vec{k}^2) \over z(1-z)q_+^2} +
2xp_-\right]\nonumber\\ &&=s^2{1 \over Q^2}{z \over 1-z} \left[m^2
+ \vec{k}^2 -z(1-z)Q^2 \right]^2\,. \label{photon29}
\end{eqnarray}
Therefore,
\begin{equation}
{1-z \over z}I^{(a)} = - {1 \over Q^2} \left[m^2 + \vec{k}^2
-z(1-z)Q^2\right]^2\,.\label{photon30}
\end{equation}
Obviously, for Diagr.(d) one has
\begin{equation}
{z \over 1-z}I^{(d)} = - {1 \over Q^2} \left[m^2 + \vec{k}^2
-z(1-z)Q^2\right]^2\,.\label{photon31}
\end{equation}

The expressions for Diagrs.(b) and (c) can be obtained in the same
way and give
\begin{eqnarray}
I^{(b)} &=& {1 \over Q^2} \left[m^2 + \vec{k}^2 - z(1-z)Q^2\right]
\left[m^2 + (\vec{k} + \vec{\kappa})^2 -
z(1-z)Q^2\right]\,;\nonumber\\ I^{(c)} &=& {1 \over Q^2} \left[m^2
+ \vec{k}^2 - z(1-z)Q^2\right] \left[m^2 + (\vec{k} - \vec{\kappa})^2
- z(1-z)Q^2\right]\,. \label{photon32}
\end{eqnarray}
Therefore Eq.(\ref{photon21}) now reads
\begin{eqnarray}
A^0 &=& is {32 \over (2\pi)^2}e_f^2 \cdot \alpha_s^2
\alpha_{em}\cdot \int dz d^2\vec{k} { [m^2 + \vec{k}^2 -
z(1-z)Q^2]\over\left[\vec{k}^2 + m^2 + z(1-z)Q^2\right]} \int {d^2
\vec{\kappa} V(\kappa)\over (\vec{\kappa}^2 + \mu^2)^2} \nonumber\\
&&\times {2\over Q^2} \Biggl[ - { [m^2 + \vec{k}^2 -
z(1-z)Q^2]\over [\vec{k}^2 + m^2 + z(1-z)Q^2]}
 +  {[m^2 + (\vec{k} - \vec{\kappa})^2 - z(1-z)Q^2]
\over [(\vec{k}-\vec{\kappa})^2 +m^2 + z(1-z)Q^2]}
\Biggr]\,.
\end{eqnarray}
In fact, this expression in brackets can be greatly simplified.
First, trivial algebra leads to
\begin{eqnarray}
&&- { [m^2 + \vec{k}^2 - z(1-z)Q^2]\over [\vec{k}^2 + m^2 +
z(1-z)Q^2]}
 + {[m^2 + (\vec{k} - \vec{\kappa})^2 - z(1-z)Q^2]
\over [(\vec{k}-\vec{\kappa})^2 +m^2 + z(1-z)Q^2]} \nonumber\\ &&=
-1+ { 2 z(1-z)Q^2]\over [\vec{k}^2 + m^2 + z(1-z)Q^2]} +1 -
{2z(1-z)Q^2 \over [(\vec{k}-\vec{\kappa})^2 +m^2 +
z(1-z)Q^2]}\nonumber\\ &&= 2z(1-z)Q^2 \left[{1\over [\vec{k}^2 +
m^2 + z(1-z)Q^2]} -{1\over [(\vec{k}-\vec{\kappa})^2 +m^2 +
z(1-z)Q^2]} \right]\,.\nonumber
\end{eqnarray}

Second, in we replace $e_0 \propto (q' +xp')$ by $q = (q' -xp')$,
we will get zero due to the gauge invariance of the totaf sum of
all diagrams. Effectively, it means that residual $[m^2 + \vec{k}^2
- z(1-z)Q^2]$ can be replaced by $-2z(1-z)Q^2$. Of course, it does
not mean that these two expressions are identical by themselves.
Only after integration over quark momenta they give the same
answer. Thus, the final answer for scalar photons reads
\begin{eqnarray}
A^0 &=& is {128 Q^2\over (2\pi)^2}e_f^2 \cdot \alpha_s^2
\alpha_{em}\cdot 2\cdot \int dz z^2(1-z)^2 d^2\vec{k} {
1\over\left[\vec{k}^2 + m^2 + z(1-z)Q^2\right]} \int {d^2
\vec{\kappa} V(\kappa)\over (\vec{\kappa}^2 + \mu^2)^2} \nonumber\\
&&\times \Biggl[ { 1\over [\vec{k}^2 + m^2 + z(1-z)Q^2]}
 -  {1\over [(\vec{k}-\vec{\kappa})^2 +m^2 + z(1-z)Q^2]}
\Biggr]\,.\label{photon33}
\end{eqnarray}

\section{Photon vertex amplitudes}

Notation is given in Fig.\ref{main3}.
We start with transverse photon.
$$
\bar u' \hat e_T v = \bar u' ( - \vec \gamma \vec e) v\,.
$$
Equal $q\bar q$ helicities give
\begin{equation}
- { \sqrt{z(1-z)} q_+ \over z(1-z)q_+}\cdot
(-m) e(-\lambda) = {1 \over \sqrt{z(1-z)}}m e(-\lambda)\label{amp1}
\end{equation}
Opposite $q\bar q$ helicities give
\begin{eqnarray}
&&- { \sqrt{z(1-z)} q_+ \over z(1-z)q_+}\cdot
\left[(\vec{e}\vec k_1)(1-z) - (\vec{e}\vec k_1)z - i\lambda
\left([\vec{e}\vec k_1](1-z) + [\vec{e}\vec k_1]z\right)
\right]\nonumber\\
&&= - {1 \over \sqrt{z(1-z)}}\left[(\vec{e}\vec k_1)(1-2z)- i\lambda[\vec{e}\vec k_1]\right]
\label{amp2}
\end{eqnarray}

In the case of scalar photon
$$
\bar u' \hat e_0 v = \bar u' {1 \over Q}(q_+\gamma_- + xp_-\gamma_+ ) v
$$
the same helicities give exactly zero while opposite helicities
result in
\begin{eqnarray}
&&{ \sqrt{z(1-z)} q_+ \over z(1-z)q_+}\cdot{1 \over Q}\cdot
\left[{q_+ \over q_+} (-m^2 -\vec k_1^2) + x 2z(1-z)p_-q_+\right]\nonumber\\
&&= - {1 \over \sqrt{z(1-z)}}{1 \over Q}
\left[m^2 + \vec k_1^2 - z(1-z)Q^2\right]
\label{amp3}
\end{eqnarray}

\section{Vector meson vertex amplitudes}
This case is more tricky due to the nonzero vector meson
transverse momentum $\vec \Delta$.
We start with the  transverse vector meson polarization:
$$
\hat V^*_T = - \vec \gamma \vec V^* + {2 (\vec V^*\vec \Delta) \over s}
p_-\gamma_+\,.
$$
The same $q\bar q$ helicities give again
\begin{equation}
- { \sqrt{z(1-z)} q_+ \over z(1-z)q_+}\cdot
m V^*(\lambda) = -{1 \over \sqrt{z(1-z)}}m V^*(\lambda)\label{amp4}
\end{equation}
while opposite helicities give
\begin{eqnarray}
&&{ \sqrt{z(1-z)} q_+ \over z(1-z)q_+}
\Biggl\{-\bigl[(\vec{V}^*\vec{k}_3)z + (\vec{V}^*\vec{k}_2)(1-z)
- i\lambda\left([\vec{V}^*\vec{k}_3]z - [\vec{V}^*\vec{k}_2](1-z)\right)\bigr]\nonumber\\
&&+ {2 (\vec{V}^*\vec\Delta) \over s}p_-2z(1-z)q_+
\Biggr\} \nonumber\\
 &&= - {1 \over \sqrt{z(1-z)}}\left\{
(\vec{V}^*\vec{k})(1-2z) + i\lambda[V\vec{k}]\right\}\label{amp5}
\end{eqnarray}
Here we used definitions and properties (see also Fig.\ref{main3}):
\begin{eqnarray}
&&\vec{k}_2 = \vec{k} + z\vec\Delta\,; \quad \vec{k}_3 = -\vec{k}+(1-z)\vec\Delta \label{d12}\\
&\Rightarrow& (1-z) \vec{k}_2 - z \vec{k}_3 = \vec{k}\,; \quad
(1-z) \vec{k}_2 + z \vec{k}_3 = (1-2z)\vec{k} + 2 z(1-z) \vec\Delta\,;\nonumber\\
&&(\vec{k}_2\vec{k}_3) = -\vec{k}^2 + (1-2z)(\vec{k}\vec\Delta) + z(1-z) \vec\Delta^2\;
\nonumber\\
&&M^2 + \vec \Delta^2 = {\vec k_2^2 +m^2 \over z} +{\vec k_3^2 +m^2 \over (1-z)} =
 {\vec k^2 +m^2 \over z(1-z)} + \vec \Delta^2\,.
\nonumber
\end{eqnarray}

For the longitudinal vector mesons one has
for equal quark-antiquark helicities
\begin{equation}
{ \sqrt{z(1-z)} q_+ \over z(1-z)q_+}{1\over M}
\cdot\left[-m\vec\Delta(\lambda) + m{q_+\over q_+} [k_2(\lambda)+k_3(\lambda)]
\right] = 0\label{amp6}
\end{equation}
and for opposite helicities
\begin{eqnarray}
&&{ \sqrt{z(1-z)} q_+ \over z(1-z)q_+}{1\over M} \Biggl\{
-\bigl[(\vec\Delta \vec{k}_3)z + (\vec\Delta \vec{k}_2)(1-z) +i\lambda[\vec\Delta \vec{k}]\bigr]
+ {q_+ \over q_+}\bigl[-m^2 +(\vec{k}_2\vec{k}_3) +i\lambda[\vec{k}_3\vec{k}_2]\bigr]
\nonumber\\
&&+ {\vec\Delta^2 -M^2 \over s} p_- 2z(1-z)q_+\Biggr\}
\nonumber\\
&& = - {1 \over \sqrt{z(1-z)}} 2z(1-z)M\label{amp7}
\end{eqnarray}

\section{Final trace calculation}

Once we have the expressions for vertex amplitudes,
the rest is done quickly. We first note that
each gluon vertex attached to the lower or upper line gives factor
\begin{equation}
2zq_+\cdot p_- = sz\,; \quad 2(1-z)q_+\cdot p_- = s(1-z)\label{amp8}
\end{equation}
correspondingly.

So. let's start with $T \to T$ amplitude and calculate it for Diagr.(c)
at Fig.\ref{main2}.

Equal $q\bar q$ helicities give
\begin{equation}
s(1-z)\cdot sz\cdot {-1 \over \sqrt{z(1-z)}}mV^*(\lambda)
{1 \over \sqrt{z(1-z)}}me(-\lambda)
= -s^2 m^2 e(-\lambda)V^*(\lambda)\label{amp9}
\end{equation}
Summing over $\lambda$ gives
\begin{equation}
2s^2 m^2 (\vec{e}\vec{V}^*)\,.\label{amp10}
\end{equation}
The opposite helicities yield
\begin{equation}
s^2 \left[(\vec{V}^*\vec{k})(1-2z) + i\lambda[\vec{V}^*\vec{k}]\right]
\left[(\vec{e}\vec{k}_1)(1-2z) - i\lambda[\vec{e}\vec{k}_1]\right]\label{amp11}
\end{equation}
Summing over helicities and making use of identity
\begin{equation}
[\vec a\vec b][\vec c\vec d] = (\vec a\vec c)(\vec b\vec d) -
(\vec a\vec d)(\vec b\vec c)\label{amp12}
\end{equation}
one obtains
\begin{equation}
2s^2 \left[(\vec{V}^*\vec{k})(\vec{e}\vec{k}_1)(1-2z)^2 +
(\vec{e}\vec{V}^*)(\vec{k}\vec k_1) - (\vec{e}\vec{k})(\vec{V}^*\vec k_1)\right]\,.
\label{amp13}
\end{equation}

Since we factored out $2s^2$ when deriving (\ref{gv4}),
we finally get
\begin{equation}
I^{(c)}_{T\to T} =
- \left[(\vec{e}\vec{V}^*)(m^2 + \vec{k}\vec k_1) + (\vec{V}^*\vec{k})(\vec{e}\vec k_1)(1-2z)^2 - (\vec{e}\vec{k})(\vec{V}^*\vec k_1)
\right]\,.\label{amp14}
\end{equation}
We included factor $(-1)$ since in this diagram we have one antiquark propagator.

An important observation here is that all other integrands, namely
$I^{(a)}\cdot (1-z)/z, I^{(c)}, I^{(d)}\cdot z/(1-z)$ give absolutely the same
result (with their own definitions of $\vec k_1$ of course). The only
thing one should not forget is that diagrams (a,d) enter with sign
'$-$' while diagrams (b,c) enter with sign '$+$':
$$
-{1-z \over z}I^{(a)} = I^{(b)} = I^{(c)} = -{z \over 1-z}I^{(d)}\,.
$$

For $L \to L$ amplitude one has immediately
\begin{equation}
I^{(c)}_{L\to L} =
-{1 \over Q}[m^2 + \vec k_1^2 - z(1-z)Q^2]\cdot {1 \over M}2z(1-z)M^2\label{amp15}
\end{equation}
In fact, using simple relation
\begin{equation}\label{trick}
  {m^2 + \vec k_1^2 - z(1-z)Q^2 \over m^2 + \vec k_1^2 + z(1-z)Q^2} =
  1 + { - 2z(1-z)Q^2 \over m^2 + \vec k_1^2 + z(1-z)Q^2}
\end{equation}
and noting that all unity terms will eventually cancel out in
(\ref{gv17}), one can rewrite (\ref{amp15}) as
\begin{equation}
I^{(c)}_{L\to L} = -4QMz^2(1-z)^2\,.\label{amp15a}
\end{equation}

For $T \to L$ amplitude one has
\begin{equation}
I^{(c)}_{T\to L} = 2z(1-z)M(\vec{e}\vec k_1)(1-2z)\label{amp16}
\end{equation}
and for $L \to T$ amplitude one has
\begin{equation}
I^{(c)}_{L\to T} =
{1 \over Q}[m^2 + \vec k_1^2 - z(1-z)Q^2](1-2z)(\vec{V}^*\vec{k})\,.\label{amp17}
\end{equation}
The same transformation as in $L\to L$ amplitude, leads to
\begin{equation}
I^{(c)}_{L\to T} = - 2z(1-z)Q^2(1-2z)(\vec{V}^*\vec{k})\,.\label{amp17a}
\end{equation}
 Note that in the
last three amplitudes only opposite $q \bar q$ helicities
contributed.

\newpage
\thispagestyle{empty}

{\Large
\begin{center}
{\bf Acknowledgements}
\vspace{1cm}
\end{center}
}

It is an honor for me to thank Prof.~J.Speth
for his perpetual desire to see me as a member of J\"ulich
theory group and for making my entire stay at Forschungszentrum
so comfortable and scientifically fruitful.

Million of thanks must undoubtedly go to Kolya Nikolaev
for pulling me in the midst of his turbulent scientific
activity and making me immediately start contributing, 
for infecting me with his desire to look for 
a clear physical meaning behind every formula,
and for innumerable discussion we have had during these years.

I want to thank all my colleagues at IKP and in Novosibirsk, 
in particularly, Ilya Ginzburg, who discussed
with me details of my work, or just listened to what I've been saying.

I wish to thank all members of the wonderful company
that grew up in J\"ulich --- Vadim, Achot, Pasha, Fedya, Dmitro, Lena,
Tanya and many others. Thanks for all those parties and journeys together!

Anuta! Thanks to you too! For your continuous and many-sided 
help and support. Well, most of time 
it actually looked rather like distracting from studies
than supporting them, but thanks anyway!
Without relaxing at home, I would get completely crazy
with all these vector mesons!

Finally, I want to collectively thank all those people who have
helped me in various ways during my work, but whom
I failed to recollect in this acknoledgement due to
my innate forgetfulness.

\newpage 

\thispagestyle{empty}
{\Large
\begin{center}
{\bf Curriculum Vitae}
\vspace{1cm}
\end{center}
}
\begin{enumerate}
\item {\bf Name}: Igor Pierovich Ivanov
\item {\bf Date of birth}: 26 October 1976
\item {\bf Place of birth}: Petropavlovsk-Kamchatsky, Russia
\item {\bf Nationality}: Russian
\item {\bf Education outline}:
\begin{itemize}
\item Sept. 1983 -- May 1991: secondary school 7, Kamchatka, Russia
\item Aug.1991 -- June 1993:  Physics and Mathematics School 
at Novosibirsk University, Novosibirsk, Russia
\item Sept. 1993 -- July 1997: undergraduate studies 
at Novosibirsk State University, Physics Dept.;
BSc diploma defended in July 1997
\item Sept 1997 -- July 1999:  graduate studies 
at Novosibirsk State University, Physics Dept.;
MSc thesis defended in July 1999.
\item from Aug. 1999 -- postgraduate studies at 
Institut f\"ur Kernphysik, Forschungszentrum J\"ulich
\end{itemize}
\end{enumerate}

\end{document}